\documentclass[rmp,aps,nofootinbib,eqsecnum,showpacs,preprintnumbers]{revtex4}
\usepackage{epsfig}
\usepackage{bm}
\usepackage{amssymb}
\usepackage{amscd}
\usepackage{latexsym}

\newcommand{\be}{\begin{equation}}
\newcommand{\dd}{\displaystyle}
\newcommand{\ee}{\end{equation}}
\newcommand{\bea}{\begin{eqnarray}}
\newcommand{\eea}{\end{eqnarray}}

\newcommand{\nn}{\nonumber}
\newcommand{\de}{\partial}

\def\slash#1{\setbox0=\hbox{$#1$}#1\hskip-\wd0\dimen0=5pt\advance
       \dimen0 by-\ht0\advance\dimen0 by\dp0\lower0.5\dimen0\hbox
         to\wd0{\hss\sl/\/\hss}}

\def\esp#1{e^{#1}}

\def\Zint{{Z \kern -.45 em Z}}
\def\complex{{\kern .1em {\raise .47ex \hbox
{$\scriptscriptstyle |$}} \kern -.4em {\rm C}}}
\def\real{{\vrule height 1.6ex width 0.05em depth 0ex
\kern -0.06em {\rm R}}}
\newcommand{\bes}{\begin{equation*}}
\newcommand{\ees}{\end{equation*}}
\def\sv{\int\frac{d\b v}{4\pi}}
\def\inbar{\,\vrule height1.5ex width.4pt depth0pt}
\def\IR{\relax{\rm I\kern-.18em R}}
\def\IC{\relax\hbox{$\inbar\kern-.3em{\rm C}$}}

\def\b#1{{\bf #1}}
\def\cre#1#2{b^\dagger_{#2}(\b #1)}
\def\dis#1#2{b_{#2}(\b #1)}
\def\creu#1{b^\dagger_{1}(\b #1)}
\def\cred#1{b^\dagger_{2}(-\b #1)}
\def\disu#1{b_{1}(\b #1)}
\def\disd#1{b_{2}(-\b #1)}
\def\acreu#1{A^\dagger_{1}(\b #1)}
\def\acred#1{A^\dagger_{2}(\b #1)}
\def\adisu#1{A_{1}(\b #1)}
\def\adisd#1{A_{2}(\b #1)}
\def\acre#1#2{A^\dagger_{#2}(\b #1)}
\def\adis#1#2{A_{#2}(\b #1)}
\begin{document}
%\preprint{{\bf CERN-TH/2003-XXXX}}
\title{\begin{center} Lecture Notes on Superconductivity:\\
 Condensed Matter and QCD
 \\ {\small (Lectures  at the
 University of Barcelona, Spain, September-October 2003)}
 \end{center}}
\author{Roberto Casalbuoni} \email{casalbuoni@fi.infn.it} \affiliation{Department of
Physics of the University of Florence,\\ Via G. Sansone 1,  50019
Sesto Fiorentino (FI),  Italy}
\date{\today}
%%%%%%%%%%%%%%%%%%%%%%%%%%%%%%%%%%%%%%%%%%%%%%%%%%%%%%%%%%%
\begin{abstract}

\end{abstract}
%%%%%%%%%%%%%%%%%%%%%%%%%%%%%%%%%%%%%%%%%%%%%%%%%%%%%%%%%%%%%%%%
\pacs{12.38.-t, 26.60.+c, 74.20.-z, 74.20.Fg, 97.60.Gb} \maketitle
\tableofcontents
\section{Introduction}
Superconductivity is one of the most fascinating chapters of
   modern physics. It has been a continuous source of inspiration for different
   realms of physics and has shown a tremendous
capacity of cross-fertilization, to say nothing of its numerous
technological applications.
 Before giving a more accurate definition of
   this phenomenon let us however briefly sketch the historical path leading to it.
   Two were the main steps in the discovery of superconductivity. The
   former
   was due to Kamerlingh Onnes \cite{Kamerlingh:1911ab} who discovered  that
   the electrical resistance of various metals, e. g. mercury, lead, tin and many others,
   disappeared when the
   temperature was lowered below some critical
   value $T_c$. The actual values of $T_c$ varied with the metal, but
   they were all of the order of a few K, or at most of the order of tenths of a K.
   Subsequently perfect diamagnetism in superconductors was discovered
   \cite{meissner:1933ab}.
     This property not only implies that
   magnetic fields are excluded from superconductors, but also
   that any field originally present in the metal is expelled from it
  when lowering the temperature below its critical
   value. These two features were captured in the equations
   proposed by the brothers F. and H. London \cite{london:1935cd}
   who first realized the quantum character of the phenomenon.
   The decade starting in 1950 was the stage of two major theoretical
breakthroughs. First,
   Ginzburg and Landau (GL) created a theory describing the transition
   between the superconducting and the normal phases \cite{ginzburg:1950xz}.
    It can be noted that,
     when it appeared, the GL
   theory looked rather phenomenological and was not
   really appreciated in the western literature. Seven  years later
   Bardeen, Cooper and Schrieffer (BCS) created the microscopic theory
that bears their name
    \cite{Bardeen:1957kj}. Their theory was based on the fundamental
    theorem \cite{cooper:1956fz}, which states that,
     for a system of many electrons at small $T$,
    any weak
   attraction, no matter how small it is, can bind two electrons together,
   forming the so called Cooper pair. Subsequently in \cite{gorkov:1959hy}  it was realized that
   the GL theory was equivalent to the BCS theory around the critical
   point, and this result vindicated the GL theory as a masterpiece in physics.
   Furthermore Gor'kov proved that the fundamental quantities of the
two theories, i.e.  the BCS parameter gap $\Delta$ and the GL
wavefunction $\psi$, were related
   by a proportionality constant and $\psi$ can be thought of as the
Cooper pair  wavefunction in the center-of-mass frame. In
   a sense, the GL theory was the prototype of the modern effective
   theories; in spite of its limitation to the phase transition it
   has a larger field of application, as shown for example by its use in
    the inhomogeneous cases,
    when the gap is not uniform in space.
     Another remarkable advance in these years
     was the Abrikosov's theory of the type II
   superconductors \cite{abrikosov:1957jk}, a class of superconductors
   allowing a penetration of the magnetic field, within certain critical
   values.

   The inspiring power of superconductivity became soon evident in the
field of elementary   particle physics. Two pioneering papers
\cite{Nambu:1961tp,Nambu:1961fr} introduced the idea
   of generating elementary particle masses through the mechanism of
   dynamical symmetry breaking  suggested by superconductivity. This
   idea was so fruitful that it eventually was a crucial ingredient of  the
   Standard Model (SM) of the elementary particles, where the masses are
   generated by the formation of the Higgs condensate much in the same way as
   superconductivity originates from the presence of a gap. Furthermore,
   the Meissner effect, which is characterized by a penetration
   length, is the origin, in the elementary particle physics
   language, of the masses of the gauge vector bosons. These masses
   are nothing but the inverse of the penetration length.

   With the advent of QCD it was early realized that at high density,
   due to the asymptotic freedom
property   \cite{gross:1973ab,politzer:1973ab} and to the
existence of an attractive channel in the color interaction,
    diquark condensates might  be   formed
   \cite{collins:1975ab,Barrois:1977xd,Frautschi:1978rz,Bailin:1984bm}.
   Since these condensates break the color gauge symmetry, the
   subject took the name of color superconductivity. However,
   only in the last few years this has become a very active field of
   research; these developments are
   reviewed in   \cite{Rajagopal:2000wf,Hong:2000ck,Alford:2001dt,Hsu:2000sy,
   Nardulli:2002ma}. It should also be  noted that color
   superconductivity might have implications in
   astrophysics because  for some  compact stars, e.g. pulsars,
the baryon densities necessary for color superconductivity can
probably be  reached.

   Superconductivity in metals was the stage of another breakthrough in  the 1980s
   with the discovery of high $T_c$ superconductors.

Finally we want to mention another development which took place
in 1964 and which is of interest also in QCD. It originates
 in high-field superconductors where a strong magnetic field,
   coupled to the spins of the conduction electrons, gives rise to a
   separation of the Fermi surfaces corresponding to electrons with
   opposite spins. If the separation is too high the pairing is
   destroyed and there is a transition (first-order at small
   temperature) from the superconducting state to the normal one.
   In two separate and contemporary papers, \cite{LO} and  \cite{FF}, it
   was   shown that a new state could be
   formed, close to the transition line. This state that hereafter
    will be called LOFF\footnote{In the literature the LOFF state is also
    known as the FFLO state.} has the feature
   of exhibiting an order parameter, or a gap, which is not a
   constant, but has a space variation whose typical wavelength is  of the
   order of the inverse of the difference in the Fermi energies of the  pairing electrons.
   The space modulation of the gap arises because the electron pair has
   non zero total momentum and it is a rather peculiar phenomenon that
leads to the
   possibility of a non uniform or anisotropic ground state, breaking
   translational and rotational symmetries. It has been also
   conjectured that the typical inhomogeneous ground state might have a
periodic
   or, in other words, a crystalline structure. For this reason other names of this
phenomenon are inhomogeneous or anisotropic or crystalline
superconductivity.

In these lectures notes I used in particular the review papers by
\cite{Polchinski:1992ed}, \cite{Rajagopal:2000wf},
\cite{Nardulli:2002ma}, \cite{Schafer:2003ab} and
\cite{Casalbuoni:2003ab}. I found also the following books  quite
useful \cite{Schrieffer}, \cite{Tinkham}, \cite{Ginzburg},
\cite{Landau2} and \cite{abrikosov}.

\subsection{Basic experimental facts}\label{introduction}
\begin{center}
\begin{figure}[htb]
\epsfxsize=9truecm \centerline{\epsffile{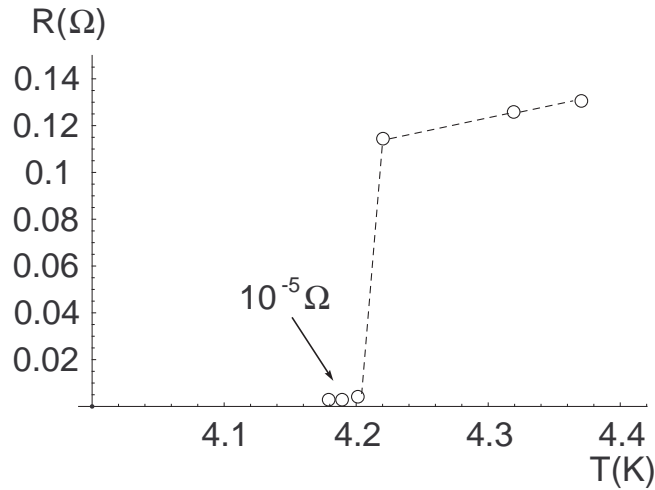}} \noindent
\caption{\it Data from Onnes' pioneering works. The plot shows the
electric resistance of the mercury vs. temperature.\label{onnes}}
\end{figure}
\end{center}

As already said, superconductivity was discovered in 1911 by
Kamerlingh Onnes in Leiden \cite{Kamerlingh:1911ab}. The basic
observation was the disappearance of electrical resistance of
various metals (mercury, lead and thin) in a very small range of
temperatures around a critical temperature $T_c$ characteristic of
the material (see Fig. \ref{onnes}).  This is particularly clear
in experiments with persistent currents in superconducting rings.
These currents have been observed to flow without measurable
decreasing up to one year allowing to put a lower bound of $10^5$
years on their decay time. Notice also that good conductors have
resistivity at a temperature of several degrees K, of the order of
$10^{-6}$ ohm cm, whereas the resistivity of a superconductor is
lower that $10^{-23}$ ohm cm. Critical temperatures for typical
superconductors range from 4.15 K for mercury, to 3.69 K for tin,
and to 7.26 K and 9.2 K for lead and niobium respectively.

 In 1933 Meissner and Ochsenfeld
\cite{meissner:1933ab} discovered the {\bf perfect diamagnetism},
that is the magnetic field ${\bf B}$ penetrates only a depth
$\lambda\backsimeq 500 $ \AA \,and is excluded from the body of
the material.

\begin{center}
\begin{figure}[ht]
\epsfxsize=10truecm \centerline{\epsffile{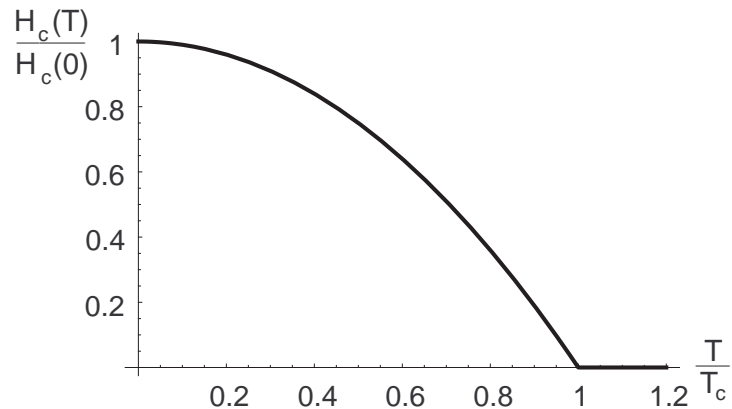}}
\noindent \caption{\it The critical field vs.
temperature.\label{critical_field}}
\end{figure}
\end{center}

One could think that due to the vanishing of the electric
resistance the electric field is zero within the material and
therefore, due to the Maxwell equation \be {\bm{\nabla}}\wedge{\bf
E}=-\frac 1 c\frac{\de{\bf B}}{\de t}\,,\ee the magnetic field is
frozen, whereas it is expelled. This implies that
superconductivity will be destroyed by a critical magnetic field
$H_c$ such that \be
f_s(T)+\frac{H_c^2(T)}{8\pi}=f_n(T)\,,\label{critical}\ee where
$f_{s,n}(T)$ are the densities of free energy in the the
superconducting phase at zero magnetic field and the density of
free energy in the normal phase. The behavior of the critical
magnetic field with temperature was found empirically to be
parabolic (see Fig. \ref{critical_field})\be H_c(T)\approx
H_c(0)\left[1-\left(\frac T{T_c}\right)^2\right]\,.\ee

The critical field at zero temperature is of the order of few
hundred gauss for superconductors as $Al$, $Sn$, $In$, $Pb$, etc.
These superconductors are said to be "soft". For "hard"
superconductors as $Nb_3Sn$ superconductivity stays up to values
of $10^5$ gauss. What happens is that up to a "lower" critical
value $H_{c1}$ we have the complete Meissner effect. Above
$H_{c1}$ the magnetic flux penetrates into the bulk of the
material in the form of vortices (Abrikosov vortices) and the
penetration is complete at $H=H_{c2}>H_{c1}$. $H_{c2}$ is called
the "upper" critical field.

At zero magnetic field a second order transition at $T=T_c$ is
observed. The jump in the specific heat is about three times the
the electronic specific heat of the normal state. In the zero
temperature limit the specific heat decreases exponentially (due
to the energy gap of the elementary excitations or
quasiparticles, see later).

An interesting observation leading eventually to appreciate the
role of the phonons in superconductivity \cite{Frolich:1950ab},
was the {\bf isotope effect}. It was found
\cite{Maxwell:1950,Reynolds:1950} that the critical field at zero
temperature and the transition temperature $T_c$ vary as \be
T_c\approx H_c(0)\approx \frac 1{M^\alpha}\,,\ee with the isotopic
mass of the material. This makes the critical temperature and
field larger for lighter isotopes. This shows the role of the
lattice vibrations, or of the phonons. It has been found that
\be\alpha\approx 0.45\div 0.5\ee for many superconductors,
although there are several exceptions as $Ru$, $Mo$, etc.

\begin{center}
\begin{figure}[ht]
\epsfxsize=10truecm \centerline{\epsffile{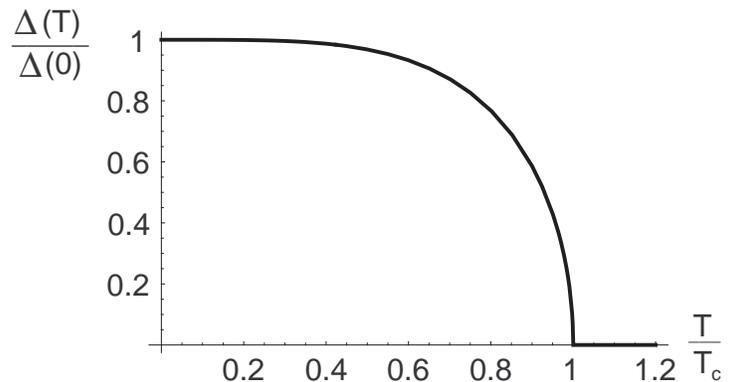}} \noindent
\caption{\it The gap vs. temperature in $Al$ as determined by
electron tunneling.\label{fig_gap}}
\end{figure}
\end{center}

The presence of an energy gap in the spectrum of the elementary
excitations has been observed directly in various ways. For
instance, through the threshold for the absorption of e.m.
radiation, or through the measure of the electron tunnelling
current between two films of superconducting material separated by
a thin ($\approx 20$ \AA) oxide layer. In the case of $Al$ the
experimental result is plotted in Fig. \ref{fig_gap}. The presence
of an energy gap of order $T_c$ was suggested by Daunt and
Mendelssohn \cite{Daunt:1946ab} to explain the absence of
thermoelectric effects, but it was also postulated theoretically
by Ginzburg \cite{Ginzburg:1953ab} and Bardeen
\cite{Bardeen:1956ab}. The first experimental evidence is due to
Corak et al. \cite{Corak:1954ab,Corak:1956ab} who measured the
specific heat of a superconductor. Below $T_c$ the specific heat
has an exponential behavior \be c_s\approx a\,\gamma\, T_c e^{-b
T_c/T}\,,\ee whereas in the normal state \be c_n\approx \gamma
T\,,\ee with $b\approx 1.5 $. This implies a minimum excitation
energy per particle of about $1.5 T_c$. This result was confirmed
experimentally by measurements of e.m. absorption
\cite{Glover:1956ab}.

\subsection{Phenomenological models}

In this Section we will describe some early phenomenological
models trying to explain superconductivity phenomena. From the
very beginning it was clear that in a superconductor a finite
fraction of electrons forms a sort of condensate or
"macromolecule" (superfluid) capable of motion as a whole. At zero
temperature the condensation is complete over all the volume, but
when increasing the temperature part of the condensate evaporates
and goes to form a weakly interacting normal Fermi liquid. At the
critical temperature all the condensate disappears.  We will start
to review the first two-fluid model as formulated by Gorter and
Casimir.

\subsubsection{Gorter-Casimir model}

This model was first formulated in 1934
\cite{Gorter:1934ab,Gorter:1934lm} and it consists in a simple
ansatz for the free energy of the superconductor. Let $x$
represents the fraction of electrons in the normal fluid and $1-x$
the ones in the superfluid. Gorter and Casimir assumed the
following expression for the free energy of the electrons \be
F(x,T)=\sqrt{x}\, f_n(T)+(1-x)\, f_s(T),\ee with \be f_n(T)=-\frac
\gamma 2 T^2,~~~f_s(T)=-\beta={\rm constant},\ee The free-energy
for the electrons in a normal metal is just $f_n(T)$, whereas
$f_s(T)$  gives the condensation energy associated to the
superfluid. Minimizing the free energy with respect to $x$, one
finds the fraction of normal electrons at a temperature $T$ \be
x=\frac 1 {16}\frac{\gamma^2}{\beta^2} T^4.\ee We see that $x=1$
at the critical temperature $T_c$ given by \be
T_c^2=\frac{4\beta}\gamma.\ee Therefore \be x=\left(\frac
T{T_c}\right)^4.\ee The corresponding value of the free energy is
\be F_s(T)=-\beta\left(1+\left(\frac{T}{T_c}\right)^4\right).\ee
Recalling  the definition (\ref{critical}) of the critical
magnetic field, and using \be F_n(T)=-\frac{\gamma}2
T^2=-2\beta\left(\frac T{T_c}\right)^2\ee we find easily \be
\frac{H_c^2(T)}{8\pi}=F_n(T)-F_s(T)=
\beta\left(1-\left(\frac{T}{T_c}\right)^2\right)^2,\ee from which
\be H_c(T)=H_0\left(1-\left(\frac{T}{T_c}\right)^2\right),\ee with
\be H_0=\sqrt{8\pi\beta}.\ee The specific heat in the normal phase
is \be c_n=-T\frac{\de^2 F_n(T)}{\de T^2}=\gamma T,\ee whereas in
the superconducting phase \be c_s=3\gamma T_c \left(\frac
T{T_c}\right)^3.\ee This shows that there is a jump in the
specific heat and  that, in general agreement with experiments,
the ratio of the two specific heats at the transition point is 3.
Of course, this is an "ad hoc" model, without any theoretical
justification but it is interesting because it leads to nontrivial
predictions and in reasonable account with the experiments.
However the postulated expression for the free energy has almost
nothing to do with the one derived from the microscopical theory.

\subsubsection{The London theory}
The brothers H. and F. London \cite{london:1935cd} gave a
phenomenological description of the basic facts of
superconductivity by proposing a scheme based on a two-fluid type
concept with superfluid and normal fluid densities $n_s$ and $n_n$
associated with velocities ${\bf v}_s$ and ${\bf v}_n$. The
densities satisfy \be n_s+n_n=n,\ee where $n$ is the average
electron number per unit volume. The two current densities satisfy
\bea \frac{\de {\bf J}_s}{\de t}&=&\frac {n_s e^2}{m} {\bf
E}~~~~~~\left({\bf J}_s=-e n_s {\bf v}_s\right),\\{\bf
J}_n&=&\sigma_n{\bf E}~~~~~~~~~\left({\bf J}_n=-e n_n {\bf
v}_n\right).\eea The first equation is nothing but the Newton
equation for particles of charge $-e$ and density $n_s$. The other
London equation is \be{\bm{\nabla}}\wedge{\bf J}_s=-\frac{n_s
e^2}{mc}{\bf B}.\label{london2}\ee From this equation the Meissner
effect follows. In fact consider the following Maxwell equation
\be {\bm{\nabla}}\wedge{\bf B}=\frac{4\pi} c{\bf J_s},\ee where we
have neglected displacement currents and the normal fluid current.
By taking the curl of this expression and using
\be{\bm{\nabla}}\wedge{\bm{\nabla}}\wedge {\bf
B}=-{\bm\nabla}^2{\bf B},\ee in conjunction with Eq.
(\ref{london2}) we get \be{\bm{\nabla}}^2{\bf B}=\frac{4\pi n_s
e^2}{mc^2}{\bf B}=\frac 1{\lambda_L^2}{\bf B},\label{meissner}\ee
with the {\it penetration depth} defined by
\be\lambda_L(T)=\left(\frac{mc^2}{4\pi n_s
e^2}\right)^{1/2}.\label{penetration}\ee Applying Eq.
(\ref{meissner}) to a plane boundary located at $x=0$ we get \be
B(x)=B(0)e^{-x/\lambda_L},\ee showing that the magnetic field
vanishes in the bulk of the material. Notice that for $T\to T_c$
one expects $n_s\to 0$ and therefore $\lambda_L(T)$ should go to
$\infty$ in the limit. On the other hand for $T\to 0$, $n_s\to n$
and we get \be\lambda_L(0)=\left(\frac{mc^2}{4\pi n
e^2}\right)^{1/2}.\ee In the two-fluid theory of Gorter and
Casimir \cite{Gorter:1934ab,Gorter:1934lm} one has \be \frac{n_s}n
=1-\left(\frac T{T_c}\right)^4,\ee and \be
\lambda_L(T)=\frac{\lambda_L(0)}{\left[1-\dd{\left(\frac
T{T_c}\right)^4}\right]^{1/2}}.\ee
\begin{center}
\begin{figure}[ht]
\epsfxsize=10truecm \centerline{\epsffile{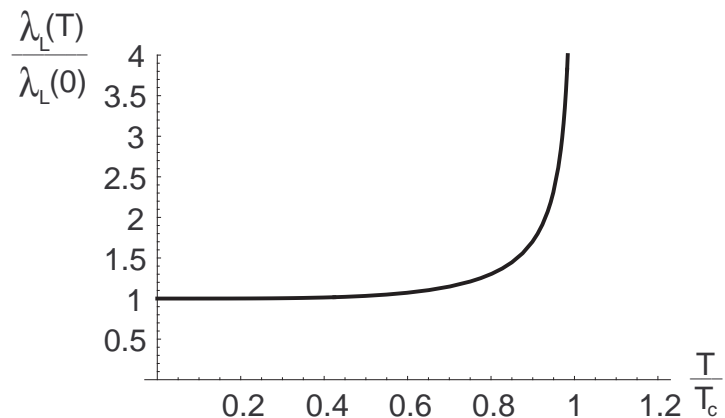}}
\noindent \caption{\it The penetration depth vs.
temperature.\label{figI2}}
\end{figure}
\end{center} This agrees very well with the
experiments. Notice that at $T_c$ the magnetic field penetrates
all the material since $\lambda_L$ diverges. However, as shown in
Fig. \ref{figI2}, as soon as the temperature is lower that $T_c$
the penetration depth goes very close to its value at $T=0$
establishing the Meissner effect in the bulk of the
superconductor.

The London equations can be justified as follows: let us assume
that the wave function describing the superfluid is not changed,
at first order, by the presence of an e.m. field. The canonical
momentum of a particle is \be {\bf p}=m{\bf v}+\frac e c{\bf
A}.\ee Then, in stationary conditions, we expect \be \langle{\bf
p}\rangle=0,\ee or \be \langle{\bf v}_s\rangle=-\frac e{mc}{\bf
A},\ee implying \be {\bf J}_s=e n_s\langle{\bf
v}_s\rangle=-\frac{n_se^2}{mc}{\bf A}.\label{london3}\ee By taking
the time derivative and the curl of this expression we get the two
London equations.
\subsubsection{Pippard non-local electrodynamics}
Pippard \cite{Pippard:1953ab} had the idea that the local relation
between ${\bf J}_s$ and ${\bf A}$ of Eq. (\ref{london3}) should be
substituted by a non-local relation. In fact the wave function of
the superconducting state is not localized. This can be seen as
follows: only electrons within $T_c$ from the Fermi surface can
play a role at the transition. The corresponding momentum will be
of order \be \Delta p\approx \frac {T_c}{v_F}\ee and \be\Delta
x\gtrsim \frac{1}{\Delta p}\approx \frac{ v_F}{T_c}.\ee This
define a characteristic length (Pippard's coherence length) \be
\xi_0=a\frac{v_F}{T_c},\label{coherence}\ee with $a\approx 1$. For
typical superconductors $\xi_0\gg\lambda_L(0)$. The importance of
this length arises from the fact that impurities increase the
penetration depth $\lambda_L(0)$. This happens because the
response of the supercurrent to the vector potential is smeared
out in a volume of order $\xi_0$. Therefore the supercurrent is
weakened. Pippard was guided by a work of
Chamber\footnote{Chamber's work is discussed in \cite{Ziman}}
studying the relation between the electric field and the current
density in normal metals. The relation found by Chamber is a
solution of Boltzmann equation in the case of a scattering
mechanism characterized by a mean free path $l$. The result of
Chamber generalizes the Ohm's law ${\bf J}({\bf r})=\sigma{\bf
E}({\bf r})$ \be {\bf J}({\bf r})=\frac{3\sigma}{4\pi
l}\int\frac{{\bf R}({\bf R}\cdot{\bf E}({\bf r}'))
e^{-R/l}}{R^4}\,d^3{\bf r}',~~~{\bf R}={\bf r}-{\bf r}'.\ee If
${\bf E}({\bf r})$ is nearly constant within a volume of radius
$l$ we get \be \b E(\b r)\cdot \b J(\b r)=\frac{3\sigma}{4\pi
l}|\b E(\b r)|^2\int\frac{\cos^2\theta\, e^{-R/l}}{R^2}\,d^3\b
r'=\sigma|\b E(\b r)|^2,\ee implying the Ohm's law. Then Pippard's
generalization of \be \b J_s(\b r)=-\frac 1 {c\Lambda(T)}\b A(\b
r),~~~~\Lambda(T)=\frac{e^2 n_s(T)}{m},\ee is \be {\bf J}({\bf
r})=-\frac{3\sigma}{4\pi \xi_0\Lambda(T) c}\int\frac{{\bf R}({\bf
R}\cdot{\bf A}({\bf r}')) e^{-R/\xi}}{R^4}\,d^3{\bf r}',\ee with
an effective coherence length defined as \be\frac 1 \xi=\frac
1{\xi_0}+\frac 1 l,\ee and $l$ the mean free path for the
scattering of the electrons over the impurities. For almost
constant field one finds as before \be  \b J_s(\b r)=-\frac 1
{c\Lambda(T)}\frac\xi{\xi_0}\b A(\b r).\ee Therefore for pure
materials ($l\to\infty$) one recover the local result, whereas for
an impure material the penetration depth increases by a factor
$\xi_0/\xi>1$. Pippard has also shown that a good fit to the
experimental values of the parameter $a$ appearing in Eq.
(\ref{coherence}) is 0.15, whereas from the microscopic theory one
has $a\approx 0.18$, corresponding to \be
\xi_0=\frac{v_F}{\pi\Delta}.\ee This is obtained using $T_c\approx
.56\,\Delta$, with $\Delta$ the energy gap (see later).

\subsubsection{The Ginzburg-Landau theory}\label{sectionGL}

In 1950 Ginzburg and Landau \cite{ginzburg:1950xz} formulated
their theory of superconductivity introducing a complex wave
function as an order parameter. This was done in the context of
Landau theory of  second order phase transitions and as such this
treatment is strictly valid only around the second order critical
point. The wave function is related to the superfluid density by
\be n_s=|\psi(\b r)|^2.\ee Furthermore it was postulated a
difference of free energy between the normal and the
superconducting phase of the form \be F_s(T)-F_n(T)=\int d^3{\b
r}\left(-\frac 1{2m^*}\psi^*(\b r)|(\bm{\nabla}+ie^*\b
A)|^2\psi(\b r)+\alpha(T)|\psi(\b r)|^2+\frac 1 2\beta(T)|\psi(\b
r)|^4\right),\label{1.35}\ee where $m^*$ and $e^*$ were the
effective mass and charge that in the microscopic theory  turned
out to be $2m$ and $2e$ respectively. One can look for a constant
wave function minimizing the free energy. We find \be
\alpha(T)\psi+\beta(T)\psi|\psi|^2=0,\ee giving
\be|\psi|^2=-\frac{\alpha(T)}{\beta(T)},\ee and for the free
energy density \be f_s(T)-f_n(T)=-\frac 12
\frac{\alpha^2(T)}{\beta(T)}=-\frac{H_c^2(T)}{8\pi},\label{GL1}\ee
where the last equality follows from Eq. (\ref{critical}).
Recalling that in the London theory (see Eq.
(\ref{penetration}))\be n_s=|\psi|^2\approx \frac
1{\lambda_L^2(T)},\ee we find \be
\frac{\lambda_L^2(0)}{\lambda_L^2(T)}=\frac{|\psi(T)|^2}
{|\psi(0)|^2}=\frac 1 n |\psi(T)|^2=-\frac 1
n\frac{\alpha(T)}{\beta(T)}.\label{GL2}\ee From Eqs. (\ref{GL1})
and (\ref{GL2}) we get \be
n\alpha(T)=-\frac{H_c^2(T)}{4\pi}\frac{\lambda_L^2(T)}{\lambda_L^2(0)}\ee
and \be
n^2\beta(T)=\frac{H_c^2(T)}{4\pi}\frac{\lambda_L^4(T)}{\lambda_L^4(0)}.\ee
The equation of motion at zero em field  is \be -\frac
1{2m^*}\bm{\nabla}^2\psi+\alpha(T)\psi+\beta(T)|\psi|^2\psi=0.\ee
We can look at solutions close to the constant one by defining
$\psi=\psi_e+f$ where \be
|\psi_e|^2=-\frac{\alpha(T)}{\beta(T)}.\ee We find, at the lowest
order in $f$ \be \frac{1}{4m^*|\alpha(T)|}\bm{\nabla}^2f-f=0.\ee
This shows an exponential decrease which we will write as \be
f\approx e^{-\sqrt{2}r/\xi(T)},\ee where we have introduced the
Ginzburg-Landau (GL) coherence length \be\xi(T)=\frac
1{\sqrt{2m^*|\alpha(T)|}}.\label{1.47}\ee Using the expression
(\ref{GL1}) for $\alpha(T)$ we have also \be
\xi(T)=\sqrt{\frac{2\pi
n}{m^*H_c^2(T)}}\frac{\lambda_L(0)}{\lambda_L(T)}.\ee Recalling
that ($t=T/T_c$)\be H_c(T)\approx \left(1-t^2\right),~~~~~~~
\lambda_L(T)\approx \frac 1{\left(1-t^4\right)^{1/2}},\ee we see
that also the GL coherence length goes to infinity for $T\to T_c$
\be \xi(T)\approx\frac 1{H_c(T)\lambda_L(T)}\approx \frac
1{(1-t^2)^{1/2}}.\ee It is possible to show that \be \xi(T)\approx
\frac{\xi_0}{(1-t^2)^{1/2}}.\ee Therefore the GL coherence length
is related but not the same as the Pippard's coherence length. A
useful quantity is \be\kappa=\frac{\lambda_L(T)}{\xi(T)},\ee which
is finite for $T\to T_c$ and approximately independent on the
temperature. For typical pure superconductors $\lambda\approx 500$
\AA, $\xi\approx 3000$ \AA, and $\kappa\ll 1$.

\subsection{Cooper pairs}

\begin{center}
\begin{figure}[ht]
\epsfxsize=10truecm \centerline{\epsffile{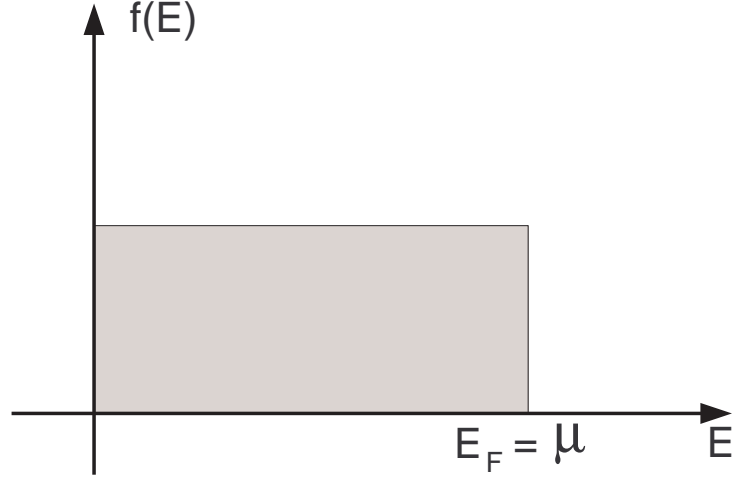}}
\noindent \caption{\it The Fermi distribution at zero
temperature.\label{figfermi}}
\end{figure}
\end{center}

One of the pillars of the microscopic theory of superconductivity
is that electrons close to the FErmi surface can be bound in pairs
by an attractive arbitrary weak interaction \cite{cooper:1956fz}.
First of all let us remember that the Fermi distribution function
for $T\to 0$ is nothing but a $\theta$-function \be f(E,T)=\frac
1{e^{(E-\mu)/T} +1},~~~~~~\lim_{T\to 0} f(E,T)= \theta(\mu-E),\ee
meaning that all the states are occupied up to the Fermi energy
\be E_F=\mu,\ee where $\mu$ is the chemical potential, as shown in
Fig. \ref{figfermi}.
The key point is that the problem has an enormous degeneracy at
the Fermi surface since there is no cost in free energy for adding
or subtracting a fermion at the Fermi surface (here and in the
following we will be quite liberal in speaking about thermodynamic
potentials; in the present case the relevant quantity is the grand
potential) \be \Omega=E-\mu N\to (E\pm E_F)-(N\pm 1)=\Omega.\ee
This observation suggests that a condensation phenomenon can take
place if two fermions are bounded. In fact, suppose that the
binding energy is $E_B$, then adding a bounded pair to the Fermi
surface we get \be \Omega\to (E+2E_F-E_B)-\mu(N+2)=-E_B.\ee
Therefore we get more stability adding more bounded pairs to the
Fermi surface.  Cooper proved that two fermions can give rise to a
bound state for an arbitrary attractive interaction by considering
the following simple model. Let us add two fermions at the Fermi
surface at $T=0$ and suppose that the two fermions interact
through an attractive potential. Interactions among this pair and
the fermion sea in the Fermi sphere are neglected except for what
follows from Fermi statistics. The next step is to look for a
convenient two-particle wave function. Assuming that the pair has
zero total momentum one starts with \be\psi_0(\b r_1-\b
r_2)=\sum_{\b k}g_{\b k}e^{i\b k\cdot(\b r_1-\b
r_2)}.\label{1.57}\ee Here and in the following we will switch
often back and forth from discretized momenta to continuous ones.
We remember that the rule to go from one notation to the other is
simply \be \sum_{\b k}\to\frac{L^3}{(2\pi)^3}\int d^3\b k,\ee
where $L^3$ is the quantization volume. Also often we will omit
the volume factor. This means that in this case we are considering
densities. We hope that from the context it will be clear what we
are doing. One has also to introduce the spin wave function and
properly antisymmetrize. We write \be\psi_0(\b r_1-\b
r_2)=(\alpha_1\beta_2-\alpha_2\beta_1)\sum_{\b k}g_{\b k}\cos(\b
k\cdot(\b r_1-\b r_2)),\ee where $\alpha_i$ and $\beta_i$ are the
spin functions. This wave function is expected to be preferred
with respect to the triplet state, since the "$\cos$" structure
gives a bigger probability for the fermions to stay together.
Inserting this wave function inside the Schr\"odinger equation \be
\left[-\frac
1{2m}\left(\bm{\nabla}_1^2+\bm{\nabla}_2^2\right)+V(\b r_1-\b
r_2)\right]\psi_0(\b r_1-\b r_2)=E\psi_0(\b r_1-\b r_2),\ee we
find \be(E-2\epsilon_{\b k})g_{\b k}=\sum_{\b k'>\b k_F} V_{\b
k,\b k'}\, g_{\b k'},\ee where $\epsilon_{\b k}=|\b k|^2/2m$ and
\be V_{\b k,\b k'}=\frac 1 {L^3}\int  V(\b r)\,e^{i(\b k'-\b
k)\cdot\b r)}\,d^3{\b r}.\ee Since one looks for solutions with
$E<2\epsilon_{\b k}$, Cooper made the following assumption on the
potential:\be V_{\b k,\b k'}=\left\{\matrix{ -G & k_F\le |\b k|\le
k_c\cr 0 & {\rm otherwise}\cr}\right. \ee with $G>0$ and
$\epsilon_{\b k_F}=E_F$. Here a cutoff $k_c$ has been introduced
such that \be \epsilon_{k_c}=E_F+\delta\ee and $\delta\ll E_F$.
This means that one is restricting the physics to the one
corresponding to degrees of freedom close to the Fermi surface.
The Schr\"odinger equation reduces to \be(E-2\epsilon_{\b k})
g_{\b k}=-G\sum_{\b k'> \b k_F}g_{\b k'}.\ee Summing over $\b k $
we get \be \frac 1 G =\sum_{\b k>\b k_F}\frac 1 {2\epsilon_{\b
k}-E}.\ee Replacing the sum with an integral we obtain \be \frac 1
G=\int_{k_F}^{k_c}\frac{d^3{\b k}}{(2\pi)^3}\frac 1 {2\epsilon_{\b
k}-E}=\int_{E_F}^{E_F+\delta} \frac{d\Omega}{(2\pi)^3}k^2
\frac{dk}{d\epsilon_{\b k}}\frac {d\epsilon}{2\epsilon-E}.\ee
Introducing the density of states at the Fermi surface for  two
electrons with  spin up and down\be
\rho=2\int\frac{d\Omega}{(2\pi)^3}k^2 \frac{dk}{d\epsilon_{\b
k}},\ee we obtain \be \frac 1 G=\frac 1 4\,
\rho\log\frac{2E_F-E+2\delta}{2E_F-E}.\label{bound}\ee Close to
the Fermi surface we may assume $k\approx k_F$ and \be\epsilon_{\b
k}=\mu+(\epsilon_{\b k}-\mu)\approx \mu+\frac{\de\epsilon_{\b
k}}{\de\b k}\Big|_{\b k=\b k_F}\cdot(\b k-\b k_F)=\mu+\b v_F(\b
k)\cdot{\bm\ell},\ee where \be{\bm \ell}=\b k-\b k_F\ee is the
"residual momentum".  Therefore \be \rho=\frac{k_F^2}{\pi^2
v_F}.\ee Solving  Eq. (\ref{bound}) we find \be
E=2E_F-2\delta\frac{e^{-4/\rho G}}{1-e^{-4/\rho G}}.\ee For most
classic superconductor \be \rho G<0.3,\ee In this case (weak
coupling approximation. $\rho G\ll 1$) we get \be E\approx
2E_F-2\delta e^{-4/\rho G}.\ee We see that a bound state is formed
with a binding energy \be E_B=2\delta e^{-4/\rho G}.\ee The result
is not analytic in $G$ and cannot be obtained by a perturbative
expansion in $G$. Notice also that the bound state exists
regardless of the strength of $G$. Defining \be N=\sum_{\b k>\b
k_F} g_{\b k},\ee we get the wave function \be \psi_0(\b
r)=N\sum_{\b k>\b k_F}\frac{\cos(\b k\cdot\b r)}{2\epsilon_{\b
k}-E}.\ee Measuring energies from $E_F$ we introduce \be \xi_{\b
k}=\epsilon_{\b k}-E_F.\ee from which \be \psi_0(\b r)=N\sum_{\b
k>\b k_F}\frac{\cos(\b k\cdot\b r)}{2\xi_{\b k}+E_B}.\ee We see
that the wave function in momentum space has a maximum for
$\xi_{\b k}=0$, that is for the pair being at the Fermi surface,
and falls off with $\xi_{\b k}$. Therefore the electrons involved
in the pairing are the ones within a range $E_B$ above $E_F$.
Since for $\rho G\ll 1$ we have $E_B\ll\delta$, it follows that
the behavior of $V_{\b k,\b k'}$ far from the Fermi surface is
irrelevant. \underline{Only the degrees of freedom close to the
Fermi surface are important}. Also using the uncertainty principle
as in the discussion of the Pippard non-local theory we have that
the size of the bound pair is larger than $v_F/E_B$. However the
critical temperature turns out to be of the same order as $E_B$,
therefore the size of the Cooper pair is of the order of the
Pippard's coherence length $\xi_0=av_F/T_c$.

\subsubsection{The size of a Cooper pair}

It is an interesting exercise to evaluate the size of a Cooper
pair  defined in terms of the mean square radius of the pair wave
function \be \bar R^2=\frac{\int|\psi_0(\b r)|^2 |\b r|^2d^3\b
r}{\int|\psi_0(\b r)|^2 d^3\b r}.\ee Using the expression
(\ref{1.57}) for $\psi_0$ we have \be |\psi_0(\b r)|^2=\sum_{\b
k,\b k'} g_{\b k}g_{\b k'}^*e^{i(\b k-\b k')\cdot\b r}\ee and \be
\int|\psi_0(\b r)|^2 d^3\b r=L^3\sum_{\b k} |g_{\b k}|^2.\ee Also
\be\int|\psi_0(\b r)|^2|\b r|^2 d^3\b r=\int \sum_{\b k\b
k'}\left[-i\b\nabla_{\b k'}g_{\b k'}^*\right]\left[i\b\nabla_{\b
k}g_{\b k}\right]e^{i(\b k-\b k')\cdot\b r}d^3\b r=L^3\sum_{\b k}
|\b\nabla_{\b k}g_{\b k}|^2.\ee Therefore \be \bar
R^2=\frac{\sum_{\b k}|\b\nabla_{\b k} g_{\b k}|^2}{\sum_{\b
k}|g_{\b k}|^2}.\ee Recalling that \be g_{\b k}\approx \frac
1{2\epsilon_{\b k} -E}=\frac 1{2\xi_{\bf k} +E_B},\ee we obtain
\be\sum_{\b k}|\b\nabla_{\b k} g_{\b k}|^2\approx \sum_{\b
k}\dd{\frac 1{(2\xi_{\b k}+E_B)^4}}\left|2\frac{\de\epsilon_{\b
k}}{\de \b k}\right|^2= 4v_F^2\sum_{\b k}\frac 1{(2\xi_{\b
k}+E_B)^4}.\ee Going to continuous variables and noticing that the
density of states cancel in the ratio we find \be \bar
R^2=4v_F^2\frac{\dd{\int_0^\infty\frac{d\xi}{(2\xi+E_B)^4}}}
{\dd{\int_0^\infty\frac{d\epsilon}{(2\xi+E_B)^2}}}=4v_F^2\frac{\dd{-\frac
1 3\frac 1{(2\xi+E_B)^3}\Big|_0^\infty}}{\dd{-\frac
1{2\xi+E_B}}\Big|_0^\infty}=\frac 43\frac{v_F^2}{E_B^2},\ee where,
due to the convergence we have extended the integrals up to
infinity. Assuming $E_B$ of the order of the critical temperature
$T_c$, with $T_c\approx 10$ K and $v_F\approx 10^8$ cm/s, we get
\be\bar R\approx 10^{-4}~{\rm cm}\approx 10^4~{\rm \AA}.\ee The
order of magnitude of $\bar R$ is the same as the coherence length
$\xi_0$. Since one electron occupies a typical size of about  (2
\AA)$^3$, this means that in a coherence volume there are about
$10^{11}$ electrons. Therefore it is not reasonable to construct a
pair wavefunction, but we need a wave function taking into account
all the electrons. This is made in the BCS theory.

\subsection{Origin of the attractive interaction}

The problem of getting an attractive interaction among electrons
is not an easy one. In fact the Coulomb interaction is repulsive,
although it gets screened in the medium by a screening length of
order of $1/k_s\approx$ 1 \AA. The screened Coulomb potential is
given by \be V(\b q)=\frac{4\pi e^2}{q^2+k_s^2}.\ee To get
attraction is necessary to consider the effect of the motion of
the ions. The rough idea is that one electron polarizes the medium
attracting positive ions. In turn these attract a second electron
giving rise to a net attraction between the two electrons. To
quantify this idea is necessary to take into account the
interaction among the electrons and the lattice or, in other
terms, the interactions among the electrons and the phonons as
suggested by \cite{Frolich:1952ab}. This idea was confirmed by the
discovery of the \b {isotope effect}, that is the dependence of
$T_c$ or of the gap from the isotope mass (see Section
\ref{introduction}). Several calculations were made by
\cite{Pines:1958ab} using the "jellium model". The potential in
this model is  \cite{deGennes} \be V(\b q,\omega)=\frac{4\pi
e^2}{q^2+k_s^2}\left(1+\frac{\omega_{\b q}^2}{\omega^2-\omega_{\b
q}^2}\right).\ee Here $\omega_{\b q}$ is the phonon energy that,
for a simple linear chain, is given by \be \omega_{\b
q}=2\sqrt{\frac k M}\sin(q a/2),\ee where $a$ is the lattice
distance, $k$ the elastic constant of the harmonic force among the
ions and $M$ their mass. For $\omega<\omega_{\b q}$ the phonon
interaction is attractive at it may overcome the Coulomb force.
Also, since the cutoff to be used in the determination of the
binding energy, or for the gap, is essentially the Debye frequency
which is proportional to $\omega_{\b q}$ one gets naturally the
isotope effect.

%%%%%%%%%%%%%%%%%%%%%%%%%%%%%%%%%%%%%%%%%%
 \section{Effective theory at the Fermi surface}
\subsection{Introduction}\label{IIA}
\begin{center}
\begin{figure}[bht]
\epsfxsize=6truecm \centerline{\epsffile{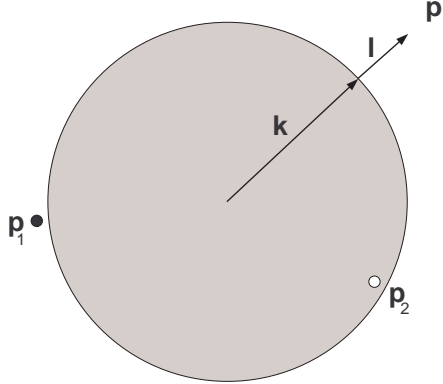}}
\noindent \caption{\it A spherical Fermi surface. Low lying
excitations are shown: a particle at ${\bf p}_1$ and a hole at
${\bf p}_2$. The decomposition of a momentum as the Fermi momentum
${\bf k}$, and the residual momentum ${\bf l}$ is also
shown.\label{6fig}}
\end{figure}
\end{center}

It turns out that the BCS theory can be derived within the Landau
theory of Fermi liquids, where a conductor is treated as a gas of
nearly free electrons. This is because one can make use of the
idea of quasiparticles, that is electrons dressed by the
interaction. A justification of this statement has been given in
\cite{benfatto:1990ab,Polchinski:1992ed,Shankar:1994pf}. Here we
will follow the treatment given by \cite{Polchinski:1992ed}. In
order to define an effective field theory one has to start
identifying a scale which, for ordinary superconductivity (let us
talk about this subject to start with) is of the order of tens of
$eV$. For instance, \be E_0={ m\alpha^2}\approx 27 ~eV\ee is the
typical energy in solids. Other possible scales as the ion masses
$M$ and velocity of light can be safely considered to be infinite.
In a conductor a current can be excited with an arbitrary small
field, meaning that the spectrum of the charged excitations goes
to zero energy. If we are interested to study these excitations we
can try to construct our effective theory at energies much smaller
than $E_0$ (the superconducting gap turns out to be of the order
of $10^{-3}~eV$). Our first problem is then to identify the
quasiparticles. The natural guess is that they are spin 1/2
particles as the electrons in the metal. If we measure the energy
with respect to the Fermi surface the most general free action can
be written as \be S_{\rm free}=\int dt\,d^3{\bf p}
\left[i\psi_{\sigma}^\dagger({\bf p})i\de_t\psi_\sigma({\bf
p})-(\epsilon({\bf p}) -\epsilon_F)\psi^\dagger_\sigma({\bf
p})\psi_\sigma({\bf p})\right]\label{free_action}\ .\ee Here
$\sigma$ is a spin index and $\epsilon_F$ is the Fermi energy. The
ground state of the theory is given by the Fermi sea with all the
states $\epsilon({\bf p})<\epsilon_F$ filled and all the states
$\epsilon({\bf p})>\epsilon_F$ empty. The Fermi surface is defined
by $\epsilon({\bf p})=\epsilon_F$. A simple example is shown in
Fig. \ref{6fig}.
%%%%%%%%%%%%%%%%%%%%%%%%%%%%%%%%%%%%%%%%%
%%%%%%%%%%%%%%%%%%%%%%%%%%%%%%%%%%%%%%%%%%%

 The free action defines the scaling properties of
the fields. In this particular instance we are interested at the
physics very close to the Fermi surface and therefore we are after
the scaling properties for $\epsilon\to\epsilon_F$. Measuring
energies with respect to the Fermi energy we introduce a scaling
factor $s<1$. Then, as the energy scales to zero the momenta must
scale toward the Fermi surface. It is convenient to decompose the
momenta as follows (see also Fig. \ref{6fig})\be\bf {\bf p}={\bf
k}+{\bf l}\ .\ee Therefore we get \be E\to s\,E,~~{\bf k}\to {\bf
k},~~{\bf l}\to s{\bf l}\ .\ee We can expand the second term in
Eq. (\ref{free_action}) obtaining \be \epsilon({\bf
p})-\epsilon_F=\frac{\de\epsilon({\bf p})}{\de{\bf p}}\Big|_{{\bf
p}={\bf k}}\cdot({\bf p}-{\bf k})=l v_F({\bf k})\ ,\ee where \be
{\bf v}_F({\bf k})=\frac{\de\epsilon({\bf p})}{\de{\bf
p}}\Big|_{{\bf p}={\bf k}}\ .\ee Notice that ${\bf v}_F({\bf k})$
is a vector orthogonal to the Fermi surface. We get \be S_{\rm
free}=\int dt\,d^3{\bf p} \left[\psi_{\sigma}^\dagger({\bf
p})\left(i\de_t-l v_F({\bf k})\right)\psi_\sigma({\bf
p})\right]\label{free_action2}\ .\ee The various scaling laws are
\bea &&dt \to s^{-1} dt,~~~~d^3{\bf p}=d^2{\bf k} dl\to s d^2{\bf
k} dl\cr &&\de_t \to s\de_t,~~~~~~~~~~~~~l\to s l\ .\eea
Therefore, in order to leave the free action invariant the fields
must scale as \be\psi_\sigma({\bf p})\to s^{-1/2}\psi_\sigma({\bf
p})\ .\ee Our analysis goes on considering all the possible
interaction terms compatible with the symmetries of the theory and
looking for the relevant ones. The symmetries of the theory are
the electron number and the spin $SU(2)$, since we are considering
the non-relativistic limit. We ignore also possible complications
coming from the real situation where one has to do with crystals.
The possible terms are:
\begin{enumerate}
\item {\underline{Quadratic terms}}:\\
\be\int dt\, d^2{\bf k}\, dl\, \mu({\bf k})
\psi^\dagger_\sigma({\bf p})\psi_\sigma({\bf p})\ .\ee This is a
relevant term since it scales as $s^{-1}$ but it can be absorbed
into the definition of the Fermi surface (that is by
$\epsilon({\bf p})$. Further terms with time derivatives or powers
of $l$ are already present or they are irrelevant.
\item {\underline{Quartic terms}}:\\
\be \int\prod_{i=1}^4\left(d^2{\bf k}_i\,
dl_i\right)\left(\psi^\dagger({\bf p}_1)\psi({\bf
p}_3\right)\left(\psi^\dagger({\bf p}_2)\psi({\bf p}_4\right)
V({\bf k}_1,{\bf k}_2,{\bf k}_3,{\bf k}_4)\delta^3({\bf p}_1+{\bf
p}_2-{\bf p}_3-{\bf p}_4).\label{quartic}\ee This scales as
$s^{-1}\, s^{4-4/2}=s$ times the scaling of the $\delta$-function.
For a generic situation the $\delta$-function does not scale (see
Fig. \ref{7fig}). However consider a scattering process $1+2\to
3+4$ and decompose the momenta as follows: \bea{\bf p}_3&=&{\bf
p}_1+\delta{\bf k}_3+\delta{\bf l}_3\ ,\\ {\bf p}_4&=&{\bf
p}_2+\delta{\bf k}_4+\delta{\bf l}_4\ .\eea This gives rise to \be
\delta^3(\delta{\bf k}_3+\delta{\bf k}_4+\delta{\bf
l}_3+\delta{\bf l}_4)\ .\ee When ${\bf p}_1=-{\bf p}_2$ and ${\bf
p}_3=-{\bf p}_4$ we see that the $\delta$-function factorizes \be
\delta^2(\delta{\bf k}_3+\delta{\bf k}_4)\delta(\delta{\bf
l}_3+\delta{\bf l}_4)\ee scaling as $s^{-1}$. Therefore, in this
kinematical situation the term (\ref{quartic}) is marginal (does
not scale). This means that its scaling properties should be
looked at the level of quantum corrections.
\begin{center}
\begin{figure}[ht]
\epsfxsize=10truecm \centerline{\epsffile{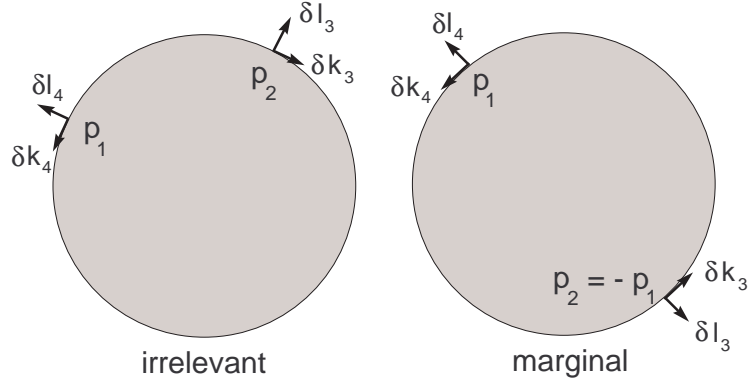}} \noindent
\caption{\it The kinematics for the quartic coupling is shown in
the generic (left) and in the special (right) situations discussed
in the text\label{7fig}}
\end{figure}
\end{center}
\item {\underline{Higher order terms}} Terms with $2n$ fermions
($n>2$) scale as $s^{n-1}$ times the scaling of the
$\delta$-function and therefore they are irrelevant.
\end{enumerate}
We see that the only potentially dangerous term is the quartic
interaction with the particular kinematical configuration
corresponding to a Cooper pair. We will discuss the one-loop
corrections to this term a bit later. Before doing that let us
study the free case.

\subsection{Free fermion gas}\label{IIB}

The statistical properties of free fermions were discussed by
Landau who, however, preferred to talk about fermion liquids. The
reason, as quoted in \cite{Ginzburg}, is that Landau thought that
"Nobody has abrogated Coulomb's law".

Let us consider the free fermion theory we have discussed before.
The fermions are described by the  equation of motion
\be{(i\de_t-\ell v_F)\psi_\sigma({\bf p}, t}=0.\ee The Green
function, or the propagator of the theory is defined by \be
(i\de_t-\ell v_F)G_{\sigma\sigma'}({\bf
p},t)=\delta_{\sigma\sigma'}\delta(t).\ee It is easy to verify
that a solution is given by \be G_{\sigma\sigma'}({\bf
p},t)=\delta_{\sigma\sigma'} G({\bf
p},t)=-i\delta_{\sigma\sigma'}\left[\theta(t)
\theta(\ell)-\theta(-t)\theta(-\ell)\right]e^{-i\ell v_F t}.\ee By
using the integral representation for the step function \be
\theta(t)=\frac i{2\pi}\int d\omega\frac{e^{-i\omega
t}}{\omega+i\epsilon},\ee we get \be G({\bf
p},t)=\frac{1}{2\pi}\int d\omega \frac{e^{-i\ell v_F
t}}{\omega+i\epsilon}\left[e^{-i\omega t}\theta(\ell)- e^{i\omega
t}\theta(-\ell)\right].\ee By  changing the variable $\omega\to
\omega'=\omega\pm\ell v_F$ in the two integrals and sending
$\omega'\to -\omega'$ in the second integral we obtain \be G({\bf
p},t)=\frac{1}{2\pi}\int d\omega e^{-i\omega
t}\left[\frac{\theta(\ell)}{\omega-\ell
v_F+i\epsilon}+\frac{\theta(-\ell)}{\omega-\ell
v_F-i\epsilon}\right].\ee We may also write \be G({\bf
p},t)\equiv\frac{1}{2\pi}\int d p_0 G(p_0,{\bf p})e^{-ip_0t},\ee
with \be G(p) =\frac 1{(1+i\epsilon)p_0 -\ell
v_F}.\label{fermipropagator}\ee Notice that this definition of
$G(p)$ corresponds to the standard Feynman propagator since it
propagates ahead in time positive energy solutions
$\ell>0~(p>p_F)$ and backward in time negative energy solutions
$\ell<0~(p<p_F)$ corresponding to holes in the Fermi sphere. In
order to have contact with the usual formulation of field quantum
theory we introduce Fermi fields \be \psi_\sigma(x)=\sum_{\bf p}
b_\sigma({\bf p},t)e^{i{\bf p}\cdot{\bf x}}=\sum_{\bf p}
b_\sigma({\bf p})e^{-ip\cdot{x}},\ee where $x^\mu=(t,{\bf x})$,
$p^\mu=\ell v_F,\b p)$ and \be p\cdot x =\ell v_F t-{\bf
p}\cdot{\bf x}.\ee Notice that within this formalism fermions have
no antiparticles, however the fundamental state is described by
the following relations \bea b_\sigma({\bf p})|0\rangle =0&&~~{\rm
for}~~ |{\bf p}|>p_F\cr b_\sigma^\dagger({\bf p})|0\rangle
=0&&~~{\rm for}~~ |{\bf p}|<p_F.\eea One could, as usual in
relativistic field theory, introduce a re-definition for the
creation operators for particles with $p<p_F$ as annihilation
operators for holes but we will not do this here. Also we are
quantizing in a box, but we will shift freely from this
normalization to the one in the continuous according to the
circumstances. The fermi operators satisfy the usual
anticommutation relations \be [b_\sigma({\bf
p}),b_{\sigma'}^\dagger({\bf p}')]_+=\delta_{{\bf p}{\bf
p}'}\delta_{\sigma\sigma'}\ee from which \be [\psi_\sigma({\bf
x},t),\psi_{\sigma'}^\dagger({\bf
y},t)]_+=\delta_{\sigma\sigma'}\delta^3({\bf x}-\b y).\ee We can
now show that the propagator is defined in configurations space in
terms of the usual $T$-product for Fermi fields \be
G_{\sigma\sigma'}(x)=-i\langle
0|T(\psi_\sigma(x)\psi_{\sigma'}(0))|0\rangle.\ee In fact we have
\be G_{\sigma\sigma'}(x)=-i\delta_{\sigma\sigma'}\sum_{\bf p}
\langle 0|T(b_\sigma({\bf p},t)b_\sigma^\dagger({\bf
p},0))|0\rangle e^{i\b p\cdot\b x}\equiv
\delta_{\sigma\sigma'}\sum_{\b p} G({\b p},t),\ee where we have
used \be \langle 0|T(b_\sigma({\bf p},t)b{_\sigma'}^\dagger({\bf
p}',0))|0\rangle=\delta_{\sigma\sigma'}\delta_{{\bf p}{\bf p}'}
\langle 0|T(b_\sigma({\bf p},t)b_\sigma^\dagger({\bf
p},0))|0\rangle.\ee Since \bea \langle 0|b_\sigma^\dagger({\bf
p})b_\sigma({\bf p})|0\rangle&=&\theta(p_F-p)=\theta(-\ell),\cr
\langle 0|b_\sigma({\bf p})b_\sigma^\dagger({\bf
p})|0\rangle&=&1-\theta(p_F-p)=\theta(p-p_F)=\theta(\ell),\eea we
get \be G({\bf p},t)=\left\{\matrix{-i\theta(\ell)e^{-i\ell v_F t}
& t>0\cr i\theta(-\ell)e^{-i\ell v_F t} & t<0.\cr}\right.\ee We
can also write \be G(x)=\int\frac{d^4p}{(2\pi)^4}e^{-ip\cdot x}
G(p),\ee with $G(p)$ defined in Eq. (\ref{fermipropagator}). It is
interesting to notice that the fermion density can be obtained
from the propagator. In fact, in the limit $\delta\to 0$ for
$\delta>0$ we have \be G_{\sigma\sigma'}({\b 0},-\delta)=-i\langle
0|T(\psi_\sigma({\b
0},-\delta)\psi^\dagger_{\sigma'}(0)|\rangle\Rightarrow i\langle
0|\psi^\dagger_{\sigma'}\psi_{\sigma}|\rangle\equiv i\rho_F.\ee
Therefore \be \rho_F=-i\lim_{\delta\to 0^+} G_{\sigma\sigma}(\b
0,-\delta)=-2i\int\frac{d^4 p}{(2\pi)^4} e^{ip_0\delta}\frac
1{(1+i\epsilon) p_0-\ell v_F}.\label{2.36}\ee The exponential is
convergent in the upper plane of $p_0$, where we pick up the pole
for $\ell<0$ at \be p_0=\ell v_F+i\epsilon.\ee Therefore
\be\rho_F=2\int\frac{d^3\b
p}{(2\pi)^3}\theta(-\ell)=2\int\frac{d^3\b
p}{(2\pi)^3}\theta(p_F-|\b p|)=\frac
{p_F^3}{3\pi^2}.\label{2.38}\ee

\subsection{One-loop corrections}

\begin{center}
\begin{figure}[ht] \epsfxsize=10truecm
\centerline{\epsffile{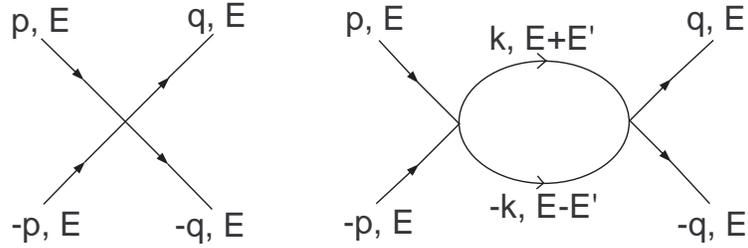}} \noindent \caption{\it The two
diagrams contributing to the one-loop four-fermi scattering
amplitude \label{3fig}}
\end{figure}
\end{center}

\begin{center}
\begin{figure}[ht] \epsfxsize=10truecm
\centerline{\epsffile{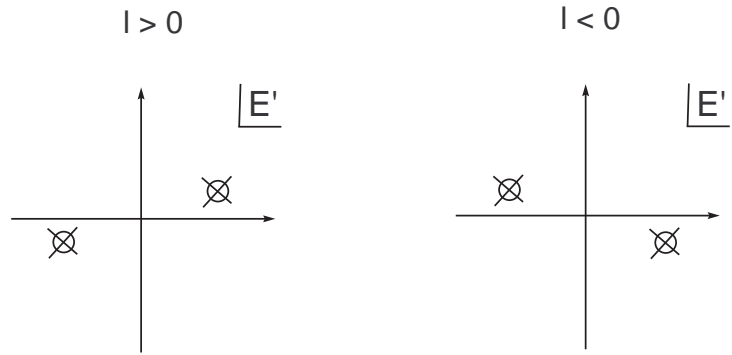}} \noindent \caption{\it The
position of the poles in the complex plane of $E'$ in the
one-loop amplitude, in the two cases $\ell\gtrless 0$
\label{poli}}
\end{figure}
\end{center}

\begin{center}
\begin{figure}[ht] \epsfxsize=10truecm
\centerline{\epsffile{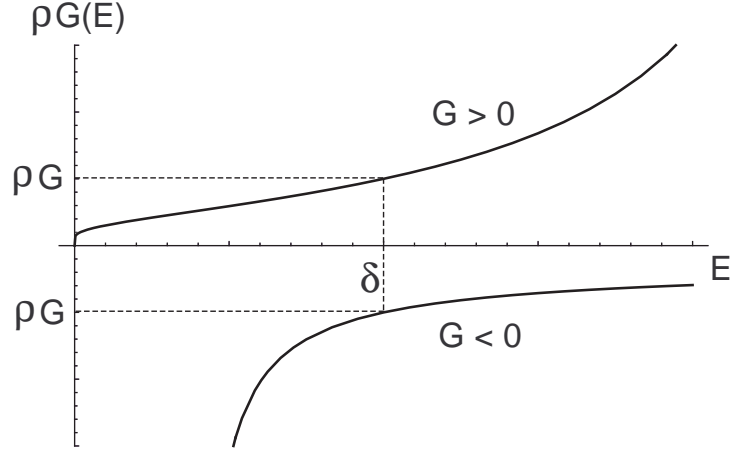}} \noindent \caption{\it The
behavior of $G(E)$  for $G>0$ and $G<0$. \label{4fig}}
\end{figure}
\end{center}

We  now  evaluate the one-loop corrections to the four-fermion
scattering. These are given in Fig. \ref{3fig}, and we get \be
G(E)=G-G^2\int\frac{dE'\,d^2{\bf k}\,dl}{(2\pi)^4}\frac
1{((E+E')(1+i\epsilon)-v_F({\bf k})
l)((E-E')(1+i\epsilon)-v_F({\bf k}) l)},\label{loopamplitude}\ee
where we have assumed the vertex $V$ as a constant $G$. The poles
of the integrand are shown in Fig. \ref{poli}
The integrand of Eq. (\ref{loopamplitude}) can be written as \be
\frac 1{2(E-\ell v_F)}\left[\frac 1{E'+E-(1-i\epsilon)\ell
v_F}-\frac 1{E'-E+(1-i\epsilon)\ell v_F}\right].\ee Therefore
closing the integration path in the upper plane we find \be
iG(E)=iG-G^2\int\frac{d^2\b k d\ell}{(2\pi)^4}\frac 1{2(E- \ell
v_F)}\left[(-2\pi i)\theta(\ell)+(2\pi i)\theta(-\ell)\right].\ee
By changing  $\ell\to -\ell$ in the second integral we find\be
iG(E)=iG+iG^2\int\frac{d^2\b k d\ell}{(2\pi)^4}\frac{\ell
v_F}{E^2-(\ell v_F)^2}\theta(\ell).\ee By putting an upper cutoff
$E_0$ on the integration over $\ell$ we get \be G(E)=G-\frac 1
2G^2\rho\log(\delta/E),\ee where $\delta$ is a cutoff on $v_F l$
and \be \rho=2\int\frac{d^2{\bf k}}{(2\pi)^3}\frac 1 {v_F({\bf
k})}\ee is the density of states at the Fermi surface for for the
two paired fermions. For a spherical surface \be
\rho=\frac{p_F^2}{\pi^2 v_F},\ee where the Fermi momentum is
defined by \be \epsilon(p_F)=\epsilon_F=\mu.\ee From the
renormalization group equation (or just at the same order of
approximation) we get easily \be G(E)\approx \frac
G{\dd{1+\frac{\rho G}2\log(\delta/E)}},\ee showing that for $E\to
0$ we have
\begin{itemize}
\item $G>0$ (repulsive interaction), $G(E)$ becomes weaker
(irrelevant interaction) \item $G<0$ (attractive interaction),
$G(E)$ becomes stronger (relevant interaction)
\end{itemize}
This is illustrated in Fig \ref{4fig}.
Therefore an attractive four-fermi interaction is unstable and one
expects a rearrangement of the vacuum. This leads to the formation
of Cooper pairs. In metals the physical origin of the four-fermi
interaction is the phonon interaction. If it happens that at some
intermediate scale $E_1$, with \be E_1\approx \left(\frac m
M\right)^{1/2}\delta,\ee with $m$ the electron mass and $M$ the
nucleus mass, the phonon interaction is stronger than the Coulomb
interaction, then we have the superconductivity, otherwise we have
a normal metal. In a superconductor we have a non-vanishing
expectation value for the difermion condensate \be
\langle\psi_\sigma({\bf p})\psi_{-\sigma}(-{\bf p})\rangle.\ee

\subsection{Renormalization group analysis}

RG analysis indicates the possible existence of instabilities at
the scale where the couplings become strong. A complete study for
QCD with 3-flavors has been done in
\cite{Evans:1998ek,Evans:1998nf}. One has to look at the
four-fermi coupling with bigger coefficient $C$ in the RG equation
\be\frac{dG(E)}{d\log E}=C G^2 \to
G(E)=\frac{G}{1-CG\log(E/E_0)}.\ee The scale of the instability is
set by the corresponding Landau pole.
\begin{center}
\begin{figure}[ht]
\epsfxsize=10truecm \centerline{\epsffile{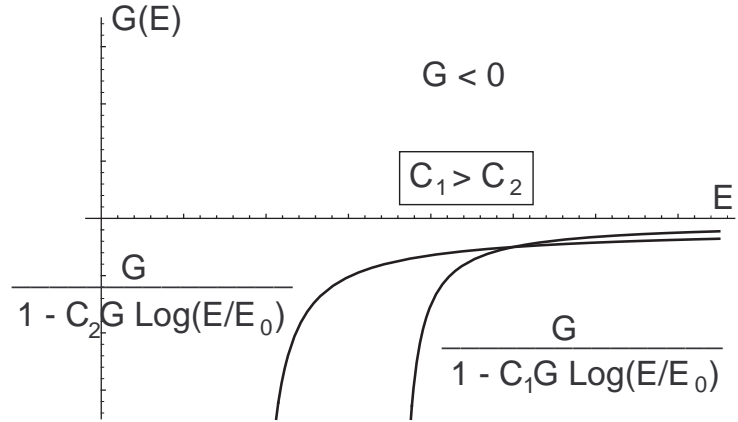}} \noindent
\caption{\it The figure shows that the instability is set in
correspondence with the bigger value of the coefficient of $G^2$
in the renormalization group equation.\label{fig_fermi}}
\end{figure}
\end{center}

In the case of 3-flavors $QCD$ one has 8 basic four-fermi
operators originating from one-gluon exchange \be
O^0_{LL}=(\bar\psi_L\gamma_0\psi_L)^2,~O^0_{LR}=
(\bar\psi_L\gamma_0\psi_L)(\bar\psi_R\gamma_0\psi_R),\ee \be
O^i_{LL}=(\bar\psi_L\gamma_i\psi_L)^2,~O^i_{LR}=
(\bar\psi_L\gamma_i\psi_L)(\bar\psi_R\gamma_i\psi_R),\ee in two
different color structures, symmetric and anti-symmetric \be
(\bar\psi^a\psi^b)(\bar\psi^c\psi^d)(\delta_{ab}\delta_{cd}\pm
\delta_{ad}\delta_{bc}).\ee The coupling with the biggest {$C$}
coefficient in the RG equations is given by the following operator
(using Fierz) \be
(\bar\psi_L\gamma_0\psi_L)^2-(\bar\psi_L\vec\gamma\psi_L)^2=
2(\psi_LC\psi_L)(\bar\psi_L C\bar\psi_L).\ee This shows that the
{dominant operator} corresponds to a scalar diquark channel. The
subdominant operators lead to vector diquark channels. A similar
analysis can be done for 2-flavors $QCD$. This is somewhat more
involved since there are new operators \be{\rm
det}_{flavor}(\bar\psi_R\psi_L),~~~ {\rm
det}_{flavor}(\bar\psi_R\vec\Sigma\psi_L).\ee The result is that
the dominant coupling is (after Fierz)\be{\rm
det}_{flavor}[(\bar\psi_R\psi_L)^2-(\bar\psi_R\vec\Sigma\psi_L)^2]=
2(\psi_L^{i\alpha}C\psi_L^{j\beta}\epsilon_{ij})\epsilon_{\alpha\beta
I}
(\psi_R^{k\gamma}C\psi_R^{l\delta}\epsilon_{kl})\epsilon_{\gamma\delta
I}.\ee The dominant operator corresponds to a flavor singlet and
to the antisymmetric color representation {$\bar 3$}.

\section{The gap equation}

In this Section we will study in detail the gap equation deriving
it within the BCS approach. We will show also how to get it from
the Nambu Gor'kov equations and  the functional approach. A
Section will be devoted to the determination of the critical
temperature.

\subsection{A toy model}

The physics of fermions at finite density and zero temperature
 can be treated in a systematic way by using  Landau's idea of
quasi-particles. An example is the Landau theory of Fermi liquids.
A  conductor is treated as a gas of almost free electrons. However
these electrons are dressed by the interactions. As we have seen,
according to Polchinski \cite{Polchinski:1992ed}, this procedure
just works because the interactions can be integrated away in the
usual sense of the effective theories. Of course, this is a
consequence of the special nature of the Fermi surface, which is
such that there are practically no relevant or marginal
interactions. In fact, all the interactions are irrelevant except
for the four-fermi couplings between pairs of opposite momentum.
Quantum corrections make  the attractive ones relevant, and   the
repulsive ones irrelevant. This explains the instability of the
Fermi surface of almost free fermions against any attractive
four-fermi interactions, but we would like to understand better
the physics underlying the formation of the condensates and how
the idea of quasi-particles comes about. To this purpose we will
make use of a toy model involving two Fermi oscillators
describing, for instance, spin up and spin down. Of course, in a
finite-dimensional system there is no spontaneous symmetry
breaking, but this model is useful just to illustrate  many points
which are common to the full treatment, but avoiding a lot of
technicalities. We assume our dynamical system to be described by
the following Hamiltonian containing a quartic coupling between
the oscillators \be H=\epsilon(a_1^\dagger a_1+a_2^\dagger
a_2)+Ga_1^\dagger a_2^\dagger a_1 a_2=\epsilon(a_1^\dagger
a_1+a_2^\dagger a_2)-Ga_1^\dagger a_2^\dagger a_2 a_1. \ee We will
study this model by using a variational principle. We start
introducing the following normalized trial wave-function
$|\Psi\rangle$\be |\Psi\rangle=\left(\cos\theta+\sin\theta\,
a_1^\dagger a_2^\dagger\right)|0\rangle. \ee The di-fermion
operator, $a_1a_2$, has the following expectation value \be
\Gamma\equiv\,\langle\Psi|\,a_1 a_2|\Psi\rangle =
-\sin\theta\cos\theta\label{gamma}. \ee Let us write the
hamiltonian $H$ as the sum of the following two pieces \be
H=H_0+H_{\rm res},\ee with \be H_0=\epsilon(a_1^\dagger
a_1+a_2^\dagger a_2)-G\Gamma(a_1 a_2- a_1^\dagger
a_2^\dagger)+G\Gamma^2,\ee and \be H_{\rm res}=G(a_1^\dagger
a_2^\dagger+\Gamma)\left(a_1 a_2-\Gamma\right),\ee Our
approximation will consist in neglecting $H_{\rm res}$. This is
equivalent to the mean field approach, where the operator $a_1a_2$
is approximated by its mean value $\Gamma$. Then we determine the
value of $\theta$ by looking for the minimum of the expectation
value of $H_0$ on the trial state \be
\langle\Psi|H_0|\Psi\rangle=2\epsilon\sin^2\theta- G\Gamma^2. \ee
We get \be 2\epsilon\sin 2\theta+2G\Gamma \cos 2\theta=0\,
\longrightarrow \, \tan 2\theta=-\frac{G\Gamma} \epsilon. \ee By
using the expression (\ref{gamma}) for $\Gamma$ we obtain the gap
equation \be \Gamma=-\frac 1 2 \sin 2\theta=\frac 1 2
\frac{G\Gamma}{\sqrt{\epsilon^2+G^2\Gamma^2}}, \ee or \be 1=\frac
1 2 \frac{G}{\sqrt{\epsilon^2+\Delta^2}},\ee where
$\Delta=G\Gamma$. Therefore the gap equation can be seen as the
equation determining the ground state of the system, since it
gives the value of the condensate. We can now introduce the idea
of quasi-particles in this particular context. The idea is to look
for  for a transformation on the Fermi oscillators such that $H_0$
acquires a canonical form (Bogoliubov transformation) and to
define a new vacuum annihilated by the new annihilation operators.
We write the  transformation in the form \be
A_1=a_1\cos\theta-a_2^\dagger\sin\theta,~~~A_2=a_1^\dagger
\sin\theta +a_2\cos\theta,\ee  Substituting this expression into
$H_0$ we find \bea H_0&=&2\epsilon\sin^2\theta+G\Gamma \sin
2\theta+G\Gamma^2+(\epsilon \cos 2\theta-G\Gamma \sin
2\theta)(A_1^\dagger A_1+A_2^\dagger A_2)\nn\\&+&(\epsilon\sin
2\theta+G\Gamma \cos 2\theta)(A_1^\dagger A_2^\dagger -A_1
A_2).\eea Requiring the cancellation of the bilinear terms in the
creation and annihilation operators we find \be \tan
2\theta=-\frac{G\Gamma}{\epsilon}=-\frac{\Delta}{\epsilon}.\label{condition}\ee
We can verify immediately that the new vacuum state annihilated by
$A_1$ and $A_2$ is \be |0\rangle_N=(\cos\theta+a_1^\dagger
a_2^\dagger
\sin\theta)|0\rangle,~~~~A_1|0\rangle_N=A_2|0\rangle_N=0.\ee The
constant term in $H_0$ which is equal to
$\langle\Psi|H_0|\Psi\rangle$ is given by\be
\langle\Psi|H_0|\Psi\rangle=
2\epsilon\sin^2\theta-G\Gamma^2=\left(\epsilon-
\frac{\epsilon^2}{\sqrt{\epsilon^2+\Delta^2}}\right)-\frac{\Delta^2}G.\ee
The first term in this expression arises from the kinetic energy
whereas the second one from the interaction. We define the weak
coupling limit by taking $\Delta\ll\epsilon$, then the first term
is given by \be \frac 1 2 \frac
{\Delta^2}{\epsilon}=\frac{\Delta^2}G,\ee where we have made use
of the gap equation at the lowest order in $\Delta$. We see that
in this limit the expectation value of $H_0$ vanishes, meaning
that the normal vacuum and the condensed one lead to the same
energy. However we will see that in the realistic case of a
3-dimensional Fermi sphere the condensed vacuum has a lower energy
by an amount which is proportional to the density of states at the
Fermi surface. In the present case there is no condensation since
there is no degeneracy of the ground state contrarily to the
realistic case. Nevertheless this case is interesting due to the
fact that the algebra is  simpler than  in the full discussion of
the next Section.

Therefore we get \be H_0=\left(\epsilon-
\frac{\epsilon^2}{\sqrt{\epsilon^2+\Delta^2}}\right)-\frac{\Delta^2}G+
\sqrt{\epsilon^2+\Delta^2}(A_1^\dagger A_1+A_2^\dagger A_2).\ee
The gap equation is recovered by evaluating $\Gamma$ \be \Gamma=
{_N\langle} 0|a_1a_2|0\rangle_N=-\frac 1 2\sin 2\theta\ee and
substituting inside Eq. (\ref{condition}). We find again
\be\Gamma=-\frac 1 2\sin 2\theta=\frac 1 2 \frac
{G\Gamma}{\sqrt{\epsilon^2+\Delta^2}},\ee or \be 1= \frac 1 2
\frac {G}{\sqrt{\epsilon^2+\Delta^2}}.\ee  From the expression of
$H_0$ we see that  the operators $A_i^\dagger$ create out of the
vacuum quasi-particles of energy \be
E=\sqrt{\epsilon^2+\Delta^2}.\ee The condensation gives rise to
the fermionic energy gap, $\Delta$. The Bogoliubov transformation
realizes the dressing of the original operators $a_i$ and
$a_i^\dagger$ to the quasi-particle ones $A_i$ and $A_i^\dagger$.
Of course, the interaction is still present, but part of it has
been absorbed in the dressing process getting a better starting
point for a perturbative expansion. As we have said this point of
view has been very fruitful in the Landau theory of conductors.

\subsection{The BCS theory}

We now proceed to the general case. We start with the following
hamiltonian containing a four-fermi interaction term of the type
giving rise to one-loop relevant contribution \be\tilde H= H-\mu
N=\sum_{\b k\sigma}\xi_{\b k} \cre k\sigma\dis k\sigma+\sum_{\b
k\b q}V_{\b k \b q}\creu k  \cred k\disd q \disu q,\ee where \be
\xi_{\b k}=\epsilon_{\b k}-E_F=\epsilon_{\b k}-\mu.\ee Here the
indices 1 and 2 refer to spin up and dow respectively. As before
we write \be \tilde H=H_0+H_{\rm res},\ee where \be H_0=\sum_{\b
k\sigma}\xi_{\b k} \cre k\sigma\dis k\sigma+ \sum_{\b k\b q} V_{\b
k\b q}\left[\creu k \cred k\Gamma_{\b q}+\disd q\disu q\Gamma_{\b
k}^*-\Gamma_{\b k}^*\Gamma_{\b q}\right]\ee and \be H_{\rm
res}=\sum_{\b k\b q} V_{\b k \b q}\Big(\creu k \cred k-\Gamma_{\b
k}^*\Big)\Big(\disd q\disu q-\Gamma_{\b q}\Big),\ee with
\be\Gamma_{\b k}=\langle \disd k\disu k\rangle\ee  the expectation
value of the difermion operator $\disd k\disu k$ in the BCS ground
state, which will be determined later. We will neglect $H_{\rm
res}$ as in the toy model. We then define \be \Delta_{\b
k}=-\sum_{\b q}V_{\b k\b q}\Gamma_{\b q},\label{4.28}\ee from
which \be H_0=\sum_{\b k\sigma}\xi_{\b k}\cre k\sigma\dis
k\sigma-\sum_{\b k}\left[\Delta_k \creu k\cred k+\Delta_{\b
k}^*\disd k \disu k-\Delta_{\b k}\Gamma_{\b k }^*\right].\ee
 Then, we look for new operators $A_i(\b k)$ \bea
\disu k&=&u_{\b k}^* A_1(\b k)+v_{\b k} A_2^\dagger(\b k),\nn\cr
\cred k&=&-v_{\b k}^* A_1(\b k)+u_{\b k} A_2^\dagger(\b k),\eea
with \be |u_{\b k}|^2+|v_{\b k}|^2=1,\ee in order to get canonical
anticommutation relations among the $A_i(\b k)$ oscillators.
Expressing $H_0$ through the new operators  we obtain \bea
H_0&=&\sum_{{\b k}\sigma}\xi_{\b k}\left[(|u_{\b k}|^2-|v_{\b
k}|^2)\acre k\sigma\adis k\sigma\right]\nn\\&+& 2\sum_{\b
k}\xi_{\b k}\left[|v_{\b k}|^2+u_{\b k} v_{\b k} \acreu k\acred
k-u_{\b k}^* v_{\b k}^* \adisu k\adisd k\right]\nn\\&+&\sum_{\b
k}\Big[\left(\Delta_{\b k}u_{\b k}v_{\b k}^*+\Delta_{\b k}^*u_{\b
k}^*v_{\b k}\right)\left(\acreu k\adisu k+\acred k\adisd
k-1\right)\nn\\ &+&\left(\Delta_{\b k}^* u_{\b k}^{*2}-\Delta_{\b
k} v_{\b k}^{*2}\right) \adisu k\adisd k- \left(\Delta_{\b k}
u_{\b k}^{2}-\Delta_{\b k}^* v_{\b k}^{2}\right)\acreu k\acred
k+\Delta_{\b k}\Gamma_{\b k}^*\Big].\eea In order to bring $H_0$
to a canonical form we must cancel the terms of the  type $\acreu
k\acred k$ and $\adisu k\adis2 k$. This can be done by choosing
\be 2\xi_{\b k} u_{\b k}v_{\b k}-(\Delta_{\b k} u_{\b
k}^2-\Delta_{\b k}^* v_{\b k}^2)=0.\ee Multiplying this Equation
by $\Delta_{\b k}^*/u_{\b k}^2$ we get \be \Delta_{\b
k}^{*2}\frac{v_{\b k}^2}{ u_{\b k}^2}+2\xi_{\b k}\Delta_{\b
k}^*\frac{v_{\b k}}{u_{\b k}}-|\Delta_{\b k}|^2=0,\ee or \be
\left(\Delta_{\b k}^*\frac{v_{\b k}}{u_{\b k}}+\xi_{\b
k}\right)^2=\xi_{\b k}^2+|\Delta_{\b k}|^2.\ee Introducing \be
E_{\b k}=\sqrt{\xi_{\b k}^2+|\Delta_{\b k}|^2},\ee which, as we
shall see, is the energy of the quasiparticles we find \be
\Delta_{\b k}^*\frac{v_{\b k}}{u_{\b k}}=E_{\b k}-\xi_{\b
k},\label{4.36}\ee or \be \left|\frac{v_{\b k}}{u_{\b
k}}\right|=\frac{E_{\b k}-\xi_{\b k}}{|\Delta_{\b k}|}.\ee This
equation together with \be |v_{\b k}|^2+|u_{\b k}|^2=1,\ee gives
\be|v_{\b k}|^2=\frac 12\left(1-\frac{\xi_{\b k}}{E_{\b
k}}\right),~~~~~ |u_{\b k}|^2=\frac 12\left(1+\frac{\xi_{\b
k}}{E_{\b k}}\right).\label{4.39}\ee Using these relations we can
easily evaluate the coefficients of the other terms in $H_0$. As
far as the bilinear term in the creation and annihilation
operators we get \bea &&\xi_{\b k}\left(|u_{\b k}|^2-|v_{\b
k}|^2\right)+\Delta_{\b k} u_{\b k} v_{\b k}^* + \Delta_{\b k}^*
u_{\b k}^* v_{\b k}\nn\\&&=\xi_{\b k}\left(|u_{\b k}|^2-|v_{\b
k}|^2\right)+2|u_{\b k}|^2\left(E_{\b k}-\xi_{\b k}\right)=E_{\b
k},\eea showing that $E_{\b k}$ is indeed the energy associated to
the new creation and annihilation operators. Therefore we get \be
H_0=\sum_{\b k\sigma} E_{\b k} A^\dagger_\sigma(\b k)A_\sigma(\b
k) + \langle H_0\rangle, \ee with \be\langle H_0\rangle=\sum_{\b
k}\left[2\,\xi_{\b k}|v_{\b k}|^2-\Delta_{\b k}^* u_{\b k}^* v_{\b
k}-\Delta_{\b k} u_{\b k} v_{\b k}^*+\Delta_{\b k} \Gamma_{\b
k}^*\right].\label{4.42}\ee We now need the BCS ground state. This
is obtained by asking for a state annihilated by the operators
$A_{\sigma}(\b k)$: \bea \adisu k&=&u_{\b k}\disu k-v_{\b k}\cred
k,\nn\\ \adisd k&=&v_{\b k}\creu k+u_{\b k}\disd k.\eea It is easy
to check that the required state is \be |0\rangle_{BCS}=\prod_{\b
k}\left(u_{\b k}+ v_{\b k} \creu k\cred k\right)|0\rangle.\ee Let
us check for $A_1(\b k)$ \bea &&A_1(\b
q)|0\rangle_{BCS}=\nn\\&=&\prod_{\b k\not=\b q}\left(u_{\b k}+
v_{\b k} \creu k\cred k\right)\left(u_{\b q}\disu q-v_{\b q}\cred
q\right)\left(u_{\b q}+ v_{\b q} \creu q\cred
q\right)|0\rangle=\nn\\&=&\prod_{\b k\not=\b q}\left(u_{\b k}+
v_{\b k} \creu k\cred k\right)\left(u_{\b q} v_{\b q }\cred
q-v_{\b q} u_{\b q} \cred q\right)|0\rangle=0.\eea We can now
evaluate  $\Gamma_{\b k}$. We have \bea\Gamma_{\b k}&=&\langle
\disd k\disu k\rangle=\left\langle\left(-v_{\b k}\acreu k+u_{\b
k}^*\adisd k\right)\left(u_{\b k}^* \adisu k+v_{\b k}\acred
k\right)\right\rangle\nn\\&=&u_{\b k}^* v_{\b
k}\left\langle\left(1-\acreu k\adisu k-\acred k\adisd
k\right)\right\rangle,\label{expectation}\eea from which \be
\Gamma_{\b k} = u_{\b k}^* v_{\b k}.\ee Therefore we can write Eq.
(\ref{4.42}) as \be\langle H_0\rangle=\sum_{\b k}\left[2\,\xi_{\b
k}|v_{\b k}|^2-\Delta_{\b k}^* u_{\b k}^* v_{\b k}\right].\ee By
Eq. (\ref{4.39}) we have  \be\langle H_0\rangle=\sum_{\b
k}\left[\xi_{\b k}-\frac{\xi_{\b k}^2}{E_{\b k}}-\Delta_{\b k}^*
u_{\b k}^* v_{\b k}\right].\ee Before proceeding we now derive the
gap equation. Starting from the complex conjugated of  Eq.
(\ref{4.36}) we can write \be \Delta_{\b k} \frac{u_{\b k}v_{\b
k}^*}{|u_{\b k}|^2}=E_{\b k}-\xi_{\b k},\ee and using (\ref{4.39})
we get \be u_{\b k}v_{\b k}^*=\frac 1 2\frac{\Delta_{\b
k}^*}{E_{\b k}}\ee and \be \Gamma_{\b k}=\frac 1 2\frac{\Delta_{\b
k}}{E_{\b k}}.\ee By the definition of $\Delta_{\b k}$ given in
Eq. (\ref{4.28}) we finally obtain the \b {gap equation} \be
\Delta_{\b k}=-\frac 1 2\sum_{\b q} V_{\b k\b q}\frac{\Delta_{\b
q}}{E_{\b q}}\label{3.53}.\ee We can now proceed to the evaluation
of the expectation value of $H_0$. Notice that if the interaction
matrix $V_{\b k \b q}$ is invertible we can write \be \langle
H_0\rangle=\sum_{\b k}\left[\xi_{\b k}-\frac{\xi_{\b k}^2}{E_{\b
k}}+\sum_{\b q} \Delta_{\b k} V^{-1}_{\b k \b q}\Delta_{\b
q}^*\right].\ee By choosing $V_{\b k \b q}$ as in the discussion
of the Cooper pairs: \be V_{\b k,\b k'}=\left\{\matrix{ -G &
|\xi_{\b k}|,~|\xi_{\b q}|<\delta\cr 0, & {\rm
otherwise}\cr}\right. \ee with $G>0$, we find \be \langle
H_0\rangle=\sum_{\b k}\left(\xi_{\b k}-\frac{\xi_{\b k}^2}{E_{\b
k}}\right)-\frac{\Delta^2}{G},\ee since the gap equation has now
solutions for $\Delta_{\b k}$ independent on the momentum. In a
more detailed way the sum can be written as \be\langle
H_0\rangle=\sum_{|\b k|>k_F}\left(\xi_{\b k}-\frac{\xi_{\b
k}^2}{E_{\b k}}\right)+\sum_{|\b k|<k_F}\left(-\xi_{\b
k}-\frac{\xi_{\b k}^2}{E_{\b k}}\right)-\frac{\Delta^2} G,\ee or
\be \langle H_0\rangle=2\sum_{|\b k|>k_F}\left(\xi_{\b
k}-\frac{\xi_{\b k}^2}{E_{\b k}}\right)-\frac{\Delta^2} G.\ee
Converting the sum in an integral we get \bea\langle
H_0\rangle&=&2\frac{p_F^2}{2\pi^2 v_F}\int_0^\delta
d\xi\left(\xi-\frac{\xi^2}{\sqrt{\xi^2+\Delta^2}}\right)-\frac{\Delta^2}
G\nn\\&=&
\rho\left[\delta^2-\delta\sqrt{\delta^2+\Delta^2}+\Delta^2
\log\frac{\delta+\sqrt{\delta^2+\Delta^2}}\Delta\right]-\frac{\Delta^2}
G\label{4.59}.\eea Let us now consider the gap equation \be
\Delta=\frac 1 2\frac{p_F^2}{2\pi^2 v_F}2G\int_0^\delta
d\xi\frac\Delta{\sqrt{\xi^2+\Delta^2}}=\frac 1 2\rho
G\Delta\log\frac{\delta+\sqrt{\delta^2+\Delta^2}}\Delta,\ee from
which \be 1=\frac 1 2\rho
G\log\frac{\delta+\sqrt{\delta^2+\Delta^2}}\Delta.\label{3.61}\ee
Using this equation in Eq. (\ref{4.59}) we find \be \langle
H_0\rangle=\frac\rho
2\left[\delta^2-\delta\sqrt{\delta^2+\Delta^2}+\frac{2\Delta^2}{\rho
G}\right]-\frac{\Delta^2} G.\ee The first term in this expression
arises from the kinetic energy whereas the second one from the
interaction. Simplifying the expression we find \be\langle
H_0\rangle=\frac\rho
2\left[\delta^2-\delta\sqrt{\delta^2+\Delta^2}\right].\ee By
taking the weak limit, that is $\rho G\ll 1$, or
$\Delta\ll\delta$, we obtain from the gap equation
\be\Delta=2\delta e^{-2/G\rho}\label{3.67}\ee and \be \langle
H_0\rangle=-\frac 1 4 \rho\Delta^2\label{3.65}.\ee All this
calculation can be easily repeated at $T\not=0$. In fact the only
point where the temperature comes in is in evaluating $\Gamma_{\b
k}$ which must be taken as a thermal average\be \langle{\cal
O}\rangle_T=\frac{Tr\left[e^{-H/T}{\cal
O}\right]}{Tr\left[e^{-H/T}\right]}.\ee The thermal average of a
Fermi oscillator of hamiltonian $H=E b^\dagger b$ is obtained
easily since\be Tr[e^{-E \,b^\dagger b/T}]=1+e^{-E/T}\ee and \be
Tr[b^\dagger b e^{-E b^\dagger b/T}]=e^{-E/T}.\ee Therefore \be
\langle b^\dagger b\rangle_T=f(E)=\frac
1{e^{E/T}+1}\label{fermidistr}.\ee It follows from Eq.
(\ref{expectation}) \be\Gamma_{\b k}(T)=\langle \disd k\disu
k\rangle_T=u_{\b k}^* v_{\b k}\left\langle\left(1-\acreu k\adisu
k-\acred k\adisd k\right)\right\rangle_T=u_{\b k}^* v_{\b
k}(1-2f(E_{\b k})).\ee Therefore the gap equation is given by \be
\Delta_{\b k}=-\sum_{\b q}V_{\b k \b q}u_{\b q}^* v_{\b q}
(1-2f(E_{\b q}))=-\sum_{\b q}V_{\b k \b q}\frac{\Delta_{\b q}}{2
E_{\b q}}\tanh\frac{E_{\b q}}{2 T},\label{3.71}\ee and in the BCS
approximation \be 1=\frac 1 4 \rho G\int_{-\delta}^{+\delta}\frac
{d\xi_{\b p}}{E_{\b p}}\tanh\frac{E_{\b p}}{2T}, ~~~E_{\b
p}=\sqrt{\xi_{\b p}^2+\Delta^2}.\ee

\subsection{The functional approach to the gap equation}
\label{IIIC}

We will now show how to derive the gap equation by using the
functional approach to field theory. We start assuming the
following action \be S[\psi,\psi^\dagger]=\int d^4
x\left[\psi^\dagger(i\de_t-\epsilon(|\bm{\nabla}|)+\mu)\psi+\frac
G 2\left(\psi^\dagger(x)\psi(x)\right)
\left(\psi^\dagger(x)\psi(x)\right)\right].\label{3.73}\ee We can
transform the interaction term in a more convenient way
(Fierzing): \be \psi_a^\dagger\psi_a\psi_b^\dagger\psi_b=-\
\psi_a^\dagger\psi_b^\dagger\psi_a\psi_b=-\frac 1 4
\epsilon_{ab}\epsilon_{ab}\psi^\dagger C\psi^*\psi^TC\psi=-\frac 1
2\psi^\dagger C\psi^*\psi^TC\psi,\ee with \be C=i\sigma_2\ee the
charge conjugation matrix. We obtain \be
S[\psi,\psi^\dagger]\equiv S_0+S_I=\int d^4
x\left[\psi^\dagger(i\de_t-\epsilon(|\bm{\nabla}|)+\mu)\psi- \frac
G 4\left(\psi^\dagger(x)C\psi^*(x)\right)
\left(\psi^T(x)C\psi(x)\right)\right],\label{3.76}\ee . The
quantum theory is defined in terms of the functional integral\be
Z=\int{\cal D}(\psi,\psi^\dagger)
e^{\dd{iS[\psi,\psi^\dagger]}}.\label{3.77}\ee The four-fermi
interaction can be eliminated by inserting inside the functional
integral the following identity \be {\rm const}=\int{\cal
D}(\Delta,\Delta^*)e^{\dd{-\frac i G\int d^4 x\left[\Delta-\frac G
2(\psi^T C\psi)\right]\left[\Delta^*+\frac G 2(\psi^\dagger
C\psi^*)\right]}}.\ee Normalizing at the free case ($G=0$) we get
\be \frac Z{Z_0}=\frac 1 {Z_0}\int{\cal D}(\psi,\psi^\dagger){\cal
D}(\Delta,\Delta^*)e^{\dd{iS_0[\psi,\psi^\dagger]+i\int
d^4x\left[-\frac{|\Delta|^2}G-\frac 1 2\Delta(\psi^\dagger
C\psi^*)+\frac 1 2\Delta^*(\psi^T C\psi)\right]}}.\label{funct}\ee
It is convenient to introduce the Nambu-Gorkov basis \be
\chi=\frac 1{\sqrt{2}}\left(\matrix{\psi\cr C\psi^*}\right),\ee in
terms of which the exponent appearing in Eq. (\ref{funct}) can be
written as \be S_0+\cdots=\int d^4 x \left(\chi^\dagger S^{-1}\chi
-\frac{|\Delta|^2}G\right),\ee where in momentum space \be
S^{-1}(p)=\left[\matrix{p_0-\xi_{\b p} &-\Delta\cr-\Delta^* &
p_0+\xi_{\b p}}\right].\ee We can now perform the functional
integral over the Fermi fields. Clearly it is convenient to
perform this integration over the Nambu-Gorkov field, but this
corresponds to double the degrees of freedom, since inside $\chi$
we count already once the fields $\psi^*$. To cover this aspect we
can use the "replica trick" by integrating also over
$\chi^\dagger$ as an independent field and taking the square root
of the result.We obtain \be\frac Z{Z_0}=\frac 1{Z_0}\left[{\rm
det}S^{-1}\right]^{1/2}e^{\dd{-i\int d^4 x \frac{|\Delta|^2}
G}}\equiv e^{i\dd{S_{\rm eff}}},\ee where \be S_{\rm
eff}(\Delta,\Delta^*)=-\frac i 2 Tr[\log(S_0 S^{-1})]-\int d^4 x
\frac{|\Delta|^2} G,\ee with $S_0$  the free propagator
($\Delta=0$). The saddle point equation for $\Delta^*$ gives \be
\frac{\delta S_{\rm eff}}{\delta\Delta^*}=-\frac\Delta G-\frac i 2
Tr\left[S\frac{\delta S^{-1}}{\delta\Delta^*}\right]=-\frac\Delta
G+\frac i 2Tr\left(\left[\matrix{\Delta & 0\cr p_0+\xi_{\b p} &
0}\right]\frac 1 {p_0^2-\xi_{\b p}^2-|\Delta|^2}\right),\ee where
we have used \be S= \frac 1 {p_0^2-\xi_{\b
p}^2-|\Delta|^2}\left[\matrix{p_0+\xi_{\b p} & \Delta\cr \Delta^*
& p_0-\xi_{\b p}}\right].\ee Therefore we get  the gap equation
(the trace gives a factor 2 from the spin)\be \Delta= i G\int
\frac {d^4p}{(2\pi)^4}\frac\Delta{p_0^2-\xi_{\b
p}^2-|\Delta|^2},\ee and performing the integration over $p_0$ we
obtain \be\Delta=\frac  G
2\int\frac{d^3p}{(2\pi)^3}\frac\Delta{\sqrt{\xi_{\b
p}^2+|\Delta|^2}},\ee in agreement with Eq. (\ref{3.53}). By
considering the case $T\not=0$ we have only to change the
integration over $p_0$ to a sum over the Matsubara frequencies \be
\omega_n=(2 n+1)\pi T,\ee obtaining \be
\Delta=GT\sum_{n=-\infty}^{+\infty}\int\frac{d^3p}{(2\pi)^3}
\frac\Delta{\omega_n^2+\xi_{\b p}^2 +|\Delta|^2}.\ee The sum can
be easily done with the result \be \sum_{n=-\infty}^{+\infty}\frac
1{\omega_n^2+\xi_{\b p}^2 +|\Delta|^2}=\frac 1 {2 E_{\b p}
T}\left(1-2f(E_{\b p}\right),\ee where $f(E)$ is the Fermi
distribution defined in Eq. (\ref{fermidistr}). From \be
1-2f(E)=\tanh\frac E{2T},\ee we get the gap equation for $T\not
=0$ \be \Delta=\frac G 2\int
\frac{d^3p}{(2\pi)^3}\frac{\Delta}{E_{\b p}}\tanh(E_{\b p}/2
T),\ee which is the same as Eq. (\ref{3.71}).

\begin{center}
\begin{figure}[ht]
\epsfxsize=12truecm \centerline{\epsffile{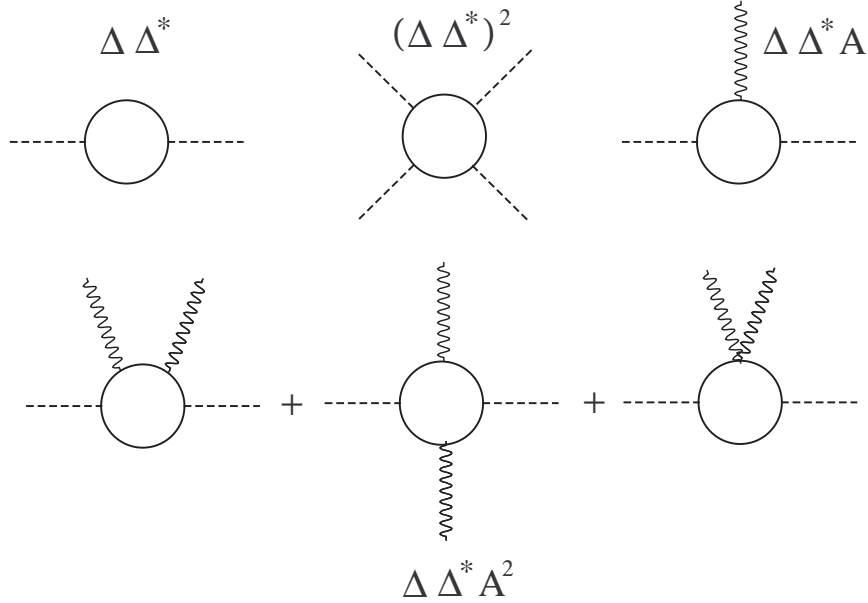}}
\noindent \caption{\it The diagrams contributing to the
Ginzburg-Landau expansion. The dashed lines represent the fields
$\Delta$ and $\Delta^*$, the solid lines the Fermi fields and the
wavy lines the photon field.\label{ginzburglandau}}
\end{figure}
\end{center}

If we consider the functional $Z$ as given by Eq. (\ref{funct}) as
a functional integral over $\psi$, $\psi^\dagger$, $\Delta$ and
$\Delta^*$, by its saddle point evaluation we see that the
classical value of $\Delta$ is given by \be \Delta=\frac G
2\langle\psi^TC\psi\rangle\label{3.95}.\ee Also, if we introduce
the em interaction in the action (\ref{3.76}) we see that $Z$, as
given by Eq. (\ref{3.77}) is gauge invariant under\be \psi\to
e^{i\alpha(x)}\psi\ee Therefore the way in which the em field
appear in $S_{\rm eff}(\Delta,\Delta^*)$ must be  such  to make it
gauge invariant. On the other side we see from Eq. (\ref{3.95})
that $\Delta$ must transform as \be \Delta\to
e^{2i\alpha(x)}\Delta,\ee meaning that $\Delta$ has charge $-2e$
and that the effective action for $\Delta$ has to contain
space-time derivatives in the form \be D^\mu=\de^\mu+2ie A^\mu.\ee
This result was achieved for the first time by
\cite{gorkov:1959hy} who derived the Ginzburg-Landau expansion of
the free energy from the microscopic theory. This calculation can
be easily repeated by inserting the em interaction and matching
the general form of the effective action against the microscopic
calculation. We will see an example of this kind of calculations
later. In practice one starts from the form (\ref{funct}) for $Z$
and, after established the Feynman rules, one evaluate the
diagrams of Fig. \ref{ginzburglandau} which give the coefficients
of the terms in $|\Delta|^2$, $|\Delta|^4$, $|\Delta|^2 A$ and
$|\Delta|^2 A^2$ in the effective lagrangian.
An explicit evaluation of these diagrams in the static case
${\dot\b A}=0$ can be found, for instance, in the book of
\cite{Sakita}. One gets an expression of the type
 \be H=\int d^3{\b r}\left( -c\,\frac
1{4m}\Delta^*(\b r)|(\bm{\nabla}+2ie\b A)|^2\Delta(\b
r)+a|\Delta(\b r)|^2+\frac 1 2 b|\Delta(\b r)|^4\right).\ee By
defining $\psi=\sqrt{c}\Delta$ we obtain \be H=\int d^3{\b
r}\left(- \frac 1{4m}\psi^*(\b r)|(\bm{\nabla}+2ie\b A)\psi(\b
r)|^2\psi(\b r)+\alpha|\psi(\b r)|^2+\frac 1 2 \beta|\psi(\b
r)|^4\right),\ee with \be \alpha =\frac a c,~~~\beta=\frac
b{c^2}.\ee This expression is the same as the original proposal
made by Ginzburg and Landau (see Eq. (\ref{1.35})) with \be
e^*=2e, ~~~~m^*=2m.\ee However, notice that contrarily to $e^*$
the value of $m^*$ depends on the normalization chosen for $\psi$.
Later we will evaluate the coefficients $a$ and $b$ directly from
the gap equation.

\subsection{The Nambu-Gor'kov equations}
\label{IIID}

We will present now a different approach, known as Nambu-Gor'kov
equations \cite{gorkov:1959hy,Nambu:1960cs} which is completely
equivalent to the previous ones and strictly related to the
effective action approach of the previous Section. We start again
from the action (\ref{3.73}) in three-momentum space
 \bea S&=&S_0+S_{BCS}\ , \\
S_0&=& \int dt\, \frac{d{\bf p}}{(2\pi)^3}\ \psi^\dagger({\bf
p})\left(i\partial_t-E({\bf p})+\mu \right)\psi({\bf p})\ ,
\label{a0}\\&&\cr S_{BCS}&=& \frac{G}2\int dt
\prod_{k=1}^4\frac{d{\bf p}_k}{(2\pi)^3}\left( \psi^\dagger({\bf
p}_1)\psi({\bf p}_4)\right)\,\left(\psi^\dagger({\bf
p}_2)\psi({\bf p}_3)\right)\nn\\&\times&(2\pi)^3 \, \delta({\bf
p}_1+{\bf p}_2-{\bf p}_3-{\bf p}_4)\,.\label{abcs}\eea Here and
below, unless explicitly stated, $\psi({\bf p})$  denotes the 3D
Fourier transform of the Pauli spinor $\psi({\bf r},t)$, i.e.
$\psi({\bf p})\equiv\psi_\sigma({\bf p},t)$. For non relativistic
particles the functional dependence of the energy is $\dd E({\bf
p})={\bf p}^{\,2}/2m$, but we prefer to leave it in the more
general form (\ref{a0}).

The BCS interaction (\ref{abcs}) can be written as follows \be
S_{BCS}=S_{cond}+S_{int}\,,\ee with \bea S_{cond}&=& -\frac{G
}4\int dt\,  \prod_{k=1}^4\frac{d{\bf p}_k}{(2\pi)^3}
\Big[\,\tilde\Xi ({\bf p}_3,\,{\bf p}_4)\psi^\dagger({\bf p}_1)
C\psi^\dagger({\bf p}_2) \cr&-& \tilde\Xi^*({\bf p}_1,\,{\bf
p}_2) \psi({\bf p}_3) C\psi({\bf p}_4)\Big](2\pi)^3 \,
\delta({\bf p}_1+{\bf p}_2-{\bf p}_3-{\bf p}_4) \ ,\cr
 S_{int}&=&-\frac{G}4
\int dt\, \prod_{k=1}^4\frac{d{\bf p}_k}{(2\pi)^3}
 \left[\psi^\dagger({\bf p}_1)
C\psi^\dagger({\bf p}_2)+\tilde\Xi^* ({\bf p}_1,\,{\bf
p}_2)\right]\times\cr&\times&\left[ \psi({\bf p}_3) C\psi({\bf
p}_4)-\tilde\Xi ({\bf p}_3,\,{\bf p}_4)\right](2\pi)^3 \,
\delta({\bf p}_1+{\bf p}_2-{\bf p}_3-{\bf p}_4)\,,
\label{interaction}\eea where $C=i\sigma_2$ and\be \tilde\Xi({\bf
p},\,{\bf p}^{\prime})=<\psi({\bf p}) C\psi({\bf p}^{\prime})>\,
.\ee In the mean field approximation the interaction term can be
neglected while the gap term $S_{cond}$ is added to $S_0$. Note
that the spin 0 condensate $\tilde\Xi({\bf p},\,{\bf
p}^{\prime})$ is simply related to the condensate wave function
  \be \Xi({\bf r})= <\psi({\bf r},t) C\psi({\bf r},t)>\label{deltax}
\ee  by the formula \be \Xi({\bf r})=\int\frac{d{\bf
p}}{(2\pi)^3}\frac{d{\bf p}^{\prime}}{(2\pi)^3}\,e^{-i({\bf
p}+{\bf p}^{\prime})\cdot{\bf r}}\
 \tilde\Xi({\bf p},\,{\bf p}^{\prime})\ .\ee
  In general the condensate wavefunction  can depend on ${\bf r}$;
only for homogeneous
materials it does not depend on the space coordinates; therefore
in this case $ \tilde\Xi({\bf p},\,{\bf p}^{\prime})$ is
proportional to $\delta({\bf p}+{\bf p}^\prime)$.

In order to write down the Nambu-Gor'kov (NG) equations we define
 the NG spinor \be \chi ({\bf p})=\frac{1}{\sqrt 2}
 \left(\matrix{\psi({\bf p}) \cr \psi^c(-{\bf p})}\right)
 \,,\label{ngspinor}\ee where we have introduced the
charge-conjugate field
 \be \psi^c=C\psi^*\ .\ee
 We also define
\be\Delta({\bf p},-{\bf p}^{\prime})=\,\frac{G}2 \int\,\frac{d{\bf
p}^{\prime\prime}}{(2\pi)^6}\tilde\Xi({\bf p}^{\prime\prime}, {\bf
p}+{\bf p}^{\prime} -{\bf p}^{\prime\prime})\ .\label{3.113}\ee
Therefore the free action can be  written as follows:
 \be S_0=\int dt\,
\frac{d{\bf p}}{(2\pi)^3}
 \frac{d{\bf p^\prime}}{(2\pi)^3}\,
\chi^\dagger({\bf p})\, S^{-1}({\bf p},\,{\bf p^\prime} )
\chi({\bf p^\prime})\\ ,\ee
 with
\bea S^{-1}({\bf p},\,{\bf p}^{\prime})=(2\pi)^3
\left(\matrix{(i\partial_t-\xi_{{\bf p}})
 \delta({\bf p}-{\bf p}^{\prime})& -
 \Delta({\bf p},{\bf p^\prime})\cr
 -\Delta^*({\bf p},{\bf p^\prime})&
(i\partial_t+\xi_{{\bf p}})\delta({\bf p}-{\bf
p}^{\prime})}\right)\,.\label{smenouno}\eea
 Here
 \be \xi_{{\bf p}}=E({\bf p})-\mu\approx {\bf v}_F
 \cdot({\bf p}-{\bf p}_F)\,,\label{vf}\ee
 where
 \be {\bf v}_F=\frac{\partial E({\bf p})}{\partial {\bf p}}
 \Big|_{{\bf p}\, =\,{\bf p}_F}\ee
 is the Fermi velocity. We have used the fact that we are considering
 only degrees of freedom near the Fermi surface, i.e. \be
 p_F-\delta<p <p_F+\delta\ ,\ee
 where
$\delta$ is the ultraviolet cutoff, of the order of the Debye
frequency. In particular in the non relativistic case
 \bea
\xi_{{\bf p}}=\frac{{{\bf
p}\,}^2}{2m}-\frac{p_F^{\,2}}{2m},~~~{\bf v}_F=\frac{ {\bf
p}_F}{m}\,. \label{diciannove}\eea $S^{-1}$ in (\ref{smenouno}) is
the 3D Fourier transform of the inverse propagator. We can make
explicit the energy dependence  by Fourier transforming the time
variable as well. In this way we get for the inverse propagator,
written as an operator: \be S^{-1}= \left(\matrix{
{\mathbf{(G_0^+)^{-1}}} & -{\mathbf\Delta}\cr -{\mathbf\Delta^*} &
-{\mathbf{(G_0^-)^{-1}}}}\right)\,, \ee and \bea [{\bf
G_0^+]^{-1}}&=& E-\xi_{{\bf P}} +\,i\,\epsilon\, {\rm sign\,}
E\,,\cr&&\cr {\bf [G_0^-]^{-1}}&=&-E-\xi_{{\bf P}}
-\,i\,\epsilon\, {\rm sign\,} E \,,\eea with $\epsilon=0^+$ and
${\bf P}$ the momentum operator. The $i\epsilon$ prescription is
the same discussed in Section \ref{IIB}. As for the NG propagator
$S$, one gets \be S=\left(\matrix{ {\bf G}&- {\bf \tilde F} \cr-
{\bf F}&{\bf \tilde G }} \right)\ . \ee $S$ has both spin,
$\sigma,\,\sigma^\prime$, and $a,b$ NG indices, i.e.
$S^{ab}_{\sigma\sigma^\prime}$\footnote{We note that the presence
of the factor $1/{\sqrt 2}$ in (\ref{ngspinor}) implies an extra
factor of 2 in the
 propagator:
 $
S(x,x^\prime)=2\,<T\left\{\chi(x)\chi^\dagger(x^\prime)\right\}>$,
as it can be seen considering e.g.  the matrix element $S^{11}$:
$<T\left\{\psi(x)\psi^\dagger(x^\prime)\right\}>=
\left(i\partial_t-\xi_{-i\vec \nabla}-\delta\mu\sigma_3
\right)^{-1}\delta(x-x^\prime),$ with  $(x\equiv (t,{\bf r}))$.}.
The NG equations in compact form are
 \be S^{-1}S=1\ ,\ee or, explicitly,
 \bea [{\bf G_0^+}]^{-1}{\bf G}+{\bf \Delta F}&=&{\bf 1}\ ,\cr
  -[{\bf G_0^-}]^{-1} {\bf F}
 +\bf{\Delta^* G}&=&{\bf 0}\label{NG}\ .\eea
 Note that we will use
 \be
 <{\bf r}\,|{\bf \Delta}|{\bf r}^{\,\prime}>=
\frac{G}2\,\Xi({\bf r})\,\delta({\bf r}-{\bf
r}^{\,\prime})=\Delta({\bf r})\,\delta({\bf r}-{\bf
r}^{\,\prime})\ ,\label{dra}\ee or \be <{\bf p}\,|{\bf
\Delta}|{\bf p}^{\,\prime}>= \Delta
 ({\bf p},{\bf p}^{\,\prime})\ee
 depending on our choice of the coordinate or momenta representation.
 The
formal solution of the system (\ref{NG}) is \bea {\bf F}&=&{\bf
G_0^-\Delta^*G}\ ,\cr
 {\bf G}&=&{\bf G_0^+}-{\bf G_0^+\Delta F}\ ,\eea so that ${\bf F}$
satisfies the equation \be {\bf
F=G_0^-\Delta^*\left(G_0^+-G_0^+\Delta F\right)}\ \label{18}\ee
and is therefore given by \be {\bf F=\frac{1}{ \Delta^*
[G_0^+]^{-1} [\Delta^{*}]^{-1} [G_0^-]^{-1} \,+\, \Delta^*
\Delta}\,\Delta^*} \label{18bis}\ .\ee

In the configuration  space the NG Eqs. (\ref{NG}) are as follows
\bea \left(E-E(-i\bm{\nabla})+\mu \right)G({\bf r},{\bf
r}^{\,\prime},E)+\,\Delta({\bf r}) F({\bf r},{\bf
r}^{\,\prime},E) &=&\delta({\bf r}-{\bf r}^{\,\prime})\ ,\cr
\left(-E-E(-i\bm{\nabla})+\mu  \right)F({\bf r},{\bf
r}^{\,\prime},E)- \Delta^*({\bf r}) G ({\bf r},{\bf
r}^{\,\prime},E)&=&0\,.\eea

The gap equation at $T=0$ is the following consistency condition
\be \Delta^*({\bf r})=-i\ \frac{G}2\int\frac{dE}{2\pi}\, {\text
Tr} F({\bf r},{\bf r},E)\,,\label{consistency}\ee where $F$ is
given by (\ref{18bis}). To derive the gap equation  we observe
that \bea \Delta^*({\bf r})&=&\,\frac{G}2\, \Xi^*({\bf
r})=\,\frac{G}2\,\int\frac{d{\bf p}_1}{(2\pi)^3} \frac{d{\bf
p}_2}{(2\pi)^3}\ e^{i({\bf p}_1+{\bf p}_2)\cdot{\bf r}}\
\tilde\Xi^*({\bf p}_1,\,{\bf p}_2)\cr &=&-\,\frac{G
}2\int\frac{dE}{2\pi}\frac{d{\bf p}_1}{(2\pi)^3}\frac{d{\bf
p}_2}{(2\pi)^3}\ e^{i({\bf p}_1+{\bf p}_2)\cdot{\bf r}}\
<\psi^\dagger({\bf p}_1,E)\psi^c({\bf p}_2,E)>\cr
&=&+\,i\,\frac{G}2\sum_\sigma\int\frac{dE}{2\pi}\frac{d{\bf
p}_1}{(2\pi)^3}\frac{d{\bf p}_2}{(2\pi)^3}e^{i({\bf p}_1-{\bf
p}_2)\cdot {\bf r}} S^{21}_{\sigma\sigma} ({\bf p}_2,{\bf
p}_1)\cr &=&+\,i\,\frac{G
}2\,\sum_\sigma\int\frac{dE}{2\pi}S^{21}_{\sigma\sigma}({\bf
r},{\bf r})\ , \eea which gives (\ref{consistency}).

 At finite temperature, introducing the Matsubara frequencies
 $\omega_n=(2n+1)\pi T$, the gap equation reads \be \Delta^*({\bf
 r})\,=\,\frac{G}2\, T\,\sum_{n=-\infty}^{+\infty} {\text
 Tr}F({\bf r},{\bf r},E)\Big|_{E=i\omega_n}\label{gap2}\ .\ee

It is useful to specialize these relations to the case of
homogeneous materials. In this case
we have \bea \Xi({\bf r})&=&{\rm const.}\equiv\frac{2\Delta}{G}\ ,\label{xir}\\
\tilde\Xi({\bf p}_1,\,{\bf
p}_2)&=&\frac{2\Delta}{G}\frac{\pi^2}{p_F^2\delta}\,(2\pi)^3\delta({\bf
p}_1+{\bf p}_2)\ \label{xip} .\eea
 Therefore one gets \be
\Delta({\bf p}_1,{\bf p}_2)= \Delta\, \delta({\bf p}_1-{\bf p}_2)
\label{deltaP}\ee and from (\ref{dra}) and (\ref{xir}) \be
\Delta({\bf r})=\Delta^*({\bf r})=\Delta\ .\label{homogeneous}\ee
Therefore $F({\bf r},{\bf r},E)$ is independent of ${\bf r}$ and,
from Eq. (\ref{18bis}), one gets \be TrF({\bf r},{\bf
r},E)=-2\,\Delta\int\frac{d^3p}{(2\pi)^3}\frac{1} {E^2-\xi_{{\bf
p}}^2-\Delta^2}\label{trf} \ee which gives the gap equation at
$T= 0$:
 \be \Delta=i\, G\Delta\int\frac{dE}{2\pi}\frac{d^3p}
 {(2\pi)^3}\frac{1}{E^2-\xi_{{\bf p}}^2-\Delta^2}\,,
\label{gapt0}\ee and at $T\neq 0$: \be \Delta\,=\, \,G
T\,\sum_{n=-\infty}^{+\infty}\int\frac{d^3p}
{(2\pi)^3}\frac{\Delta}{\omega_n^2+ \epsilon({\bf
p},\Delta)^2}\,,\label{gaptne0}\ee where \be \epsilon({\bf
p},\Delta)=\sqrt{\Delta^2+\xi^2_{{\bf p}}}\, . \ee is the same
quantity that we had previously defined as $E_{\b p}$. We now use
the identity \be \frac 1 2\left[1-n_u-n_d \right] =\epsilon({\bf
p},\Delta) T
\sum_{n=-\infty}^{+\infty}\frac{1}{\omega_n^2+\epsilon^2({\bf
p},\Delta)}\,, \label{28}\ee where \be n_u({\bf p})= n_d({\bf
p}=\frac{1}{\dd e^{\epsilon/T}+1} \, .\ee The gap equation can be
therefore written as \be\Delta=\frac{G\,\Delta}2\,\int
\frac{d^3p}{(2\pi)^3}\, \frac{1}{\epsilon({\bf
p},\Delta)}\,\left(1-n_u({\bf p})-n_d({\bf
p})\right)\,.\label{gap1} \ee In the Landau theory of the Fermi
liquid $n_u,\,n_d$ are interpreted as
 the equilibrium distributions for the quasiparticles of type $u,d$.
It can be noted  that the last two terms act as blocking factors,
reducing the phase space, and producing eventually $\Delta\to 0$
when  $T$ reaches a critical value $T_c$ (see below).

\subsection{The critical temperature
\label{IIIE}}

We are now in the position to evaluate the critical temperature.
This can be done by deriving the Ginzburg-Landau expansion, since
we are interested to the case of $\Delta\to 0$. The  free energy
(or rather in this case the grand potential),  as measured from
the normal state, near a second order phase transition is given by
\be\Omega=\frac 1 2 \alpha\Delta^2+\frac 1
4\beta\Delta^4\,.\label{eq:potential1}\ee Minimization gives the
gap equation \be \alpha\Delta+\beta\Delta^3=0\,.\ee

Expanding the gap equation (\ref{gaptne0}) up to the third order
in the gap, $\Delta$, we can obtain the coefficients $\alpha$ and
$\beta$ up to a normalization constant. One gets \be
\Delta=2\,G\,\rho\,T \,Re\,\sum_{n=0}^{\infty}\int_0^\delta
d\xi\left[\frac{\Delta}{(\omega_n^2+\xi^2)}-
\frac{\Delta^3}{(\omega_n^2+\xi^2)^2}+ +\cdots\,\right]\,,
\label{eq:gap_expan}\ee with\be\omega_n=(2n+1)\pi
T\,.\label{eq:70}\ee The grand potential can be obtained, up to a
normalization factor, integrating in $\Delta$ the gap equation.
The normalization can be obtained by the simple BCS case,
considering the grand potential as obtained, in the weak coupling
limit, from Eqs. (\ref{3.65}) \be \Omega=-\frac{\rho}
4\Delta^2\,.\label{eq:70b}\ee The same result can be obtained
multiplying the gap equation (\ref{3.61}) in the weak coupling
limit \be 1-\frac{G\rho}
2\log\frac{2\delta}\Delta=0\label{3.150}\ee by $\Delta$ and
integrating over $\Delta$ starting from $\Delta=0$, that is the
normal state. We find \be \frac 1 2\Delta^2-\frac {G\rho}8
\Delta^2-\frac{G\rho}4\Delta^2\log\frac{2\delta}\Delta=-\frac
{G\rho}8 \Delta^2+\frac 1 2\Delta^2\left(1-\frac
{G\rho}2\log\frac{2\delta}\Delta\right).\ee Using again the gap
equation to cancel the second term, we see that the grand
potential is recovered if we multiply the result of the
integration by $2/G$. Therefore the coefficients $\alpha$ and
$\beta$ appearing the grand potential are obtained by multiplying
by $2/G$ the coefficients in the expansion of the gap equation. We
get
  \bea \alpha&=&\frac 2 G\left(1-2\, G\,\rho\,T\,
Re\sum_{n=0}^\infty\int_0^\delta\frac{d\xi}{(\omega_n^2+\xi^2)}\right)\,,
\label{eq:alpha}\\
\beta&=&4\rho\,T\,Re\sum_{n=0}^\infty
\int_0^\infty\frac{d\xi}{(\omega_n^2+\xi^2)^2}\,,\label{eq:beta}
\eea In the coefficient $\beta$  we have extended the integration
in $\xi$ up to infinity since both the sum and the integral are
convergent. To evaluate $\alpha$ is less trivial. One can proceed
in two different ways.  One can sum over the Matsubara frequencies
and then integrate over $\xi$ or one can perform the operations in
the inverse order. Let us begin with the former method. We get \be
\alpha=\frac 2
G\left[1-\frac{gG\,\rho}2\int_0^\delta\frac{d\xi}{\xi}
\tanh\left(\frac{\xi}{2T}\right)\right]\,.\ee Performing an
integration by part we can extract the logarithmic divergence in
$\delta$. This can be eliminated using the result (\ref{3.150})
valid for $\delta\mu=T=0$ in the weak coupling limit ($\Delta_0$
is the gap at $T=0$)\be
1=\frac{G\,\rho}2\log\frac{2\delta}{\Delta_0}\,.\ee We find \be
\alpha=\rho\left[\log\frac{2T}{\Delta_0}+\frac 12\int_0^\infty dx
{\rm\, ln } x\, \frac  1{\cosh^2\frac{x} 2}
\right]\,.\label{eq:alpha2}\ee  Defining \be\log\frac{\Delta_0}{2
T_c}=\frac 12 \int_0^\infty dx {\rm\, ln } x\, \frac
1{\cosh^2\frac{x} 2} \,,\label{eq:78}\ee we get \be
\alpha(T)=\rho\log\frac T{T_c}\,,\label{eq:79}\ee  Performing the
calculation in the reverse we first integrate over $\xi$ obtaining
a divergent series which can be regulated cutting the sum at a
maximal value of $n$ determined by \be\omega_N=\delta \Rightarrow
N\approx \frac{\delta}{2\pi T}\,.\ee We obtain \be \alpha=\frac 2
G\left(1-\pi\,G\,\rho\,T\,\sum_{n=0}^N\frac{1}{\omega_n}\right)\,.\ee
The sum can be performed in terms of the Euler's function
$\psi(z)$:\be \sum_{n=0}^N\frac{1}{\omega_n}=\frac{1}{2\pi T}\,
\left[\psi\left(\frac 3 2+N\right)-\psi\left(\frac 1
2\right)\right]\approx \frac 1 {2\pi T}\left(
\log\frac{\delta}{2\pi T}-\,\psi\left(\frac 1
2\right)\right)\,.\ee Eliminating the cutoff as we did before we
get \be \alpha(T)=\rho\left(\log\left(4\pi \frac
T\Delta_0\right)+\,\psi\left(\frac 1 2\right)\right)\,.\ee By
comparing with Eq. (\ref{eq:alpha2}) we get the following identity
 \be \,\psi\left(\frac 1 2\right)
 =-\log(2\pi)+
\frac 12 \int_0^\infty dx {\rm\, ln } x\, \left(\frac
1{\cosh^2\frac{x} 2}  \right)\,.\ee The equation (\ref{eq:78}) can
be re-written as \be \log\frac{\Delta_0}{4\pi T_c}=\psi\left(\frac
1 2\right)\,.\ee Using ($C$ the Euler-Mascheroni
constant)\be\psi\left(\frac 1 2\right)=-\log(4\gamma),
~~~~\gamma=e^C,~~~~C=0.5777\dots\,,\ee we find
\be\alpha(T)=\rho\log\frac{\pi T}{\gamma\Delta_0}\,,\ee Therefore
the critical temperature, that is the value of $T$ at which
$\alpha=0$, is\be T_c=\frac\gamma\pi\Delta_0\approx
0.56693\,\Delta_0\,.\label{eq:90}\ee The other terms in the
expansion of the gap equation are easily evaluated integrating
over $\xi$ and  summing over the Matsubara frequencies. We get \be
\beta(T)=\pi\,\rho\, T\,\sum_{n=0}^\infty\frac
1{\omega_n^3}=-\frac\rho{16\,\pi^2\, T^2} \,\psi^{(2)}\left(\frac
1 2\right)\,,\label{eq:beta2}\ee where \be\psi^{(n)}(z)=\frac
{d^n}{dz^n}\psi(z)\,.\ee Using \be \psi^{(2)}\left(\frac 1
2\right)=-14\zeta(3),\ee where $\zeta(3)$ is the function zeta of
Riemann \be\zeta(s)=\sum_{k=1}^\infty k^{-s}\ee we get \be
\beta(T)=\frac 7 8\frac \rho{\pi^2 T^2}\zeta(3)\ee Close to the
critical temperature we have \be \alpha(T)\approx
-\rho\left(1-\frac T {T_c}\right),~~~\beta(T)\approx \frac
{7\rho}{8\pi^2 T_c^2}\zeta(3).\ee From the grand potential (or
from the gap equation) we
obtain\be\Delta^2(T)=-\frac{\alpha(T)}{\beta(T)}\to
\Delta(T)=\frac{2\sqrt{2}\pi T_c}{\sqrt{7\zeta(3)}}\left(1-\frac
T{T_c}\right)^{1/2},\ee in agreement with the results of Section
\ref{sectionGL}.

\section{The role of the broken gauge symmetry}

Superconductivity appears to be a fundamental phenomenon and
therefore we would like to understand it from a more fundamental
way than doing a lot of microscopical calculations. This is in
fact the case if one makes the observation that the
electromagnetic $U(1)$ symmetry is spontaneously broken. We will
follow here the treatment given in \cite{WeinbergII}. We have seen
that in the ground state of a superconductor the following
condensate is formed \be
\langle\epsilon_{\alpha\beta}\psi^\alpha\psi^\beta\rangle.\ee This
condensate breaks the em $U(1)$ since the difermion operator has
charge $-2e$. As a matter of fact this is the only thing that one
has to assume, i.e., the breaking of $U(1)$ by an operator of
charge $-2e$. Thinking in terms of an order parameter one
introduces a scalar field, $\Phi$, transforming as the condensate
under a gauge transformation \be A_\mu\to
A_\mu+\de_\mu\Lambda,~~\psi\to e^{ie\Lambda}\psi\Rightarrow
\Phi\to e^{2ie\Lambda}\Phi,\ee where $\psi$ is the electron field.
We then introduce the Goldstone field $\phi$ as the phase of the
field $\Phi$  \be \Phi=\rho\, e^{\dd{2ie\phi}}.\ee Therefore
$\phi$ transforms as the phase of the condensate under a gauge
transformation \be \phi\to\phi+\Lambda.\ee In the case of constant
$\Lambda$ this implies that the theory may depend only on
$\de_\mu\phi$. Notice also that the gauge invariance is broken but
a subgroup $Z_2$ remains unbroken, the one corresponding to
$\Lambda =0$ and $\Lambda= \pi/e$. In particular $\phi$ and
$\phi+\pi/e$ should be identified.

It is also convenient to introduce gauge invariant Fermi fields
\be \tilde\psi=e^{\dd{-ie\phi}}\,\psi.\ee The system will be
described by a gauge invariant lagrangian depending on
$\tilde\psi$, $A_\mu$ and $\de_\mu\phi$. Integrating out the Fermi
fields one is left with a gauge invariant lagrangian depending on
$A_\mu$ and $\de_\mu\phi$. Gauge invariance requires that these
fields should appear only in the combinations \be
F_{\mu\nu}=\de_\mu A_\nu-\de_\nu A_\mu,~~~A_\mu-\de_\mu\phi.\ee
Therefore the lagrangian has the form \be L=-\frac 1 4\int d^3\b
x\,F_{\mu\nu}F^{\mu\nu}+L_s(A_\mu-\de_\mu\phi).\ee The equation of
motion for the scalar field is \be 0=\de_\mu\frac{\delta
L_s}{\delta\de_\mu\phi}=-\de_\mu\frac{\delta L_s}{\delta
A_\mu}=-\de_\mu J^\mu,\ee where $J_\mu$ is the em current defined
as  \be J_\mu=\frac{\delta L_s}{\delta A^\mu}.\ee Therefore the
equation of motion for $\phi$ is nothing but the conservation of
the current. The only condition on $L_s$ is that it gives rise to
a stable state of the system in the absence of $A_\mu$ and $\phi$.
In particular this amounts to say that the point
$A_\mu=\de_\mu\phi$ is a local minimum of the theory. Therefore
the second derivative of $L_s$ with respect to its argument should
not vanish at that point.

\begin{center}
\begin{figure}[ht]
\epsfxsize=8truecm \centerline{\epsffile{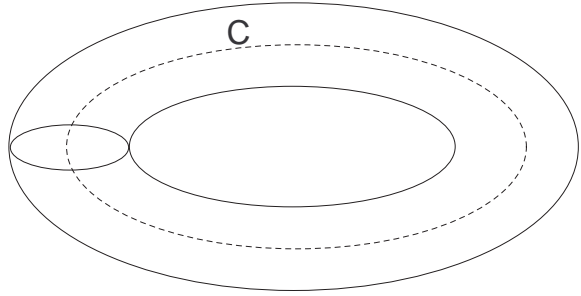}} \noindent
\caption{\it Inside the toroidal superconductor the line $C$ of
integration is shown\label{torus}}
\end{figure}
\end{center}

The Meissner effect follows easily from the previous
considerations. In fact, if we go deep inside the superconductor
we will be in the minimum $A_\mu=\de_\mu\phi$, implying that
$A_\mu$ is a pure gauge since \be
F_{\mu\nu}(\de_\lambda\phi)=0.\ee In particular the magnetic field
inside the superconductor is vanishing, $\b B=0$. We may refine
this analysis by doing some considerations about the energy. Close
to the minimum we have \be L_s(A_\mu-\de_\mu\phi)\approx
L_s(0)+\frac 1 2\frac{\delta^2
L_s}{\delta(A_\mu-\de_\mu\phi)^2}(A_\mu-\de_\mu\phi)^2.\ee Notice
that the dimensions of the second derivative are $[E\times
E^{-2}]=[E^{-1}]=[L]$. Therefore, in the static case, up to a
constant \be L_s\approx \frac {L^3}{\lambda^2}|\b
A-{\bm\nabla}\phi|^2,\ee with $L^3$ the volume of the
superconductor and $\lambda$ some length  typical of the material.
If a magnetic field $\b B$ penetrates inside the material, we
expect \be |\b A-{\bm\nabla}\phi|\approx BL,\ee from which \be
L_s\approx \frac {B^2 L^5}{\lambda^2}.\ee For the superconductor
to remain in that state the magnetic field must be expelled with
an energy cost of \be B^2L^3.\ee Therefore there will be
convenience in  expelling $\b B$ if \be
\frac{B^2L^5}{\lambda^2}\gg B^2L^3,\ee or \be L\gg\lambda.\ee
$\lambda$ is the penetration depth, in fact from its definition it
follows that it is the region over which the magnetic field is non
zero. Repeating the same reasoning made in the Introduction one
can see the existence of a critical magnetic field. Notice that a
magnetic field smaller than the critical one penetrates inside the
superconductor up to a depth $\lambda$ and in that region the
electric current will flow, since \be \b J\propto
{\bm\nabla}\wedge\b B.\ee Consider now a thick superconductor with
a shape of  a torus. Along the internal line $C$ (see Fig.
\ref{torus}) the quantity $|\b A-{\bm \nabla}\phi|$ vanishes but
the two fields do not need to be zero. However, going around the
path $C$ $\phi$ has to go to an equivalent value, $\phi+n\pi/e$.
Therefore \be \int_{\cal A}\b B\cdot d\b S=\oint_{C}\b A\cdot d\b
x=\oint_C{\bm \nabla}\phi\cdot d\b x=\frac{n\pi}e,\ee where ${\cal
A}$ is an area surrounded by $C$. We see that the flux of $\b B$
inside the torus is quantized.

Notice also that the electric current sustaining $\b B$ flows in a
layer of thickness $\lambda$ below the surface of the torus. It
follows that the current cannot decay smoothly but it must jump in
such a way that the variation of the magnetic flux is a multiple
of $\pi/e$. Therefore the electric resistance of a superconductor
is rather peculiar. In order to understand better the resistance
let us consider the following equation \be \frac{\delta
L_s}{\delta\dot\phi}=-\frac{\delta L_s}{\delta A_0}=-J_0.\ee This
shows that $-J_0$ is the canonical momentum density conjugated to
$\phi$. Therefore the Hamilton equations of motion give \be
\dot\phi(x)=\frac{\delta H_s}{\delta(-J_0(x))}=-V(x),\ee where
$V(x)$, being the variation of energy per change in the current
density, is the "voltage" at $x$. If in the superconductor there
is a stationary current (that is with time-independent fields),
the previous equation shows that the voltage is zero. But a
current with zero voltage means that the electric resistance is
zero.

We are now in the position of explaining the Josephson effect
\cite{Josephson:1962ab,Josephson:1965ab}. The effect arises at the
junction of two superconductors separated by a thin insulating
barrier. At zero voltage difference between the two
superconductors a continuous current flows, depending on the phase
difference due to the two different Goldstone fields. Furthermore,
if a constant voltage difference is maintained between the two
superconductor an alternate current flows. These two effects are
known as the {\it dc} and the {\it ac} {\it Josephson effects}.
Consider first the case of zero voltage difference. By gauge
invariance the lagrangian at the junction may depend only on the
phase difference\be L_{\rm junction}=F(\Delta\phi).\ee The
function $F$ must be periodic since the Goldstone fields in the
two superconductors are defined mod $\pi/e$, that is \be
F(\Delta\phi)=F(\Delta\phi+n\pi/e).\ee To evaluate the current let
us introduce a vector potential $\b A$. Then \be \Delta_{\b
A}\phi=\int_\ell d\b x\cdot({\bm\nabla}\phi-\b A),\ee where the
line $\ell$ is taken across the junction. Therefore we get \be
J^k=\frac{\delta L_{\rm junction}}{\delta A_k}=n^kF'(\Delta_{\b
A}\phi),\ee where $n^k$ is the normal unit vector at the junction
surface. By putting $\b A=0$ we get \be \b J=\b n
F'(\Delta\phi),\ee showing the {\it dc} Josephson effect. To get
the second one, consider a constant voltage difference. From
\be\dot\phi=-V,\ee we get \be\Delta\phi(t)=|\Delta
V|t+\Delta\phi(0).\ee Since $F$ has a period $\pi/e$ it follows
that the current oscillates with a frequency \be
\nu=\frac{e|\Delta V|}\pi.\ee Using this relation one can get a
very accurate measure of $e/\hbar$ (going back to standard units,
one has $\nu=e|\Delta V|/\pi\hbar$).

The current in the {\it dc} Josephson effect can be of several
milliamperes for conventional superconductors. In the case of the
{\it ac} effect for voltages of the order of millivolts, the
frequency can rise to hundreds and thousands of gigahertz.

When close to the phase transition the description of the theory
in terms of the Goldstone boson is not enough. In fact there is
another long wave-length mode associate to the order parameter.
This is because the $U(1)$ symmetry gets restored and its minimal
description is in terms of a complex field. Therefore one
introduces \be \Phi=\rho\, e^{\dd{2ie\phi}}.\ee Expanding $L_s$
for small values of $\Phi$ we get (with an appropriate
normalization for $\Phi$) \bea L_s&\approx&\int d^3\b x
\left[-\frac 1 2\Phi^*|({\bm\nabla}-2ie\b A)|^2\Phi-\frac 1
2\alpha|\Phi|^2-\frac 1 4\beta|\Phi|^4\right]\nn\\&=&\int d^3\b
x\left[-2e^2\rho^2({\bm\nabla}\phi-e\b A)^2-\frac 1
2({\bm\nabla}\rho)^2-\frac 1 2 \alpha\rho^2-\frac 14
\beta\rho^4\right].\eea In this Section we have defined the
penetration depth as the inverse square root of the coefficient of
$-({\bm\nabla}\phi-\b A)^2/2$. Therefore \be \lambda=\frac 1
{\sqrt{4e^2\langle\rho^2\rangle}}.\ee Using
\be\langle\rho^2\rangle=-\frac\alpha\beta,\ee we get
\be\lambda=\frac 1{2e}\sqrt{-\frac\beta\alpha},\ee in agreement
with Eq. (\ref{GL2}) (notice that $\alpha$ and $\beta$ are not
normalized in the same way). Another length is obtained by
studying the behavior of the fluctuations of the field $\rho$.
Defining \be \rho=\langle\rho\rangle+\rho',\ee
 we get
 \be{\bm\nabla}^2\rho'=-2\alpha\rho'.\ee This allows us to
 introduce the coherence length $\xi$ as
 \be\xi=\frac 1{\sqrt{-2\alpha}},\ee in agreement with Eq.
 (\ref{1.47}). Using the definitions of $\lambda$ and $\xi$ we
 get
 \be\alpha=-\frac 1
 {2\xi^2},~~~\beta=2\frac{e^2\lambda^2}{\xi^2}.\ee
Therefore the energy density of the superconducting state is lower
than the energy density of the normal state by \be \frac 1
4\frac{\alpha^2}\beta=\frac 1 {32}\frac 1 {e^2\lambda^2\xi^2}.\ee

The relative size of $\xi$ and $\lambda$ is very important, since
vortex lines can be formed inside the superconductor and their
stability depends on this point. More precisely inside the
vortices the normal state is realized, meaning $\rho=0$ and a
flux-quantized magnetic field. The superconductors are therefore
classified according to the following criterium:

\begin{itemize}
\item Type I superconductors: $\xi>\lambda$. The vortices are not
stable since the penetration of the magnetic field is very small.
\item Type II superconductors: $\xi<\lambda$. The vortices are
stable and the magnetic field penetrates inside the
superconductor exactly inside the vortices. This may happen since
the core of the vortex is much smaller than the region where the
magnetic field goes to zero. In this cases there are two critical
magnetic fields, $H_{c1}$, where for $H<H_{c1}$ the state is
superconducting, whereas for $H>H_{c_1}$ vortices are formed.
Increasing  the magnetic field, more and more vortex lines are
formed, up to a value $H_{c_2}$ where the magnetic field
penetrates all the superconductor and the transition to the
normal state arises.
\end{itemize}

\section{Color superconductivity}

\begin{center}
\begin{figure}[ht]
\epsfxsize=12truecm \centerline{\epsffile{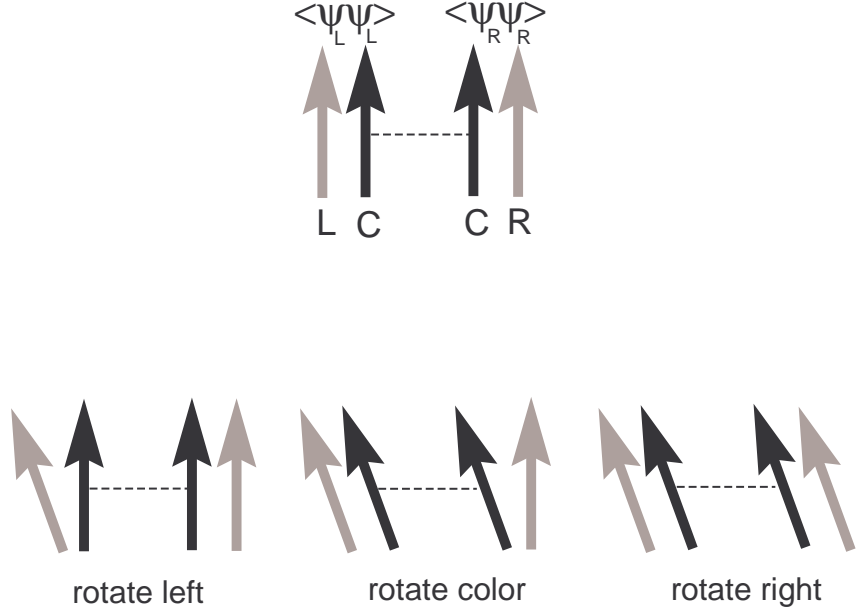}} \noindent
\caption{\it Color and flavor indices are represented by black and
grey arrows respectively. In the lower part it is shown that we
can leave the left-handed condensate invariant if we perform a
flavor rotation followed by a related color rotation. However this
rotates the color index in the right-handed condensate, so in
order to have both  condensates  invariant we have to rotate also
the right- flavor index. In conclusion we have invariance under
the diagonal product of the three $SU(3)$ groups.\label{CFLeps}}
\end{figure}
\end{center}

Ideas about color superconductivity go back to almost 25 years ago
\cite{collins:1975ab,Barrois:1977xd,Frautschi:1978rz,Bailin:1984bm},
but only recently this phenomenon has received a lot of attention
(for recent reviews see ref.
\cite{Rajagopal:2000wf,Hsu:2000sy,Hong:2000ck,Alford:2001dt,Nardulli:2002ma,
Schafer:2003ab}). The naive expectation is that at very high
density, due to the asymptotic freedom, quarks would form a Fermi
sphere of almost free fermions. However, as we know, Bardeen,
Cooper and Schrieffer proved that the Fermi surface of free
fermions is unstable in presence of an attractive, arbitrary
small, interaction.  Since in QCD the gluon exchange in the $\bar
3$ channel is attractive one expects the formation of a coherent
state of Cooper pairs.  The phase structure of QCD at high density
depends on the number of flavors and there are two very
interesting cases, corresponding to two massless flavors (2SC)
\cite{Barrois:1977xd,Alford:1998zt,Rapp:1998zu} and to three
massless flavors (CFL) \cite{Alford:1998mk,Schafer:1998ef}
respectively.  The two cases give rise to very different patterns
of symmetry breaking. If we denote left- and right-handed quark
fields by $\psi_{iL(R)}^\alpha=\psi_{ia(\dot a)}^\alpha$ by their
Weyl components, with $\alpha=1,2,3$, the $SU(3)_c$ color index,
$i=1,\cdots,N_f$ the flavor index ($N_f$ is the number of massless
flavors) and $a(\dot a)=1,2$ the Weyl indices, the structure of
the condensate at very high density can be easily understood on
the basis of the following considerations. Consider the matrix
element
\begin{equation}
  \langle 0|\psi_{ia}^\alpha\psi_{jb}^\beta|0\rangle.
\end{equation}
its color, spin and flavor structure is completely fixed by the
following considerations:
\begin{itemize}
\item The condensate should be antisymmetric in color indices
$(\alpha,\beta)$ in order to have attraction; \item The condensate
should be antisymmetric in spin indices $(a,b)$ in order to get a
spin zero condensate. The isotropic structure of the condensate is
generally favored since it allows a better use of the Fermi
surface
\cite{Evans:1998ek,Schafer:1998na,Brown:1999yd,Schafer:2000tw};
\item given the structure in color and spin, Pauli principles
requires antisymmetry in flavor indices.
\end{itemize}
Since the momenta in a Cooper pair are opposite, as the spins of
the quarks (the condensate has spin 0), it follows that the
left(right)-handed quarks can pair only with left(right)-handed
quarks. In the case of 3 flavors the favored condensate is
\begin{equation}
\langle 0|\psi_{iL}^\alpha\psi_{jL}^\beta|0\rangle=-\langle
0|\psi_{iR}^\alpha\psi_{jR}^\beta|0\rangle=
\Delta\sum_{C=1}^3\epsilon^{\alpha\beta C}\epsilon_{ijC}.
\end{equation}
This gives rise to the so-called color--flavor--locked (CFL)
phase \cite{Alford:1998mk,Schafer:1998ef}. The reason for the
name is that simultaneous transformations in color and in flavor
leave the condensate invariant as shown in Fig. \ref{CFLeps}.
More generally  the following condensate is formed
\cite{Alford:1998mk,Schafer:1998ef} \be \langle q_{iL(R)}^\alpha C
q_{jL(R)}^\beta\rangle\propto \sum_{C=1}^3\epsilon^{ijC}
\epsilon_{\alpha\beta
C}+\kappa(\delta^i_\alpha\delta^j_\beta+\delta_\beta^i\delta_\alpha^j)
\label{CFL}. \ee Due to the Fermi statistics, the condensate must
be symmetric in color and flavor. As a consequence the two terms
appearing in Eq. (\ref{CFL}) correspond to the $(\mathbf{\bar
3},\mathbf{\bar 3})$ and $(\mathbf{6},\mathbf{6})$ channels of
$SU(3)_c\otimes SU(3)_{L(R)}$. It turns out that $\kappa$ is small
\cite{Alford:1998mk,Schafer:1999fe,Shovkovy:1999mr} and therefore
the condensation occurs mainly in the $(\mathbf{\bar
3},\mathbf{\bar 3})$ channel. Also in this case  the ground state
is left invariant by a simultaneous transformation of $SU(3)_c$
and $SU(3)_{L(R)}$.  The symmetry breaking pattern is
%\newpage
\bea &SU(3)_c\otimes SU(3)_L\otimes SU(3)_R\otimes U(1)_B\otimes
U(1)_A&\nn\\ &\downarrow&\\ &SU(3)_{c+L+R}\otimes Z_2\otimes
Z_2.&\nn \eea The $U(1)_A$ symmetry is broken at the quantum level
by the anomaly, but it gets restored at very high density since
the instanton contribution is suppressed \cite{Rapp:1999qa,
Schafer:1999fe,Son:1999cm,Son:2000tu}. The $Z_2$ symmetries arise
since the condensate is left invariant by a change of sign of the
left- and/or right-handed fields. The electric charge is broken
but a linear combination with the broken color generator $T_8$
annihilates the ground state. On the contrary the baryon number is
broken. Therefore there are $8+2$ broken global symmetries giving
rise to 10 Goldstone bosons. The one associated to $U(1)_A$ gets
massless only at very high density. The color group is completely
broken and all the gauge particles acquire mass. Also all the
fermions are gapped. We will show in the following how to
construct an effective lagrangian describing the Goldstone bosons,
and how to compute their couplings in the high density limit where
the QCD coupling gets weaker.

 The previous one  is the typical situation when the chemical
potential is much bigger than the quark masses $m_u$, $m_d$ and
$m_s$ (here the masses to be considered are in principle density
depending). However we may ask what happens  decreasing the
chemical potential. At intermediate densities we have no more the
support of asymptotic freedom, but all the model calculations
show that one still has a sizeable color condensation. In
particular if the chemical potential $\mu$ is much less than the
strange quark mass one expects that the strange quark decouples,
and the corresponding condensate should be
\begin{equation}
\langle 0|\psi_{iL}^\alpha\psi_{jL}^\beta|0\rangle=
\Delta\epsilon^{\alpha\beta 3}\epsilon_{ij}.\label{2SC}
\end{equation}
In fact, due to the antisymmetry in color the condensate must
necessarily choose a direction in color space. Notice that now the
symmetry breaking pattern is completely different from the
three-flavor case: \bea
  SU(3)_c\otimes SU(2)_L\otimes SU(2)_R\otimes U(1)_B\to
  SU(2)_c\otimes SU(2)_L\otimes
  SU(2)_R\otimes U(1)\otimes Z_2.\nn
\eea The condensate breaks the  color group $SU(3)_c$ down to the
subgroup $SU(2)_c$ but it does not break any flavor symmetry.
Although the baryon number, $B$, is broken, there is a combination
of $B$ and of the broken color generator, $T_8$, which is unbroken
in the 2SC phase. Therefore no massless Goldstone bosons are
present in this phase. On the other hand, five gluon fields
acquire mass whereas three are left massless. We notice also that
for the electric charge the situation is very similar to the one
for the baryon number. Again a linear combination of the broken
electric charge and of the broken generator $T_8$ is unbroken in
the 2SC phase. The condensate (\ref{2SC})  gives rise to a gap,
$\Delta$, for quarks of color 1 and 2, whereas the two quarks of
color 3 remain un-gapped (massless). The resulting effective
low-energy theory has been described in
\cite{Casalbuoni:2000jn,Casalbuoni:2000cn,Rischke:2000cn}.

 A final problem we will discuss has to do with the fact
that when quarks (in particular the strange quark) are massive,
their chemical potentials cannot be all equal. This situation has
been modelled out in \cite{Alford:2000ze}.  If the Fermi surfaces
of different flavors are too far apart, BCS pairing does not
occur. However it might be favorable for different quarks to pair
each of one lying at its own Fermi surface and originating a pair
of non-zero total momentum. This is the LOFF  state first studied
by the authors of ref. \cite{LO,FF} in the context of electron
superconductivity in the presence of magnetic impurities. Since
the Cooper pair has non-zero momentum the condensate breaks space
symmetries and we will show that  in the low-energy spectrum a
massless particle, a phonon, the Goldstone boson of the broken
translational symmetry, is present. We will construct the
effective lagrangian also for this case (for a general review of
the LOFF phase see \cite{Casalbuoni:2003ab,Bowers:2003ye}.

Of course it would be very nice if we could test all these ideas
on the lattice. However the usual sampling method, which is based
on a positive definite measure, does not work in presence of a
chemical potential since the fermionic determinant turns out to be
complex in euclidean space. The argument is rather simple. We
define euclidean variables through the following substitutions
\bea x_0\to -ix_E^4,&&~~x^i\to x^i_E,\\
\gamma_0\to \gamma_E^4,&&~~\gamma^i\to -i \gamma_E^i.\eea Then the
Dirac operator, in presence of a chemical potential,
becomes\footnote{We neglect the mass term, since it correspond to
an operator multiple of the identity, and all the following
considerations can be trivially extended.}\be
D(\mu)=\gamma^\mu_ED_E^\mu +\mu\gamma_E^4,\ee where
$D^\mu_E=\de_E^\mu+iA_E^\mu$ is the euclidean covariant
derivative. In absence of a chemical potential the operator has
the following properties\be D(0)^\dagger=-D(0),~~~\gamma_5
D(0)\gamma_5=-D(0).\ee Therefore the eigenvalues of $D(0)$ are
pure imaginary. Also if $|\lambda\rangle$ is an eigenvector of
$D(0)$, the same is for $\gamma_5|\lambda\rangle$ but with
eigenvalue $-\lambda$. This follows from \be \gamma_5
D(0)|\lambda\rangle=\lambda\gamma_5
|\lambda\rangle=-D(0)\gamma_5|\lambda\rangle.\ee Therefore \be{\rm
det}[D(0)]=\prod_\lambda(\lambda)(-\lambda)>0.\ee For $\mu\not=0$
this argument does not hold and the determinant is complex. Notice
that this argument depends on the kind of chemical potential one
has to do. For instance in the case of two degenerate flavored
quarks, $u$ and $d$, if we consider the isospin chemical
potential, which is coupled to the conserved charge $\tau_3$ in
flavor space, we may still prove the positivity by using, for
instance, $\tau_1\gamma_5$ instead of $\gamma_5$. Therefore these
cases can be treated on the lattice.

\subsection{Hierarchies of effective lagrangians}\label{VA}
QCD at high density is conveniently studied through a hierarchy of
effective field theories. The starting point is the fundamental
QCD lagrangian. The way of obtaining a low energy effective
lagrangian is to integrate out high-energy degrees of freedom. As
we have seen Polchinski \cite{Polchinski:1992ed} has shown  that
the physics is particularly simple for energies close to the Fermi
energy. He has shown that all the interactions are irrelevant
except for a four-fermi interaction coupling pair of fermions with
opposite momenta. This is nothing but the interaction giving rise
to the BCS condensation. This physics can be described using the
High Density Effective Theory (HDET)
\cite{Hong:1998tn,Hong:1999ru,Beane:2000ms,Casalbuoni:2000na,Casalbuoni:2000nb}.
\begin{center}
\begin{figure}[ht]
\epsfxsize=8truecm \centerline{\epsffile{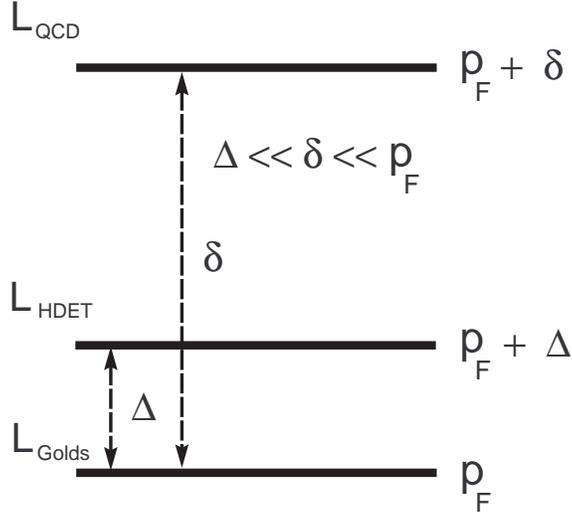}} \noindent
\caption{\it The hierarchy of effective lagrangians entering in
the discussion of high density QCD.\label{hierarchy}}
\end{figure}
\end{center}
In this theory the condensation effects are taken into account
through the introduction of a Majorana mass term. The degrees of
freedom are quasi-particles (dressed fermions), holes and gauge
fields. This description is supposed to hold up to a cutoff
$p_F+\delta$, with $\delta$  smaller than the Fermi momentum but
bigger than the gap $\Delta$, $\Delta\ll\delta\ll p_F$. Going at
momenta much smaller than the gap energy $\Delta$ all the gapped
particles decouple and one is left with the low energy modes as
Goldstone bosons, ungapped fermions and holes and massless gauged
fields according to the breaking scheme. The corresponding
effective theory in the Goldstone sector can be easily formulated
using standard techniques. In the case of CFL and 2SC such
effective lagrangians have been given in refs.
\cite{Casalbuoni:1999wu} and
\cite{Casalbuoni:2000cn,Rischke:2000cn}. The parameters of the
effective lagrangian can be evaluated at each step of the
hierarchy by matching the Greens functions with the ones evaluated
at the upper level.

\subsection{The High Density Effective Theory (HDET)}\label{HDET}
We will present here the High Density Effective Theory
\cite{Hong:1998tn,Hong:1999ru,Beane:2000ms,Casalbuoni:2000na,Casalbuoni:2000nb}
in the context of QCD with $N_f$ massless flavors. As already
discussed we will integrate out all the fermionic degrees of
freedom corresponding to momenta greater than $p_F+\delta$ with
$\delta$ a cutoff such that $\Delta\ll\delta\ll p_F$.  The QCD
lagrangian at finite density is given by\be {\cal L}_{QCD}=
\bar\psi i\slash D\psi-\frac 1 4 F_{\mu\nu}^a
F^{a\mu\nu}+\mu\bar\psi\gamma_0\psi,~~~a=1,\cdots, 8,\ee \be
D_\mu=\partial_\mu+ig_s A_\mu^a T^a,~~~~ \slash D=\gamma_\mu
D^\mu,\ee where $T^a=\lambda^a/2$, with $\lambda^a$ the Gell-Mann
matrices. At asymptotic values of $\mu\gg \Lambda_{QCD}$ quarks
can be considered as almost free particles due to the asymptotic
freedom. The corresponding Dirac equation in momentum space is \be
(\slash p+\mu\gamma_0)\psi(p)=0,\ee or \be
(p^0+\mu)\psi(p)={\bm\alpha}\cdot\b p\psi(p), \ee where
${\bm\alpha}=\gamma_0{\bm\gamma}$. From this we get immediately
the dispersion relation \be (p^0+\mu)^2=|\b p|^2,\ee or \be
p^0=E_{\pm}=-\mu\pm |\b p|.\ee For the following it is convenient
to write the Dirac equation in terms of the following projectors
\be P_\pm=\frac{1\pm{\bm\alpha}\cdot\b v_F}2,\ee where \be\b
v\equiv\b v_F=\frac{\de E(\b p)}{\de \b p}=\frac{\de|\b p|}{\de \b
p}=\b{\hat p}.\ee Decomposing $\b p=\mu\b v_F+{\bm \ell}$ we get
\be H\psi_+={\bm\alpha}\cdot{\bm \ell}\,\psi_+,~~~~
H\psi_-=\left(-2\mu+{\bm\alpha}\cdot{\bm \ell}\right)\,\psi_-.\ee
Therefore only the states $\psi_+$ with energies  close to the
Fermi surface, $|\b p|\approx\mu$ can be easily excited
($E_+\approx 0$). On the contrary, the states $\psi_-$ have
$E_-\approx -2\mu$. They are well inside the Fermi sphere and
decouple for high values of $\mu$. In the limit $\mu\to\infty$ the
only physical degrees of freedom are $\psi_+$ and the gluons.
After this quantum mechanical introduction, let us consider the
field theoretical version of the previous argument. The main idea
of the effective theory is the observation that the quarks
participating in the dynamics have large ($\sim \mu$) momenta.
Therefore one can introduce velocity dependent fields by
extracting the large part $\mu\b v$ of this momentum. We start
with the Fourier decomposition of the quark field $\psi(x)$: \be
\psi(x)=\int\frac{d^4p}{(2\pi)^4}e^{-i\,p\cdot
x}\psi(p),\,\label{4.0.dec}\ee and then we introduce the Fermi
velocity as
 \be p^\mu=\mu
v^\mu+\ell^\mu\ , \label{4.1.dec} \ee where \be v^\mu=(0,\b v ),
\ee with $|\b v|=1$. The four-vector \be
\ell^\mu=(\ell^0,\vec\ell),\ee is called the residual momentum. We
define also \be {\bm\ell}=\b v \ell_\parallel+{\bm\ell}_\perp,\ee
with \be {\bm\ell}_\perp= {\bm\ell}-({\bm\ell}\cdot\b v )\b v.\ee
Since we can always choose the velocity $\b v$ parallel to $\b p$,
we have ${\bm\ell}_\perp=0$. We will now separate in Eq.
(\ref{4.0.dec}) the light and heavy fermion degrees of freedom.
These are defined by the following restrictions in momentum space:
\bea &{\rm light~
d.o.f.}& ~~~\mu-\delta\le|\b p|\le\mu +\delta,\nn\\
&{\rm heavy~d.o.f.}&~~~|\b p|\le\mu-\delta,~~|\b
p|\ge\mu+\delta,\eea or, in terms of the residual momentum \bea
&{\rm light~
d.o.f.}& ~~~-\delta\le\ell_\|\le +\delta,\nn\\
&{\rm heavy~d.o.f.}&~~~\ell_\|\le-\delta,~~\ell_\|\ge+\delta.\eea
Here $\delta$ is the cutoff that we choose for defining the
effective theory around the Fermi sphere. We will assume that
$\delta\gg\Delta$, with $\Delta$ the gap. We will assume also that
the shell around the Fermi sphere is a narrow one and therefore
$\delta\ll\mu$.  Let us start our discussion with the light
fields. We will discuss how to integrate out the heavy d.o.f later
on. In this case we can write the integration in momentum space in
the form \be \int \frac{d^4
p}{(2\pi)^4}=\frac{\mu^2}{(2\pi)^4}\int
d\Omega\int_{-\delta}^{+\delta}
d\ell_\parallel\int_{-\infty}^{+\infty}d\ell_0= \int\frac{d\b
v}{4\pi}\frac{\mu^2}\pi\int\frac{d^2\ell}{(2\pi)^2},\label{4.0.8}\ee
where we have taken into account that we are interested in the
degrees of freedom in a shell of amplitude 2$\delta$ around the
Fermi sphere, and  that in the shell we can assume a constant
radius $\mu$. We have also substituted the angular integration
with the integration over the Fermi velocity taking into account
that it is a unit vector. That is \be \int d\b v\equiv\int d^3{\b
v} \delta(|\b v|-1)=\int |\b v|^2\, d|\b v|\,d\Omega\,\delta(|\b
v|-1)=\int d\Omega.\ee
 In this way the original 4-dimensional integration in momentum has been
 factorized in the product of two 2-dimensional integrations. In
 particular
\be \int_{|\b p|\in {\rm shell}} \frac{d^3\b p}{(2\pi)^3}=\int
\frac{d\b v}{4\pi}\frac{\mu^2}\pi\int d\ell_\|\label{5.25}.\ee The
Fourier decomposition (\ref{4.0.dec}) for the light d.o.f.  takes
the form \be \psi(x) = \int\frac{d\b v}{4\pi} e^{-i\mu v\cdot
x}\psi_{\b v}(x)\ ,\label{6.95}\ee where \be \psi_{\b
v}(x)=\frac{\mu^2}\pi\int\frac{d^2\ell}{(2\pi)^2}e^{-i\ell\cdot
x}\,\psi_{\b v }(\ell),\label{6.96}\ee  with $\psi_{\b
v}(\ell)\equiv\psi(p)$. Notice that the fields $\psi_{\b v}(x)$
are velocity-dependent and they include only the degrees of
freedom corresponding to the shell around the Fermi sphere.
 Projecting with the operators $P_\pm$ we get
 \be\psi(x)= \int\frac{d\b v}{4\pi}e^{-i\mu v\cdot x}
 \left[\psi_+(x)+\psi_-(x)\right]\ ,\ee where
\be  \psi_\pm(x)=\ P_\pm \psi_{\b v}(x)=P_\pm\frac{\mu^2}\pi\int
\frac{d^2\ell}{(2\pi)^2}e^{-i\ell\cdot x} \psi_{\b v}(\ell).\ee

Let us now define \bea V^\mu&=&(1,\,\b v)\ ,~~~~~~~~~~~~~~~\tilde
V^\mu=(1,\,-\b v)\, ,\cr&&\cr
\gamma^\mu_\parallel&=&(\gamma^0,\,(\b v\cdot\bm \gamma)\,\b v)\
,~~~~~~ \gamma_\perp^\mu=\gamma^\mu-\gamma_\parallel^\mu.\eea We
can then prove  the following relations\bea
\bar\psi_+\gamma^\mu\psi_+&=&V^\mu\bar\psi_+\gamma^0\psi_+\ ,\cr
\bar\psi_-\gamma^\mu\psi_-&=&\tilde V^\mu\bar\psi_-\gamma^0\psi_-\
,\cr
\bar\psi_+\gamma^\mu\psi_-&=&\bar\psi_+\gamma_\perp^\mu\psi_-\
,\cr
\bar\psi_-\gamma^\mu\psi_+&=&\bar\psi_-\gamma^\mu_\perp\psi_+.
\label{6.102}\eea Now we want to evaluate the effective action in
terms of the velocity-dependent fields defined in Eqs.
(\ref{6.95}) and (\ref{6.96}).  Substituting the expression
(\ref{6.96}) in terms of the type $ \int d^4x \psi^\dagger\psi$ we
find \bea&\displaystyle{\int} d^4x\,
\psi^\dagger\psi=\nn\\&\displaystyle{\left(\frac{\mu^2}{\pi}\right)^2\int}
\displaystyle{\frac{d \vec v_F}{4\pi} \frac{d\vec
v'_F}{4\pi}\frac{d^2\ell}{(2\pi)^2}\frac{d^2\ell'}{(2\pi)^2}}
(2\pi)^4\delta^4(\ell'-\ell+\mu v'-\mu v)\psi_{\b
v'}^\dagger(\ell') \psi_{\b v}(\ell).\eea For large values of
$\mu$ the integral is different from zero only if $v=v'$ and the
$\delta$-function factorizes in the product of two
$\delta$-functions one for the velocity and one for the residual
momenta. Both these $\delta$-functions are two-dimensional.
Therefore we obtain \be \displaystyle{\int} d^4x\,
\psi^\dagger\psi=\frac{\mu^2}{\pi}\int \frac{d\b
v}{4\pi}\frac{d^2\ell}{(2\pi)^2}\psi^\dagger_{\b v}(\ell)\psi_{\b
v}(\ell).\ee Analogously one could start from (\ref{6.95}) finding
\be \int d^4 x\,\psi^\dagger\psi=\int d^4x\,\int\frac{d\b
v}{4\pi}\psi_{\b v}^\dagger(x)\psi_{\b v}(x).\ee

In the following we will need also the expression $\int
d^4x\psi^TC\psi$. We find \be \int d^4x\psi^TC\psi=\int
d^4x\,\int\frac{d\b v}{4\pi}\psi_{\b v}^T(x)C\psi_{-\b
v}(x)=\frac{\mu^2}\pi\int\frac{d\b
v}{4\pi}\frac{d^2\ell}{(2\pi)^2}\psi^T_{\b v}(\ell)C\psi_{-\b
v}(-\ell).\ee These expressions show that there is a {\it
superselection rule for the Fermi velocity}.

Now we are in the position to evaluate the effective lagrangian
for the fields $\psi_+$ which should be the relevant degrees of
freedom in the  high density limit. Since we have \be
i\partial_\mu \esp{-i\mu v\cdot x}=\mu v_\mu\esp{-i\mu v\cdot
x},\ee we get \be\int d^4 x\,\bar\psi\left(i\slash
D+\mu\gamma_0\right)\psi=\int d^4 x\,\int\frac{d\b
v}{4\pi}(\bar\psi_++\bar\psi_-)\left[\mu(\slash
v+\gamma_0)+i\slash D\right](\psi_++\psi_-).\ee Expanding and
using Eqs. (\ref{6.102}) we find \bea
 &\int d^4 x\,\bar\psi\left(i\slash D+\mu\gamma_0\right)\psi=\cr&\int d^4 x\,\left(\psi_+^\dagger
iV\cdot D\psi_++\psi_-^\dagger i\tilde V\cdot
D\psi_-+2\mu\psi_-^\dagger\psi_-+\bar\psi_+i\slash D_\perp\psi_-
+\bar\psi_-i\slash D_\perp\psi_+\right). \eea The lagrangian is
given by \be {\cal L}_D=\int\frac{d\b v}{4\pi}\left[\psi_+^\dagger
iV\cdot D\psi_++\psi_-^\dagger(2\mu+ i\tilde V\cdot
D)\psi_-+(\bar\psi_+i\slash D_\perp\psi_- + h.c.)\right]
\label{5.42}.\ee From which we get the equations of motion \bea
iV\cdot D\psi_++i\gamma^0\slash D_\perp\psi_-=0,\cr (2\mu+i\tilde
V\cdot D)\psi_-+i\gamma^0\slash D_\perp\psi_+=0. \eea At the
leading order in $1/\mu$ \be \psi_-=0,~~~ iV\cdot D\psi_+=0, \ee
proving the decoupling of $\psi_-$. At this order the effective
lagrangian is simply \be {\cal L}_D=\int\frac{d\b
v}{4\pi}\psi^\dagger_+ i V\cdot D\psi_+\label{5.45}.\ee Therefore
the free propagator, $\langle T(\psi_+\psi_+^\dagger)\rangle$, is
given by \be \frac 1{ V\cdot\ell}.\ee This can be seen directly
starting from the Dirac propagator  \be \frac 1{p^0\gamma^0-\vec
p\cdot\vec \gamma+\mu\gamma^0}= \frac{(p^0+\mu)\gamma^0-\vec
p\cdot\vec \gamma}{(p^0+\mu)^2-|\vec p\,|^2}. \ee By putting
$p=\mu v+\ell$ and expanding at the leading order in $1/\mu$ we
find \be \frac 1{\slash p+\mu\gamma_0}\approx \frac{\slash V}
2\frac {1}{V\cdot \ell}, \ee where $V\cdot
\ell=\ell^0-{\bm\ell}\cdot\b v$. Therefore the propagator depends
(at the leading order) only on the energy $\ell^0$ and on the
momentum perpendicular to the Fermi surface
$\ell_\parallel={\bm\ell}\cdot\b v$. Notice also that \be
\frac{\slash V} 2=\frac 1 2 \gamma^0(1-{\bm\alpha}\cdot \b
v)=P_+\gamma_0. \ee Recalling that the Dirac propagator is the
$\psi\bar\psi$ $T$-product, we see  that the propagator $\langle
T(\psi_+\psi_+^\dagger)\rangle$ at this order is just
$1/V\cdot\ell$.

\subsubsection{Integrating out the heavy degrees of freedom}

\begin{figure}[ht]
\epsfxsize=9truecm \centerline{\epsffile{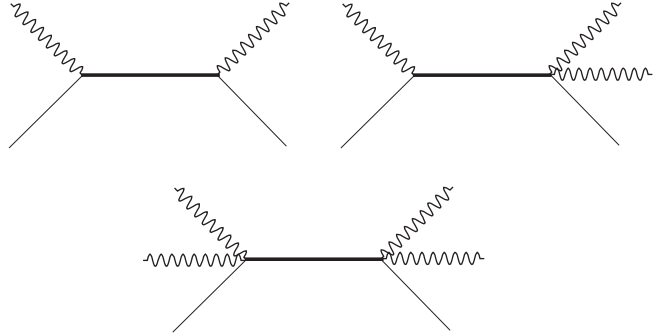}}
\noindent \caption{\it Some of the operators arising from
integrating out the heavy degrees of freedom in ${\cal L}_D^{lh}$.
The wavy lines represent the gluons, the thin lines the light
fermionic d.o.f. and the thick lines the heavy
ones\label{light_heavy}}
\end{figure}

\begin{center}
\begin{figure}[ht]
\epsfxsize=5truecm \centerline{\epsffile{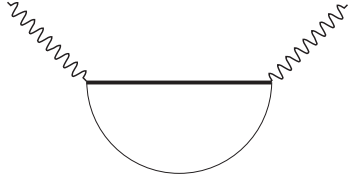}}
\noindent \caption{\it A contribution to hard gluon terms obtained
by closing the light fermion lines in the diagrams of Fig.
\ref{light_heavy}\label{heavy_loop}}
\end{figure}
\end{center}

All the steps leading to Eq. (\ref{5.42}) can be formally
performed both for light and heavy degrees of freedom. However, in
order to integrate out the heavy fields we need to write the
effective lagrangian
 at all orders in $\mu$ in the following non-local form
: \be {\cal L}=-\frac{1}{4}F_{\mu\nu}^a F^{a\mu\nu}+{\cal L}_D\
,\ee where\be {\cal L}_D=\int\frac{d\b v}{4\pi}
\left[\psi^\dagger_+ iV\cdot D\psi_+ -
\psi^\dagger_+\frac{1}{2\mu+i\tilde V\cdot D}\slash
D_\perp^2\psi_+\right]\ .\label{4.2} \ee \vskip.5cm \noindent
Using the identity \be\psi^\dagger_+
\gamma_\perp^\mu\gamma_\perp^\nu\psi_+=\psi^\dagger_+P^{\mu\nu}\psi_+,
\ee where\vskip0.1cm \be P^{\mu\nu}=g^{\mu\nu}-\frac 1
2\left[V^\mu\tilde V^\nu+V^\nu\tilde V^\mu\right]\ ,\label{219}\ee
we can write (\ref{4.2}) as:
 \be{\cal L}_D=\int\frac{d\b
v}{4\pi}\left[\psi^\dagger_+ iV\cdot D\psi_+  -
P^{\mu\nu}\psi^\dagger_+\frac{1}{2\mu+i\tilde V\cdot D}D_\mu
D_\nu\psi_+\right]\ .\label{4.2nuovo} \ee Notice that the Eq.
(\ref{6.96}) does not hold in this case. We need now to decompose
the fields in light and heavy contributions \be
\psi_+=\psi_+^l+\psi_+^h.\ee Substituting inside Eq.
(\ref{4.2nuovo}) we get \be {\cal L}_D={\cal L}_D^l+{\cal
L}_D^{lh}+{\cal L}_D^h,\ee where ${\cal L}_D^l$ is nothing but
(\ref{5.45}) at the leading order in $\mu$: \be {\cal
L}_D^l=\int\frac{d\b v}{4\pi}\psi^{l\dagger}_+ i V\cdot
D\psi_+^l,\ee whereas \be {\cal L}_D^{lh}=\int\frac{d\b
v}{4\pi}\left(\psi_+^{l\dagger}iV\cdot
D\psi_+^{h}-P^{\mu\nu}\psi^{l\dagger}_+\frac{1}{2\mu+i\tilde
V\cdot D}D_\mu D_\nu\psi_+^h ~~~+~~~(l\leftrightarrow h)\right)\ee
and \be {\cal L}_D^{h}=\int\frac{d\b
v}{4\pi}\left(\psi_+^{h\dagger}iV\cdot
D\psi_+^h-P^{\mu\nu}\psi^{h\dagger}_+\frac{1}{2\mu+i\tilde V\cdot
D}D_\mu D_\nu\psi_+^h\right).\ee When we integrate out the heavy
fields, the terms arising from ${\cal L}_D^{lh}$ produce diagrams
with two light fermions external lines and a bunch of external
gluons with a heavy propagator (see, for instance Fig.
\ref{light_heavy}).
%%%%%%%%%%%%%%%%%%%%%%%%
%\begin{center}
%\end{center}
Due to the momentum conservation they can contribute only if some
of the gluon momenta are harder than $\delta$. However the hard
gluons are suppressed by asymptotic freedom. Notice that these
terms may also give rise to pure gluon terms by closing the light
line in the loop, see Fig. \ref{heavy_loop}.
%%%%%%%%%%%%%%%%%%%%%%%%
%%%%%%%%%%%%%%%%%%%%%%%%%%%%%%%%%%%
\begin{center}
\begin{figure}[ht]
\epsfxsize=2.5truecm \centerline{\epsffile{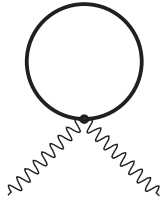}}
\noindent \caption{\it The two gluon operator arising from
integrating out the heavy degrees of freedom in ${\cal L}_D^{h}$.
The wavy lines represent the gluons, the thin lines the light
fermionic d.o.f. and the thick lines the heavy
oneslight\label{meissner_heavy}}
\end{figure}
\end{center}
%%%%%%%%%%%%%%%%%%%%%%%%%%%%%%%%%%%
On the other hand the second term in ${\cal L}_D^h$ can contribute
to an operator containing only soft gluon external lines and a
loop of heavy fermions at zero momentum (see for instance in Fig.
\ref{meissner_heavy} the contribution to the two gluon operator).
Notice that the heavy fermion  propagator comes from the first
term in ${\cal L}_D^h$ and it coincides with the propagator
evaluated for the free Fermi gas in Section \ref{IIA}. This is so
also in case of condensation since we are assuming
$\delta\gg\Delta$. Furthermore, from Eq. (\ref{2.38}) we know that
the propagator at zero momentum is nothing but the density of
states inside the Fermi surface. Therefore we expect
 an order $\mu^3$ contribution\footnote{To be precise we should
take out the volume of the portion of the shell inside the Fermi
surface that goes in the light fields definition, but since we are
taking the limit of large $\mu$ this is negligible.}. At the
leading order in $\mu$  this gives a term of the type $\mu^2 A^2$,
where $A$ is the gluon field. We will evaluate this contribution
later in Section \ref{evaluating}. Since this is a mass term for
the gluons and, as we shall see, gives a contribution only to the
spatial gluons, it will be referred to as {\bf bare Meissner
mass}. The result of this discussion is that the effective
lagrangian for the light fields is simply given by ${\cal L}^l_D$,
plus terms containing powers of soft gluon fields.

\subsubsection{The HDET in the condensed phase}
\label{VB2}

 Let us now see what happens in the case
of condensation after having integrated out the heavy fields. We
will omit from now on the superscript identifying the light fields
since these are the only fields we will deal with from now on. We
will describe color and flavor with a collective index
$A=1,\cdots, N$ ($N=N_c N_f$). The general structure of the
condensate, both for left- and right- handed fields is  (we will
neglect from now on the Weyl indices since we assume the
difermions  in the spin 0 state) \be \langle\psi^A
C\psi^B\rangle\approx \Delta_{AB},\ee with $\Delta_{AB}$ a complex
symmetric matrix.

We will now consider a four-fermi interaction of the BCS type
\be{\cal L}_I=-\frac G 4\epsilon_{ab}\epsilon_{\dot a\dot
b}V_{ABCD}\psi_a^A\psi_b^B\psi_{\dot a}^{C\dagger}\psi_{\dot
b}^{D\dagger}.\label{5.54}\ee We require ${\cal L}_I$ to be
hermitian, therefore \be V_{ABCD}=V_{CDAB}^*\ee and furthermore
\be V_{ABCD}=V_{BACD}=V_{ABDC}.\ee As we have done in Section
\ref{IIID} we write  \be {\cal L}_I={\cal L}_{cond}+{\cal
L}_{int},\ee with \be {\cal L}_{cond}=\frac  G
4V_{ABCD}\Gamma^{CD*}\psi^{AT}C\psi^B-\frac G 4
V_{ABCD}\Gamma^{AB}\psi^{C\dagger}C\psi^{D*}\ee and \be{\cal
L}_{int}=-\frac G 4 V_{ABCD}
\left(\psi^{AT}C\psi^B-\Gamma^{AB}\right)
\left(\psi^{C\dagger}C\psi^{D*}+\Gamma^{CD*}\right).\ee We now
define \be \Delta_{AB}=\frac
G2\,V_{CDAB}\Gamma^{CD},~~~\Delta^*_{AB}=\frac G
2V_{CDAB}^*\Gamma^{CD*}=\frac G 2\, V_{ABCD}\Gamma^{CD*}.\ee
Clearly $\Delta_{AB}$ is a symmetric matrix. We will assume also
that it can be diagonalized, meaning that \be
[{\bm\Delta},{\bm\Delta}^\dagger]=0,\ee where
${\bm\Delta}_{AB}=\Delta_{AB}$. Therefore ($C$ is the charge
conjugation matrix, $C=i\sigma_2$)\be{\cal L}_{cond}=\frac 1
2\Delta_{AB}^*\psi^{AT}C\psi^B-\frac 1 2\Delta_{AB}
\psi^{A\dagger}C\psi^{B*}.\ee

For the following it is convenient to introduce the following
notation for the positive energy fields: \be
\psi_\pm(x)=\psi_+(\pm\b v,x).\ee One should be careful in not
identifying $\psi_-$ with the negative energy solution with
velocity $\b v$. The condensation is taken into account by adding
${\cal L}_{cond}$ to the effective lagrangian of the previous
Section (here we consider only the leading term), see Sections
\ref{IIIC} and \ref{IIID}. Therefore we will assume the following
Lagrangian \be {\cal L}_D=\int \frac{d\b
v}{4\pi}\sum_{A,B}\left[\psi_+^{A\dagger}(iV\cdot
D)_{AB}\psi_+^B+\frac 1 2\psi_-^AC\psi_+^B\Delta_{AB}^*- \frac 1 2
\psi_+^{A\dagger}C\psi_-^{B*}\Delta_{AB}\right].\ee Using the
symmetry $\b v\to -\b v$ of the velocity integration we may write
the previous expression in the form \be {\cal L}_D=\int \frac{d\b
v}{4\pi}\sum_{A,B}\frac 1 2\left[\psi_+^{A\dagger}(iV\cdot
D)_{AB}\psi_+^B+\psi_-^{A\dagger}(i\tilde V\cdot
D)_{AB}\psi_-^B+\psi_-^AC\psi_+^B\Delta_{AB}^*-
\psi_+^{A\dagger}C\psi_-^{B*}\Delta_{AB}\right].\ee We then
introduce the Nambu-Gor'kov basis \be \chi^A=\frac
1{\sqrt{2}}\left(\matrix{\psi_+^A\cr C\psi_-^{A*}}\right)\ee in
terms of which \be {\cal L}_D=\int\frac{d\b
v}{4\pi}\chi^{A\dagger}\left[\matrix{iV\cdot D_{AB}
&-\Delta_{AB}\cr -\Delta_{AB}^*&i\tilde V\cdot
D_{AB}^*}\right]\chi^B.\ee The lagrangian we have derived here
coincides with the one that we obtained in Section \ref{IIID},
that is the lagrangian giving rise to the Nambu-Gor'kov equations.

The inverse free propagator in operator notations is (notice that
since ${\bm\Delta}$ is symmetric we have
${\bm\Delta}^*={\bm\Delta}^\dagger$) \be
S^{-1}(\ell)=\left(\matrix{V\cdot \ell & -{\bm\Delta}\cr
-{\bm\Delta}^\dagger & \tilde V\cdot \ell}\right).\ee From which
\be S(\ell)= \frac{1}{(V\cdot \ell)(\tilde V\cdot
\ell)-{\bm\Delta}{\bm\Delta}^\dagger}\left(\matrix{\tilde V\cdot
\ell & {\bm\Delta}\cr {\bm\Delta}^\dagger & V\cdot
\ell}\right).\ee

Let us now consider the relation \be\Delta_{AB}^*=-\frac G 2
V_{ABCD}\langle\psi^{C\dagger} C\psi^{D*}\rangle.\ee This Equation
is the analogous of Eq. (\ref{3.113}) in configuration space.
Repeating the same steps leading to Eq. (\ref{consistency})  we
find \be \Delta_{AB}^*=i\,  2\times\frac G 2 V_{ABCD}\int
\frac{d\b v}{4\pi}\frac{\mu^2}{\pi}\int
\frac{d^2\ell}{(2\pi)^2}\Delta^*_{CE}\frac 1{D_{ED}},\ee where we
have made use of Eq. (\ref{5.25}) and \be \frac
1{D_{AB}}=\left(\frac{1}{(V\cdot \ell)(\tilde V\cdot
\ell)-{\bm\Delta}{\bm\Delta}^\dagger}\right)_{AB}.\ee Notice that
the factor 2 arises from the sum over the Weyl indices. The final
result is  \be \Delta_{AB}^*=i  G
 V_{ABCD}\int \frac{d\b v}{4\pi}\frac{\mu^2}{\pi}\int
\frac{d^2\ell}{(2\pi)^2}\Delta^*_{CE}\frac
1{D_{ED}}\label{5.73},\ee in agreement with the result we found
from the Nambu-Gor'kov equations by identifying $E$ with $\ell_0$
and $d^3\b p$ with the integration over the velocity and
$\ell_\|$. The same result can be found via the functional
approach, see  Appendix \ref{Appendix1}.

Let us apply this formalism to the case of an effective four-fermi
interaction due to one gluon exchange \be {\cal L}_I=\frac
3{16}G\bar\psi\gamma_\mu\lambda^a\psi\bar\psi\gamma^\mu\lambda^a\psi,\ee
where $\lambda^a$ are the Gell-Mann matrices. Using the following
identities
\be\sum_{a=1}^8(\lambda^a)_{\alpha\beta}(\lambda^a)_{\delta\gamma}=\frac
2 3\left(3\delta_{\alpha\gamma}\delta_{\beta\delta}-
\delta_{\alpha\beta}\delta_{\gamma\delta}\right)\ee and
\be(\sigma_\mu)_{\dot a b}(\tilde\sigma^\mu)_{d\dot
c}=2\epsilon_{\dot a \dot c}\epsilon_{bd}\ee where
\be\sigma^\mu=(1,{\bm\sigma}),~~~~~~
\tilde\sigma^\mu=(1,-{\bm\sigma}),\ee with ${\bm\sigma}$ the Pauli
matrices, we find \be {\cal L}_I=-\frac G 4 V_{(\alpha i)(\beta
j)(\gamma k)(\delta
\ell)}\psi_i^\alpha\psi_j^\beta\psi_k^{\gamma\dagger}\psi_\ell^{\delta\dagger}\ee
with \be V_{(\alpha i)(\beta j)(\gamma k)(\delta
\ell)}=-(3\delta_{\alpha\delta}\delta_{\beta\gamma}-
\delta_{\alpha\gamma}\delta_{\beta\delta})\delta_{ik}\delta_{j\ell}.\ee
Let us now consider the simple case of the 2SC phase. Then, as we
have seen, \be \Delta_{(\alpha i)(\beta j)}=\epsilon_{\alpha\beta
3}\epsilon_{ij}\Delta.\ee We get at once
\be\Delta=4iG\int\frac{d\b v}{4\pi}\frac{\mu^2}\pi\int
\frac{d^2\ell}{(2\pi)^2}\frac{\Delta}{\ell_0^2-\ell_\|^2-\Delta^2},
\ee with $\ell_\|=\b v\cdot{\bm\ell}$. Performing the integration
over $\ell_0$ we obtain \be\Delta=\frac G 2\,\rho\int_0^\delta
d\xi\frac{\Delta}{\sqrt{\xi^2+\Delta^2}}.\ee Here we have defined
the density of states as \be \rho=\frac {4\mu^2}{\pi^2},\ee which
is the appropriate one for this case. In fact remember that in the
BCS case the density is defined as $p_F^2/v_F\pi^2$. In the actual
case $p_F=E_F=\mu$ and $v_F=1$. The factor 4 arises since  there
are 4 fermions, $\psi_i^\alpha$ with $\alpha=1,2$, which are
pairing.

It is worth to note that an alternative approach is to work
directly with the Schwinger-Dyson equation
\cite{Rajagopal:2000wf}. In that case there is no necessity to
Fierz transform the interaction term. The reason we have not
adopted this scheme, although technically more simple, is just to
illustrate more the similarities with the condensed matter
treatment.

In order to make an evaluation of the gap one can fix the coupling
by the requirement that this theory reproduces the chiral
phenomenology in the limit of zero density and temperature. It is
not difficult to show that the chiral gap equation is given by \be
1=8 G\int_0^\Lambda \frac{d^3\b p}{(2\pi)^3}\frac 1
{\sqrt{p^2+M^2}}.\ee Here $\Lambda$ is the Nambu Jona-Lasinio
cutoff and $M$ is the constituent mass. Correspondingly one
chooses $\delta=\Lambda-\mu$ in the gap equation at finite
density. By choosing typical values of $\Lambda=800~MeV$,
$M=400~MeV$ and $\mu=400\div 500~MeV$ one finds respectively
$\Delta=39\div 88~MeV$\footnote{For a different way of choosing
the cutoff in NJL models, see \cite{Casalbuoni:2003cs}}. Similar
values are also found in the CFL case.

\subsection{The gap equation in QCD}\label{VD}

Having discussed the gap equation in the context of a four-fermi
interaction we will now discuss the real QCD case. However this
calculation has real meaning only at extremely high densities much
larger than 10$^8$ $MeV$, see \cite{Rajagopal:2000rs}. Therefore
we will give only a brief sketch of the main results. We will work
in the simpler case of the 2SC phase.

%%%%%%%%%%%%%%%%%%%%%%%%%%%%%%%%%%%
\begin{center}
\begin{figure}[ht]
\epsfxsize=10truecm \centerline{\epsffile{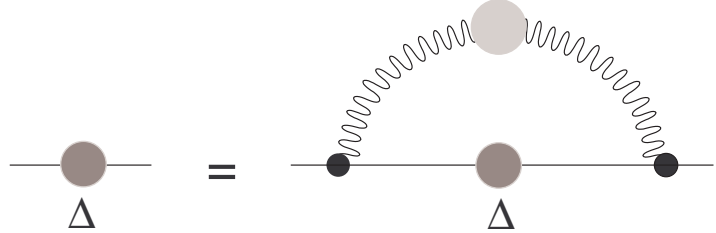}}
\noindent \caption{\it The gap equation in QCD at finite density.
The light grey circle denotes the gluon self-energy insertion. The
dark grey ones a gap insertions. At the leading order there is no
need of vertex corrections \label{schafer}}
\end{figure}
\end{center}
%%%%%%%%%%%%%%%%%%%%%%%%%%%%%%%%%%%%%%%%%%%%%%%%%%%%%%%%%%%%%%%%%%%%

  As shown in Appendix \ref{Appendix1} the gap equation can be
  obtained simply by writing down the Schwinger-Dyson equation, as
  shown in Fig. \ref{schafer}. The diagram has been evaluated in
\cite{Son:1998uk,Schafer:1999jg,Pisarski:1999tv,Hong:1999fh,Brown:1999aq}.
The result in euclidean space is \bea  \Delta(p_0) &=&
\frac{g^2}{12\pi^2} \int dq_0\int d\cos\theta\,
 \left(\frac{\frac{3}{2}-\frac{1}{2}\cos\theta}
            {1-\cos\theta+G/(2\mu^2)}\right.\nn\\
    &+&\left.\frac{\frac{1}{2}+\frac{1}{2}\cos\theta}
            {1-\cos\theta+F/(2\mu^2)} \right)
 \frac{\Delta(q_0)}{\sqrt{q_0^2+\Delta(q_0)^2}}.\label{eliash}
\eea Here, $\Delta(p_0)$ is the energy dependent gap, $g$ is the
QCD coupling constant and $G$ and $F$ are the self energies of
magnetic and electric gluons.  The terms in the curly brackets
arise from the magnetic and electric components of the gluon
propagator. The numerators are the on-shell matrix elements ${\cal
M}_{ii,00}=[\bar{u}_h(p_1)
\gamma_{i,0}u_h(p_3)][\bar{u}_h(p_2)\gamma_{i,0}u_h(p_4)]$ for the
scattering of back-to-back fermions on the Fermi surface. The
scattering angle is $\cos\theta=\b{p}_1\cdot\b{p}_3$. In the case
of a spin zero order parameter, the helicity $h$ of all fermions
is the same (see \cite{Schafer:1999jg}).

The important difference between Eq.~(\ref{eliash}) and the case
of a contact four-fermi interaction is due to the fact that the
self-energy of the magnetic gluons vanishes at zero energy.
Therefore the gap equation contains a collinear $\cos\theta \sim
1$ divergence.  One makes use of the hard-loop approximation
\cite{LeBellac:1996ab} and for $q_0\ll |\b{q}|\to 0$ and to
leading order in perturbation theory we have \be \label{pi_qcd}
 F = m_D^2, \hspace{1cm}
 G = \frac{\pi}{4}m_D^2\frac{q_0}{|\vec{q}|},
\ee with \be m_D^2=N_f\frac{g^2\mu^2}{2\pi^2}.\ee In the electric
part, $m_D^2$ is the  Debye screening mass. In the magnetic part,
there is no screening of static modes, but non-static modes are
dynamically screened due to Landau damping.  For small energies
dynamic screening of magnetic modes is much weaker than Debye
screening of electric modes. As a consequence, perturbative color
superconductivity is dominated by magnetic gluon exchanges. We are
now able to perform the angular integral in Eq.~(\ref{eliash})
finding\be \label{eliash_mel} \Delta(p_0) = \frac{g^2}{18\pi^2}
\int dq_0
 \log\left(\frac{b\mu}{|p_0-q_0|}\right)
    \frac{\Delta(q_0)}{\sqrt{q_0^2+\Delta(q_0)^2}},
\ee with \be b=256\pi^4(2/N_f)^{5/2}g^{-5}.\ee This equation has
been derived for the first time in QCD in Ref. \cite{Son:1998uk}.
In ordinary superconductivity it was realized by
\cite{Eliashberg:1960ab} that  the effects of retardation of
phonons (taking the place of the gluons) are important and that
they produce the extra logarithmic term in the gap equation. The
integral equation we have obtained can be converted to a
differential equation \cite{Son:1998uk} and in the weak coupling
limit an approximate solution is \cite{Son:1998uk,Pisarski:1999av,
Pisarski:1999bf,Pisarski:1999tv,Schafer:1999jg,Brown:1999aq,Brown:1999yd,
Brown:2000eh} \be\Delta(p_0)\approx\Delta_0\sin\left(\frac
g{3\sqrt{2}\pi}\log\left(\frac{b}{p_0}\right)\right),\ee with
\be\Delta_0=2b\mu\exp\left(-\frac{3\pi^2}{\sqrt{2}g}\right).\ee
This result shows why it is important to keep track of the energy
dependence of $\Delta$. In particular neglecting the energy
dependence would give a wrong coefficient in the exponent
appearing in $\Delta_0$. Also we see that the collinear divergence
leads to a gap equation with a double-log behavior. Qualitatively
\be \label{dlog}
 1 \sim \frac{g^2}{18\pi^2}
 \left[\log\left(\frac{\mu}{\Delta}\right)\right]^2,
\ee from which we conclude that $\Delta\sim\exp(-c/g)$. The
prefactor in the expression for $\Delta_0$ is not of easy
evaluation. By writing \be \label{gap_oge} \Delta_0 \simeq
512\pi^4(2/N_f)^{5/2}b_0'\mu g^{-5}
   \exp\left(-\frac{3\pi^2}{\sqrt{2}g}\right).
\ee we have $b_0'=1$ in the previous case, whereas with different
approximations in \cite{Brown:1999yd,Wang:2001aq} it has been
found\be b_0'=\exp\left(-\frac{4+\pi^2}{8}\right).\ee Numerically
one finds at $\mu=10^{10}~MeV$, $g=.67$ and $b_0'=2/5$,
$\Delta_0\approx 40~MeV$. Extrapolating at $\mu=400~MeV$,
$g=3.43$, one finds $\Delta_0\approx 90~MeV$. It turns out that
$\Delta_0$ decreases from 90 $MeV$ to about 10 $MeV$ for $\mu$
increasing from 400 $MeV$ to $10^6$ $MeV$. Continuing to increase
$\mu$ $\Delta$ increases as it should be according to its
asymptotic value. In fact we see that for increasing $\mu$,
$\Delta$ increases although $\Delta/\mu\to 0$. In particular we
notice that neither the four-fermi interaction approach and the
present one from first principles can be trusted at
phenomenologically interesting chemical potentials of the order
$400\div 500~MeV$. Still it is of some interest to observe that
both methods lead to gaps of the same order of magnitude.

\subsection{The symmetries of the superconductive
phases}\label{VC}

\subsubsection{The CFL phase}

As we have already discussed we expect that at very high density
the condensate $\langle\psi_{iL}^\alpha C\psi_{j L}^\beta\rangle$
is antisymmetric in color and in flavor. Also, if we require
parity invariance we have \be\langle\psi_{iL}^\alpha \psi_{j
L}^\beta\rangle=-\langle\psi_{iR}^\alpha \psi_{j
R}^\beta\rangle.\ee In fact in Dirac notation we notice that
\be\langle\psi^TC\gamma_5\psi\rangle\ee is parity invariant ($\psi
\to\eta_P\gamma_0\psi$) and its L and R components are precisely
$\langle\psi_L\psi_L\rangle$ and $\langle\psi_R\psi_R\rangle$. The
required antisymmetry implies \be\langle\psi_{iL}^\alpha \psi_{j
L}^\beta\rangle=\epsilon^{\alpha\beta\gamma}\epsilon_{ijk}
A^k_{\gamma},\ee with $A$ a $3\times 3$ matrix. In
\cite{Evans:1999at} it has been shown that $A$ can be diagonalized
by using an $SU(3)_c\otimes SU(3)_L$ global rotation \be A\to
g_{SU(3)_c} A g_{SU(3)_L}=A_D,\ee with \be A_D=\left(\matrix{a & 0
& 0\cr 0 & b & 0\cr 0&0& c}\right).\ee Studying the gap equation
(this calculation has been done in full QCD) it can be seen that
three cases are possible \bea (1,1,1)&:&~~~~A_D=
\left(\matrix{1 & 0 & 0\cr 0 & 1 & 0\cr 0&0& 1}\right)\nn\\
(1,1,0)&:&~~~~A_D=
\left(\matrix{1 & 0 & 0\cr 0 & 1 & 0\cr 0&0& 0}\right)\nn\\
(1,0,0)&:&~~~~A_D= \left(\matrix{1 & 0 & 0\cr 0 & 0 & 0\cr
0&0&0}\right).\eea The corresponding gaps satisfy \be
\Delta_{(1,1,1)}<\Delta_{(1,1,0)}<\Delta_{(1,0,0)},\ee but for the
free energies the result is \be
F_{(1,1,1)}<F_{(1,1,0)}<F_{(1,0,0)}.\ee The reason is that
although the gaps for the less symmetric solution is bigger, there
are more fermions paired in the more symmetric configurations.

The analysis of the gap equation shows that there is also a
component from the color channel ${\bm 6}$. In fact, under the
group $SU(3)_c\otimes SU(3)_L\otimes SU(3)_R$ we have
\be\psi_{iL(R)}^\alpha\in({\bm 3}_c,{\bm 3}_{L(R)}).\ee Therefore
\be \left[({\bm 3}_c,{\bm 3}_{L(R)})\otimes({\bm 3}_c,{\bm
3}_{L(R)})\right]_S=({\bm 3}_c^*,{\bm 3}_{L(R)}^*)\oplus ({\bm
6}_c,{\bm 6}_{L(R)}).\ee Here the index $S$ means that we have to
take the symmetric combination of the tensor product since
$\langle\psi_L^T C\psi_L\rangle$ is already antisymmetric in spin.
From the previous argument and in the absence of the $({\bm
6}_c,{\bm 6}_{L(R)})$ component we have
\be\langle\psi_{iL}^\alpha\psi_{jL}^{\beta}\rangle=\Delta\epsilon^{\alpha\beta
I}\epsilon_{ijI}.\ee The presence of the $({\bm 6}_c,{\bm
6}_{L(R)})$ implies
\be\langle\psi_{iL}^\alpha\psi_{jL}^{\beta}\rangle=\Delta\epsilon^{\alpha\beta
I}\epsilon_{ijI}+\Delta_6
(\delta^\alpha_i\delta^\beta_j+\delta^\beta_i\delta^\alpha_j),\ee
or
\be\langle\psi_{iL}^\alpha\psi_{jL}^{\beta}\rangle=\Delta(\delta^\alpha_i\delta^\beta_j-
\delta^\beta_i\delta^\alpha_j)+\Delta_6
(\delta^\alpha_i\delta^\beta_j+\delta^\beta_i\delta^\alpha_j)=(\Delta+\Delta_6)
\delta^\alpha_i\delta^\beta_j+(\Delta_6-\Delta)\delta_i^\beta\delta_j^\alpha.\ee
Notice that the $({\bm 6}_c,{\bm 6}_{L(R)})$ term does not break
any further symmetry other than the ones already broken by the
$({\bm 3}_c^*,{\bm 3}_{L(R)}^*)$. Numerically $\Delta_6$ turns out
to be quite small. The analysis has been done in
\cite{Alford:1998mk} using  the parameters $\Delta_8$ and
$\Delta_1$, defined as \be \Delta+\Delta_6=\frac 1
3\left(\Delta_8+\frac 1 8\Delta_1\right),~~~\Delta-\Delta_6=-\frac
1 8 \Delta_1.\ee Therefore \be \Delta=\frac 1
3\left(\Delta_8-\Delta_1\right),~~~\Delta_6=\frac 1
6\left(\Delta_8+\frac 1 2\Delta_1\right).\ee The absence of the
sextet is equivalent to require $\Delta_1=-2\Delta_8$. The result
found in \cite{Alford:1998mk}, for the choice of parameters,
$\Lambda=800~MeV$, $M=400~MeV$ and $\mu =400~MeV$ is \be
\Delta_8=80~MeV,~~~~\Delta_1=-176~ MeV,\ee implying
\be\Delta=85.3~MeV,~~~\Delta_6=-1.3~MeV.\ee

As already discussed the original symmetry of the theory is \be
G_{QCD}=SU(3)_c\otimes SU(3)_L\otimes SU(3)_R\otimes U(1)_B.\ee
The first factor is local whereas the other three are global
since, for the moment, we are neglecting the em interaction which
makes local a $U(1)$ subgroup of $SU(3)_L\otimes SU(3)_R$. The
condensates lock together the transformations of $SU(3)_c$,
$SU(3)_L$ and $SU(3)_R$, therefore the symmetry of the CFL phase
is \be G_{\rm CFL}=SU(3)_{c+L+R}\otimes Z_2.\ee In fact, also the
$U(1)_B$ group is broken leaving a $Z_2$ symmetry corresponding to
the multiplication of the quark fields by -1. Also, locking
$SU(3)_L$ and $SU(3)_R$ to $SU(3)_c$ makes $SU(3)_L$ and $SU(3)_R$
lock together producing the breaking of the chiral symmetry. The
breaking of $G_{QCD}$ to $G_{\rm CFL}$ gives rise to \be 3\times
8+1-8=8+8+1\ee Nambu-Goldstone (NB) bosons. However 8 of the NG
bosons disappear from the physical spectrum through the Higgs
mechanism, giving masses to the 8 gluons, whereas 8+1 massless NG
bosons are left in the physical spectrum. Of course, these NG
bosons take mass due to the explicit breaking of the symmetry
produced by the quark masses. Since all the local symmetries are
broken (but see later as far the electromagnetism is concerned)
all the gauge bosons acquire a mass.

Although the baryon number is broken no dramatic event takes
place. The point is that we are dealing with a finite sample of
superconductive matter. In fact, applying the Gauss' law to a
surface surrounding the sample we find that changes of the baryon
number inside the superconductor must be accompanied by
compensating fluxes. In other words, inside the sample there might
be large fluctuations and transport of baryonic number. Things are
not different from what happens in ordinary superconductors where
the quantum number, number of electrons (or lepton number), is not
conserved. The connection between the violation of quantum numbers
and phenomena of supertransport as superfluidity and
superconductivity are very strictly related.

The condensate $\langle\psi_{iL}^\alpha\psi_{jL}^{\beta}\rangle$
is not gauge invariant and we may wonder if it is possible to
define gauge invariant order parameters. To this end let us
introduce the matrices \be
X_\gamma^k=\langle\psi_{iL}^\alpha\psi_{jL}^\beta\rangle^*\epsilon_{\alpha\beta\gamma}
\epsilon^{ijk},~~~Y_\gamma^k=\langle\psi_{iR}^\alpha\psi_{jR}^\beta\rangle^*\epsilon_{\alpha\beta\gamma}
\epsilon^{ijk}.\label{5.100}\ee The conjugation has been
introduced for convenience reasons (see in the following). The
matrix \be (Y^\dagger
X)_j^i=\sum_\alpha(Y_\alpha^j)^*X_\alpha^i\ee is gauge invariant,
since color indices are saturated and breaks $SU(3)_L\otimes
SU(3)_R$ to $SU(3)_{L+R}$. Analogously the 6-fermion operators \be
{\rm det}(X),~~~{\rm det}(Y)\ee are gauge invariant flavor singlet
and break $U(1)_B$.

So far we have neglected $U(1)_A$. This group is broken by the
anomaly. However, the anomaly is induced by a 6-fermion operator
(the 't Hooft determinant) which becomes irrelevant at the Fermi
surface. On the other hand this operator is qualitatively
important since it is the main cause of the breaking of $U(1)_A$.
If there was no such operator we would have a further NG boson
associated to the spontaneous breaking of $U(1)_A$ produced by the
di-fermion condensate. Since the instanton contribution is
parametrically small at high density, we expect the NG boson to be
very light. Notice that the   $U(1)_A$ approximate symmetry is
broken by the condensate to a discrete group $Z_2$.

\vskip0.5cm \noindent{\bf The spectrum of the CFL phase}
\vskip0.5cm

Let us spend some word about the spectrum of QCD in the CFL phase.
We start with the fermions. Since all the fermions are paired they
are all gapped. To understand better this point let us notice that
under the symmetry group of the CFL phase $SU(3)_{c+L+R}$ quarks
transform as ${\bm 1}\oplus{\bm 8}$. Therefore it is useful to
introduce the basis \be\psi_i^\alpha=\frac 1
{\sqrt{2}}\sum_{A=1}^9(\lambda_A)_i^\alpha\psi^A,\label{5.103}\ee
where $\lambda_A$, $A=1,\cdots, 8$ are the Gell-Mann matrices and
\be\lambda_9=\lambda_0=\sqrt{\frac 2 3}\times {\bm 1},\ee with
${\bm 1}$ the identity matrix in the $3\times 3$ space. With this
normalization \be {\rm Tr}(\lambda_A\lambda_B)=2\delta_{AB}.\ee
Inverting Eq. (\ref{5.103}) \be \psi^A=\frac 1
{\sqrt{2}}\sum_{\alpha i}({\lambda_A})_\alpha^i\psi_i^\alpha=\frac
1 {\sqrt{2}}{\rm Tr}(\lambda_A\psi),\ee we obtain
\be\langle\psi^A\psi^B\rangle=\frac 1
2\sum({\lambda_A})_\alpha^i(\lambda_B)_\beta^j\Delta\epsilon^{\alpha\beta
I}\epsilon_{ijI}=\frac \Delta 2{\rm
Tr}\sum_I\left(\lambda_A\epsilon_I\lambda_B^T\epsilon_I\right),\ee
where we have defined the following three, $3\times 3$, matrices
\be (\epsilon_I)_{\alpha\beta}=\epsilon_{\alpha\beta
I}\label{5.125},\ee which have the following property valid for
any $3\times 3$ matrix $g$: \be\sum_I\epsilon_I
g^T\epsilon_I=g-{\rm Tr }[g]\label{5.126}.\ee This is a simple
consequence of the definition of the $\epsilon_I$ matrices. Using
this equation it follows \be
\langle\psi^A\psi^B\rangle=\Delta_A\delta_{AB},\ee with \be
\Delta_{A}=\left\{\matrix{A=1,\cdots, 8 &\Delta_A=\Delta,\cr A=9 &
\Delta_9=-2\Delta,}\right.\ee There are two gaps, one for the
octet ($A=1,\cdots,8$) and  a larger one for the singlet ($A=9$).
In each of these cases the dispersion relation for the
quasi-fermions is \be\epsilon_A(\b p)=\sqrt{(\b
v\cdot{\bm\ell})^2+\Delta_A^2}. \ee

The next category of particles are the gluons which acquire a mass
through the Higgs mechanism. Therefore we expect \be m_g^2\approx
g_s^2 F^2,\ee with $g_s$ the coupling constant of QCD and $F$ the
coupling constant of the NG bosons of the CFL phase. However the
situation is more complicated than this due to two different
effects. First there is a very large wave-function renormalization
making the mass of the gluons proportional to $\Delta$. Second,
the theory is not relativistically invariant. In fact being at
finite density implies a breaking of the Lorentz group down to
$O(3)$, the invariance under spatial rotations.

The last category of particles are the NG bosons. As already
discussed we expect 9 massless NG bosons from the breaking of
$G_{QCD}$ to $G_{\rm CFL}$, plus a light NG boson from the
breaking of $U(1)_A$. Again, this is not the end of the story,
since quarks are not massless. As a consequence the  NG bosons of
the octet acquire mass. The NG boson associated to the breaking of
$U(1)_B$ remain massless since this symmetry is not broken by
quark masses. However the masses of the NG bosons of the octet are
parametrically small since their square turns out to be quadratic
in the quark masses. The reason comes from the approximate
symmetry $(Z_2)_L\otimes(Z_2)_R $ defined by \be \matrix{(Z_2)_L &
\psi_L\to -\psi_L\cr (Z_2)_R & \psi_R\to -\psi_R.}\ee This is an
approximate symmetry since only the diagonal $Z_2$ is preserved by
the axial anomaly. However the quark mass term \be \bar\psi_L
M\psi_R+{\rm h.c.}\ee is such that $M\to - M$ (here we are
treating $M$ as a spurion field) under the previous symmetry.
Therefore the mass square for the NG bosons in the CFL phase must
be quadratic in $M$. The anomaly breaks $(Z_2)_L\otimes(Z_2)_R$
through the instantons. As a result a condensation of the type
$\langle\bar\psi_L\psi_R\rangle$ is produced.  This is because 4
of the  6 fermions in the 't Hooft determinant
\cite{'tHooft:1976up,Shifman:1980uw,Schafer:1998wv} condensate
through the operators $X$ and $Y$ (see Eq. (\ref{5.100})), leaving
a condensate of the form $\langle\bar\psi_L\psi_R\rangle$
\cite{Alford:1998mk,Rapp:1999qa,Schafer:1999fe}. However it has
been shown that this contribution is very small, of the order
$(\Lambda_{QCD}/\mu)^8$ \cite{Manuel:2000wm,Schafer:2000ew}.
\vskip0.5cm\noindent{\bf In-medium electric charge}\vskip0.5cm

{The em interaction is included noticing that {$U(1)_{\rm
em}\subset SU(3)_L\otimes SU(3)_R$} and extending the covariant
derivative}\be D_\mu \psi=\de_\mu \hat \psi-ig_\mu^a T_a\psi
-i\psi Q A_\mu\nn,\ee where $T_a=\lambda_a/2$ with $\lambda_a$ the
Gell-Mann matrices. {The condensate breaks $U(1)_{\rm em}$ but
leaves invariant a combination of $Q$ and of the color generator.
The result can be seen immediately in terms of the matrices $X$
and $Y$ introduced before. In fact the CFL vacuum is defined by
\be X_\alpha^i=Y_\alpha^i=\delta_\alpha^i.\ee Defining
($T_8=(1,1,-2)/2\sqrt{3}$) \be Q_{SU(3)_c}\equiv -\frac
2{\sqrt{3}}\,T_8={\rm diag}(-1/3,-1/3,+2/3)=Q,\ee we see that the
combination \be Q_{SU(3)_c}\otimes {\bm 1}-{\bm 1}\otimes Q=
Q\otimes {\bm 1}-{\bm 1}\otimes Q\ee leaves invariant the
condensates \be Q\langle \hat X\rangle -\langle \hat X\rangle Q
~{\to}~ Q_{\alpha\beta}\delta_{\beta i}-\delta_{\alpha j}
Q_{ji}=0.\ee Therefore the in-medium conserved electric charge is
\be\tilde Q={\bf 1}\otimes Q-Q\otimes {\bf 1}.\ee {The eigenvalues
of $\tilde Q$ are {$0,\pm 1$} as in the old Han-Nambu model}.

{The in-medium em field $A_\mu$ and the gluon field $g_\mu^8$ get
rotated to new fields {$\tilde A_\mu$} and {$\tilde G_\mu$}} \bea
A_\mu=\tilde
A_\mu\cos\theta -\tilde G_\mu\sin\theta,\nn\\
g_\mu^8=\tilde A_\mu\sin\theta+\tilde G_\mu\cos\theta,\eea with
new interactions  \be g_s g_\mu^8 T_8\otimes  1+e A_\mu 1\otimes
Q~~{\to}~~{\tilde e}\tilde Q\tilde A_\mu+g'_s \tilde G\tilde T,\ee
{where} \bea \tan\theta=\dd\frac 2{\sqrt{3}}\frac e{g_s},~~
{\tilde e
=e\cos\theta},~~g'_s=\dd\frac{g_s}{\cos\theta},\nn\\
\tilde T=-\dd\frac{\sqrt{3}}2\left[(\cos^2\theta) \,Q\otimes
1+(\sin^2\theta)\,1\otimes Q\right].\eea Therefore the photon
associated with the field $\tilde A_\mu$ remains massless, whereas
the gluon associated to $g_\mu^8$ becomes massive due to the
Meissner effect.

We shall show that also gluons and NG bosons have integer charges,
$0,\pm 1$. Therefore all the elementary excitations are integrally
charged.

It is interesting to consider a sample of CFL material. If quarks
were massless there would be charged massless NG bosons and the
low-energy em response would be dominated by these modes. This
would look as a "bosonic metal". However quarks are massive and so
the charged NG bosons. Therefore, for quarks of the same mass, the
CFL material would look like as a transparent insulator with no
charged excitations at zero temperature (transport properties of
massive excitations are exponentially suppressed at zero
temperature). However making quark masses different makes the
story somewhat complicated since for the equilibrium one needs a
non zero density of electrons or a condensate of charged kaons
\cite{Schafer:2000ew}. In both cases there are massless or almost
massless excitations. \vskip0.5cm \noindent{\bf Quark-Hadron
Continuity} \vskip0.5cm

The main properties we have described so far of the CFL phase are:
confinement (integral charges), chiral symmetry breaking to a
diagonal subgroup and baryon number superfluidity (due to the
massless NG boson). If not for the $U(1)_B$ NG boson these
properties are the same as the hadronic phase of three-flavor QCD:
\bea\text{{CFL phase}:} ~~~U(1)_B~~~\text{broken} \to
{NGB}\nn\\
\text{{hadr. phase at} }~ T=\mu=0: ~U(1)_B~\text{unbroken}\nn\eea

\begin{table}[htb]
\begin{center}
\begin{tabular}{|c|c|}
\hline &\\{CFL phase} & {Hypernuclear phase}\\&\\
\hline &\\{$\psi^i_\alpha{\langle D^\alpha_k\rangle}$} &
{$B^i_k=\psi^i_\alpha D^\alpha_k$}
\\&\\ \hline &\\ {${\langle
(D^*)^i_\alpha\rangle} g^\alpha_\beta {\langle
D^\beta_k\rangle}$}&{$(D^*)^i_\alpha g^\alpha_\beta D^\beta_k$}\\
&\\ \hline &\\~~ {Mesons = phases of {$(D^*)^i_{\alpha L}
D^\alpha_{j R}$}}~~ &~~{Mesons = phases of} {$\bar\psi^\alpha_{j
L}\psi^i_{\alpha R}$}~~
\\&\\
 \hline
\end{tabular}
\end{center}
\caption {\label{complementarity}{\it This table shows the
complementarity of the CFL phase and of the hypernuclear phase.
The diquark fields $D_i^\alpha$ are defined as
$D_i^\alpha=\epsilon_{ijk}
\epsilon^{\alpha\beta\gamma}\psi_\beta^j\psi_\gamma^k$.}}
\end{table}

\noindent{ The NGB makes the CFL phase a superfluid}. For
3-flavors a dibaryon condensate, {H}, of the type $(udsuds)\approx
\det(X)$ is possible \cite{Jaffe:1977ab}. This may arise {{at
$\mu$ such that the Fermi momenta of the baryons in the octet are
similar}} allowing pairing in strange, isosinglet dibaryon states
of the type ($p\Xi^-$, $n\Xi^0$, $\Sigma^+\Sigma^-$,
$\Sigma^0\Sigma^0$, $\Lambda\Lambda$) (all of the type $udsuds$).
This would be again a {superfluid phase}. The symmetries of this
phase, called {hypernuclear matter phase}, are the same as the
ones in CFL. Therefore there is {no need of phase transition
between hypernuclear matter and CFL phase} \cite{Schafer:1998ef}.
This is strongly suggested by complementarity idea.
Complementarity refers to gauge theories with a one-to-one
correspondence between the spectra of the physical states in the
Higgs and in the confined phases, see
\cite{Banks:1979ab,Fradkin:1979ab} for $U(1)$ theories and
\cite{Hooft:1979ab,Dimopoulos:1980ab,Dimopoulos:1980cd,Abbott:1980ab}
for $SU(2)$. Specific examples \cite{Fradkin:1979ab} show that the
two phases are rigorously indistinguishable.{ No phase transition
but a smooth variation of the parameters characterizes the
transition between  the two phases}.

\begin{table}[h]
\begin{center}
\begin{tabular}{|c|c|c|c|}
\hline {$\psi^i_\alpha$}& {$u$} & {$d$} & {$s$}\\
\hline {$R$} & 2/3 & -1/3 & -1/3\\
\hline {$B$} & 2/3 & -1/3 & -1/3\\
\hline {$W$} & 2/3 & -1/3 & -1/3\\
\hline
\end{tabular}
\end{center}
\caption{\label{quarks}{\it The electric charges of the quark
fields $\psi^i_\alpha$.}}
\end{table}

\begin{table}[h]
\begin{center}
\begin{tabular}{|c|c|c|c|}
\hline {$D^\gamma_k$}& {$R$} & {$B$} &{$W$} \\
\hline {$u$} & -2/3 & -2/3 & -2/3 \\
\hline {$d$} &  1/3& 1/3 & 1/3\\
\hline {$s$} & 1/3& 1/3 & 1/3\\
\hline
\end{tabular}
\end{center}
\caption{\label{diquarks}{\it The electric charges of the diquark
fields $D^\gamma_k=\epsilon_{ijk}
\epsilon^{\alpha\beta\gamma}\psi^i_\alpha\psi^j_\beta$. However
their $\tilde Q$ charges are integers and equal t0 $0,\pm1$. This
follows since the charges $\tilde Q$ of the fermions enjoy  the
same property.}}
\end{table}

\begin{table}[h]
\begin{center}
\begin{tabular}{|c|c|c|c|}
\hline{$B^i_k$}&{$u$}  &{$d$}  & {$s$}\\
\hline {$u$} & 0& -1 & -1 \\
\hline {$d$} &  1& 0 & 0\\
\hline {$s$} & 1& 0 & 0\\
\hline
\end{tabular}
\end{center}
\caption{\label{baryons}{\it The electric charges of the baryon
fields $B^i_k=\psi^i_\gamma
D^\gamma_k=\psi^i_\gamma\left(\epsilon_{rsk}
\epsilon^{\alpha\beta\gamma}\psi^r_\alpha\psi^s_\beta\right)$.}}
\end{table}

\begin{table}[h]
\begin{center}
\begin{tabular}{|c|c|c|c|}
\hline {$G^i_k$}&{$u$}  &{$d$}  & {$s$}\\
\hline {$u$} & 0& -1 & -1 \\
\hline {$d$} &  1& 0 & 0\\
\hline {$s$} & 1& 0 & 0\\
\hline
\end{tabular}
\end{center}
\caption{\label{vectors}{\it The electric charges of the vector
meson fields $G^i_k={(D^*)}^i_\alpha g^\alpha_\beta D^\beta_k$.}}
\end{table}

 A way to implement complementarity is the following
\cite{Casalbuoni:1981nd}. Suppose to have the following spectrum
of fields \bea\psi_i\in R~\text{of}~G, ~~\text{{elementary
states}}\nn\\Q_\alpha\in \tilde R~\text{of}~\tilde G,
~~\text{{composite states}}\nn\eea with $G$ the gauge group in the
unbroken phase and $\tilde G$ the global symmetry group of the
broken phase. We assume that $R$ and $G$ are isomorphic to $\tilde
R$ and $\tilde G$. We assume also that the breaking is such that
the the effective Higgs fields ($\phi^i_\alpha$) are such to map
the two set of states \be
Q_\alpha(x)=\psi_i(x)\phi^i_\alpha(x),\alpha\in \tilde G, i\in
G\nn.\ee In the broken phase,
$\langle\phi_\alpha^i\rangle\propto\delta_\alpha^i$ implying that
the states in the two phases are the same, except for a necessary
redefinition of the conserved quantum numbers following from the
requirement that the Higgs fields should be neutral in the broken
vacuum The gauge fields $(g_\mu)_\alpha^\beta$ go into the vector
mesons of the confined phase \be
(Z_\mu)_j^i=-{\phi^{*\beta}_j}\left[
\de_\mu-(g_\mu)_\beta^\alpha\right]\phi^i_\alpha.\ee

In the case of CFL phase and hypernuclear matter, we have
{$G=SU(3)_c$ and $\tilde G=SU(3)$ with the fermion fields in three
copies  in both phases. The effective Higgs field is given by the
diquark field \be D^\gamma_k=\epsilon_{ijk}
\epsilon^{\alpha\beta\gamma}\psi^i_\alpha\psi^j_\beta,\ee, with
the property \be {\langle D_k^\gamma\rangle\propto
\delta_k^\gamma}.\ee

 The two phases, as shown in Table \ref{complementarity} are very similar but
 there are also
several differences \cite{Schafer:1998ef}
\begin{itemize} \item In the hypernuclear phase
there is a nonet of vector bosons. However if the dibaryon H
exists the singlet vector becomes unstable and does not need to
appear in the effective theory  \item In the CFL phase there are
nine ($\mathbf{8\oplus 1}$) quark states, but the gap of the
singlet is bigger than for the octet. The baryonic singlet has the
structure \be \epsilon_{ijk}\epsilon_{\alpha\beta\gamma}
\psi_i^\alpha\psi_j^\beta\psi_k^\gamma,\ee and it is precisely the
condensation of this baryon with itself which may produce the
state $H$ discussed before.
\end{itemize}

{{The CFL phase is a concrete example of complementarity}}. In the
tables \ref{quarks}, \ref{diquarks}, \ref{baryons}, \ref{vectors}
we show the electric charges of the various states. In the CFL
phase the charge $\tilde Q$ of diquarks is zero whereas for
quarks, $\psi_\alpha^i$, and gluons, $g_{\alpha\beta}$, coincides
with the charge $Q$ of baryons, $B^i_k$, and of vector mesons,
$G^i_k$.

\subsubsection{The 2SC phase}
\label{2SCphase}

We remember that in this case
\be\langle\psi_{iL}^\alpha\psi_{jL}^\beta\rangle=\Delta\epsilon^{\alpha\beta
3}\epsilon_{ij}.\ee Only 4 out of the 6 quarks are gapped, the
ones with color 1 and 2 whereas the 2 quarks of color 3 remain
ungapped. The symmetry breaking pattern is \[ SU(3)_c\otimes
SU(2)_L\otimes SU(2)_R\otimes U(1)_B\to SU(2)_c\otimes
SU(2)_L\otimes SU(2)_R\otimes U(1)_{\tilde B}.\] In this phase the
baryon number is not broken but there is a combination of $B$ and
of the color generator $T_8$, acting upon the color indices,
defined as \be \tilde B=B-\frac 2{\sqrt{3}} T_8=\left(\frac 1
3,\frac 1 3,\frac 1 3\right)-\frac 1
3\left(1,1,-2\right)=(0,0,1),\ee which is conserved by the
condensate. In fact, although both $B$ and $T_8$ are broken, the
condensate, involving only quarks of color 1 and 2, is neutral
under $\tilde B$. Also the electric charge is rotated. In fact
consider the following combination \be \tilde Q=Q\otimes {\bm
1}-\frac 1{\sqrt{3}}{\bm 1}\otimes T_8=\left(\frac 2  3,-\frac 1
3\right)\otimes {\bm 1}-{\bm 1}\otimes \frac 1 6(1,1,-2).\ee The
$\tilde B$ and $\tilde Q$ quantum numbers of the quarks are given
in Table \ref{rotated}
\begin{table}[h]
\begin{center}
\begin{tabular}{|c|c|c|}
\hline {}&{~~~$\tilde Q$~~~}  &{~~~$\tilde B$~~~}  \\
\hline {$u^\alpha$,~~$\alpha=1,2$} & $\frac 1 2$&0  \\
\hline {$d^\alpha$,~~$\alpha=1,2$} & $-\frac 1 2$& 0 \\
\hline {$u^3$} & 1& 1 \\
\hline {$d^3$} & 0& 1\\
 \hline
\end{tabular}
\end{center}
\caption{\label{rotated}The electric charge  ($\tilde Q)$ and the
baryon number ($\tilde B)$ for the quarks in the CFL phase}
\end{table}
We see that the condensate is neutral under $\tilde Q$ since it
pairs together up and down quarks of color 1 and 2. Notice that
quarks $u^3$ and $d^3$ have integer values of both $\tilde B$ and
$\tilde Q$. They look like proton and neutron respectively. We can
understand why these quarks are ungapped by looking at the 't
Hooft anomaly condition \cite{Hooft:1979ab}. In fact, it has been
shown in \cite{Sannino:2000kg,Hsu:2000by} that the anomaly
coefficient does not change at finite density. The theory in the
confined phase at zero density has an anomaly $SU(2)_{L(R)}\otimes
U(1)_B$ which is given by \be \frac 14\times\frac 1 3\times
3=\frac 1 4,\ee whereas in the broken CFL phase at finite density
there is an anomaly $SU(2)_{L(R)}\otimes U(1)_{\tilde B}$ given by
\be \frac 1  4\times 1=\frac 1 4\ee This anomaly is due entirely
to the states $\psi_i^3$, therefore they should remain massless as
the quarks in the zero density phase. In the 2SC phase there are
no broken global symmetries therefore we expect a first order
phase transition to the nuclear matter phase with a competition
between chiral and di-fermion condensates. Notice that there is no
superfluidity in the 2SC phase. \vskip0.5cm\noindent {\bf The
spectrum of the 2SC phase}\vskip0.5cm

We have already discussed the fermionic part, the modes
$\psi_i^\alpha$ with $\alpha=1,2$ are gapped, whereas $\psi_i^3$
remain massless.

Since the gauge group $SU(3)_c$ is broken down to $SU(2)_c$ we get
\be 8-3=5\ee massive gluons. However we have still 3 massless
gluons belonging to the confining gauge group $SU(2)_c$. With
respect to this group the gapped fermions $\psi_i^\alpha$ with
$\alpha=1,2$ are confined, whereas the massless fermions
$\psi_i^3$ are un-confined. Notice also that the electric charges
of the confined states are not integers whereas the ones of the
un-confined states are integers.

As in 2SC no global symmetries are broken, there are no massless
NG bosons. In conclusion the only light degrees of freedom are 3
gluons and 2 fermions.

\subsubsection{The case of 2+1 flavors}\label{VD3}

In nature the strange quark is much heavier than the other two and
it may happen that $\mu\approx m_s$ with $\mu\gg m_{u,d}$. In this
situation neither of the discussions above applies. In practice we
expect that decreasing $\mu$ from $\mu\gg m_{u,d,s}$ the system
undergoes a phase transition from CFL to 2SC for values of $\mu$
in between $m_s$ and $m_{d,u}$. This is because the Fermi momenta
of the different Fermi spheres get separated. In fact remember
that for massive quarks the Fermi momentum is defined by the
equation \be E_F=\mu=\sqrt{p_F^2+M^2}\to p_F=\sqrt{\mu^2-M^2}.\ee
As a consequence the radius of the Fermi sphere of a given quark
decreases increasing its mass. To see why such a transition is
expected, let us consider a simplified model with two quarks, one
massless and the other one with mass $m_s$ at the same chemical
potential $\mu$. The Fermi momenta are
\begin{equation}
  p_{F_1}=\sqrt{\mu^2-m_s^2},~~~~p_{F_2}=\mu.
\end{equation}
The grand potential for the two unpaired fermions is (factor 2
from the spin degrees of freedom)
\begin{equation}
 \Omega_{\rm unpair.}=2\int_{0}^{p_{F_1}}\frac{d^3p}{(2\pi)^3}\left(\sqrt{{\vec p\,}^2+m_s^2}-\mu\right)+
 2\int_{0}^{p_{F_2}}\frac{d^3p}{(2\pi)^3}\left(|\vec
 p\,|-\mu\right).
\end{equation}
In fact, the grand potential is given in general by the expression
\be \Omega=-T\sum_{\b k}\log\sum_{n_{\b
k}}\left(e^{(\mu-\epsilon_{\b k})/T}\right)^{n_{\b k}}=-T\sum_{\b
k}\log\left(1+e^{(\mu-\epsilon_{\b k})/T}\right),\ee where
$\epsilon_{\b k}$ is the energy per particle and $n_{\b k}$ is the
occupation number of the mode ${\b k}$. The expression above
refers to fermions. Also \be \sum_{\b k}\to gV\int\frac{d^3\b
p}{(2\pi)^3},\ee with $g$ the degeneracy factor. In the limit of
$T\to 0$ and in the continuum we get \be\Omega=gV\int\frac{d^3\b
p}{(2\pi)^3}(\epsilon_{\b p}-\mu)\theta(\mu-\epsilon_{\b
p})\label{5.164}.\ee In order to pair the two fermions must reach
some common momentum $p_{\rm comm}^F$, and the corresponding grand
potential can be written as
 \bea
 \Omega_{\rm pair.}&=&2\int_{0}^{p_{\rm comm}^F}\frac{d^3p}{(2\pi)^3}
 \left(\sqrt{{\b p\,}^2+m_s^2}-\mu\right)+
 2\int_{0}^{p_{\rm comm}^F}\frac{d^3p}{(2\pi)^3}\left(|\b
 p\,|-\mu\right)\nonumber\\&-&\frac{\mu^2\Delta^2}{4\pi^2},
 \label{omega_pair}\eea
where the last term is the energy necessary for the condensation
of a fermion pair, that we recall from Eq. (\ref{3.65}) to be
given by \be -\frac 1 4\rho \Delta^2=-\frac 1 4\frac{p_F^2}{\pi^2
v_F}\Delta^2=-\frac{\mu^2\Delta^2}{4\pi^2}.\ee This expression can
be adapted to the present case by neglecting terms of order
$m_s^2\Delta^2$. The common momentum $p^F_{\rm comm}$ can be
determined by minimizing $\Omega_{\rm pair.}$ with respect to
$p^F_{\rm comm}$, with the result \be p^F_{\rm
comm}=\mu-\frac{m_s^2}{4\mu}.\ee It is now easy to evaluate the
difference $\Omega_{\rm unpair.}-\Omega_{\rm pair.}$ at the order
$m_s^4$, with the result \be \Omega_{\rm pair.}-\Omega_{\rm
unpair.}\approx\frac
1{16\pi^2}\left(m_s^4-4\Delta^2\mu^2\right).\ee We see that in
order to have condensation the condition \be
\mu>\frac{m_s^2}{2\Delta}\ee must be realized.
 Therefore if $m_s$ is too large the pairing does not occur. It is
 easy to see that the transition must be first order. In fact if
 it were second order the gap $\Delta$ should vanish at the
 transition, but in order this to happen we must have
 $\Delta> m_s^2/2\mu$. It should be noticed that the value of
 $m_s$ appearing in this equations should be considered density
 dependent. For  $m_s\approx 200\div 300~MeV$ and chemical
 potentials interesting for compact stellar objects (see later)
 the situation is shown in Fig. \ref{stress}. We see that the gap
 should be larger than $40\div 110~MeV$ in order to get the CFL
 phase, whereas for smaller value we have the 2SC phase. These
 values of $\Delta$ are very close to the ones that one gets from
 the gap equation. Therefore we are not really able to judge if
 quark matter with the values of the parameters is in the CFL or
 in the 2SC phase. The situation is far more complicated, because
 differences in the Fermi momenta can be generated also from
 different chemical potentials (arising from the requirement of
 weak equilibrium, see later) and/or from the requirement of
 electrical and color neutrality as appropriated for compact
 stars. Furthermore, at the border of this transition a
 crystalline phase, the so-called LOFF phase, can be formed. We
 will discuss all these questions later on.

 \begin{center}
\begin{figure}[ht]
\epsfxsize=8truecm \centerline{\epsffile{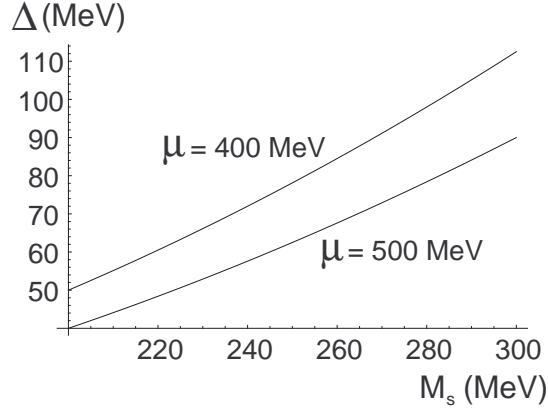}} \noindent
\caption{\it The minimal values of the gap in order to get pairing
for two given values of $\mu$ vs. $m_s$.\label{stress}}
\end{figure}
\end{center}

\subsubsection{Single flavor and single color}

According to the values of the parameters  single flavor and/or
single color condensation could arise. A complete list of
possibilities has been studied in \cite{Alford:2002rz} (for a
review see \cite{Bowers:2003ye}).  In general a single flavor
condensate will occur  in a state of angular momentum $J=1$. In
this case rotational invariance is not broken since the condensate
is in an antisymmetric state of color ${\bm 3}^*$ and and
rotations and color transformations are locked together \be
\langle s^\alpha\sigma^\beta
s^\gamma\rangle=\Delta\epsilon^{\alpha\beta\gamma},\ee where
$\sigma^\alpha$ is a spin matrix. This condensate breaks
$SU(3)\otimes O(3)$ down to $O(3)$. However, it is a general
statement that condensation in higher momentum states than $J=0$
gives rise to smaller gaps, mainly due to  a less efficient use of
the Fermi surface. The typical gaps may range from 10 to 100
$keV$. We will not continue here this discussion.

In  \cite{Rajagopal:2000wf} it is possible to find a discussion of
many other possibilities as $N_f>3$, the case of two colors (of
interest because it can be discussed on the lattice) and the limit
$N_f\to\infty$.

\section{Effective lagrangians}

In this Section we will derive the effective lagrangians relevant
to the two phases CFL and 2SC. The approach will be the classical
one, that is looking at the relevant degrees of freedom at low
energy (with respect to the Fermi energy) and constructing the
corresponding theory on the basis of the symmetries.

\subsection{Effective lagrangian for the CFL phase}\label{VIA}

We will now derive the effective lagrangian for the light modes of
the CFL phase. As we have seen such modes are the NG bosons. We
will introduce the Goldstone fields as the phases of the
condensates in the $(\mathbf{\bar 3},\mathbf{\bar 3})$ channel
\cite{Casalbuoni:1999wu,Hong:1999tz} (remember also the discussion
made in Section \ref{VC}) \be X_\alpha^i\approx
\epsilon^{ijk}\epsilon_{\alpha\beta\gamma}\langle \psi^j_{\beta
L}\psi^k_{\gamma L}\rangle^*,~~~ Y_\alpha^i\approx
\epsilon^{ijk}\epsilon_{\alpha\beta\gamma}\langle \psi^j_{\beta
R}\psi^k_{\gamma R}\rangle^*. \label{6.1}\ee Since quarks belong
to the representation $(\mathbf{3},\mathbf{3})$ of $SU(3)_c\otimes
SU(3)_{L(R)}$ they transform as ($g_c\in SU(3)_c$, $g_{L(R)}\in
SU(3)_{L(R)}$)\be \psi_L\to e^{i(\alpha+\beta)} g_c\psi_L
g_L^T,~~~\psi_R\to e^{i(\alpha-\beta)} g_c\psi_R
g_R^T,~~~e^{i\alpha}\in U(1)_B,~~~e^{i\beta}\in U(1)_A,
\label{6.2}\ee whereas the transformation properties of the fields
$X$ and $Y$ under the total symmetry group $G=SU(3)_c\otimes
SU(3)_L\otimes SU(3)_R\otimes U(1)_B\otimes U(1)_A$ are  \be X\to
g_cXg_L^T e^{-2i(\alpha+\beta)},~~~Y\to g_cYg_R^T
e^{-2i(\alpha-\beta)}. \label{6.3}\ee The fields $X$ and $Y$ are
$U(3)$ matrices and as such they describe $9+9=18$ fields. Eight
of these fields are eaten up by the gauge bosons, producing eight
massive gauge particles. Therefore we get the right number of
Goldstone bosons, $10 =18-10$. In this Section we will treat the
field associated to the breaking of $U(1)_A$ as a true NG boson.
However, remember that this is a massive particle with a light
mass at very high density. These fields correspond to the breaking
of the global symmetries in $G=SU(3)_L\otimes SU(3)_L\otimes
U(1)_L\otimes U(1)_R$ (18 generators) to the symmetry group of the
ground state $H=SU(3)_{c+L+R}\otimes Z_2\otimes Z_2$ (8
generators). For the following it is convenient to separate the
$U(1)$ factors in $X$ and $Y$ defining fields, $\hat X$ and $\hat
Y$, belonging to $SU(3)$ \be X=\hat X
e^{2i(\phi+\theta)},~~~Y=\hat Y e^{2i(\phi-\theta)},~~~\hat X,\hat
Y\in SU(3). \label{6.4}\ee The fields $\phi$ and $\theta$ can also
be described  through the determinants of $X$ and $Y$ \be d_X={\rm
det}(X)=e^{6i(\phi+\theta)},~~~d_Y={\rm
det}(Y)=e^{6i(\phi-\theta)}, \label{6.5}\ee The transformation
properties under $G$ are\be \hat X\to g_c\hat X g_L^T,~~~\hat Y\to
g_c\hat Y g_R^T,~~~\phi\to\phi-\alpha,~~~\theta\to\theta-\beta.
\label{6.6}\ee The breaking of the global symmetry can   be
discussed in terms of gauge invariant fields given by $d_X$, $d_Y$
and\be\Sigma^i_j=\sum_\alpha (\hat Y_\alpha^j)^* \hat
X_\alpha^i\to \Sigma=\hat Y^\dagger \hat X. \label{6.7}\ee The
$\Sigma$ field describes the 8 Goldstone bosons corresponding to
the breaking of the chiral symmetry $SU(3)_L\otimes SU(3)_R$, as
it is made clear by the transformation properties of $\Sigma^T$,
$\Sigma^T\to g_L \Sigma^T g_R^\dagger$. That is $\Sigma^T$
transforms  as the usual chiral field. The other two fields $d_X$
and $d_Y$ provide the remaining  Goldstone bosons related to the
breaking of the $U(1)$ factors.

In order to build up an invariant lagrangian, it is convenient to
define the following currents \bea J_X^\mu&=&\hat X D^\mu \hat
X^\dagger=\hat X(\partial^\mu\hat X^\dagger+\hat X^\dagger
g^\mu)=\hat X\partial^\mu\hat X^\dagger+ g^\mu,\nn\\J_Y^\mu&=&\hat
Y D^\mu \hat Y^\dagger=\hat Y(\partial^\mu\hat Y^\dagger+\hat
Y^\dagger g^\mu)=\hat Y\partial^\mu\hat Y^\dagger+ g^\mu, \eea
with \be g_\mu=ig_s g_\mu^a T^a\ee the gluon field and \be
T^a=\frac{\lambda_a}2\ee the $SU(3)_c$ generators. These currents
have simple transformation properties under the full symmetry
group $G$:\be J^\mu_{X,Y}\to g_c J^\mu_{X,Y}g_c^\dagger.\ee The
most general lagrangian, up to two derivative terms, invariant
under $G$, the rotation group $O(3)$ (Lorentz invariance is broken
by the chemical potential term) and the parity transformation,
defined as: \be P:~~~~\hat X\leftrightarrow \hat Y,~~~
\phi\to\phi,~~~ \theta\to -\theta,\ee is \cite{Casalbuoni:1999wu}
\bea {\cal L}&=&-\frac{F_T^2}4{\rm
Tr}\left[\left(J^0_X-J^0_Y)^2\right)\right]-\alpha_T\frac{F_T^2}4{\rm
Tr}\left[\left(J^0_X+J^0_Y)^2\right)\right]+\frac 12
(\de_0\phi)^2+\frac
12(\de_0\theta)^2\nn\\
&&+\frac{F_S^2}4{\rm Tr}\left[\left|\b J_X-\b
J_Y\right|^2\right]+\alpha_S\frac{F_S^2}4{\rm Tr}\left[\left|\b
J_X+\b J_Y\right|^2\right]-\frac{v_\phi^2}2|{\bm\nabla}\phi|^2-
\frac{v_\theta^2}2|{\bm\nabla}\theta|^2.\nn\label{lagrangian1}
 \eea or
\bea {\cal L}&=&-\frac{F_T^2}4{\rm Tr}\left[\left(\hat X\de_0\hat
X^\dagger-\hat Y\de_0\hat
Y^\dagger)^2\right)\right]-\alpha_T\frac{F_T^2}4{\rm
Tr}\left[\left(\hat X\de_0 \hat X^\dagger+\hat Y\de_0\hat
Y^\dagger+2 g_0)^2\right)\right]\nn\\
&&+\frac{F_S^2}4{\rm Tr}\left[\left| \hat X{\bm\nabla}\hat
X^\dagger-\hat Y{\bm\nabla}\hat
Y^\dagger\right|^2\right]+\alpha_S\frac{F_S^2}4{\rm
Tr}\left[\left|\hat X{\bm\nabla}\hat X^\dagger+\hat
Y{\bm\nabla}\hat Y^\dagger+2 \b g\right|^2\right] \nn\\&&+\frac 12
(\de_0\phi)^2+\frac
12(\de_0\theta)^2-\frac{v_\phi^2}2|{\bm\nabla}\phi|^2-
\frac{v_\theta^2}2|{\bm\nabla}\theta|^2.\label{lagrangian1}
 \eea
 Using $SU(3)_c$ color gauge invariance
we can choose $\hat X=\hat Y^\dagger$, making 8 of the Goldstone
bosons  disappear and giving mass to the gluons. The properly
normalized Goldstone bosons, $\Pi^a$,  are given in this gauge by
\be \hat X=\hat Y^\dagger =e^{i\Pi^a T^a/F_T}, \ee and expanding
Eq. (\ref{lagrangian1}) at the lowest order in the fields we get
\be {\cal L}\approx\frac 12 (\de_0\Pi^a)^2+\frac 12
(\de_0\phi)^2+\frac 12(\de_0\theta)^2 -\frac{v^2}
2|{\bm\nabla}\Pi^a|^2-\frac{v_\phi^2}2|{{\bm\nabla}}\phi|^2-
\frac{v_\theta^2}2|{\bm\nabla}\theta|^2,\ee with \be
v=\frac{F_S}{F_T}.\ee The gluons $g_0^a$ and $g_i^a$ acquire Debye
and Meissner masses given by \be m_D^2=\alpha_Tg_s^2
F_T^2,~~~m_M^2=\alpha_Sg_s^2 F_S^2=\alpha_Sg_s^2v^2 F_T^2.
\label{masses}\ee It should be stressed that these are not the
true rest masses of the gluons, since there is a large wave
function renormalization effect making the gluon masses of order
of the gap $\Delta$, rather than $\mu$ (see later)
\cite{Casalbuoni:2000na,Casalbuoni:2000nb}. Since this description
is supposed to be valid at low energies (we expect much below the
gap $\Delta$), we could also decouple the gluons solving their
classical equations of motion neglecting the kinetic term. The
result from Eq. (\ref{lagrangian1}) is \be g_\mu=-\frac 12
\left(\hat X\de_\mu\hat X^\dagger+\hat Y\de_\mu\hat
Y^\dagger\right). \label{6.18}\ee It is easy to show that
substituting this expression in Eq. (\ref{lagrangian1}) one gets
\be {\cal L}=\frac{F_T^2}4\left({\rm
Tr}[\dot\Sigma\dot\Sigma^\dagger]-v^2{\rm
Tr}[\vec\nabla\Sigma\cdot\vec\nabla\Sigma^\dagger]\right)+ \frac
12\left(\dot\phi^2-v_\phi^2|\vec\nabla\phi|^2\right)+ \frac
12\left(\dot\theta^2-v_\phi^2|\vec\nabla\theta|^2\right).
\label{6.17}\ee Notice that the first term is nothing but the
chiral lagrangian except for the breaking of the Lorentz
invariance. This is a way of seeing the quark-hadron continuity,
that is the continuity between the CFL and the hypernuclear matter
phase in three flavor QCD discussed previously. Furthermore one
has to identify the NG $\phi$ associated to the breaking of
$U(1)_B$ with the meson $H$ of the hypernuclear phase
\cite{Schafer:1998ef}.

\subsection{Effective lagrangian for the 2SC phase}
\label{General}

In Section \ref{2SCphase} we have seen that the only light degrees
of freedom in the 2SC phase are the $u$ and $d$ quarks of color 3
and the gluons belonging to the unbroken $SU(2)_c$. The fermions
are easily described in terms of ungapped quasi-particles at the
Fermi surface and we will not elaborate about these modes further.
In this Section we will discuss only the massless gluons. An
effective lagrangian describing the  5 would-be Goldstone bosons
and their couplings to the gluons has been given in
\cite{Casalbuoni:2000cn}. The effective lagrangian for the
massless gluons has been given in \cite{Rischke:2000cn} and it can
be obtained simply by noticing that it should be gauge invariant.
Therefore it  depends only on the field strengths \be
E_i^a=F_{0i}^a,~~~B_i^a=\frac 1 2\epsilon_{ijk}F_{jk}^a,\ee where
$F_{\mu\nu}^a$ is the usual non-abelian curvature. As we know,
Lorentz invariance is broken, but rotations are good symmetries,
therefore the most general effective lagrangian we can write at
the lowest order in the derivatives of the gauge fields is \be
{\cal L}_{\rm eff}=\frac
1{g^2}\sum_{a=1}^3\left(\frac{\epsilon}2\b E^a\cdot\b E^a-\frac
1{2\lambda}\b B^a\cdot \b B^a\right).\ee The constants $\epsilon$
and $\lambda$ have the meaning of the dielectric constant and of
the magnetic permeability. The speed of the gluons turns out to be
given by \be v=\frac 1{\sqrt{\epsilon\lambda}}.\ee We will see
later how to evaluate these constants starting from the HDET but
we can see what are the physical consequences of  this
modification of the usual relativistic lagrangian. The most
interesting consequence has to do with the Coulomb law and the
effective gauge coupling. In fact, as we shall see, at the lowest
order in $1/\mu$ expansion, $\lambda=1$. The Coulomb potential
between two static charges get modified \be V_{\rm
Coul}=\frac{g^2}{\epsilon
r}=\frac{g_{eff}^2}{r},~~~g_{eff}^2=\frac{g^2}{\epsilon}.\ee
Furthermore, the value of $\alpha_s^{eff}$, reintroducing the
velocity of light $c$, and keeping $\slash h=1$ is
\be\alpha_s=\frac{g^2}{4\pi c}\to \alpha_s'=\frac{g_{eff}^2}{4\pi
v}=\frac{g^2}{4\pi c\sqrt{\epsilon}}\label{6.24}.\ee If, as it is
the case, $\epsilon$ is much bigger than 1, the gluons move slowly
in the superconducting medium and \be\frac
{\alpha_s'}{\alpha_s}=\frac 1{\sqrt{\epsilon}}\ll 1.\ee Therefore
the effect of the medium is to weaken the residual strong
interaction. The same result can be obtained by performing the
following transformation in time, fields and coupling: \be
x^{0'}=\frac{x^0}{\sqrt{\epsilon}},~~~A_0^{a'}=\sqrt{\epsilon}A_0^a,~~~g'=\frac
g{{\epsilon}^{1/4}}.\ee The effective lagrangian takes the usual
form for a relativistic gauge theory in terms of the new
quantities \be{\cal L}_{\rm eff}=-\frac{1}{4g^{'2}}
F_{\mu\nu}^{a'} F^{\mu\nu a'},\ee with \be
F_{\mu\nu}^{a'}=\de_\mu^{'}A_{\nu}^{a'}-\de_\nu^{'}
A_\mu^{a'}+f^{abc}A_\mu^{b'} A_\nu^{c'}.\ee The calculation shows
\cite{Rischke:2000cn,Rischke:2000qz,Casalbuoni:2001ha} that \be
\epsilon=
1+\frac{g^2\mu^2}{18\pi^2\Delta^2}\approx\frac{g^2\mu^2}{18\pi^2\Delta^2}
\label{6.29},\ee since typically $\Delta\ll\mu$. We recall also
that at asymptotic values of $\mu$ the calculations from QCD show
that (see Section \ref{VD}) \be \Delta=c\mu
g^{-5}e^{\dd{-3\pi^2/\sqrt{2}g}}.\ee Using Eqs. (\ref{6.29}) and
(\ref{6.24}) we obtain
\be\alpha_s'=\frac{3}{2\sqrt{2}}\frac{g\Delta}{\mu}.\ee This is
the way in which the coupling gets defined at the matching scale,
that is at $\Delta$. Since $SU(2)$ is an asymptotically free gauge
theory, going  at lower energies makes the coupling to increase.
The coupling gets of order unity at a scale $\Lambda'_{QCD}$.
Since the coupling at the matching scale is rather small it takes
a long way before it gets of order one. As a consequence
$\Lambda_{QCD}'$ is expected to be small. From one-loop beta
function we get \be \Lambda_{QCD}'\approx \Delta
e^{\dd{-2\pi/\beta_0\alpha_s'}}\approx
\Delta\exp\left(-\frac{2\sqrt{2}\pi}{11}\frac{\mu}{g\Delta}\right),\ee
where  $\beta_0$ is the first coefficient of the beta function. In
SU(2) we have $\beta_0=22/3$. Notice that it is very difficult to
give a good estimate of $\Lambda_{QCD}'$ since it depends
crucially on the value of $c$ which is very poorly known. In fact
different approximations give different values of $c$. In
\cite{Schafer:1999pb,Pisarski:1999tv} it has been found \be
c=512\pi^4,\ee whereas in \cite{Brown:1999yd,Wang:2001aq} \be
c=512\pi^4 \exp\left(-\frac{4+\pi^2}{8}\right).\ee Using
$\Lambda_{QCD}=200~MeV$ to get $g$ and $\mu=600~MeV$ one finds
$\Lambda_{QCD}'=10~MeV$ in the first case and
$\Lambda_{QCD}'=0.3~keV$ in the second case. Although a precise
determination of $\Lambda_{QCD}'$ is lacking, it is quite clear
that \be\Lambda_{QCD}'\ll\Lambda_{QCD}.\ee Notice also that
increasing the density, that is $\mu$, $\Lambda_{QCD}'$ decreases
exponentially. Therefore the confinement radius $1/\Lambda_{QCD}'$
grows exponentially with $\mu$. This means that looking at physics
at some large, but fixed, distance and  increasing the density,
there is a crossover density when the color degrees of freedom
become deconfined.

\section{NGB and their parameters}

In this Section we will evaluate the parameters appearing in the
effective lagrangian for the NG bosons in the CFL phase.
Furthermore we will determine  the properties of the gluons in the
2SC phases. A nice way of organizing this calculation is to make
use of the HDET which holds for residual momenta such that $\Delta
\ll\ell\ll\delta$ and to match, at the scale $\Delta$, with the
effective lagrangian supposed to hold for momenta $\ell\ll\Delta$.
This calculation can be extended also to  massive gluons both in
CFL and in 2SC. However the expansion in momenta is not justified
here, since the physical masses turn out to be of order $\Delta$.
On the other hand by a numerical evaluation we can show that the
error is not more than 30\%, and therefore this approach gives
sensible results also in this case. The way of evaluating the
propagation parameters  for NG bosons and for gluons is rather
simple, one needs to evaluate the self-energy in both cases. The
self-energy for the gluons is not a problem since we know the
couplings of the gluons to fermions. However  we need the explicit
expression of the interaction between NG bosons and quarks. This
interaction can be easily obtained since we know the coupling of
the fermions to the gap and the phases of the gap are the NG boson
fields. Therefore we need only to generalize the Majorana mass
terms (the coupling with the gap) to an interaction with the NG
bosons such to respect the total symmetry of the theory. We will
illustrate the way of performing this calculation in the specific
example of the NG boson corresponding to the breaking of $U(1)_B$
and we will illustrate the results for the NG bosons. Before doing
that we will derive first the lagrangians for the HDET in the CFL
and in the 2SC phases. We have also to mention that the parameters
of the NG bosons have been derived by many authors
\cite{Son:1999cm,Son:2000tu,Rho:1999xf,Hong:1999ei,Manuel:2000wm,
Rho:2000ww,Beane:2000ms,Manuel:2000xt}, although using different
approaches from the one considered here \cite{Casalbuoni:2000na,
Casalbuoni:2000nb,Casalbuoni:2001ha}.

\subsection{HDET for the CFL phase} In the CFL
phase the symmetry breaking is induced by the condensates
 \be
\langle\psi_{\alpha i}^{L\,T} C\psi_{\beta  j}^L\rangle=
-\langle\psi_{\alpha i}^{R\,T}C\psi_{\beta
j}^R\rangle\approx{\Delta}\, \epsilon_{\alpha\beta
I}\epsilon_{ijI}~,\label{3.2.1}\ee where $\psi^{L,\,R}$ are Weyl
fermions and $C=i\sigma_2$. The corresponding Majorana mass term
is given by ($\psi\equiv \psi_L$): \be \frac{\Delta} 2
\sum_{I=1}^3\psi^T_- C\epsilon_I\psi_+\epsilon_I~+(L\to R)\, +
h.c.,\ee with the $3\times 3$ matrices $\epsilon_I$ defined as in
Eq. (\ref{5.125}) \be \left(\epsilon_I\right)_{ab}=\epsilon_{Iab}\
.\ee Expanding the quark fields in the basis (\ref{5.103})
\be\psi_\pm=\frac{1}{\sqrt{2}}\sum_{A=1}^9
\lambda_A\psi^A_\pm~.\ee and following Section \ref{HDET} we get
the effective lagrangian at the Fermi surface  \be {\cal L}_D=
\int\frac{d\b v}{4\pi}\sum_{A,B=1}^9
\chi^{A\dagger}\left(\matrix{\frac i2{\rm Tr}[\lambda_A\,V\cdot
D\,\lambda_B] & -\Delta_{AB}\cr -\Delta_{AB} &\frac i 2{\rm
Tr}[\lambda_A\,\tilde V\cdot D^*\,\lambda_B]}\right)\chi^B + (L\to
R), \label{cflcomplete0} \ee where \be \Delta_{AB} =\,\frac 1
2\sum_{I=1}^3\Delta\,{\rm
Tr}[\epsilon_I\lambda_A^T\epsilon_I\lambda_B],\label{2.31}\ee and
we recall from Section \ref{HDET} that  \be
\Delta_{AB}\,=\,\Delta_A\delta_{AB}, \label{7.8}\ee with \be
\Delta_1=\cdots=\Delta_8=\Delta, \label{2.39}\ee  \be
\Delta_9=-2\Delta. \label{2.40}\ee The CFL free fermionic
lagrangian assumes therefore the form: \be {\cal
L}_D=\int\frac{d\b v}{4\pi}\sum_{A=1}^9
\chi^{A\dagger}\left(\matrix{iV\cdot
\partial & -\Delta_A\cr -\Delta_A &i\tilde V\cdot \de}\right)\chi^A\
+\ (L\to R)\ .\label{cflcomplete}\ee  From this equation one can
immediately obtain the free fermion propagator   in momentum space
\be S_{AB}(p)=\frac{\delta_{AB}}{V\cdot \ell\,\tilde V\cdot
\ell-\Delta_A^2}\left(\matrix{\tilde V\cdot \ell &
\Delta_A\cr\Delta_A & V\cdot \ell}\right).\label{propagatorcfl}\ee

We recall that the NG bosons have been described in terms of  the
fields $X$ and $Y$ of Eq. (\ref{6.1}). Since we want to couple the
NG bosons with fermions in a $G$ invariant way let us look at the
transformation properties of the gap term. We have \be
\sum_{I=1}^3{\rm Tr}[\psi_L^TC\epsilon_I\psi_L\epsilon_I]\to
\sum_{I=1}^3{\rm Tr}[g_L\psi_L^T g_c^T\epsilon_I g_c\psi_L
g_L^T\epsilon_I].\ee Using the following property,  holding for
any unitary $3\times 3$ matrix , $g$ \be g^T\epsilon_I
g=\sum_{J=1}^3\epsilon_J g_{JI}^\dagger {\rm det}[g],\ee which
follows from \be \epsilon_{ijk}
g_{ii'}g_{jj'}g_{kk'}=\epsilon_{i'j'k'}{\rm det}[g],\ee we get \be
\sum_{I=1}^3{\rm Tr}[\psi_L^TC\epsilon_I\psi_L\epsilon_I]\to
\sum_{I,J,K=1}^3(g_c)_{JI}^\dagger(g_L)_{KI}^\dagger{\rm
Tr}[\psi^T_L\epsilon_J\psi_L\epsilon_K].\ee This expression shows
not only that  the gap term is invariant under the locked
transformations $g_L^*=g_c$, but also that it could be made
invariant modifying the transformation properties of the fermions
by changing $g_L^T$ to $g_c^\dagger$. This is simply done by using
the matrices $X$ and $Y$ respectively for the left- and right-
handed quarks  \be  (\psi_L X^\dagger)\to g_c(\psi_L X^\dagger)
g_c^\dagger,~~~~(\psi_R Y^\dagger)\to g_c(\psi_R Y^\dagger)
g_c^\dagger.\ee Therefore,
 the coupling of the left-handed Weyl spinors
$\psi$'s to the octet of NG boson fields is:
 \be
\frac{\Delta}2\sum_{I=1,3}Tr[(\psi_- X^\dagger)^T
C\epsilon_I(\psi_+ X^\dagger)\epsilon_I]+~(L\to R) +{ h.c.},\ee
 Both $X$ and $Y$ have
 v.e.v. given by   \be\langle
X\rangle=\langle Y\rangle=1\ee and we shall use the gauge \be
X=Y^\dag\ . \ee  The lagrangian giving the the coupling of the
quarks to the external fields can be obtained from
(\ref{cflcomplete0}) and is given by \be {\cal L}_D=\int\frac{d\b
v}{4\pi}\sum_{A,B=1}^9 \chi^{A\dagger}\left(\matrix{i\,V\cdot
\partial\delta_{AB} & -\Xi_{BA}^*
\cr -\Xi_{AB} &i \,\tilde V\cdot
\partial\delta_{AB}}\right)\chi^B+ (L\to R)
\label{7.20},\ee
 where
\bea \Xi_{AB} &=&\,\sum_{I=1}^3\frac 1 2\Delta\,{\rm
Tr}\left[\epsilon_I\left(\lambda_AX^\dag\right)^T\epsilon_I
\left(\lambda_BX^\dag\right)\right]\,=\cr&=&\,\sum_{I=1}^3\frac 1
2\Delta\left(\,{\rm Tr}[\lambda_AX^\dag \lambda_BX^\dag]\,-\,Tr
[\lambda_AX^\dag]\,{\rm Tr}[\lambda_BX^\dag]\right)\, . \eea One
can now expand $X$  in terms of the NGB fields \be X=
 \exp{\displaystyle{ i\left(
\frac{\lambda_a\Pi^a}{2F}\right)}},~~~~~~a=1,\cdots,8~,\ee and
obtain  the 3-point $\chi\chi\Pi$, and the 4-point
$\chi\chi\Pi\Pi$  couplings. The result is, for the 3-point
coupling  \bea {\cal L}_{\chi\chi\Pi}& =-\,i\ &\int\frac{d\b
v}{4\pi} \frac{\Delta} {F}
 \Big\{\sum_{a=1}^8\frac{\Pi^a}{\sqrt 6}
\Big[\chi^{9\dagger}\Gamma_0\,\chi^a + \chi^{a\dagger}
\Gamma_0\,\chi^9 \Big] - \cr && - \sum_{a,b,c=1}^8
d_{abc}\chi^{a\dagger} \Gamma_0\chi^b\Pi^c\Big\}
~.\label{trilin}\eea and for the 4-point coupling  \bea {\cal
L}_{\chi\chi\Pi\Pi}&=&\int\frac{d\b v}{4\pi}
 \Big\{~\frac{4}{3} \sum_{a=1}^8 \frac{\Delta}{8F^2}
\chi^{9\dagger}\Gamma_1\, \chi^9~\Pi^a\Pi^a~ \nonumber
\\&& + 3~\sqrt{\frac{2}{3}}
\sum_{a,b,c=1}^8\left(\frac{\Delta}{8F^2} d_{abc}
\chi^{c\dagger}\Gamma_1\,\chi^9~\Pi^a\Pi^b + {\rm
h.c.}\right)~\nonumber
\\ &&+~ \sum_{a,b,c,d=1}^8
\frac{\Delta}{8F^2}h_{abcd}\chi^{c\dagger} \Gamma_0\,
\chi^d\Pi^a\Pi^b
 \Big\}~,\label{quadrilin}\eea where we have defined
 \be h_{abcd}=2 \sum_{p=1}^8
\left(g_{cap}g_{dbp}+d_{cdp}d_{abp}\right) \, -\frac 8 3
\delta_{ac}\delta_{db} +\frac 4 3 \delta_{cd}\delta_{ab},~\ee with
\be g_{abc}=d_{abc}+if_{abc}\ee and \be \Gamma_0=\left(\matrix{0
&+1\cr -1 &0}\right) \ ,\hskip1cm \Gamma_1=\left(\matrix{0 &1\cr 1
&0}\right) \ .\ee In the CFL case we have also the NG bosons
associated to the breaking of the $U(1)$ factors. Let us consider
the following fields \be
U=e^{i\sigma/f_\sigma},~~~~V=e^{i\tau/f_\tau},\ee with
\be\sigma=f_\sigma\phi,~~~~\tau=f_\tau\theta,\ee where $\phi$ and
$\theta$ are the dimensionless $U(1)$ fields transforming
according to Eq. (\ref{6.6})\footnote{The couplings $f_\sigma$ and
$f_\tau$ have dimension 1 in units of mass.}. Clearly under
$U(1)_B$ and $U(1)_A$ groups we have \be U\to e^{-i\alpha}
U,~~~V\to e^{-i\beta}V,\ee whereas \be\psi_L\to
e^{i(\alpha+\beta)}\psi_L,~~~ \psi_R\to
e^{i(\alpha-\beta)}\psi_R.\ee We can make invariant couplings with
the fermions simply by taking the combinations \be
UV\psi_L,~~~UV^\dagger\psi_R.\ee As a result we get full invariant
couplings to the NG bosons modifying the quantity $\Xi_{AB}$
appearing in Eq. (\ref{7.20}}) as follows:\be \Xi_{AB}\to\Xi_{AB}
U^2V^2.\label{7.31}\ee

\subsection{HDET for the 2SC phase}

For the two flavour case, which encompasses both the 2SC model and
the existing calculation in the LOFF phase (see later) we follow a
similar approach.

The symmetry breaking is induced by the condensates
 \be
\langle\psi_{\alpha i}^{L\,T}C\psi_{\beta  j}^L\rangle=
-\langle\psi_{\alpha i}^{R\,T}C\psi_{\beta
j}^R\rangle\approx{\Delta}\, \epsilon_{\alpha\beta
3}\epsilon_{ij3}~,\ee
  corresponding to the
invariant coupling ($\psi\equiv \psi^L$): \be \frac{\Delta}2\,
\psi^T_- C\epsilon\,\psi_+\epsilon\ -(L\to R)+{\rm h.c.}~,\ee
where \be\epsilon=i\sigma_2\ .\ee As in the previous Section we
use a different basis for the fermion fields
 by writing \bea \psi_{+,\alpha i}&=&
\sum_{A=0}^3\frac{(\sigma_A)_{\alpha i}}{\sqrt 2}\psi_{+}^A
~~~~~~~~(i,\,\alpha=1,\,2)~\cr \psi_{+,3 1}&=&\psi_{+}^4\cr
\psi_{+,3 2}&=&\psi_{+}^5\ ,\eea where $\sigma_A$ are the Pauli
matrices for $A=1,2,3$ and $\sigma_0=1$.

A different, but also convenient notation for the fields
$\psi_{+,\,\alpha i}$, makes use of the following combination of
$\lambda$ matrices, as follows \be \psi_{+,\alpha i}=
\sum_{A=0}^5\frac{(\tilde\lambda_A)_{\alpha i}}{\sqrt 2}\psi_{+}^A
~.\label{7.37}\ee The $\tilde\lambda_A$ matrices are defined in
terms of the usual $\lambda$ matrices as follows: \be\dd
\tilde\lambda_0=\frac 1{\sqrt 3}\lambda_8\,+\, {\sqrt\frac 2
3}\lambda_0,~~~ \dd \tilde\lambda_A=\lambda_A\,~~(A=1,2,3),~~~ \dd
\tilde\lambda_4=\frac{\lambda_{4-i5}}{\sqrt 2},~~~ \dd
\tilde\lambda_5=\frac{\lambda_{6-i7}}{\sqrt 2}.\ee

 Proceeding as before the 2SC fermionic lagrangian assumes the
form:  \be {\cal L}_D= \int\frac{d\b v}{4\pi}\sum_{A,B=0}^5
\chi^{A\dagger}\left(\matrix{\frac i2{\rm Tr}[\tilde
\lambda_A\,V\cdot D\,\tilde \lambda_B ]& -\Delta_{AB}\cr
-\Delta_{AB} &\frac i2 {\rm Tr}[\tilde \lambda_A\,\tilde V\cdot
D^*\,\tilde \lambda_B ]}\right)\chi^B+ (L\to R)\ .
\label{2sccomplete0} \ee Here \bea \Delta_{AB}
&=&\,\frac{\Delta}{2}\,{\rm Tr}[\,\epsilon \sigma_A^T\epsilon
\sigma_B] \hskip1cm(A,B=0,...3),\cr  \Delta_{AB}
&=&\,0\hskip3.3cm(A,B=4,5)\ . \eea

 We now use the identity ($g$ any $2\times 2$
matrix), analogous to (\ref{5.126}): \be \epsilon
g^T\epsilon\,=\,g\,-\,{\rm Tr}[g] \ ; \ee and we obtain \be
\Delta_{AB}\,=\,\Delta_A\delta_{AB} \ee where
  $\Delta_{A}$ is defined as follows:
\be \Delta_A=\left(
-\,\Delta,\,+\Delta,\,+\Delta,\,+\Delta,\,0,\,0\right)\
.\label{delta2sc}\ee Therefore the effective lagrangian for free
quarks in the 2SC model
 can be written as follows
 \be {\cal L}_D=\int\frac{d\b v}{4\pi}\sum_{A=0}^5
\chi^{A\dagger}\left(\matrix{iV\cdot \partial & -\Delta_A\cr
-\Delta_A &i\tilde V\cdot \de}\right)\chi^A\ +\ (L\to R)\
.\label{2sccomplete}\ee From this equation one can immediately
obtain the free fermion propagator that  in momentum space is
still given by (\ref{propagatorcfl}),
 with the $\Delta_A$ given by (\ref{delta2sc}).

\subsection{Gradient expansion for the $U(1)$ NGB in the CFL model and in the 2SC
model\label{gradient}}

In order to illustrate the procedure of evaluating the self-energy
of the NG bosons we will consider here the case of the NG boson
associated with the breaking of $U(1)_B$. In particular, we will
get the first terms in the effective action for the NG bosons by
performing an expansion in momenta, $\ell\ll \Delta$. An expansion
of this kind is also called gradient expansion (see
\cite{Eguchi:1976iz} for a very clear introduction).

To start with we consider the $U(1)_B$ NG boson within the CFL
model. Therefore we can put to zero all the NG fields in the HDET
lagrangian except for the field $U=\exp{i\sigma/f_\sigma}$.
Correspondingly the quark lagrangian of Eq. (\ref{7.20})with the
modification (\ref{7.31}) becomes
  \be {\cal L}_D =
 \sv~\sum_{A=1}^9
\chi^{A\dagger}\left(\matrix{iV\cdot \partial &
-\Delta_A\,U^{\dag\,2}\cr -\Delta_A \,U^{2} &i\tilde
V\cdot\partial}\right)\chi^A\ +\ (L\to R)\
.\label{cflcompleteexternal0}\ee At the lowest order in the field
$\sigma$ we have \be {\cal L}_\sigma\approx \sv~\sum_{A=1}^9
\chi^{A\dagger} \Delta_A\left(\matrix{0 &\dd
\frac{2i\sigma}{f_\sigma}\,+\, \frac{2\sigma^2}{f^2_\sigma}\cr \dd
-\frac{2i\sigma}{f_\sigma}\,+\, \frac{2\sigma^2}{f^2_\sigma}
&}\right)\chi^A\ +\ (L\to R)\ ,\label{cflcompleteexternal}\ee
which contains the couplings $\sigma\chi\chi$ and
$\sigma\sigma\chi\chi$. Notice that $\sigma$ does not propagate at
tree level, however a non trivial kinetic term is generated by
quantum corrections. To show this we consider the generating
functional  where we take only left-handed
 fields for simplicity. Also in this case we will make use of
 the replica trick to take into account the fact that $\chi$ and
$\chi^\dagger$ are not independent variables. Therefore  we will
have to take the square root of the fermion determinant. In
\cite{Casalbuoni:2000na,Casalbuoni:2000nb} this was taken into
account by dividing by a factor 2 the sum over the velocities
appearing in the fermionic loops \be\sv\to\int\frac{d\b
v}{8\pi}.\ee The generating functional is
 \be
 Z[\sigma]\,=\,\int{\cal D}\chi{\cal D}\chi^\dag
 \exp\left\{i\int\chi^\dag
 A\chi\right\}
\ee where we have introduced (omit the indices of the fields for
simplicity)\be
A=S^{-1}+\frac{2i\sigma\Delta}{f_\sigma}\,\Gamma_0\,+\,
\frac{2\sigma^2\Delta}{f^2_\sigma}\,{\Gamma_1} \ee and \be
\Gamma_0=\left(\matrix{0 &+1\cr -1 &0}\right) \ ,\hskip1cm
\Gamma_1=\left(\matrix{0 &1\cr 1 &0}\right) \ .\ee with $S^{-1}$
the free propagator. Performing the integration over the Fermi
fields we get \be Z[\sigma]=({\rm det}[A])^{1/2}=e^{\dd{\frac 1
2{\rm Tr}[\log A]}}.\ee Therefore \be S_{eff}(\sigma)=-\frac
i2{\rm Tr}[\log A].\ee Evaluating the trace we get
 \bea &&-\,i\,Tr\log A\,=\,- i \,Tr\log
S^{-1}\left(1\,+\, S\,\frac{2i\sigma\Delta}{f_\sigma}\Gamma_0
\,+\,S\,\frac{2\sigma^2\Delta}{f^2_\sigma}{\Gamma_1}\right)\cr&&=
-i\,Tr\log S^{-1}-i\sum_{n=1}^\infty\frac{(-1)^{2n-1}}{n}
\left(iS\,\frac{2i\sigma\Delta}{f_\sigma}i\Gamma_0
\,+\,iS\frac{2\,\sigma^2\Delta}{f^2_\sigma}{i\Gamma_1} \right)^n\
.\eea This is a loop expansion. At the lowest order  it produces
the effective action \bea {\cal S}_{eff}&=&\frac i 4 \,Tr\int dx
dy\ \left[
\frac{i\,S(y,x)2i\sigma(x)\Delta}{f_\sigma}\,i\,\Gamma_0\,
\frac{i\,S(x,y)2i\sigma(y)\Delta}{f_\sigma}\,i\,\Gamma_0\right]\cr&&\cr
&+&\,\frac i 2\,Tr\,\int dx \,\left[
\frac{i\,S(x,x)\,2\,\Delta\,\sigma^2(x)}{f^2_\sigma}\,i\,\Gamma_1\right]\
.\eea
\begin{center}
\begin{figure}[b]
\epsfxsize=8truecm \centerline{\epsffile{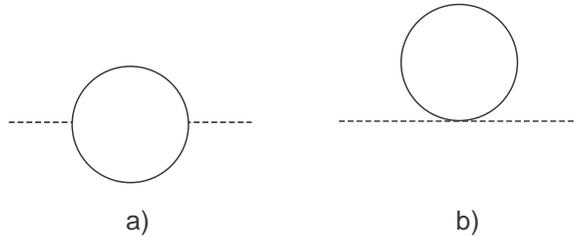}} \noindent
\caption{\it One-loop diagrams. External lines represent the NG
boson field $\sigma$. Full lines are fermion
propagators.}\label{grafici}
\end{figure}
\end{center}
The two terms correspond to the diagrams in Fig. \ref{grafici},
i.e. the self-energy, Fig. \ref{grafici} a), and the tadpole, Fig.
\ref{grafici} b). They can be computed by the following set of
Feynman rules to provide $S_{eff}(\sigma)$ In momentum space the
Feynman rules are as follows:
\begin{enumerate}
  \item For each fermionic internal line with momentum $p$, associate  the propagator
  \be iS_{AB}(p)\,=\,i\delta_{AB}S(p)\,=
  \,\frac{i\,\delta_{AB}}{V\cdot \ell\,\tilde V\cdot
\ell-\Delta_A^2+i\epsilon}\left(\matrix{\tilde V\cdot \ell &
\Delta_A\cr \Delta_A & V\cdot \ell}\right)\label{propagator}\ ;\ee
  \item Each vertex introduces a term $i{\cal L}_{int}$ that
  can be obtained from the effective lagrangian;
for example the $\sigma$ couplings to quarks can be derived from
(\ref{cflcompleteexternal}); \item For each internal momentum not
constrained by the momentum conservation perform the integration
 \be \frac{4\pi\mu^2}{(2\pi)^4}\int d^2\ell\,=\,\frac{\mu^2}{4\pi^3}\int_{-\delta}^{+\delta}
 d\ell_\parallel\,
 \int_{-\infty}^{+\infty}d\ell_0\ ;\ee
  \item A factor $(-1)$ and a factor of 2 for
  each fermion loop to take into account the spin ($L+R$). Also a
  factor 1/2 is necessary due to the replica trick. This 1/2 will
  be associated to the sum over the velocities
  \item A statistical factor arising from the Wick theorem if needed.
\end{enumerate}The result of the calculation
of the effective lagrangian in momentum space is as follows: \bea
i{\cal L}_I&=&-2\frac 1 2\frac{\mu^2}{4\pi^3}\frac 1
2\sv\sum_{A,B} \int d^2\ell\,\cr&\times& Tr \left[iS_{AB}(\ell+p)
\frac{2i\Delta_B\sigma}{f_\sigma} i\Gamma_0 iS_{BA}(\ell)
\frac{2i\Delta_A\sigma}{f_\sigma}i\Gamma_0\right]\ ,\cr&&\cr
i\,{\cal L}_{II}&=&-2\,\frac{\mu^2}{4\pi^3}\frac 1 2\sv\sum_{A,B}
\int d^2\ell\, Tr \left[iS_{AB}(\ell)
\frac{2\,\Delta_B\sigma^2}{f^2_\sigma}\,i\,\Gamma_1 \right], \eea
 corresponding  to the two diagrams of
Figs. \ref{grafici} a) and \ref{grafici} b) respectively.  After
some computation one has \bea &&i\,{\cal L}_{eff}(p)\,=\,i\,{\cal
L}_{I}(p)\,+ \,i\,{\cal L}_{II}(p)\,=\,-\frac 1 2\sv\sum_{
A}\frac{\mu^2\Delta^2_A}{\pi^3f^2_\sigma} \cr&\times& \int
d^2\ell\,\Big[\frac{\tilde V\cdot(\ell+p)\sigma
V\cdot\ell\,\sigma+ V\cdot(\ell+p)\sigma \tilde
V\cdot\ell\sigma-2\Delta^2_A\sigma^2}{D_A(\ell+p)D_A(\ell)}
-\,\frac{2\sigma^2}{D_A(\ell)} \Big], \eea where we have defined
\be D_A(p)\,=\,V\cdot p\,\tilde V\cdot p-\Delta_A^2+i\epsilon\ .
\ee One can immediately notice that \be \,{\cal
L}_I(p=0)\,+\,{\cal L}_{II}(p=0)\,=\,0\ . \ee This result implies
that the scalar $\sigma$ particle has no mass, in agreement with
Goldstone's theorem.
 To get the effective lagrangian in the CFL model at the lowest order in the $\sigma$ momentum
 we expand the function in $p$ ($|p|\ll|\Delta$) to get, in momentum space \be
i\,{\cal L}_{eff}(p)\,=\,-\frac 1
2\sv\sum_{A}\,\frac{2\mu^2\Delta^4_A}{\pi^3f^2_\sigma}(V\cdot
p)\sigma(\tilde V\cdot p )\sigma\ I_2\ ,\ee where we have defined
\be I_2=\int\,\frac{ d^2\ell}{D_A^3(\ell)} .\ee This and other
integrals can be found in Appendix \ref{appendix2}. In getting
this result we have used also the fact
\be\int\frac{(V\cdot\ell)^2}{D_A(\ell)^3}=\int\frac{(\tilde
V\cdot\ell)^2}{D_A(\ell)^3}=0.\ee In fact, since the integral is
convergent we can send the cutoff $\delta$ to $\infty$ and to use
the Lorentz invariance in 2 dimensions to prove that these
integrals are proportional to $V^2=\tilde V^2=0$. We have \be
I_2=-\frac{i\,\pi}{2\Delta_A^4}\ .\ee Therefore we get, in
configuration space, \be {\cal
L}_{eff}(x)\,=\,\frac{9\mu^2}{\pi^2f^2_\sigma}\frac 1
2\sv(V\cdot\partial \sigma)(\tilde V\cdot\partial\sigma)\ .\ee
Since
\begin{equation}
\frac 1 2 \int\frac{d\b v}{4\pi} V^\mu \tilde V^\nu=\frac 1 2
  \left(\matrix{
    1 & 0&0&0 \cr 0 & -\frac 1 3 &0&0 \cr 0 & 0&-\frac 1 3 &0
    \cr 0 & 0&0&-\frac 1 3}\right)_{\mu\nu}\ ,
\end{equation}we obtain
 \be
{\cal L}_{eff}(x)\,=\,\frac{9\mu^2}{\pi^2f^2_\sigma}\frac 1
2\left((\partial_0\sigma)^2 \,-\,v^2_\sigma(\vec\nabla\sigma)^2
\right) \ ,\ee with \be v^2_\sigma\,=\,\frac 1 3 \
.\label{velocity}\ee This kinetic lagrangian has the canonical
 normalization factor provided
 \be f_\sigma^2=\frac{9\mu^2}{\pi^2}\hskip 1cm(\rm CFL)\ ,\label{fsigma}\ee
 Therefore the effective lagrangian for the NGB $\sigma$ particle
is:
 \be
{\cal L}_{eff}=\frac{1} 2 \left((\partial_0\sigma)^2 \,-\,\frac 1
3(\vec\nabla\sigma)^2 \right)=\frac 12f_\sigma^2\left(\dot U{\dot
U}^\dagger-\frac 1 3{\bm\nabla}U\cdot{\bm\nabla}U^\dagger\right) \
,\label{efflagsigma}\ee
 We note that
the value of the velocity (\ref{velocity}) is a consequence of the
average over the Fermi velocities and reflects the number of the
space dimensions, i.e. 3. Therefore it is universal and we expect
  the same value in all the calculations of this type.

In the case of the 2SC model there is no $\sigma$ field since the
baryon number is not broken. However, neglecting the mass of the
$\tau$ field one can perform the same kind of calculation by using
the invariant coupling derived for the CFL case. The final result
differs from the CFL case only in the coefficient in front of
(\ref{fsigma}) which reflects the number of color-flavor gapped
degrees of freedom,
 9 in the CFL case and 4 in the 2SC case; therefore one has
 \be f_\tau^2=\frac{4\mu^2}{\pi^2}\hskip 1cm(\rm 2SC)\ ,\label{fsigma2sc}\ee whereas
 the result (\ref{velocity}), being universal, holds also in this case. The NGB effective lagrangian
 is still of the form
 (\ref{efflagsigma}). The NGB  boson is, in this case, only a
 would-be NGB because the axial $U(1)$ is explicitly broken, though
 this breaking is expected to be small at high $\mu$ since the
 instanton density vanishes for increasing $\mu$.
 \cite{Son:2000fh}.

    \subsection{The parameters of the NG bosons of the CFL phase\label{ngbcfl}}

Using the Feynman rules given above and the interaction
lagrangians (\ref{trilin}) and (\ref{quadrilin}) we get the
effective lagrangian as follows:
 \be  {\cal
L}_{eff}^{kin}=  \frac {\mu^2(21-8\ln 2)}{36\pi^2F^2}\,\frac
1{2}\sum_{a=1}^8 \left(\dot\Pi^a\dot\Pi^a-\frac 1
3|\vec\nabla\Pi_a|^2\right) \ .\ee Comparing with the effective
lagrangian in Eq. (\ref{6.17}) we see that \be F_T^2=F^2= \frac
{\mu^2(21-8\ln 2)}{36\pi^2},~~~~v^2=\frac 1 3.\ee Therefore the
pion satisfy the dispersion relation \be E=\frac 1{\sqrt{3}}|\b
p|.\ee We notice that the evaluation of $F_T$ could be done just
computing the coupling of the pions to their currents. To this end
one has to compute  the diagram of Fig. \ref{NGcoupling}
\cite{Casalbuoni:2000na,Casalbuoni:2000nb}. The result is
\be\langle 0|J_\mu^a|\Pi^b\rangle=iF\delta_{ab}\tilde
p_\mu,~~~\tilde p_\mu=\left(p^0,\frac 1 3\b p\right),\ee with $F$
the same evaluated before. This result is particularly interesting
since it shows how the conservation of the color currents is
realized through   the dispersion relation of the pions. In fact
\be p\cdot\tilde p=E^2-\frac 1 3 |\b p|^2=0.\ee
\begin{center}
\begin{figure}[ht]
\epsfxsize=6truecm \centerline{\epsffile{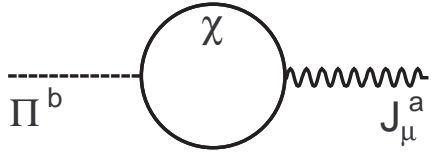}}
\noindent \caption{\it The one-loop contribution to the coupling
of the NG bosons (dotted line)  to their currents (wavy line). The
fermions in the loop are represented by a solid
line.\label{NGcoupling}}
\end{figure}
\end{center}

\subsection{The masses of the NG bosons in the CFL
phase}\label{VIIE}

Up to now we have considered massless quarks. In this Section we
want to determine the influence of mass terms at the level of NG
bosons. Remember that the QCD mass terms have the structure \be
\bar\psi_LM\psi_R~~+~~h.c.\label{7.77}\ee where $M$ is the quark
mass matrix. In order to determine the structure of the NG boson
masses it is convenient to think to $M$ as a field (usually called
a spurion field) transforming in such a way that the expression
(\ref{7.77}) is invariant. Recalling the transformation properties
(\ref{6.2}) of quark fields \be \psi_L\to e^{i(\alpha+\beta)}
g_c\psi_L g_L^T,~~~\psi_R\to e^{i(\alpha-\beta)} g_c\psi_R
g_R^T,~~~e^{i\alpha}\in U(1)_B,~~~e^{i\beta}\in U(1)_A,
\label{7.78}\ee it follows that $M$ transforms as\be M\to
e^{2i\beta} g_L Mg_R^\dagger.\ee In this Section we will make use
of the $U(3)$ fields \be X\to g_cXg_L^T
e^{-2i(\alpha+\beta)},~~~Y\to g_cYg_R^T e^{-2i(\alpha-\beta)}.\ee
In this discussion the $U(1)_A$ symmetry will play an important
role. We recall again that this symmetry is broken by the anomaly
but this breaking goes to zero in the high density limit. If, in
particular,  we consider the subgroup $(Z_2)_A$ we see that NG
boson mass terms can be only even in the matrix $M$. Linear terms
in $M$ may arise only from this breaking which is suppressed by
$(\Lambda_{QCD}/\mu)^8$ (see discussion in Section \ref{VC}). In
order to build up the most general mass term for the NG bosons let
us introduce the gauge invariant $U(3)$ field \be
\tilde\Sigma=(Y^\dagger X)^T= e^{4i\theta}\Sigma^T,\ee where
$\theta$ is the $U(1)_A$ field and $\Sigma$ was defined in
(\ref{6.7}). We have \be \tilde\Sigma\to
e^{-4i\beta}g_L\tilde\Sigma g_R^\dagger.\ee We recall also the
other two gauge invariant fields
 defined in (\ref{6.5}) \be d_X={\rm
det}(X)=e^{6i(\phi+\theta)},~~~d_Y={\rm
det}(Y)=e^{6i(\phi-\theta)},\ee and transforming as \be d_X\to
e^{-6i(\alpha+\beta)} d_X,~~~d_Y\to e^{-6i(\alpha-\beta)}d_Y.\ee
We have also \be {\rm det}(M)\to e^{6i\beta}{\rm det}(M),~~~{\rm
det}(\tilde\Sigma)\to  e^{-12i\beta}{\rm det}(\tilde\Sigma).\ee
Notice that since $d_X$ and $d_Y$ are the only terms transforming
under $U(1)_V$, invariance requires that the effective lagrangian
should depend only on the combination \be d_Xd_{Y^\dagger}={\rm
det}(\tilde\Sigma).\ee Other quantities that can be considered are
\bea M\Sigma^\dagger,&\to& e^{6i\beta}g_L M\Sigma^\dagger
g_L^\dagger,~~~~ M^\dagger\Sigma\to e^{-6i\beta}g_R
M^\dagger\Sigma g_R^\dagger\cr M^{-1}\Sigma&\to& e^{-6i\beta}g_R
M\Sigma^\dagger g_R^\dagger,~~~~ M^{-1\dagger}\Sigma^\dagger\to
e^{6i\beta}g_L M^{-1\dagger}\Sigma^\dagger g_L^\dagger.
\label{7.87}\eea One can build terms transforming only with
respecto to $U(1)_A$ by taking traces of these quantities.
However, taking into account the Cayley identity for $3\times 3$
matrices\be A^3- Tr(A)\,A^2 +\frac 1 2 ((Tr(A))^2-Tr(A^2))\,A
-{\rm det}\,(A)=0, \ee we see that multiplying this expression by
$A^{-1}$ and taking the trace we can express the trace of terms in
the second line of (\ref{7.87}) as a combination of the trace of
the terms in the first line, their second power and the
determinants of $M$ and $\tilde\Sigma$. In the same way we see
that it is enough to consider only the first and the second power
of the terms in the first line of (\ref{7.87}). Summarizing, it is
enough to take into considerations the following quantities \bea
&{\rm
det}(\tilde\Sigma),~~{\rm det}(M),~~ {\rm det}(M^\dagger),&\nn\\
&{\rm Tr}[M\tilde\Sigma^\dagger],~~{\rm
Tr}[M^\dagger\tilde\Sigma],~~{\rm
Tr}[(M\tilde\Sigma^\dagger)^2],~~{\rm
Tr}[(M^\dagger\tilde\Sigma)^2].\eea Therefore the most general
invariant term involving the quark mass matrix $M$ is of the form
\bea &I=({\rm det}(\tilde\Sigma))^{a_1}({\rm det}(M))^{a_2}({\rm
det}(M^\dagger))^{\bar a_2}&\nn\\&\times({\rm
Tr}[M\tilde\Sigma^\dagger])^{a_3}({\rm
Tr}[M^\dagger\tilde\Sigma])^{\bar a_3}({\rm
Tr}[(M\tilde\Sigma^\dagger)^2])^{a_4}({\rm
Tr}[(M^\dagger\tilde\Sigma)^2])^{\bar a _4}&.\eea Requiring
analyticity all the exponents must be integers. The term $I$ is
invariant under the full group except for $U(1)_A$. Requiring also
$U(1)_A$ invariance we get the equation \be -2 a_1+(a_2-\bar
a_2)+(a_3-\bar a_3)+2(a_4-\bar a _4)=0.\ee If we ask $I$ to be of
a given order $n$ in the mass matrix we have also the condition
\be 3(a_2+\bar a_2)+(a_3+\bar a_3)+2(a_4+\bar
a_4)=n\label{7.92}.\ee If we subtract these two equations one by
another we find \be 2a_1+4a_2+2\bar a_2+2a_3+4
a_4=n\label{7.93},\ee which implies that $n$ must be even, as it
should be by $(Z_2)_A$ invariance alone. This argument shows also
that the it would be enough to require the invariance under the
discrete axial group. If we now select $n=2$ (the lowest power in
the mass), Eq. (\ref{7.92}) implies \be a_2=\bar a_2=0.\ee
Therefore the only solutions to our conditions are \bea &a_1=1,
~~~a_3=2,~~~\bar a_3=0,~~~a_4=0,~~~\bar a_4=0&\nn\\&a_1=1,
~~~a_3=0,~~~\bar a_3=0,~~~a_4=1,~~~\bar a_4=0&
\\&a_1=0, ~~~a_3=1,~~~\bar a_3=1,~~~a_4=0,~~~\bar a_4=0&.\nn\eea
If the matrix $M$ has no zero eigenvalues we can use the Cayley
identity to write a linear combination of the first two solutions
in the following form \be{\rm Tr}[M^{-1}\tilde \Sigma]{\rm
det}(M)=\frac 1 2{\rm det}(\tilde\Sigma)\left\{\left({\rm
Tr}[M\tilde\Sigma^\dagger]\right)^2-{\rm
Tr}[(M\tilde\Sigma^\dagger)^2]\right\}.\ee In conclusion the most
general invariant term is given by \bea &{\cal
L}_{masses}=-c\left({\rm det}(M){\rm
Tr}[M^{-1}\tilde\Sigma]+h.c.\right)-c'\left({\rm
det}({\tilde\Sigma}){\rm
Tr}[(M\tilde\Sigma^\dagger)^2]+h.c.\right)&\nn\\&-c''\left({\rm
Tr}[M\tilde\Sigma^\dagger]{\rm
Tr}[M^\dagger\tilde\Sigma]\right).&\eea The coefficients appearing
in this expression can be evaluated by using the matching
technique
\cite{Rho:1999xf,Hong:1999ei,Beane:2000ms,Son:1999cm,Son:2000tu}
(see also the review paper \cite{Schafer:2003ab}). The idea is to
evaluate the contribution of ${\cal L}_{masses}$ to the vacuum
energy. One starts evaluating an effective four-fermi interaction
due to a hard gluon exchange  and performing a matching between
QCD and HDET as illustrated in Fig. \ref{NGmasses}. The
contribution from QCD arises from a double chirality violating
process producing a contribution proportional to the square of the
masses \cite{Schafer:2001za}.
%%%%%%%%%%%%%%%%%%%%%%%%%%%%%%%%%%%%%%%%%%%
\begin{center}
\begin{figure}[ht]
\epsfxsize=9truecm \centerline{\epsffile{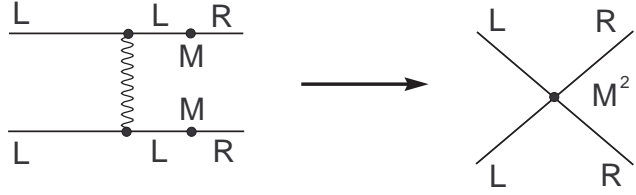}} \noindent
\caption{\it The effective four-fermi interaction produced by a
chirality violating process in QCD. There are other contributions
obtained by exchanging the points where the gluon line is attached
with the mass insertions.\label{NGmasses}}
\end{figure}
\end{center}
%%%%%%%%%%%%%%%%%%%%%%%%%%%%%%%%%%%%%%%%%
Then one uses this  effective interaction in HDET to evaluate the
contribution to the  vacuum energy to be matched against the
contribution from ${\cal L}_{masses}$ obtained by putting
$\tilde\Sigma=1$ (the vacuum state). This is illustrated in Fig.
\ref{NGvacuum}.
%%%%%%%%%%%%%%%%%%%%%%%%%%%%%%%%%%%%%%%%%%%
\begin{center}
\begin{figure}[ht]
\epsfxsize=5.5truecm \centerline{\epsffile{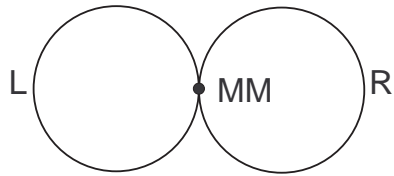}}
\noindent \caption{\it The effective four-fermi interaction of
Fig. \ref{NGmasses} is used to evaluate the contribution to the
vacuum energy of the terms proportional to $M^2$.
\label{NGvacuum}}
\end{figure}
\end{center}
%%%%%%%%%%%%%%%%%%%%%%%%%%%%%%%%%%%%%%%%%
The result of this computation is \be
c=\frac{3\Delta^2}{2\pi^2},~~~ c'=c''=0.\ee There is a simple way
of understanding why only the $c$ contribution survives at the
leading order. In fact, this is the only term which arises working
at the second order in perturbation theory and approximating the
product of Fermi fields with the corresponding NG fields $X$ and
$Y$. We find \bea &(\bar\psi_LM\psi_R)^2=(\psi_{L\alpha}^{i
\dagger}M_i^j\psi_{Rj}^\alpha)(\psi_{L\beta}^{k
\dagger}M_k^l\psi_{Rl}^\beta)\approx
\epsilon^{ikm}\epsilon_{\alpha\beta\gamma} X_m^\gamma M_i^j M_k^l
\epsilon_{jlp}\epsilon^{\alpha\beta\delta} Y_\delta^{p*}&\nn\\
&\approx \epsilon^{ikm}\epsilon_{jlp}M_i^j
M_k^l\tilde\Sigma_m^p=\epsilon^{ikm}\epsilon_{jlp}M_i^j M_k^l
M_m^a(M^{-1})_a^b\tilde\Sigma_b^p \approx {\rm det}(M){\rm
Tr}[M^{-1}\tilde\Sigma].\eea

There is also another interesting point about the quark masses.
Starting from the complete QCD lagrangian \be {\cal
L}_{QCD}=\bar\psi(i\slash D+\mu\gamma_0)\psi-\bar\psi_L
M\psi_R-\bar\psi_R M^\dagger\psi_L-\frac 1 4 G_{\mu\nu}^a
G^{\mu\nu a}\ee and repeating what we did in Section \ref{HDET}
for deriving the HDET expansion we find that at the leading order
the expression for the negative energy left-handed fields is \be
\psi_{-,L}=\frac 1{2\mu}\left(-i\gamma_0\slash
D_\bot\psi_{+,L}+\gamma_0 M\psi_{+,R}\right).\ee Substituting
inside the lagrangian one finds (we neglect here the condensate
terms for simplicity) \be {\cal L}_D=\psi_{+,L}^\dagger(iV\cdot
D)\psi_{+,L}-\frac{1}{2\mu}\psi_{+,L}^\dagger\left[(\slash
D_\bot)^2+M M^\dagger\right]\psi_{+,L}+(L\leftrightarrow R,
M\leftrightarrow M^\dagger)+\cdots.\ee The new term that we have
obtained is just what expected from the expansion of the kinetic
energy of the quark. We see that the mass terms in the effective
lagrangian behave as effective chemical potentials. Notice that in
presence of a chemical potential a Dirac lagrangian has the form
\be\bar\psi i\gamma_\nu(\de^\nu-i\mu g^{\nu 0})\psi,\ee therefore
the chemical potential acts as the fourth component of an abelian
gauge field. This implies that the lagrangian is invariant under
the gauge transformation \be\psi\to e^{i\alpha(t)}\psi,~~~~\mu\to
\mu+\dot\alpha(t).\ee This argument implies that the HDET
lagrangian has the invariance\footnote{Notice that $\slash D_\bot$
does not involve time derivatives.} \be\psi_L\to
L(t)\psi_L,~~~~\psi_R\to R(t)\psi_R,\ee where $L(t)$ and $R(t)$
are time dependent flavor transformation and with \be
X_L=\frac{1}{2\mu} M M^\dagger,~~~X_R=\frac{1}{2\mu} M^\dagger
M\ee transforming as left- and right- handed gauge fields
\cite{Bedaque:2001je}. But then also the effective lagrangian for
the filed $\Sigma$ has to satisfy this symmetry, meaning that we
should substitute the time derivative  with the covariant
derivative \be \de_0\Sigma\to\nabla_0\Sigma=\de_0\Sigma
+i\Sigma\left(\frac{M
M^\dagger}{2\mu}\right)^T-i\left(\frac{M^\dagger M
}{2\mu}\right)^T\Sigma,\ee  where we have taken into account that
\be M M^\dagger\to g_L MM^\dagger g_L^\dagger,~~~M^\dagger M\to
g_R M^\dagger M g_L^\dagger,~~~\Sigma \to g_R^*\Sigma g_L^T.\ee
This result has been confirmed by a microscopic calculation done
in \cite{Bedaque:2001je}. We can now see that a generic term in
the expansion of the effective lagrangian has the form \be
F^2\Delta^2\left(\frac{\de_0-iM
M^\dagger/(2\mu)}{\Delta}\right)^n\left(\frac{\bm\nabla}{\Delta}\right)^m
\left(\frac{M^2}{F^2}\right)^p(\Sigma)^q(\Sigma^\dagger)^r\label{7.109}.\ee
The various terms are easily understood on the basis of the fact
that the momentum expansion is a series in $p/\Delta$. The factor
$\Delta^2$ in front is just to adjust the normalization of the
kinetic terms arising for $n=2$ and $m=0$ and $n=0$, $m=2$, with
$q=r=1$. The term quadratic in the mass matrix, $M$, has also the
correct normalization. We now see that the contribution to the
square masses of the mesons of the masses originating from the
kinetic energy expansion and the ones coming from the explicit
quark mass terms might be of the same order of magnitude. In fact,
the contribution to the NG bosons square masses are (the extra
$1/F^2$ terms originate from the expansion of $\Sigma$)\bea&&{\rm
kinetic~ energy}:~~~~F^2\Delta^2\frac {m^4}{\mu^2\Delta^2}\frac
1{F^2}\approx \frac{m^4}{\mu^2},\cr &&{\rm quark~
masses}:~~~~F^2\Delta^2\frac{m^2}{F^2}\frac 1 {F^2}\approx
\frac{\Delta^2 m^2}{F^2},\eea and the two contributions are of the
same order of magnitude  for $m\approx\Delta$, since $F\approx
\mu$. Before studying the mass spectrum we will now introduce
another subject which is relevant to the physics of the NG bosons
of the CFL phase, that is the subject of Boson-Einstein
condensation.

\subsubsection{The role of the chemical potential for scalar
fields: Bose-Einstein condensation}

Let us consider a relativistic complex scalar field (however
similar considerations could be done for the non-relativistic
case) described by the lagrangian density \be
\de_\mu\phi^\dagger\de^\mu\phi-m^2\psi^\dagger\phi-\lambda(\phi^\dagger\phi)^2.\ee
The quartic term gives rise to a repulsive interaction for
$\lambda>0$ as it is required by the stability of theory. The
lagrangian has a phase symmetry $U(1)$ giving rise to a conserved
current. A simple way of introducing the chemical potential is to
notice that we can promote the global $U(1)$ to a local one by
adding a gauge field. However, by definition \be \frac{\de {\cal
L}}{\de A_\mu}=-j^\mu,\ee and therefore the charge density is
obtained by varying the lagrangian with respect to the fourth
component  of the gauge potential. But in a system at finite
density the variation with respect to the chemical potential is
just the corresponding conserved charge, therefore the chemical
potential must  enter in the lagrangian exactly as the fourth
component of a gauge field. Therefore we get \bea {\cal
L}&=&(\de_0+i\mu)\phi^\dagger(\de_0-i\mu)\phi-({\bm\nabla}\phi^\dagger)
\cdot({\bm\nabla}\phi)-m^2\phi^\dagger\phi-\lambda(\phi^\dagger\phi)^2\nn\\&=&
\de_\mu\phi^\dagger\de^\mu\phi-(m^2-\mu^2)\phi^\dagger\phi-
\lambda(\phi^\dagger\phi)^2+i\mu(\phi^\dagger\de_0\phi-\de_0\phi^\dagger\phi).
\eea Notice that the last term breaks the charge conjugation
symmetry of the theory since $\mu$ multiplies the charge density.
As a result the mass spectrum is given by \be p^2-(m^2-\mu^2)+2\mu
Q p_0=0,\ee where $Q=\pm 1$ is the charge for particles and
antiparticles. The dispersion relation can be written as \be
(E+\mu Q)^2=m^2+|\b p|^2\ee with $E=p_0$. We see that the mass of
particles and antiparticles are different and given by \be
m_{p,\bar p}=\mp\mu+m.\label{7.116}\ee This is what happens if
$\mu^2< m^2$. At $\mu^2=m^2$ the system undergoes a second order
phase transition and a condensate is formed. The condensate is
obtained by minimizing the potential \be
V(\phi)=(m^2-\mu^2)\phi^\dagger\phi+
\lambda(\phi^\dagger\phi)^2,\ee from
which\be\langle\phi^\dagger\phi\rangle=\frac{\mu^2-m^2}{2\lambda}.\ee
Notice also that at $\mu=m$  the mass of the particle mode goes to
zero. In correspondence of the condensation the system has a
charge density given by \be \rho=-\frac{\de V}{\de\mu}=2\mu
\langle\phi^\dagger\phi\rangle=\frac{\mu}{\lambda}(\mu^2-m^2).\ee
Therefore the ground state of the system is a Bose-Einstein
condensate. Defining
\be\langle\phi\rangle=\frac{v}{\sqrt{2}},~~~v^2=\frac{\mu^2-m^2}{\lambda},\ee
we can derive the physical spectrum of the system through the
replacement \be \phi(x)=\frac
1{\sqrt{2}}(v+h(x))e^{\dd{i\theta(x)/v}}.\ee The quadratic part of
the lagrangian is \be {\cal L}_2=\frac 1
2\de_\mu\theta\de^\mu\theta+\frac 1 2\de_\mu h\de^\mu h-\lambda
v^2 h^2-2\mu h\de_0\theta.\ee The mass spectrum is given by the
condition \be{\rm det}\left(\matrix{ p^2-2\lambda v^2 & 2i\mu E\cr
-2i\mu E & p^2}\right)=0.\ee At zero momentum we get
($p^\mu=(M,{\bm 0})$)\be M^2(M^2-2\lambda v^2-4\mu^2)=0.\ee
Therefore  the antiparticle remains massless after the transition,
whereas the particle gets a mass given by \be M^2=6\mu^2-2m^2.\ee
At the transition point these two masses agree with the ones in
the unbroken phase, Eq. (\ref{7.116}). The masses in the two
phases are illustrated in Fig. \ref{bose}.
%%%%%%%%%%%%%%%%%%%%%%%%%%%%%%%%%%%%%%%%%%%
\begin{center}
\begin{figure}[ht]
\epsfxsize=8truecm \centerline{\epsffile{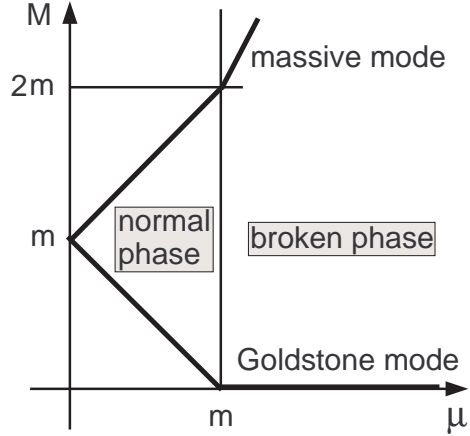}} \noindent
\caption{\it This figure shows the evolution of the particle and
antiparticle masses from the normal phase to the broken one.
\label{bose}}
\end{figure}
\end{center}
%%%%%%%%%%%%%%%%%%%%%%%%%%%%%%%%%%%%%%%%%
It is interesting to find out the energies of these particles in
the small momentum limit. We find for the Goldstone mode\be
E_{NG}\approx\sqrt{\frac{\mu^2-m^2}{3\mu^2-m^2}}|\b p|,\ee whereas
for the massive one \be
E_{massive}\approx\sqrt{6\mu^2-2m^2+\frac{9\mu^2-m^2}{6\mu^2-2m^2}|\b
p|^2}.\ee The velocity of the Goldstone mode is particularly
interesting since it is zero at the transition point and goes to
$1/\sqrt{3}$ at large densities. Remember that this is just the
velocity we have found in our effective lagrangian describing the
NG bosons in the CFL phase.

\subsubsection{Kaon condensation}

Starting from the mass terms for the NG bosons of the CFL phase
obtained in Section \ref{VIIE} one can easily understand the
results for the masses of the flavored NG bosons \bea
m_{\pi^\pm}&=&\mp\frac{m_d^2-m_u^2}{2\mu}+\sqrt{\frac{2c}{F^2}(m_u+m_d)m_s},\nn\\
m_{K^\pm}&=&\mp\frac{m_s^2-m_u^2}{2\mu}+\sqrt{\frac{2c}{F^2}(m_u+m_s)m_d},\nn\\
m_{K^0,\bar
K^0}&=&\mp\frac{m_s^2-m_d^2}{2\mu}+\sqrt{\frac{2c}{F^2}(m_d+m_s)m_u}.
\eea The two terms in this expression have their counterpart in
Eq. (\ref{7.116}). The first term arises from the "chemical
potential" terms $M M^\dagger/2\mu$ whereas the second term comes
from the "true" mass term. Also in this case a Bose-Einstein
condensate might be formed. For simplicity consider the case
$m_s\gg m_u,m_d$. Then we get \bea m_{\pi^\pm}&\approx&\frac 1
{F}\sqrt{
{2cm_s}(m_u+m_d)},\nn\\
m_{K^\pm}&\approx& \mp \frac{m_s^2}{2\mu}+\frac 1 F\sqrt{2cm_s m_d},\nn\\
m_{K^0,\bar K^0}&\approx& \mp \frac{m_s^2}{2\mu}+\frac 1
F\sqrt{2cm_s m_u}.\eea The pion masses are independent on the
chemical potential terms, however  the masses of $K^+$ and $K^0$
are pushed down (whereas the ones of $K^-$ and $\bar K^0$ are
pushed up) and therefore they become massless at \be m_s\Big|_{\rm
crit}=\left(\frac {12\mu^2}{\pi^2 F^2}\right)^{1/3}
\sqrt[3]{m_{u,d}\Delta^2}=6\left(\frac 2{21-8\log
2}\right)^{1/3}\sqrt[3]{m_{u,d}\Delta^2}\approx
3.03\sqrt[3]{m_u\Delta}.\ee The critical value of $m_s$ can vary
between $41~MeV$ for $m_u=1.5~MeV$ and $\Delta=40~MeV$ and
$107~MeV$ for $m_u=4.5~MeV$ and $\Delta=100~MeV$. For larger
values of $m_s$ the modes $K^+$ and $K^0$ become unstable. This is
the signal for condensation \cite{Schafer:2000ew,Bedaque:2001je}.
In fact if we look for a kaon condensed ground state of the
type\footnote{The most general ansatz would be to assume in the
exponential a linear combination of $\lambda^a$, $a=4,5,6,7$, but
it turns out that the effective potential depends only on the
coefficient of $\lambda_4$. This is related to the fact that, as
we shall see, there are three broken symmetries and therefore
there must be three flat directions in the potential.}
\be\Sigma=e^{i\alpha\lambda_4}=1+(\cos\alpha-1)\lambda_4^2+
i\lambda_4\sin\alpha, \ee we obtain from our effective lagrangian,
in the limit of exact isospin symmetry, (subtracting the term for
$\alpha=0$) the potential \be V(\alpha)= F^2\left(-\frac 1
2\left(\frac{m_s^2}{2\mu}\right)^2\sin^2\alpha+(m_K^0)^2(1-\cos\alpha)\right),\ee
where  $m_K^0$ is the lowest order square mass of the kaon in
$m_s$, that is \be (m_K^0)^2=\frac{2\,c\,m_s\,m}{F^2}\ee and
$m=m_u=m_d$. As in the example of a complex scalar field we see
that the "chemical potential" terms give a negative contribution,
whereas the "mass" terms gives a positive one. Therefore it is
convenient to introduce the effective chemical potential
\be\mu_{eff}=\frac{m_s^2}{2\mu}.\ee We see that \be
V(\alpha)=F^2\left(-\frac 1
2\mu_{eff}^2\sin^2\alpha+(m_K^0)^2(1-\cos\alpha)\right).\ee
Minimizing the potential we find a  solution with $\alpha\not=0$,
given by \be \cos\alpha=\frac{(m_K^0)^2}{\mu_{eff}^2}\ee if \be
{\mu_{eff}}\ge m_K^0. \ee One can also derive the hypercharge
density \be n_Y=-\frac{\de V}{\de \mu_{eff}}=\mu_{eff}
F^2\left(1-\frac{(m_K^0)^4}{\mu_{eff}^4}\right).\ee The mass terms
break the original $SU(3)_{c+L+R}$ symmetry of the CFL ground
state to $SU(2)_I\otimes U(1)_Y$. The kaon condensation breaks
this symmetry to the diagonal $U(1)$ group generated by \be
Q=\frac 1 2\left(\lambda_3-\frac 1 {\sqrt{3}}\lambda_8\right).\ee
In fact, one can easily verify that \be [Q,\Sigma]=0.\ee This
result can be simply understood by the observation that under
$SU(2)_I\otimes U(1)_Y$ the NG bosons in $\Sigma$ decompose as a
triplet, the pions, a complex doublet $(\bar K_0,K^-)$ and its
complex conjugate $(K^+,K^0)$ and a singlet, $\eta$. The
$\alpha\not=0$ solution for $\Sigma$ gives rise to an expectation
value for the doublets. We see that the symmetry breaking
mechanism is the same as for the electroweak sector of the
Standard Model (SM). Therefore $SU(2)\otimes U(1)$ is broken down
to $U(1)$

Notice that by the usual counting of NG bosons one would expect 3
massless modes. However we have seen from the unbroken phase that
only two modes become massless at the transition. This comes from
the breaking of Lorentz invariance due to the presence of the
chemical potential in the original QCD lagrangian. A theorem due
to Chada and Nielsen \cite{Chada:1976cn} gives the key for the
right counting of the NG physical bosons. The essence of the
theorem is that the number depends on the dispersion relation of
the NG bosons. If the energy is linear in the momentum (or a odd
power) the counting is normal. If the energy depends on the
momentum quadratically (or through an even power), then there are
two broken generators associated to a single NG bosons. Notice
that in relativistic theories the dispersion relations are always
of the first type. A more recent discussion of this topics is in
\cite{Miransky:2001tw,Schafer:2001bq}. In particular in
\cite{Schafer:2001bq} it is proved a theorem which helps to show
algebraically when the dispersion relations are odd or even in the
momenta. The behavior of the masses with the effective chemical
potential is represented in Fig. \ref{bose2}.

 The effective lagrangian for the NG bosons in the CFL phase has
been useful for the study of many phenomenological interesting
questions in the realm of compact stellar objects
\cite{Kaplan:2001qk,Kaplan:2001hh,Buckley:2002ur,Reddy:2002xc,
Jaikumar:2002vg,Shovkovy:2002kv}.
%%%%%%%%%%%%%%%%%%%%%%%%%%%%%%%%%%%%%%%%%%%
\begin{center}
\begin{figure}[ht]
\epsfxsize=8truecm \centerline{\epsffile{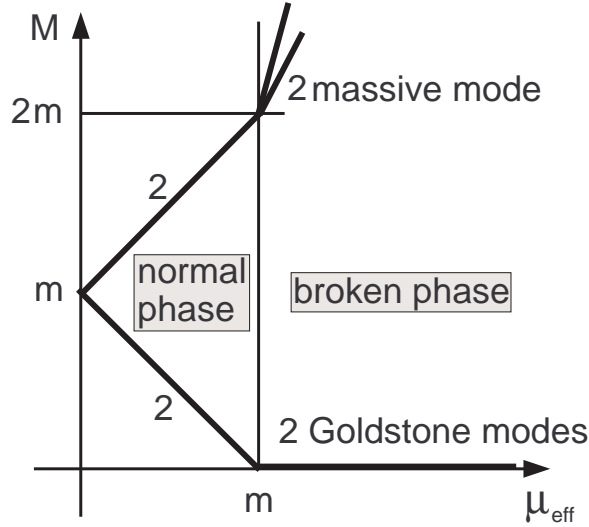}} \noindent
\caption{\it This figure shows the evolution of the masses of the
$K$ modes in going from the normal to the broken phase.
\label{bose2}}
\end{figure}
\end{center}
%%%%%%%%%%%%%%%%%%%%%%%%%%%%%%%%%%%%%%%%%

\section{The dispersion law for the gluons}

In this Section will be evaluate the dispersion relations both for
the gluons in the 2SC and in the CFL phase. In the 2SC phase we
will be able, in this way, of getting the parameters of the
effective theory introduced in Section \ref{General}. We will also
show that the low-energy expansion is accurate at about 30\% when
compared with the numerical calculations.

\subsection{Evaluating the bare gluon mass\label{evaluating}}

In deriving the HDET we have so far neglected the heavy fields
contribution to the leading correction in $1/\mu$ given by the
operator \be -P^{\mu\nu}\psi_+^\dagger\frac{D_\mu
D_\nu}{2\mu+i\tilde V\cdot D}\psi_+.\label{8.1}\ee Remember that
in order to get the effective lagrangian we have integrated out
all the fields with momenta greater that $\delta$. However these
degrees of freedom give a contribution of order $\mu$ which
compensates the $\mu$ in the denominator leaving a finite
contribution to the operator $g^2$, where $g$ is the gluon field.
This contribution is called the bare gluon mass
\cite{Son:1999cm,Son:2000tu} and it must be inserted in the HDET
when one is studying the gluon properties. In fact, as we shall
see, this contribution is needed, in the case of unbroken gauge
symmetries, to cancel a contribution to the Meissner mass coming
form the gluon polarization. To make a correct calculation of
these contribution it is necessary to evaluate the tadpole
contribution from the interaction (\ref{8.1}) for momenta greater
than $\delta$ (see Fig. \ref{meissner_heavy}). Since it  appears a
heavy propagator at zero momentum, which is essentially the
density of the states (see Eq. (\ref{2.36})), it is  easier to
derive the result in a first-quantization approach, where we sum
over all the degrees of freedom within the Fermi sphere in order
to get the two-gluon contribution. Notice also that we are
integrating over residual momenta $|\ell|>\delta>>\Delta$,
therefore we can neglect the gap. As a consequence this
calculation applies to both cases of gapped and/or ungapped
fermions.

We start with the observation  that in first-quantization we have
$H=|\b p|$. Coupling the particle to a gauge field \be H=|\b p-g\b
A|+e A_0\approx |\b p|+g A_0-g\b v\cdot\b A+\frac{g^2}{2|\b
p|}\left(|\b A|^2-(\b v\cdot \b A)^2\right).\ee The first three
terms correspond to the terms considered so far in the HDET
lagrangian. The fourth term is nothing but the operator sandwiched
among the quark fields in Eq. (\ref{8.1}) evaluated at the Fermi
surface. In fact \be P_{\mu\nu}A^\mu A^\nu\approx |\b A|^2-(\b
v\cdot\b A)^2.\ee What we have to do now is just to sum over all
the particles inside the Fermi sphere, which is equivalent to the
tadpole calculation. Notice that we should leave aside the
fermions within the shell of momentum $\mu-\delta<|\b p|<
\mu+\delta$, but this give a negligible contribution since
$\delta\ll\mu$. Therefore we get \be \frac {g^2} 2\times 2\times
N_f\times\int_{|\b p|\le\mu}\frac{d^3\b p}{(2\pi)^3}\frac 1{|\b
p|}{\rm Tr}\left[{\b A}^2-\frac{(\b p\cdot \b A)^2}{|\b
p|^2}\right],\ee where the factors 2 and $N_f$ comes from the spin
and the number of flavors. On the other hand  the trace is over
the color indices. The result is simply \be
N_f\frac{g^2\mu^2}{6\pi^2}\frac 1 2\sum_a{\b A}^a\cdot{\b A}^a.\ee
Therefore in the effective lagrangian of HDET we have to introduce
a term \be -\frac 1 2 m_D^2\sum_a{\b A}^a\cdot{\b A}^a\ee with \be
m_{BM}^2=N_f\frac{g^2\mu^2}{6\pi^2}\label{8.7}.\ee
 One could equally well perform this evaluation by using the
 Feynman rules for the heavy fields, which are easily obtained,
 to determine the contribution to the polarization
 function $\Pi_{\mu\nu}^{ab}$. The contribution is the following
 \be
\Pi_{\mu\nu}^{ab\,{\rm BM}}=(-i) 2\times 2\times
N_f\times(-1)\int\frac{d \ell_\|}{(2\pi)}
\frac{(\ell_\|+\mu)^2}\pi\int\frac{d\b v}{4\pi}\int\frac{
d\ell_0}{2\pi}\frac{ig^2\delta_{ab}}{2(2\mu+\tilde V\cdot
\ell)}\frac{i}{V\cdot\ell} P_{\mu\nu}.\ee The different factors
have the following origin. The first $(-i)$ is due to the
definition of $\Pi_{\mu\nu}$ such to reproduce the mass term in
the lagrangian. Then there is a factor 2 from the spin, a factor 2
from symmetry reasons (2 gluon fields). A factor $N_f$ from the
trace over the heavy fermions, a $(-1)$ from the loop. We have
left the residual momenta as integration variable, but the measure
cannot be any more approximated as $\mu^2$ and we have put back
the original integration factors. There is no extra 2 in the
velocity integration since we have now no necessity of introducing
the Nambu-Gor'kov fields. The next factor arises from the vertex 2
gluons-2heavy fields and it is the result after the trace over the
color indices. Finally we have the propagator of the $\psi_+^h$
field. If we define the propagator at zero momentum as in Section
\ref{IIA} we get a contribution from the integral over $\ell_0$
proportional to $\theta(-\ell_\|)$  (see Eq. (\ref{2.38})). This
restricts the integration over $\ell_\|$ between $-\mu$ and 0.
Performing the calculation one gets easily the result (\ref{8.7}).
In fact we have \bea&\Pi_{\mu\nu}^{ab\,{\rm BM}}=-2iN_f
g^2\delta_{ab}\dd{\int_{-\mu}^0\frac{d\ell_\|}{2\pi}\frac{(\ell_\|+\mu)^2}\pi\frac
1{2\pi}\frac{(2\pi i)}{2(\ell_\|+\mu)}\int \frac {d\b v}{4\pi}
P_{\mu\nu}}=\frac{g^2}{2\pi^2}\delta_{ab}\frac{\mu^2}2(-\frac 2
3\delta_{ij})&\nn\\&=-m_{BM}^2\delta_{ij}\delta_{ab}.&\eea

\subsection{The parameters of the effective lagrangian for the 2SC case}

We have now all the elements to evaluate the parameters appearing
in the effective lagrangian for the 2SC case, which amounts to
evaluate the propagation properties of the gluons belonging to the
unbroken color group $SU(2)_c$. We will need to evaluate the
vacuum polarization function $\Pi_{\mu\nu}^{ab}$, $a,b=1,2,3$,
defined by the diagram in Fig. \ref{gluons}.
 %%%%%%%%%%%%%%%%%%%%%%%%%%%%%%%%%%%%%%%%%%%%%%%
\begin{center}
\begin{figure}[ht]
\epsfxsize=6truecm \centerline{\epsffile{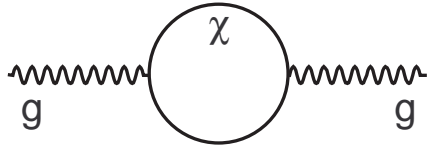}} \noindent
\caption{\it The Feynman diagram corresponding to the vacuum
polarization of the gluons.\label{gluons}}
\end{figure}
\end{center}
%%%%%%%%%%%%%%%%%%%%%%%%%%%%%%%%%%%%%%%%%%%%%%%%%%%%%%%
As usual we need the interaction term which is given by the
coupling of the gauge field with the fermionic current  \be i g
A_\mu^a J_a^\mu.\ee One finds \be J_\mu^a\,=\,\sv\sum_{A,B=0}^5
\chi^{A\dagger}\left(\matrix{i\,V_\mu K_{AaB}& 0\cr 0 &-i\,\tilde
V_\mu K^*_{AaB}}\right)\chi^B+(L\to R)\ . \label{2sccomplete0new}
\ee where we have used the basis (\ref{7.37}) for the quark
fields, leading to the coefficients  $K _{AaB}$ \be
K_{AaB}\,=\,\frac 1
4\,Tr\{\tilde\lambda_A\lambda^a\tilde\lambda_B\}=(\b K)_{AB}\ .\ee
Using our Feynman rules we can compute $\Pi_{\mu\nu}^{ab}$
\cite{Casalbuoni:2001ha}\footnote{For the same calculation using
different methods see: \cite{Rischke:2000cn,Rischke:2000qz}.}.
From the self energy diagram of Fig. \ref{gluons} ($a,b=1,2,3$) we
find:
 \bea \Pi^{\mu \nu\,{\rm self}}_{a b~}(p)
&=& 2\times(-i)(-1)   \int\frac{d\b v}{8\pi}\frac{\mu^2}\pi
 \int \! \frac{d^2\ell}{(2\pi)^2} \, \Big( \frac{ (igV^{\mu})(igV^{\nu})
 i\tilde V\cdot
 \ell\,
 i\tilde V \cdot(\ell+p)
 + (V\leftrightarrow\tilde V)}{D(\ell+p) D (\ell)} \,+\cr&&\cr&-& \Delta^2
\frac{(igV^{\mu})(ig \tilde V^{\nu}) + (igV^{\nu})(ig \tilde
V^{\mu})}{D(\ell+p) D(\ell)} \,\Big){\rm Tr}[{\b K}_a {\b K}_b]\ .
\eea where \be Tr[\b K_a \b K_b]=\delta_{ab}.\ee In order to
derive the effective lagrangian it is enough to expand this
expression up to the second order in momenta and add the
contribution from the bare Meissner mass. The result is
\cite{Casalbuoni:2001ha}
 \be \Pi^{0 0}_{ab}(p)\ =\
\delta_{ab}\frac{ \mu^2 g^2}{18 \pi^2 \Delta^2} |\vec p\,|^2\, ,
\label{glue1} \ee \be \Pi^{k l}_{ab}(p)\ =\ \Pi^{k l\,{\rm
self}}_{ab} +  \Pi^{k l\,{\rm BM}}_{ab} = \ \delta_{ab}
\delta^{kl} \frac{ \mu^2 g^2}{3 \pi^2} \left( 1 + \frac{p_0^2}{6
\Delta^2} \right)-\delta_{ab} \delta^{kl} \frac{ \mu^2 g^2}{3
\pi^2}=\delta_{ab} \delta^{kl} \frac{\mu^2 g^2}{18
\pi^2}\frac{p_0^2}{\Delta^2}\, , \label{glue2} \ee \be \Pi^{0
k}_{ab}(p)\  =\ \delta_{ab} \frac{ \mu^2 g^2}{18 \pi^2 \Delta^2}
 p^0 p^k \,.
\label{glue3} \ee We see that the bare Meissner mass contribution
$\Pi^{k l\,{\rm BM}}_{ab}(p)$ just cancels the zero momentum
contribution to the gluon self-energy. As a consequence the
unbroken gluons remain massless. These results agree with the
outcomes of \cite{Rischke:2000cn,Rischke:2000qz}.

We notice also that there is no contribution to the Debye mass
(the mass associated to the component $\Pi_{ab}^{00}$ of the
vacuum polarization).  This results reflects the fact that in the
2SC model the $SU(2)_c$ color subgroup generated by the first
three generators $T^c$ ($c=1,2,3)$ remains unbroken.

One can now compute the dispersion laws for the unbroken gluons.
From the previous formulas one gets: \be {\mathcal L}\ =\
-\frac{1}{4} F^{\mu \nu}_{a} F_{\mu \nu}^{a} + \frac{1}{2}
\Pi^{\mu \nu}_{ab} A_{ \mu}^a A_{ \nu}^b\ ~. \ee Introducing the
fields $\dd E_i^a \equiv F_{0i}^a$ and $\dd B_i^a \equiv i
\varepsilon_{ijk} F_{jk}^a$, and using (\ref{glue1}),
(\ref{glue2}) and (\ref{glue3}) these results can be written as
follows (assuming gauge invariance the terms with 3 and 4 gluons
are completely fixed, but see later): \be {\mathcal L}\ =\
\frac{1}{2} (E_i^a E_i^a - B_i^a B_i^a) + \frac{k}{2} E_i^a E_i^a
~, \label{33} \ee with \be k= \frac{g^2 \mu^2}{18 \pi^2 \Delta^2}
~. \ee These results have been first obtained in
\cite{Rischke:2000cn}. As discussed in this paper, these results
imply that the medium has a very high {\it dielectric constant}
$\dd \epsilon = k +1$ and a {\it magnetic permeability} $\dd
\lambda = 1$. The gluon speed in this medium is now \be
v=\frac{1}{\sqrt{\epsilon \lambda}} \propto \frac{ \Delta}{g \mu}
\ee and in the high density limit it tends to zero. We have
already discussed the physical consequences of these results in
Section \ref{General}.

In \cite{Casalbuoni:2001ha} we have also computed the vacuum
polarization of the gluons belonging to the broken part of the
gauge group. That is the gluons with color index $a=4,5,6,7,8$. In
Table \ref{tableVII} we give the results at zero momentum in all
the cases.
\begin{table}[ht]
\begin{center}
\begin{tabular}{|c|c|c|} \hline
  \hspace{.8in} & \hspace{.8in} & \hspace{.8in}  \\ [-.06in]
  $\dd a$     &  $\dd \Pi^{00}(0)$  & $\dd - \Pi^{ij}(0)$ \\ [.07in] \hline
                    &           &  \\ [-.09in]
$\dd 1 - 3$   &  $\dd  0$  &  $\dd 0 $ \\ [.07in] \hline
                   &           &  \\ [-.09in]
$\dd 4 - 7$ & $ \frac{3}{2} m^2_g $  &  $ \frac{1}{2} m^2_g$
\\[.07in] \hline
                   &           &  \\ [-.09in]
$\dd 8$   &  $ 3 m^2_g $& $ \frac{1}{3} m^2_g $\\[.07in] \hline
\end{tabular}\\
\caption{Debye and Meissner masses in the 2SC
  phase; $m^2_g=\mu^2 g^2/3 \pi^2$.\label{tableVII}}
\end{center}
\end{table}
These results are in agreement with a calculation performed by
\cite{Rischke:2000qz} with a different method. In
\cite{Casalbuoni:2001ha} we have also made an expansion in
momenta. A priori this expansion cannot be used to derive the
dispersion relation for the broken gluons. In fact their mass,
taking into account the  wave function renormalization (of  order
$g^2\mu^2/\Delta^2$), is of order $\Delta$. Defining the mass of
the gluon as the value of the energy at zero momentum one finds
from the expansion \be m_R=\sqrt{2}\Delta\ee for colors
$a=4,5,6,7$, whereas the numerical calculation gives\be
m_R=0.894\Delta\ee and we see that the expansion overestimates the
value of the mass. For the color 8 we find\be m_R=\frac{\mu
g}\pi.\ee The large value obtained of the mass for the gluon 8 in
the small momentum expansion can be shown to be valid at all
orders in momentum. Except for the case $a=8$ we see that
generally there is a very large wave function renormalization
making the physical gluon masses of order $\Delta$ rather than of
order $g\mu$. This result was shown for the first time in
\cite{Casalbuoni:2000na,Casalbuoni:2000nb} in the case of the CFL
phase. In \cite{Casalbuoni:2001ha}  the one-loop contributions to
the three and four gluon vertices has been evaluated. It has been
checked that the result gives rise to the correct gauge invariant
contribution when added to the tree level functions.

\subsection{The gluons of the CFL phase}

We will discuss here some of the results obtained in
\cite{Casalbuoni:2000na,Casalbuoni:2000nb} for the CFL case. We
will skip  many technical details. We will only give the coupling
of the currents to fermions which are again obtained from the
gauge coupling. Working in the basis $\chi^A$ we use Eq.
(\ref{cflcomplete0}) to write \be {\cal
L}_1\,=\,i\,g\,A^\mu_aJ_\mu^a\ , \ee where \be J_\mu^a\,=\,
\sv\sum_{A,B=1}^9 \chi^{A\dagger}\left(\matrix{i\,V_\mu h_{AaB}&
0\cr 0 &-i\,\tilde V_\mu h^*_{AaB}}\right)\chi^B+(L\to R),
\label{cflcomplete0new} \ee and \be h_{AaB} =\frac 1 4
\,Tr[T_A\lambda_a T_B]\ .\label{cflcomplete0newnew} \ee

Performing the trace we find
 \bea
J_\mu^a&=&\frac i 2 \sqrt{\frac 2 3} \sum_{\vec v}
%\sum_{a=1}^8
\left(\chi^{9\dagger}\left(\matrix{V_\mu & 0\cr 0 &-\tilde
V_\mu}\right)\chi^a+ {\rm h.c.}\right)\ +\cr &+&\frac{i}{2}
\sum_{\vec v} \sum_{b,c=1}^8\chi^{b\dagger} \left(\matrix{V_\mu
g_{bac} & 0\cr 0 &-\tilde V_\mu g_{bac}^*}\right)\chi^c,\eea where
\be g_{abc}=d_{abc}+if_{abc}\ .\label{current}\ee and $d_{abc}$,
$f_{abc}$ are the usual $SU(3)$ symbols. The result of the self
energy diagram (see Fig. \ref{gluons}) can be written as follows:
\bea &&i\,\Pi^{\mu\nu\,{\rm self}}_{ab}(p)=-2\int\frac{d\b
v}{8\pi}\sum_{A,C,D,E}\left(-ig\right)^2 \frac{\mu^2}\pi\int
\frac{d^2\ell}{(2\pi)^2}\,
Tr\Big[iS_{CD}(\ell+p)\times\cr&&\cr&\times& \left(\matrix{V_\nu
h_{DbE}& 0\cr 0 &-\tilde V_\nu h^*_{DbE}}\right)
iS_{EA}(\ell)\left(\matrix{V_\mu h_{AaC}& 0\cr 0 &-\tilde V_\mu
h^*_{AaC}}\right)\Big] \label{pimunu},\eea where the propagator is
given by Eq. (\ref{propagator}). We note the minus sign on the
r.h.s of (\ref{pimunu}), due to the presence of a fermion loop and
the factor 2 due to the spin ($L+R$).  To this result one should
add the contribution arising from the bare Meissner mass (see Eq.
(\ref{8.7})):  \be \Pi^{\mu\nu\,{\rm
BM}}_{ab}~=~-~\frac{g^2\mu^2}{2\pi^2}\delta_{ab}\delta^{j
k}\delta^{\mu j}\delta^{\nu k}\ . \label{1b}\ee
 To derive the dispersion law for the gluons,
we write the equations of motion for the gluon field $A^b_\mu$ in
momentum space and high-density limit:

\be \left[\left(-p^2 g^{\nu\mu}+p^\nu p^\mu
 \right)\delta_{ab}+\Pi^{\nu\mu}_{ab}\right]A^b_\mu~=0\ .\label{105}\ee
  We define the rotational invariant quantities $\dd \Pi_0,
\Pi_1, \Pi_2$ and $\Pi_3$ by means of the following equations, \be
\Pi^{\mu \nu}(p_0,\,\vec p) = \left\{ \begin{array}{ll}
    \Pi^{00}=\Pi_0(p_0,\,\vec p)  \\
    \Pi^{0i}=\Pi^{i0}=\Pi_1(p_0,\,\vec p) \, n^i \\
    \Pi^{ij}=\Pi_2 (p_0,\,\vec p) \,\delta^{ij}\, +\, \Pi_3(p_0,\,\vec p)\,
n^i n^j
    \end{array}
    \right.
\label{dec} \ee with $\dd \b n  = \b p/|\b p|$ . It is clear that
in deriving the dispersion laws we cannot go beyond momenta  of
order $\Delta$, as the Fermi velocity superselection rule excludes
gluon exchanges with very high momentum; it is therefore an
approximation, but nevertheless a useful one, as in most cases
hard gluon exchanges are strongly suppressed by the asymptotic
freedom property of QCD.

 By the equation
\be p_{\nu} \Pi^{\nu \mu}_{a b} A^b_{\mu} = 0 \ ,\label{108} \ee
one obtains the relation \be \left( p_0 \,\Pi_0\, -\, |\vec p|\,
\Pi_1 \right) A_0 = \vec n \cdot \vec A
 \left( p_0\, \Pi_1
\,-\,|\vec p|\, (\Pi_2+\Pi_3) \right) \ , \label{7} \ee between
the scalar and the longitudinal component of the gluon fields. The
dispersion laws for the scalar, longitudinal and transverse gluons
are respectively \bea \left(\Pi_2\,+\,\Pi_3 \,+\,p_0^2\right)
\left(|\vec p|^2\,+\,\Pi_0 \right)&=&p_0|\vec p|
\left(2\,\Pi_1\,+p_0|\vec p|\right)\ ,\cr
\left(\Pi_2\,+\,\Pi_3\,+\,p_0^2\right) \left(|\vec
p|\,p_0\,+\,\Pi_0\right)&=&p_0|\vec p|
\left(2\,\Pi_1\,+\,p_0^2\right)+\Pi_1^2 \ ,\cr p_0^2-|\vec
p|^2+\Pi_2&=&0\ .\label{cflpig} \eea The analysis of these
equations is rather complicated \cite{Gusynin:2001tt} and we will
give only the results arising from the expansion of the vacuum
polarization up to the second order in momenta. The relevant
expressions can be found in
\cite{Casalbuoni:2000na,Casalbuoni:2000nb}. The results at zero
momentum are the same for all the gluons and they are summarized
in the Debye mass (from $\Pi^{00}$) and in the Meissner mass (from
$\Pi^{ii}$ \be m^2_D\,=\,\frac{\mu^2 g^2}{36\pi^2} \left(21-8\ln 2
\right)~=\,g^2F^2\ ,\ee and \be m^2_M\,=\,\frac{\mu^2 g^2}{\pi^2}
\left(-\frac{11}{36}-\frac{2}{27} \ln 2+\frac{1}{2} \right)
\,=\,\frac{m^2_D}{3}~,
 \ee
where the first two terms are the result of the diagram of Fig.
\ref{gluons}, whereas the last one is the bare Meissner mass.
These results  agree with the findings of other authors as, for
instance, \cite{Son:1999cm,Son:2000tu,Zarembo:2000pj}. Recalling
Eqs. (\ref{masses}) \be m_D^2=\alpha_T g^2F_T^2,~~~m_M^2=\alpha_S
v^2g^2 F_T^2\ee derived from the effective lagrangian for the CFL
phase and the results already obtained for the parameters $F_T^2$
and $v^2$ we see that \be \alpha_S=\alpha_T=1.\ee This completes
the evaluation of the parameters of the effective lagrangian for
the CFL phase.

The Debye and Meissner masses do not exhaust  the analysis of the
dispersion laws for the gluons in the medium. We will give here
the result for the rest mass defined as the energy at zero
momentum. Due to the large wave function renormalization, of order
$g^2\mu^2/\Delta^2$, the rest mass turns out to be of order
$\Delta$. Precisely we find that for large values of $\mu$ one has
\be m^R_{A}\approx\frac{m_D}{\sqrt{3\alpha_1}}~.\ee with \be
\alpha_1=\frac{\mu^2
g_s^2}{216\Delta^2\pi^2}\left(7+\frac{16}{3}\ln 2\right)~.\ee
Therefore\be m^R_{A}\approx\frac{m_D}{\sqrt{3\alpha_1}}=
\sqrt{6\,\frac{21-8\ln 2}{21+16 \ln2}}\,\Delta\approx
1.70\,\Delta~. \label{127}\ee It is interesting to notice that the
gluons are below threshold for the decay $g\to q\bar q$ since in
the CFL all the quarks are gapped with rest masses $\Delta$ or
$2\Delta$. The rest mass can be also evaluated numerically without
expanding in momenta. One finds \be m\equiv m_R=\,1.36\, \, \Delta
\ . \ee A comparison with (\ref{127}) shows that the relative
error between the two procedures is of the order of $20\%$ and
this is also the estimated difference for the dispersion law  at
small $\vec p$. We notice that also in this case the momentum
expansion approximation overestimates the correct result
\cite{Gusynin:2001tt}.

The result about the large renormalization of the gluon fields can
be easily understood by using the chiral expansion of Eq.
(\ref{7.109}) \cite{Jackson:2003dk}. To this end let us first
notice that the renormalizion factor for the fields to be
canonical is \be g_\mu\to \frac{g \mu}{\Delta} g_\mu,\ee which
implies a coupling renormalization \be g\to \frac{\Delta}{g\mu}
g=\frac{\Delta}{\mu}.\ee Therefore, after renormalization the
coupling becomes $\Delta/\mu$. Now let us go back to the effective
theory for the CFL phase and precisely to the equation used to
decouple the gluon fields, Eq. (\ref{6.18}) \be g_\mu^a=-\frac
1{2g} \left(\hat X\de_\mu\hat X^\dagger+\hat Y\de_\mu\hat
Y^\dagger\right)\equiv \frac 1 g\omega_\mu^a. \label{8.48}\ee Now
consider the kinetic term for the gluon fields \be \sum_a
F_{\mu\nu}^a F^{\mu\nu a},\ee with \be F_{\mu\nu}^a=g\de_\mu
g^a_\nu-\de_\nu g^a_\mu-g f_{abc} g_\mu^b g_\nu^c\ee and $f_{abc}$
the structure constants of the gauge group. Substituting the
expression (\ref{8.48}) inside the kinetic term we find \be{\rm
kinetic ~rem}\approx \sum_a\left[\left(\frac 1
g(\de_\mu\omega^a_\nu-\de_\nu\omega^a_\mu-g\frac 1
{g^2}f_{abc}\omega^b_\mu\omega^c_{\nu}\right)^2\right]\approx
\frac 1 {g^2}\times{\rm four~derivative~operator}.\ee However the
chiral expansion gives a coefficient for a four derivative
operator given by \be F^2\Delta^2\times\frac{1}{\Delta^4}=\frac
{\mu^2}{\Delta^2}.\ee Comparing these two results, obtained both
for canonical gluon fields, we see that the coupling must be of
the order $\Delta/\mu$, recovering the  result obtained from the
explicit calculation.

\section{Quark masses and the gap equation}\label{IX}

We have discussed in Section \ref{VD3}  the more realistic case of
2 massless flavors and a third massive one, showing that for
$\mu<m_s^2/2\Delta$ the condensate of the heavy quark with the
light ones  may be disrupted. In this Section we will investigate
more carefully this problem but limiting ourselves two the case of
two flavors (one massless and one massive). We know (always from
Section \ref{VD3}) that the radii of the Fermi spheres are \be
p_{F_1}=\sqrt{\mu^2-m_s^2}, ~~~p_{F_2}=\mu.\ee In the case
$m_s\ll\mu$ we can approximate the radius of the sphere of the
massive fermion as \be p_{F_1}\approx \mu-\frac{m_s^2}{2\mu}.\ee
At the lowest order the effect of the mass is to split by a
constant term the Fermi momenta. Also, the splitting is the ratio
$m_s^2/2\mu$ that we have already encountered in Section
\ref{VIIE}, and we see from here in a simple way why this
expression is an effective chemical potential. In this
approximation we can substitute the problem with another one where
we have two massless fermions but with a split chemical potential.
This problem has received a lot of attention in normal
superconductivity  in presence of a magnetic field. The coupling
of the field to the spin of the electrons produces a splitting of
the Fermi surfaces related to spin up and spin down electrons. In
practice this coupling is completely dominated by the coupling of
the magnetic field to the orbital angular momentum, but it is
possible to conceive situations where the effect is
important\footnote{For  recent reviews about this point and the
LOFF phase that will be discussed later, see
\cite{Casalbuoni:2003ab,Bowers:2003ye}}

Measuring the splitting from the middle point we define the
chemical potentials for the two species of fermions as \be
\mu_u=\mu+\delta\mu,~~~~\mu_d=\mu-\delta\mu,\ee where we have
denoted by $u$ and $d$ the two species of fermions under study. To
describe the situation we can add the following interaction
hamiltonian \be H_I=-\delta\mu\psi^\dagger\sigma_3\psi,\ee where
$\sigma_3$ is a Pauli matrix acting in the two-dimensional space
corresponding to the two fermions. Notice that in the case of
normal superconductivity $\delta\mu$ is proportional to the
magnetic field. Let us start considering the case of normal
superconductivity. In this case we have a simple modification in
the diagonal terms of the Gor'kov equations, introduced in Section
\ref{IIID}, leading to the inverse propagator \bea S^{-1}({\bf
p})= \left(\matrix{(i\partial_t-\xi_{{\bf p}}+\delta\mu\sigma_3)
 & -
 \Delta\cr
 -\Delta^*&
(i\partial_t+\xi_{{\bf
p}}+\delta\mu\sigma_3)}\right)\,.\label{smenouno}\eea Then,
evaluating the gap equation, as given in Eq. (\ref{consistency}},
one gets easily at $T= 0$:
 \be \Delta=i\, g\Delta\int\frac{dE}{2\pi}\frac{d^3p}
 {(2\pi)^3}\frac{1}{(E-\delta\mu)^2-\xi_{{\bf p}}^2-\Delta^2}\,,
\label{gapt0}\ee and at $T\neq 0$: \be \Delta\,=\, \,g
T\,\sum_{n=-\infty}^{+\infty}\int\frac{d^3p}
{(2\pi)^3}\frac{\Delta}{(\omega_n+i\delta\mu)^2+ \epsilon({\bf
p},\Delta)^2}\,,\label{gaptne0}\ee where, we recall that \be
\epsilon({\bf p},\Delta)=\sqrt{\Delta^2+\xi^2_{{\bf p}}}\, . \ee
We now use the identity \be \frac 1 2\left[1-n_u-n_d \right]
=\epsilon({\bf p},\Delta) T
\sum_{n=-\infty}^{+\infty}\frac{1}{(\omega_n+i\delta\mu)^2+\epsilon^2({\bf
p},\Delta)}\,, \label{28}\ee where \be n_u({\bf p})=\frac{1}{\dd
e^{(\epsilon+\delta\mu)/T}+1}\ ,~~~~~~~~~~~ n_d({\bf
p}=\frac{1}{\dd e^{(\epsilon-\delta\mu)/T}+1} \, .\ee The gap
equation can be therefore written as
\be\Delta=\frac{g\,\Delta}2\,\int \frac{d^3p}{(2\pi)^3}\,
\frac{1}{\epsilon({\bf p},\Delta)}\,\left(1-n_u({\bf p})-n_d({\bf
p})\right)\,.\label{gappp} \ee In the Landau theory of the Fermi
liquid $n_u,\,n_d$ are interpreted as
 the equilibrium distributions for the quasiparticles of type $u,d$.
It can be noted  that the last two terms act as blocking factors,
reducing the phase space.

Before considering the solutions of the gap equations in the
general case let us  consider the case $\delta\mu=0$; the
corresponding gap is denoted $\Delta_0$. At $T=0$ there is no
reduction of the phase space and we know already the solution
\be\Delta_0=\frac{\delta}{\dd\sinh\frac 2{g\rho}}\label{gap0}\
.\ee Here \be \rho=\frac{p_F^2}{\pi^2 v_F}\ee is the density of
states and we have used  $\xi_{{\bf p}}\approx v_F(p-p_F)$, see
Eqs. (\ref{vf})-(\ref{diciannove}). In the weak coupling limit
(\ref{gap0}) gives \be \Delta_0=2\delta\, e^{\dd{-2/\rho
g}}\label{bcsgap}\ .\ee

Let us now consider the case $\delta\mu\neq 0$.
 By (\ref{gap1}) the gap equation is written as
 \be-1+\frac{g}2\int\frac{d^3p}{(2\pi)^3}\frac{1}{\epsilon}
 =\frac{g}2\int\frac{d^3p}{(2\pi)^3}\,\frac{n_u+n_d}{\epsilon}\ .\ee
Using the gap equation for the BCS superconductor, the l.h.s can
be written, in the weak coupling limit, as \be {\text
{l.h.s}}=\frac{g\rho}{2}\log\frac{\Delta_0}{\Delta}\ ,\ee where we
got rid of the cutoff $\delta$ by using $\Delta_0$, the  gap at
$\delta\mu=0$ and $T=0$. Let us now evaluate the r.h.s. at $T=0$.
 We get \be
\text {r.h.s.}\Big|_{T=0}\,=\,\frac{g\rho}{2}
\int_{0}^{\delta}\frac{d\xi_{{\bf p}}}{\epsilon}
\left[\theta(-\epsilon-\delta\mu)+\theta(-\epsilon+\delta\mu
)\right]\,. \label{Gap}\ee The gap equation at $T=0$ can therefore
be written as follows:
\be\log\frac{\Delta_0}{\Delta}=\theta(\delta\mu-\Delta) {\rm
arcsinh}
\frac{\sqrt{\delta\mu^2-\Delta^2}}{\Delta}\,,\label{gap5}\ee i.e.
\be
\log\frac{\Delta_0}{\delta\mu+\sqrt{\delta\mu^2-\Delta^2}}=0\,.\label{54}\ee
One can immediately see that there are no solutions for
$\delta\mu>\Delta_0$. For $\delta\mu\le \Delta_0$ one has two
solutions.
\bea a)~~~\,\,\Delta&=&\Delta_0\,,\label{d2b}\\
b)~~~ \Delta^2&=&2\,\delta\mu\Delta_0-\Delta_0^2\,.\label{d1b}
\eea The first arises since for $\Delta=\Delta_0$ the l.h.s. of
the Eq. (\ref{gap5}) is zero. But  since we may have solutions
only for $\delta\mu\le\Delta_0$ the $\theta$-function in Eq.
(\ref{gap5}) makes zero also the r.h.s..

The existence of this solution can also be seen  from Eq.
(\ref{gapt0}). In fact in this equation one can shift the
integration variable as follows: $E\to E+\delta\mu$, getting the
result that, in the superconductive phase, {\it the gap $\Delta$
is independent of} $\delta\mu$, i.e. $\Delta=\Delta_0$. Notice
that the shift is admissible only if no singularity is found.
However, since the integrand has poles at
$-\delta\mu\pm\sqrt{\xi^2_{\b p}+\Delta^2}-i\, sign(E)$, when
$\delta\mu>\Delta_0$ the pole corresponding to the sign plus
becomes negative and therefore it goes into the upper plane
together with the other pole. In this case by closing the path in
the lower half plane we find zero

We take this occasion to compute the contribution of the free
energy to the grand potential again in a different way. To compute
this contribution we make use of a theorem asserting that for
small variations of an external parameter all the thermodynamical
quantities vary in the same way \cite{landau1}. We apply this to
the grand potential to get \be \frac{\de\Omega}{\de
g}=\Big\langle\frac{\de H}{\de g}\Big\rangle\,.\ee From the
expression of the interaction hamiltonian (see Eq. (\ref{abcs})
with $G=g$) we find immediately  (cfr. \cite{abrikosov}, cap.
7)\footnote{We will use indifferently the symbol $\Omega$ for the
grand potential and its density $\Omega/V$.}: \be \Omega=-\int
\frac{dg}{g^2} |\Delta|^2\ \label{eq:59}. \ee Using the result
(\ref{bcsgap}) one can trade the integration over the coupling
constant $g$ for an integration over $\Delta_0$, the BCS gap at
$\delta\mu=0$, because ${d\Delta_0}/{\Delta_0}= {2dg}/{\rho g^2}$.
Therefore the difference in free energy between the superconductor
and the normal state is \be
\Omega_\Delta-\Omega_0=\,-\,\frac{\rho}{2}\int_{\Delta_f}^{\Delta_0}
\Delta^2\ \frac{d\Delta_0}{\Delta_0}\,. \label{d0bis}\ee Here
$\Delta_f$ is the value of $\Delta_0$ corresponding to $\Delta=0$.
$\Delta_f=0$ in the case a) of Eq. (\ref{d2b}) and
 $\Delta_f=2\delta\mu$  in the case b) of Eq. (\ref{d1b}); in the latter case
 one sees immediately that $\Omega_\Delta-\Omega_0>0$ because from Eq.
 (\ref{d1b}) it follows that $\Delta_0<2\delta\mu$.  The free energies
for $\delta\mu\neq 0$ corresponding to the cases a), b) above can
be computed substituting (\ref{d2b}) and (\ref{d1b}) in
(\ref{d0bis}). Before doing that let us derive the density of free
energy at $T=0$ and $\delta\mu\neq 0$ in the normal non
superconducting state. Let us start from the very definition of
the grand potential for free spin  1/2 particles\be \Omega_0(0,T)
\,=\,-2{VT}\int \frac{d^3p}{(2\pi)^3}\,{\rm
ln}\left(1+e^{(\mu-\epsilon({\bf p}))/T}\right)\,.\label{eq:61}\ee
Integrating by parts this expression we get, for $T\to 0$, \be
\Omega_0(0) \,=\,-\frac{V}{12\pi^3}\int d\Omega_{{\bf p}}\, p^3\,
d\epsilon\,\theta(\mu-\epsilon)\,.\ee From this expression we can
easily evaluate the grand-potential for two fermions with
different chemical potentials expanding at the first non-trivial
order in $\delta\mu/\mu$. The result is
 \be
\Omega_0 (\delta\mu)\,=\,\Omega_0 (0)-\frac{\delta\mu^2}{2}\rho
\label{om1}\,.\ee Therefore
 from  (\ref{d2b}), (\ref{d1b})  and (\ref{d0bis}) in the cases a), b)
 one has
\bea a)~~~\Omega_\Delta(\delta\mu)&=&\Omega_0(\delta\mu)
-\frac \rho 4 \,(-2\,\delta\mu^2+\Delta_0^2)\label{oma1}\ ,\\
b)~~~ \Omega_\Delta(\delta\mu)&=&\Omega_0(\delta\mu) -\frac \rho 4
\,(-4\,\delta\mu^2+4\delta\mu\Delta_0-\Delta_0^2)\ .\label{omb1}
\eea Comparing  (\ref{oma1}) and (\ref{omb1}) we see that the
solution $a)$ has lower $\Omega$. Therefore, for
$\delta\mu<\Delta_0/\sqrt 2$ the BCS superconductive state is
stable \cite{clogston,chandrasekhar}. At $\delta\mu=\Delta_0/\sqrt
2$ it becomes metastable,
 as the normal state has a lower free energy. This transition is first
 order since the gap does not depend on $\delta\mu$.

\begin{center}
\begin{figure}[ht]
\epsfxsize=9truecm \centerline{\epsffile{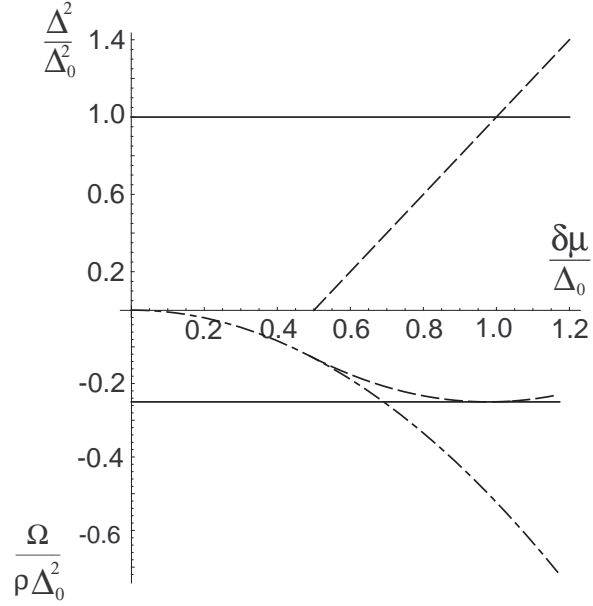}} \noindent
\caption{\it Gap and grand potential as functions of $\delta\mu$
for the two solutions a) and b) discussed in the text, see
Eqs.(\ref{d2b}), (\ref{d1b}) and (\ref{oma1}), (\ref{omb1}). Upper
solid (resp. dashed) line: Gap for solution a) (resp. solution
b)). In the lower part we plot the grand potential for the
solution a) (solid line) and solution b) (dashed line); we also
plot the grand potential for the normal gapless state with
$\delta\mu\neq 0$ (dashed-dotted line). All the grand potentials
are referred to the value $\Omega_0(0)$ (normal state with
$\delta\mu=0$). \label{fig0}}
\end{figure}
\end{center}

The grand potentials for the two cases a) and b) and for the
gapless phase, Eq. (\ref{om1}),  are given in Fig. \ref{fig0},
together with the corresponding gaps.
This analysis shows that at $\delta\mu=\delta\mu_1=\Delta_0/{\sqrt
2}$ one goes from the superconducting ($\Delta\neq 0$) to the
normal ($\Delta=0$) phase. However, as we shall discuss below,
 the real ground state  for $\delta\mu>\delta\mu_1$
 turns out to be an inhomogeneous one, where the assumption (\ref{homogeneous})
of a uniform gap is not justified.

The considerations made in this Section may be repeated for the
2SC case in color superconductivity. In fact, as we have seen, the
only difference is in the density at the Fermi surface which is
four time the one considered here with $v_F=1$, $p_F=\mu$.

\subsection{Phase diagram of homogeneous superconductors
\label{phasediagramhomogeneous}}

We will now study the phase diagram of the homogeneous
superconductor for small values of the gap parameter, which allows
to perform a Ginzburg-Landau expansion of gap equation and grand
potential. In order to perform a complete study we need to expand
the grand-potential  up to the 6$^{\text{th}}$ order in the gap.
As a matter of fact in the plane $(\delta\mu, T)$ there is a first
order transition at $(\delta\mu_1,0)$ and a second order one at
$(0,T_c)$ (the usual BCS second order transition). Therefore we
expect that a second order and a first order line start from these
points and meet at a tricritical point, which by definition is the
meeting point of a second order and a first order transition line.
A tricritical point is characterized by the simultaneous vanishing
of the $\Delta^2$ and  $\Delta^4$ coefficients in the
grand-potential expansion, which is why one needs to introduce in
the grand potential the 6$^{\text{th}}$ order term. For stability
reasons the corresponding coefficient should be positive; if not,
one should include also the $\Delta^8$ term.

We consider the  grand potential, as measured from the normal
state, near a second order phase transition \be\Omega=\frac 1 2
\alpha\Delta^2+\frac 1 4\beta\Delta^4+\frac 1
6\gamma\Delta^6\,.\label{eq:potential1}\ee Minimization gives the
gap equation: \be
\alpha\Delta+\beta\Delta^3+\gamma\Delta^5=0\,.\ee Expanding Eq.
(\ref{gaptne0}) up to the $5^{th}$ order in $\Delta$ and comparing
with the previous equation one determines the coefficients
$\alpha$, $\beta$ and $\gamma$ up to a normalization constant. One
gets \be \Delta=2\,g\,\rho\,T
\,Re\,\sum_{n=0}^{\infty}\int_0^\delta
d\xi\left[\frac{\Delta}{(\bar\omega_n^2+\xi^2)}-
\frac{\Delta^3}{(\bar\omega_n^2+\xi^2)^2}+
\frac{\Delta^5}{(\bar\omega_n^2+\xi^2)^3} +\cdots\,\right]\,,
\label{eq:gap_expan}\ee
with\be\bar\omega_n=\omega_n+i\delta\mu=(2n+1)\pi
T+i\delta\mu\,.\label{eq:70}\ee As we have discussed in Section
\ref{IIIE} the grand potential can be obtained,
 integrating in $\Delta$ the gap
equation  and integrating the result provided that we multiply it
by the factor $2/g$. Therefore \bea \alpha&=&\frac 2 g\left(1-2\,
g\,\rho\,T\,
Re\sum_{n=0}^\infty\int_0^\delta\frac{d\xi}{(\bar\omega_n^2+\xi^2)}\right)\,,
\label{eq:alpha}\\
\beta&=&4\rho\,T\,Re\sum_{n=0}^\infty
\int_0^\infty\frac{d\xi}{(\bar\omega_n^2+\xi^2)^2}\,,\label{eq:beta}\\
\gamma&=&-4\rho\,T\,Re\sum_{n=0}^\infty
\int_0^\infty\frac{d\xi}{(\bar\omega_n^2+\xi^2)^3}\,.\label{eq:gamma}
\eea Notice that for $\delta\mu=0$ we recover the expressions of
Section \ref{IIIE}. In the coefficients $\beta$ and $\gamma$ we
have extended the integration in $\xi$ up to infinity since both
the sum and the integral are convergent. To evaluate $\alpha$ we
proceed as in Section \ref{IIIE} first integrating over $\xi$ and
then summing over the Matsubara frequencies \cite{buzdin:1997ab}.
In Eq. (\ref{eq:alpha}) we first integrate over $\xi$ obtaining a
divergent series which can be regulated cutting the sum at a
maximal value of $n$ determined by \be\omega_N=\delta \Rightarrow
N\approx \frac{\delta}{2\pi T}\,.\ee We obtain \be \alpha=\frac 2
g\left(1-\pi\,g\,\rho\,T\,Re\sum_{n=0}^N\frac{1}{\bar\omega_n}\right)\,.\ee
The sum can be performed in terms of the Euler's function
$\psi(z)$:\bea
Re\sum_{n=0}^N\frac{1}{\bar\omega_n}&=&\frac{1}{2\pi T}\,
Re\left[\psi\left(\frac 3 2+i\frac
y{2\pi}+N\right)-\psi\left(\frac 1 2+i\frac
y{2\pi}\right)\right]\nn\\&\approx& \frac 1 {2\pi T}\left(
\log\frac{\delta}{2\pi T}-Re\,\psi\left(\frac 1 2+i\frac
y{2\pi}\right)\right)\,,\eea where \be y=\frac{\delta\mu}T.\ee
Eliminating the cutoff and using the gap equation at $T=0$ we find
\be \alpha(v,t)=\rho\left(\log(4\pi t)+Re\,\psi\left(\frac 1
2+i\frac v{2\pi t}\right)\right)\,.\label{eq:alpha_f}\ee with \be
v=\frac{\delta\mu}{\Delta_0},~~~t=\frac{T}{\Delta_0},~~~y=\frac v
t.\ee Let us introduce the function $T_c(y)$ defined by
\be\log\frac{\Delta_0}{4\pi T_c(y)}=Re\,\psi\left(\frac 1 2+\frac
{iy}{2\pi}\right).\ee Then we find \be \alpha(v,t)=\rho\log\frac
t{t_c(v/t)},\ee where \be t_c(y)=\frac{T_c(y)}{\Delta_0}.\ee The
line where $\alpha(v,t)=0$, that is\be t=t_c(v/t)\ee defines the
second order phase transitions (see discussion later). In
particular at $\delta\mu=0$, using ($C$ the Euler-Mascheroni
constant)\be\psi\left(\frac 1 2\right)=-\log(4\gamma),
~~~~\gamma=e^C,~~~~C=0.5777\dots\,,\ee we find from Eq.
(\ref{eq:alpha_f})\be\alpha(0,T/\Delta_0)=\rho\,\log\frac{\pi
T}{\gamma\Delta_0}\,,\ee reproducing the critical temperature for
the BCS case \be T_c=\frac\gamma\pi\Delta_0\approx
0.56693\,\Delta_0\,.\label{eq:90}\ee The other terms in the
expansion of the gap equation are easily evaluated integrating
over $\xi$ and  summing over the Matsubara frequencies. We get
\bea \beta&=&\pi\,\rho\, T\,Re\sum_{n=0}^\infty\frac
1{\bar\omega_n^3}=-\frac\rho{16\,\pi^2\, T^2}
Re\,\psi^{(2)}\left(\frac 1 2+i\frac{\delta\mu}{2\pi
T}\right)\,,\label{eq:beta2}\\
\gamma&=&-\frac 3 4\pi\,\rho\, T\,Re\sum_{n=0}^\infty\frac
1{\bar\omega_n^5}=\frac 3 4\frac\rho{768\,\pi^4 \,T^4}
Re\,\psi^{(4)}\left(\frac 1 2+i\frac{\delta\mu}{2\pi T}\right)\,,
\label{eq:gamma2}\eea where \be\psi^{(n)}(z)=\frac
{d^n}{dz^n}\psi(z)\,.\ee

\begin{figure}[ht]
%\epsfxsize=10cm   %width of figure - will enlarge/reduce the figures
%\epsfbox{fig3.eps}
%\figurebox{2cm}{3cm}{} %to have a box alone
\centerline{\epsfxsize=11cm\epsfbox{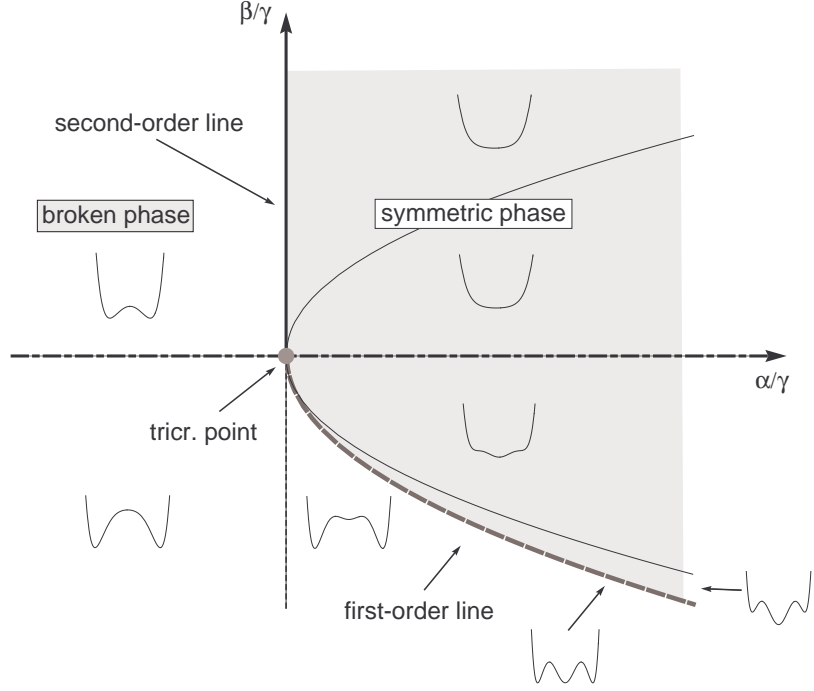}} \caption{{\it
The graph shows the first order and the second order transition
lines for the potential of Eq. (\ref{eq:potential1}). We show the
tricritical point and the regions corresponding to the symmetric
and the broken phase. Also shown is the behavior of the grand
potential in the various regions. The thin solid line is the locus
of the points $\beta^2-4\alpha\gamma=0$. In the interior region we
have $\beta^2-4\alpha\gamma<0$.\label{fig1a}}}
\end{figure}

Let us now briefly review some results on the grand potential in
the GL expansion (\ref{eq:potential1}). We will assume $\gamma>0$
in order to ensure the stability of the potential. The
minimization leads to the solutions \be
\Delta=0\,,\label{eq:94}\ee\be\Delta^2=\Delta_\pm^2=\frac
1{2\gamma}\left(-\beta\pm\sqrt{\beta^2-4\alpha\gamma}
\right)\,.\label{eq:95}\ee The discussion of the minima of
$\Omega$ depends on the signs of the parameters $\alpha$ and
$\beta$. The results are the following:
\begin{enumerate}
\item \fbox{${\alpha>0,~~\beta>0}$}\\\\In this case there is a
single minimum given by (\ref{eq:94}) and the phase is
\underline{symmetric}. \item \fbox{${\alpha>0,~~\beta<0}$}\\\\
Here there are three minima, one is given by (\ref{eq:94}) and the
other two are degenerate minima
at\be\Delta=\pm\Delta_+\,.\label{eq:96}\ee The line along which
the three minima become equal is given
by:\be\Omega(0)=\Omega(\pm\Delta_+)~~\longrightarrow~~
\beta=-4\,\sqrt{\frac{\alpha\gamma}3}\,.\label{eq:101}\ee Along
this line there is a first order transition with a discontinuity
in the gap given by \be\Delta_+^2=-\frac{4\alpha}\beta=-\frac 3
4\frac\beta\gamma\,.\label{eq:102}\ee To the right of the first
order line we have $\Omega(0)<\Omega(\pm\Delta_+)$. It follows
that to the right of this line there is the symmetric phase,
whereas the broken phase is in the left part (see Fig.
\ref{fig1a}). \item \fbox{${\alpha<0,~~\beta>0}$}\\\\In this case
Eq. (\ref{eq:94}) gives a maximum, and there are  two degenerate
minima given by Eq. (\ref{eq:96}).
 Since for $\alpha>0$ the two minima
disappear, it follows that there is a  second order phase
transition along the line $\alpha=0$. This can also be seen by
noticing that going from the broken phase to the symmetric one we
have \be\lim_{\alpha\,\to\, 0} \Delta_+^2=0\,.\ee \item
\fbox{${\alpha<0,~~\beta<0}$}\\\\ The minima and the maximum are
as in the previous case.
\end{enumerate}
Notice also that the solutions $\Delta_\pm$ do not exist in the
region $\beta^2<4\alpha\gamma$.  The situation is summarized in
Fig. \ref{fig1a}. Here we show the behavior of the grand potential
in the different sectors of the plane
$(\alpha/\gamma,\beta/\gamma)$, together with the transition
lines. Notice that in the quadrant $(\alpha>0,\beta<0)$ there are
metastable phases corresponding to non absolute minima. In the
sector included between the line $\beta=-2\sqrt{\alpha/\gamma}$
and the first order transition line the metastable phase is the
broken one, whereas in the region between the first order  and the
$\alpha=0$ lines the metastable phase is the symmetric one.

\begin{figure}[ht]
%\epsfxsize=10cm   %width of figure - will enlarge/reduce the figures
%\epsfbox{fig3.eps}
%\figurebox{2cm}{3cm}{} %to have a box alone
\centerline{\epsfxsize=12cm\epsfbox{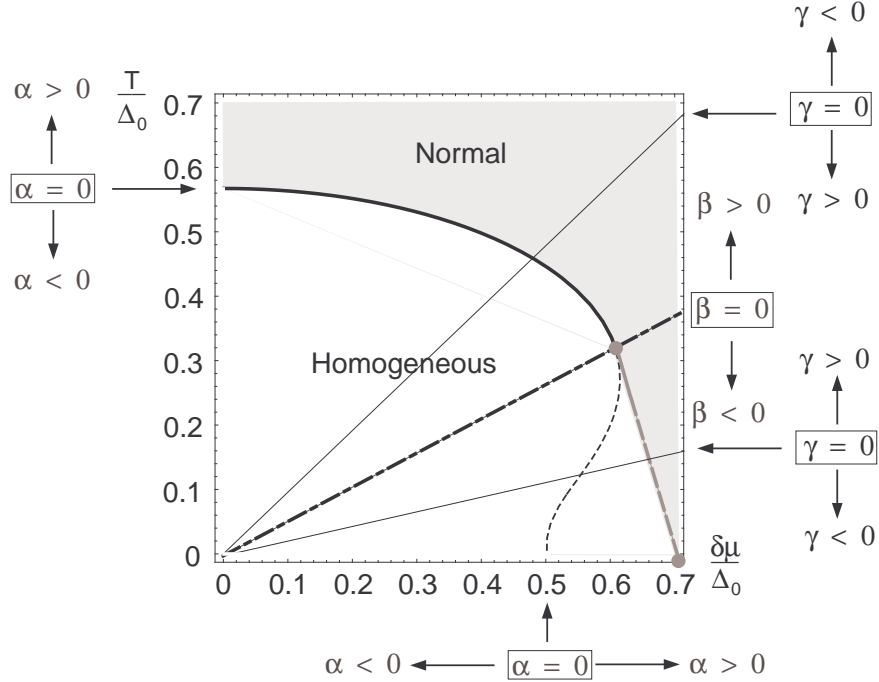}} \caption{{\it
The curve shows the points solutions of the equation $\Delta=0$ in
the plane $(v,t)=(\delta\mu/\Delta_0,T/\Delta_0)$. The tricritical
point at $(\delta\mu,T)\approx(0.61,0.32)\,\Delta_0$ is also
shown. The upper part of the curve (solid line) separates the
homogeneous phase from the normal one. Along the dashed line
$\Delta=0$ but this is not the absolute minimum of the grand
potential. \label{fig30}}}
\end{figure}

Using Eqs. (\ref{eq:alpha_f}), (\ref{eq:beta2}) and
(\ref{eq:gamma2}) which give the parameters $\alpha$, $\beta$ and
$\gamma$ in terms of the variables $v=\delta\mu/\Delta_0$ and
$t=T/\Delta_0$, we can map the plane $\alpha$ and $\beta$ into the
plane $(\delta\mu/\Delta_0,T/\Delta_0)$. The result is shown in
Fig. \ref{fig30}. From this mapping we can draw several
conclusions. First of all the region where the previous discussion
in terms of the parameters $\alpha$, $\beta$ and $\gamma$ applies
is the inner region of the triangular part delimited by the lines
$\gamma=0$. In fact, as  already stressed, our expansion does not
hold outside this region. This statement can be made quantitative
by noticing that along the first order transition line the gap
increases when going away from the tricritical point
as\be\Delta_+^2=-\,\frac{4\alpha}\beta=\sqrt{\frac{3\alpha}\gamma}\,.\ee

Notice  that the lines $\beta(v,t)=0$ and $\gamma(v,t)=0$ are
straight lines, since these zeroes are determined  by the
functions $\psi^{(2)}$ and $\psi^{(4)}$ which depend only on the
ratio $v/t$. Calculating the first order line around the
tricritical point one gets the result  plotted as a solid line in
Fig. \ref{fig30}. Since we know that
$\delta\mu=\delta\mu_1=\Delta_0/\sqrt{2}$ is a first order
transition point, the first order line must end there. In Fig.
\ref{fig30} we have simply connected the line with the point with
a grey dashed line. To get this line a numerical evaluation at all
orders in $\Delta$ would be required. This is feasible but we will
skip it since the results will not be necessary in the following,
see \cite{sarma:1963ab}. The location of the tricritical point is
determined by the intersection of the lines $\alpha=0$ and
$\beta=0$. One finds \cite{combescot:2002ab,buzdin:1997ab}\be
\frac{\delta\mu}{\Delta_0}\Big|_{\rm tric}=0.60822,~~~~
\frac{T}{\Delta_0}\Big|_{\rm tric}=0.31833\,.\ee
 We  also note that the line $\alpha=0$ should cross the temperature axis at the
 BCS point. In this way one reobtains the result in Eq. (\ref{eq:90})
 for the BCS critical temperature, and also the value for the
 tricritical temperature
\be \frac{T_{\rm tric}}{T_{\rm BCS}}=0.56149\,.\ee

The results given in this Section are valid as long as other
possible condensates are neglected. In fact, we will see that
close to the first order transition of the homogeneous phase the
LOFF phase with inhomogeneous  gap can be formed.
\subsection{Dependence of the condensate on the quark masses}

In the previous Section we have ignored all the mass corrections
but for the ones coming from the shift of the chemical potential.
However there is a small mass dependence on the condensate itself.
This is because the presence of the mass gives rise to a reduction
of the density of the states at the Fermi surface weakening the
gap. This problem has been studied in
\cite{Kundu:2001tt,Casalbuoni:zx}. We will follow here the
approach given in \cite{Casalbuoni:zx}. Let us consider the QCD
lagrangian \be {\cal L}_{QCD}=\bar{\psi}\,(i\,\slash
D+\mu\,\gamma^{0}-m)\,\psi= \bar{\psi}\,(i\,\slash D
+\mu\,\gamma^{0}-x\mu)\,\psi\label{lag}, \ee where we will keep\be
x=\frac m\mu\ ,\hskip1cm 0<x<1 \label{x}\ee fixed in the
$\mu\to\infty$ limit. This particular limit is convenient when one
discusses problems related to compact stellar objects where the
strange quark mass is not very far from the relevant chemical
potential (of order of $400~MeV$). This lagrangian and the one at
zero quark mass can be related by making use of the Cini-Touschek
transformation \cite{Cini:1958ab} that was invented to study the
ultra-relativistic limit of the Dirac equation for a massive
fermion.

 In order to describe the method used we need to discuss a little
 kinematics. In the massive case, as we have seen, the Fermi
 momentum is given by
 \be
 p_F^2=\mu^2(1-x^2)\ee and the Fermi velocity by
 \be\b v_F=\frac{\de E(\b p)}{\de \b p}\Big|_{p =p_F}=\frac{\b
 p_F}{E_F}=\frac{\b
 p_F}{\mu}.\ee
Introducing the unit vector \be \b n=\frac{\b p_F}{|\b p_F|},\ee
we get \be \b v_F=\sqrt{1-x^2}\vec n.\ee Now let us do again the
decomposition of the momentum in the Fermi momentum plus the
residual momentum \be \b p=\b p_F+{\bm \ell}=\mu\sqrt{1-x^2}\,\b
n+{\bm \ell}.\ee Substituting inside the Dirac hamiltonian \be
H=p_0={\bm\alpha}\cdot{\b p}-\mu+m\gamma_0,\ee we get \be
H=-\mu+\mu(\sqrt{1-x^2}{\bm\alpha}\cdot{\b n}+x\gamma_0)+
{\bm\alpha}\cdot{\bm\ell}.\ee We proceed now introducing the
following two projection operators \be P_\pm=\frac{1\pm
\left(x\gamma_0-\sqrt{1-x^2}{\bm\alpha}\cdot\b n\right)}2.\ee
These are projection operators since the square of the operator in
parenthesis is one. In terms of the projected wave functions the
Dirac equations splits into the two equations \be
H\psi_+={\bm\alpha}\cdot{\bm\ell}\psi_+,~~~H
\psi_-=\left(-2\mu+{\bm\alpha}\cdot{\bm\ell}\right)\psi_- .\ee
Therefore we reproduce exactly the same situation as in the
massless case (see Section \ref{HDET}) and all the formalism holds
true introducing the two four-vectors \be V^\mu=(1,\b
v_F),~~~~\tilde V^\mu=(1,-\b v_F),\ee with \be |\b
v_F|^2=1-x^2.\ee
\begin{center}
\begin{figure}[htb]
\epsfxsize=9truecm \centerline{\epsffile{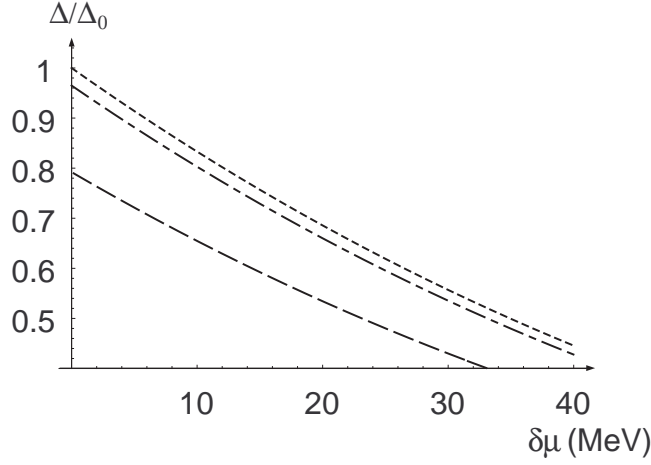}} \noindent
\caption{\it The ratio of the gap for finite masses $\Delta$ over
the gap for massless quarks $\Delta_0$ as a function of
$\delta\mu$. The lines give the ratio for values  $x_1=0$ (dotted
line),  $x_1=0.1$ (dash-dotted line),  $x_1=0.25$ (dashed one).
The values used are $\mu=400$ MeV, $\delta=400$ $MeV$,
 $G=10.2~GeV^{-2}$, corresponding to $\Delta_0\approx 40~MeV$.
\label{ruggieri}}
\end{figure}
\end{center}
Without entering in too many details we can now explain the
results  found in \cite{Casalbuoni:zx}. One starts from the same
four-fermi interaction of Section \ref{VB2} and writes the
Schwinger-Dyson equation obtaining  a  gap equation which is the
same obtained in Section \ref{VB2} with a few differences. Since
the main interest of the paper was to understand how the
condensate varies with quark masses the assumption was made that
the chemical potentials of the up and down quarks were arranged in
such a way to make equal the two Fermi momentum. This gives the
relation \be
p_F=p_{F_1}=p_{F_2}=\sqrt{\mu_1^2-m_1^2}=\sqrt{\mu_2^2-m_2^2},\ee
from which \be \delta\mu=\frac{m_1^2-m_2^2} 4.\ee Then we have the
main modifications come from a change of the density at the Fermi
surface. Precisely \be \mu^2\to
p_F^2=(\mu+\delta\mu)(\mu-\delta\mu)\alpha(x_1,x_2),\ee where \be
\alpha(x_1,x_2)=\sqrt{(1-x_1^2)(1-x_2^2)}.\ee Also there is a
factor \be \tilde V_2 \cdot V_1=\frac{1+\alpha(x_1,x_2)} 2\ee
coming from the interaction, and finally a modification of the
propagator (where the product $(\tilde V_2\cdot\ell)V_1\cdot\ell)$
appears. Putting everything together we find that the gap is given
by \be \Delta=2\sqrt{\frac{4\alpha+\beta^2}4}\delta e^{-2/\rho_N
G}\ee where\be\beta(x_1,x_2)=\sqrt{1-x_1^2}-\sqrt{1-x_2^2}\ee and
\be\rho_N=\frac
{4\mu^2}{\pi^2}\left(1-\frac{\delta\mu^2}{\mu^2}\right)\frac{\alpha(x_1,x_2)(1+
\alpha(x_1,x_2))}{\sqrt{4\alpha(x_1,x_2)+\beta(x_1,x_2)^2}}.\ee In
Fig. \ref{ruggieri} we plot the condensate normalized at its
value, $\Delta_0$, for $m_1=m_2=0$ as a function of $\delta\mu$ in
different situations. We have chosen $\mu=400~MeV$,
$\delta=400~MeV$ and $G=10.3~GeV^{-2}$ in such a way that
$\Delta_0=40~MeV$. The diagram refers to one massive quark and one
massless. Notice also that plotting $\delta\mu$ is the same as
plotting $m_2^2-m_1^2$.

\section{The LOFF phase}\label{X}

\begin{figure}[ht]
%\epsfxsize=10cm   %width of figure - will enlarge/reduce the figures
%\epsfbox{fig3.eps}
%\figurebox{2cm}{3cm}{} %to have a box alone
\centerline{\epsfxsize=3.0in\epsfbox{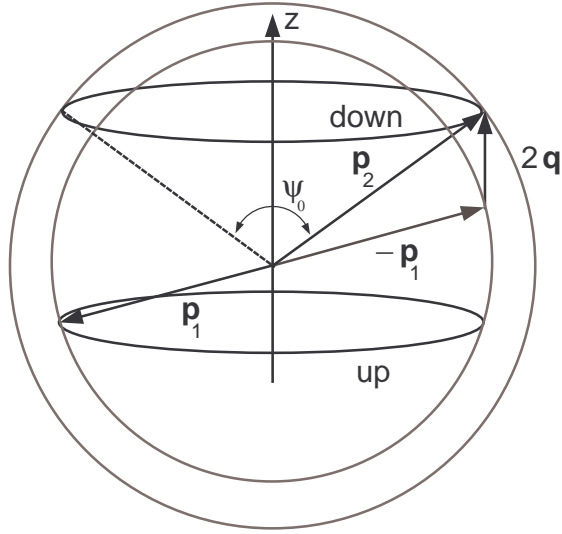}}
\caption{\it Pairing of fermions belonging to two Fermi spheres of
different radii according to LOFF. \label{fig1}}
\end{figure}

We will review here briefly the so called LOFF phase. Many more
details can be found in two recent reviews
\cite{Casalbuoni:2003ab,Bowers:2003ye}. We have already discussed
the fact that when the chemical potentials of two fermions are too
apart the condensate may break. In particular we have shown (see
Section \ref{IX}) that when the difference between the two
chemical potentials, $\delta\mu$, satisfies
\be\delta\mu=\frac{\Delta_0}{\sqrt{2}},\ee where $\Delta_0$ is the
BCS gap at $\delta\mu=0$, the system undergoes a first order phase
transition, with the gap going from $\Delta_0$ to zero. However
just close at this point something different may happen. According
to the authors \cite{LO,FF} when fermions belong to two different
Fermi spheres, they  may prefer to pair staying as much as
possible close to their own Fermi surface. When they are sitting
exactly at the surface, the pairing is as shown in Fig.
\ref{fig1}.
We see that the total momentum of the pair is ${\b p}_1+{\b
p}_2=2\b q$ and, as we shall see, $|\b q\,|$ is fixed
variationally whereas the direction of $\b q$ is chosen
spontaneously. Since the total momentum of the pair is not zero
the condensate breaks rotational and translational invariance. The
simplest form of the condensate compatible with this breaking is
just a simple plane wave (more complicated functions will be
considered later) \be \langle\psi(x)\psi(x)\rangle\approx\Delta\,
e^{2i\b q\cdot\b x}.\label{single-wave}\ee It should also be
noticed that the pairs use much less of the Fermi surface than
they do in the BCS case. In fact, in the case considered in Fig.
\ref{fig1} the fermions can pair only if they belong to the
circles drawn there. More generally there is a quite large region
in momentum space (the so called blocking region) which is
excluded from the pairing. This leads to a  condensate smaller
than the BCS one.

Let us now begin the discussion. Remember that for two fermions at
different densities we have an extra term in the hamiltonian which
can be written as \be
H_I=-\delta\mu\sigma_3.\label{interaction}\ee In the original LOFF
papers \cite{LO,FF} the case of ferromagnetic alloys with
paramagnetic impurities was considered. The impurities produce a
constant magnetic exchange field which, acting upon the electron
spin, gives rise to an effective difference in the chemical
potential of the opposite fields. In this case $\delta\mu$ is
proportional to the exchange field. In the actual case
$\delta\mu=(\mu_1-\mu_2)/2$ and $\sigma_3$ is a Pauli matrix
acting on the space of the two fermions. According to
 \cite{LO,FF} this favors the formation of pairs with momenta
\be \b p_1=\b k+\b q,~~~\b p_2=-\b k+\b q.\ee We will discuss in
detail the case of a single plane wave (see Eq.
(\ref{single-wave})) and we will give some results about the
general case. The interaction term of Eq. (\ref{interaction})
gives rise to a shift in the variable $\xi=E(\b p)-\mu$  due both
to the non-zero momentum of the pair and to the different chemical
potential \be \xi=E(\vec p)-\mu\to E(\pm\vec k+\vec
q)-\mu\mp\delta\mu\approx \xi\mp\bar\mu,\ee with \be
\bar\mu=\delta\mu-{\vec v}_F\cdot\vec q.\ee Here we have assumed
$\delta\mu\ll\mu$ (with $\mu=(\mu_1+\mu_2)/2$) allowing us to
expand $E$ at the first order in $\b q$ (see Fig. \ref{fig1}). The
gap equation has the same formal expression as Eq. (\ref{gap1})
for the BCS case \be
1=\frac{g}2\int\frac{d^3p}{(2\pi)^3}\frac{1-n_u-n_d}{\epsilon(\vec
p,\Delta)},\ee but now $n_u\not=n_d$ \be n_{u,d}=\frac
1{e^{(\epsilon(\vec p,\Delta)\pm\bar\mu)/T}+1},\ee where $\Delta$
is the LOFF gap. In the limit of zero temperature we obtain \be
1=\frac{g}2\int\frac{d^3p}{(2\pi)^3}\frac{1}{\epsilon(\vec
p,\Delta)}\left(1-\theta(-\epsilon-\bar\mu)-\theta(-\epsilon+\bar\mu)
\right)\label{gap}.\ee The two step functions can be interpreted
saying that at zero temperature  there is no pairing when
$\epsilon(\vec p,\Delta)<|\bar\mu|$. This inequality defines the
so called blocking region. The effect is to inhibit part of the
Fermi surface to the pairing giving rise a to a smaller condensate
with respect to the BCS case where all the surface is used.

\begin{figure}[ht]
%\epsfxsize=10cm   %width of figure - will enlarge/reduce the figures
%\epsfbox{fig3.eps}
%\figurebox{2cm}{3cm}{} %to have a box alone
\centerline{\epsfxsize=5.in\epsfbox{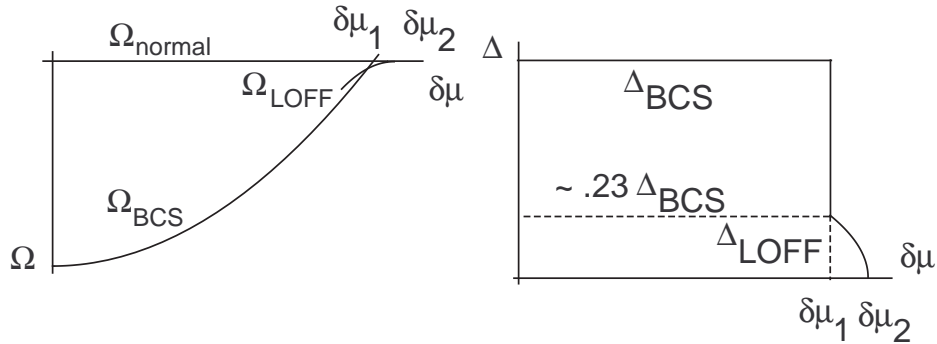}} \caption{\it
The grand potential and the condensate of the BCS and LOFF phases
vs. $\delta\mu$. \label{fig2}}
\end{figure}

We are now in the position to show that increasing $\delta\mu$
from zero we have first the BCS phase. Then  at
$\delta\mu\approx\delta\mu_1$ there is a first order transition to
the LOFF phase \cite{LO,Alford:2000ze},  and at
$\delta\mu=\delta\mu_2>\delta\mu_1$ there is a second order phase
transition to the normal phase (with zero gap)
\cite{LO,Alford:2000ze}. We start comparing the grand potential in
the BCS phase to the one in the normal phase. Their difference,
from Eq. (\ref{oma1}), is given by \be \Omega_{\rm
BCS}-\Omega_{\rm
normal}=-\frac{\rho}{4}\left(\Delta^2_{BCS}-2\delta\mu^2\right).\ee
We have  assumed $\delta\mu\ll\mu$.  The situation is represented
in Fig. \ref{fig2}.
In order to compare with the LOFF phase we will now expand the gap
equation around the point $\Delta=0$ (Ginzburg-Landau expansion)
exploring the possibility of a second order phase transition.
Using the gap equation for the BCS phase in the first term on the
right-hand side of Eq. (\ref{gap}) and integrating the other two
terms in $\xi$ we get \be \rho\log\frac{\Delta_{BCS}}{\Delta}=\rho
\int\frac{d\Omega}{4\pi}\, {\rm
arcsinh}\frac{C(\theta)}{\Delta},\ee where \be
C(\theta)=\sqrt{(\delta\mu-q v_F\cos\theta)^2-\Delta^2}.\ee For
$\Delta\to 0$ we get easily \be
-\log\frac{\Delta_{BCS}}{2\delta\mu}+\frac 1 2
\int_{-1}^{+1}\log\left(1-\frac u
z\right)=0,~~~z=\frac{\delta\mu}{qv_F}.\label{expansion}\ee This
expression is proportional to the coefficient $\alpha$ in the
Ginzburg-Landau expansion (recall the discussion in Section
\ref{phasediagramhomogeneous}). Therefore it should be minimized
with respect to $q$. The minimum is given by \be \frac 1
z\tanh\frac 1 z=1,\ee implying \be qv_F\approx 1.2\,
\delta\mu_2\label{q}.\ee Putting this value back in eq.
(\ref{expansion}) we obtain \be \delta\mu_2\approx
0.754\,\Delta_{BCS}\label{deltamu2}.\ee From the expansion of the
gap equation around $\delta\mu_2$ it is easy to obtain \be
\Delta^2\approx 1.76\,\delta\mu_2(\delta\mu_2-\delta\mu).\ee
Recalling Eq. (\ref{eq:59}), we can express the grand potential of
the LOFF phase relatively to the one of the normal phase as \be
\Omega_{\rm LOFF}-\Omega_{\rm normal}=-\int_0^g\frac
{dg}{g^2}\Delta^2\label{LOFF}.\ee Using Eq. (\ref{3.67}) for the
BCS gap and Eq.(\ref{deltamu2}) we can write \be
\frac{dg}{g^2}=\frac{\rho}2\frac{d\Delta_{BCS}}{\Delta_{BCS}}=
\frac{\rho}2\frac{d\delta\mu_2}{\delta\mu_2}.\ee Noticing that
$\Delta$ is zero for $\delta\mu_2=\delta\mu$ we are now able to
perform the integral (\ref{LOFF}) obtaining \be \Omega_{\rm
LOFF}-\Omega_{\rm normal}\approx
-0.44\,\rho(\delta\mu-\delta\mu_2)^2\label{10.20}.\ee We see that
in the window between the intersection of the BCS curve and the
LOFF curve in Fig. \ref{fig2} and $\delta\mu_2$ the LOFF phase is
favored. Furthermore at the intersection there is a first order
transition between the LOFF and the BCS phase. Notice that since
$\delta\mu_2$ is very close to $\delta\mu_1$ the intersection
point is practically given by $\delta\mu_1$. In Fig. \ref{fig2} we
show also the behaviour of the condensates. Altough the window
$(\delta\mu_1,\delta\mu_2)\simeq(0.707,0.754)\Delta_{BCS}$ is
rather narrow, there are indications that considering the
realistic case of QCD \cite{Leibovich:2001xr} the window may open
up. Also, for different structures than the single plane wave
there is the possibility that the window opens up
\cite{Leibovich:2001xr}.

\subsection{Crystalline structures}

\begin{figure}[ht]
%\epsfxsize=10cm   %width of figure - will enlarge/reduce the figures
%\epsfbox{fig3.eps}
%\figurebox{2cm}{3cm}{} %to have a box alone
\centerline{\epsfxsize=1.9in\epsfbox{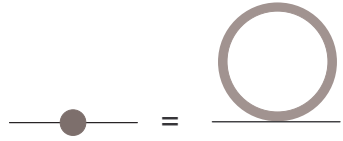}} \caption{\it Gap
equation in graphical form. The thick line is the exact
propagator. The black dot the gap insertion. \label{fig3}}
\end{figure}

\begin{figure}[ht]
%\epsfxsize=10cm   %width of figure - will enlarge/reduce the figures
%\epsfbox{fig3.eps}
%\figurebox{2cm}{3cm}{} %to have a box alone
\centerline{\epsfxsize=3.3in\epsfbox{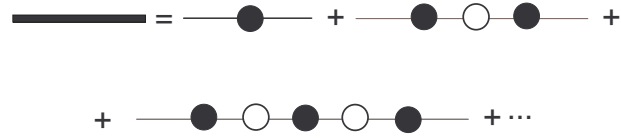}} \caption{\it
The expansion of the propagator in graphical form. The darker
boxes represent a $\Delta^*$ insertion, the lighter ones a
$\Delta$ insertion. \label{fig4}}
\end{figure}

\begin{center}
\begin{figure}[ht]
\centerline{\epsfxsize=3.3in\epsfbox{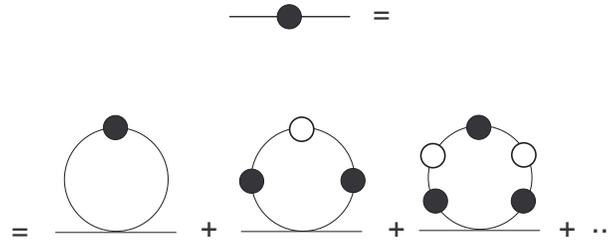}}
\caption{\it The expansion of the gap equation in graphical form.
Notations as in Fig. \ref{fig4}. \label{fig5}}
\end{figure}
\end{center}

The ground state in the LOFF phase is a superposition of states
with different   occupation numbers ($N$ even)\be
|0\rangle_{LOFF}=\sum_N c_N|N\rangle.\ee Therefore the general
structure of the condensate in the LOFF ground state will be \bea
\langle \psi(x)\psi(x)\rangle&=&\sum_N c_N^*c_{N+2}e^{2i\b
q_N\cdot\b x}\langle N|\psi(x)\psi(x)|N+2\rangle\nn\\&=& \sum_N
\Delta_N e^{2i\b q_N\cdot\vec x}.\eea The case considered
previously corresponds to all the Cooper pairs having the same
total momentum $2\b q$. A more general situation, although not the
most general, is when the vectors $\b q_N$ reduce to a set $\b
q_i$ defining a regular crystalline structure. The corresponding
coefficients $\Delta_{\b q_i}$ (linear combinations of subsets of
the $\Delta_N$'s) do not depend on the vectors $\b q_i$ since all
the vectors belong to the same orbit of the group. Furthermore all
the vectors $\b q_i$ have the same length \cite{Bowers:2002xr}
given by Eq. (\ref{q}). In this case \be\langle
0|\psi(x)\psi(x)|0\rangle=\Delta_q\sum_i e^{2i\b q_i\cdot\b x}.\ee
This more general case has been considered in
\cite{LO,Bowers:2002xr} by evaluating the grand-potential of
various crystalline structures through a Ginzburg-Landau
expansion, up to sixth order in the gap \cite{Bowers:2002xr}\be
\Omega=\alpha\Delta^2+\frac\beta 2\Delta^4+\frac\gamma
3\Delta^6.\label{potential}\ee These coefficients  can be
evaluated microscopically for each given crystalline structure by
following the procedure outlined in Sections
\ref{phasediagramhomogeneous} and \ref{IIIE}. The general
procedure  is to start from the gap equation represented
graphically in Fig. \ref{fig3}.
Then, one expands the exact propagator in a series of the gap
insertions as given in Fig. \ref{fig4}.
Inserting this expression back into the gap equation one gets the
expansion illustrated in Fig. \ref{fig5}.
On the other hand the gap equation is obtained minimizing the
grand-potential (\ref{potential}), i.e. \be
\alpha\Delta+\beta\Delta^3+\gamma\Delta^5+\cdots=0.\ee Comparing
this expression with the result  of Fig. \ref{fig5} one is able to
derive  the coefficients $\alpha$, $\beta$ and $\gamma$. Except
for an overall coefficient (the number of plane waves) the
coefficient $\alpha$ has the same expression for all kind of
crystals. Therefore the results obtained for the single plane wave
and depending on the properties of this coefficients are
universal.

\begin{center}
\begin{figure}[ht]
\centerline{\epsfxsize=3.3in\epsfbox{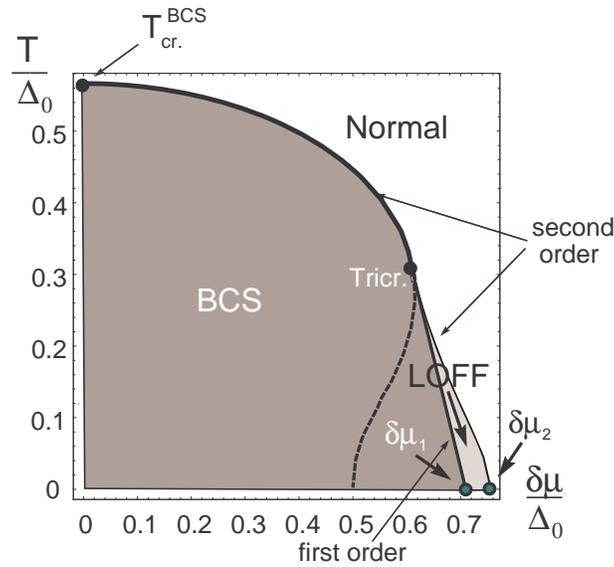}} \caption{\it
The phase diagram for the LOFF phase in the plane $(\delta\mu,T)$.
It is shown the tricritical point. Here three lines meet: the
second order transition line from the normal case to the BCS
phase, the first order transition line from the BCS phase to the
LOFF phase and the second order transition line from the LOFF
phase to the normal one. \label{LOFFphase}}
\end{figure}
\end{center}
\begin{center}
\begin{figure}[ht]
\centerline{\epsfxsize=4in\epsfbox{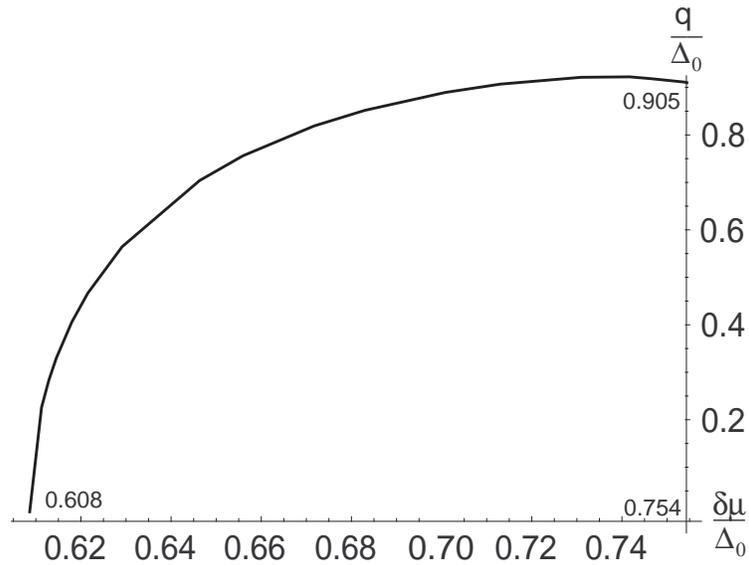}} \caption{\it The
behavior of the length of the vector $\b q$ along the second order
transition line from the LOFF phase to the normal phase is shown,
vs. $\delta\mu/\Delta_0$. \label{qvector}}
\end{figure}
\end{center}

In ref. \cite{Bowers:2002xr} more than 20 crystalline structures
have been considered, evaluating for each of them  the
coefficients of Eq. (\ref{potential}). The result of this analysis
is that the face-centered cube appears to be the favored structure
among the ones considered (for more details see ref.
\cite{Bowers:2002xr}). However it should be noticed that this
result can be trusted only at $T=0$. In fact one knows that in the
$(\delta\mu,T)$ phase space, the LOFF phase has a tricritical
point and that around this point the favored crystalline phase
corresponds to two antipodal waves (see
\cite{buzdin:1997ab,combescot:2002ab} and for a review
\cite{Casalbuoni:2003ab}). Therefore there could be various phase
transitions going down in temperature as it happens in the
two-dimensional case \cite{Shimahara:1998ab}.
 For completeness we give here the phase diagram of the LOFF phase
 in the plane $(\delta\mu,T)$, in Fig. \ref{LOFFphase}, and the
 behavior of the length of the vector $\b q$ along the second
 order critical line in Fig \ref{qvector}.

\subsection{Phonons}

As we have seen QCD at high density is conveniently studied
through a hierarchy of effective field theories, see Section
\ref{VA}. By using the same procedure in the case of the LOFF
phase one can derive the analogue of the HDET
\cite{Casalbuoni:2001gt,Casalbuoni:2002pa} and the effective
lagrangian for the Goldstone bosons (phonons) associated to the
breaking of translational and rotational symmetries in the LOFF
phase \cite{Casalbuoni:2002hr,Casalbuoni:2002my}.  The number and
the features of the phonons depend on the particular crystalline
structure. We will consider here the case of the single plane-wave
\cite{Casalbuoni:2001gt} and of the face-centered cube
\cite{Casalbuoni:2002hr}. We will introduce the phonons as it is
usual for NG bosons \cite{Casalbuoni:2001gt}, that is as the
phases of the condensate. Considering the case of a single
plane-wave we introduce a scalar field $\Phi(x)$ through the
replacement \be \Delta(\vec x)=\exp^{2i\b q\cdot\b x}\Delta\to
e^{i\Phi(x)}\Delta\label{subst1}.\ee We require that the scalar
field $\Phi(x)$  acquires the following expectation value in the
ground state \be \langle\Phi(x)\rangle=2\,\b q\cdot\b
x\label{expectation}.\ee The phonon field is defined as \be\frac 1
f\phi(x)=\Phi(x)-2\b q\cdot\b x.\label{phonon}\ee Notice that the
phonon field transforms nontrivially under rotations and
translations. From this it follows that non derivative terms for
$\phi(x)$ are not allowed. One starts with the most general
invariant lagrangian for the field $\Phi(x)$ in the low-energy
limit. This cuts the expansion of $\Phi$ to the second order in
the time derivative. However one may have an arbitrary number of
space derivative, since from Eq. (\ref{expectation}) it follows
that the space derivatives do not need to be small. Therefore \be
{\cal L}_{\rm phonon} =\frac{f^2}2\left({\dot\Phi}^2+\sum_k
c_k\Phi(\vec\nabla^2)^k\Phi\right).\ee Using the definition
(\ref{phonon}) and keeping the space derivative up to the second
order (we can make this assumption for the phonon field) we find
\be {\cal L}_{\rm phonon}=\frac 1
2\left({\dot\phi}^2-v_\perp^2\vec\nabla_\perp\phi\cdot\vec\nabla_\perp\phi-
v_\parallel^2\vec\nabla_\parallel\phi\cdot\vec\nabla_\parallel\phi\right),\ee
where \be\vec\nabla_\parallel=\vec n(\vec n\cdot\vec\nabla),~~~~
\vec\nabla_\perp=\vec\nabla-\vec\nabla_\parallel,~~~~\vec
n=\frac{\vec q}{|\vec q\,|}.\ee We see that the propagation of the
phonon in the crystalline medium is anisotropic.

\begin{figure}[ht]
%\epsfxsize=10cm   %width of figure - will enlarge/reduce the figures
%\epsfbox{fig3.eps}
%\figurebox{2cm}{3cm}{} %to have a box alone
\centerline{\epsfxsize=2.5in\epsfbox{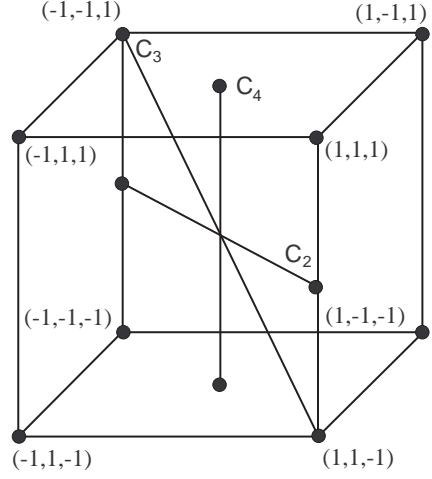}} \caption{\it
The figure shows the vertices and corresponding coordinates of the
cube described in the text. Also shown are the symmetry axes.
\label{fig6}}
\end{figure}

The same kind of considerations can be made in the case of the
cube. The cube is defined by 8 vectors $\b q_i$ pointing from the
origin of the coordinates to the vertices of the cube
parameterized as in Fig. \ref{fig6}.

The condensate is given by \cite{Bowers:2002xr} \be
\Delta(x)=\Delta\sum_{k=1}^8e^{2i\vec q_k\cdot\vec
x}=\Delta\sum_{i=1,(\epsilon_i=\pm)}^3e^{2i|\vec q\,|\epsilon_i
x_i}.\ee We introduce now three scalar fields such that
\be\langle\Phi^{(i)}(x)\rangle=2|\vec q\,|x_i,\ee through the
substitution
\be\Delta(x)\to\Delta\sum_{i=1,(\epsilon_i=\pm)}^3e^{i\epsilon_i\Phi^{(i)}(x)}
\label{phonon_cube}\ee and the phonon fields \be \frac 1 f
\phi^{(i)}(x)=\Phi^{(i)}(x)-2|\vec q\,|x_i.\ee Notice that the
expression (\ref{phonon_cube}) is invariant under the symmetry
group of the cube acting upon the scalar fields $\Phi^{(i)}(x)$.
This group has three invariants for the vector representation \bea
&I_2(\vec X)=|\vec X|^2,~~~~I_4(\vec X)=X_1^2 X_2^2 +X_2^2
X_3^3+X_3^2 X_1^2&\nn\\&I_6(\vec X)=X_1^2 X_2^2 X_3^2.&\eea
Therefore the most general invariant lagrangian invariant under
rotations, translations and the symmetry group of the cube, at the
lowest order in the time derivative, is \bea  L_{\rm
phonon}&=&\frac {f^2} 2\sum_{i=1,2,3}({\dot\Phi}^{(i)})^2\nn\\&+&
L_{\rm s}(I_2(\vec\nabla\Phi^{(i)}),
I_4(\vec\nabla\Phi^{(i)}),I_6(\vec\nabla\Phi^{(i)})).\eea
Expanding this expression at the lowest order in the space
derivatives of the phonon fields one finds
\cite{Casalbuoni:2002hr} \bea L_{\rm phonos} &=&\frac 1
2\sum_{i=1,2,3}({\dot\phi}^{(i)})^2-\frac a 2
\sum_{i=1,2,3}|\vec\nabla\phi^{(i)}|^2\nn\\&-& \frac b 2
\sum_{i=1,2,3}(\de_i\phi^{(i)})^2-
c\sum_{i<j=1,2,3}\de_i\phi^{(i)}\de_j\phi^{(j)}.\eea

The parameters appearing in the phonon lagrangian can be evaluated
following the strategy outlined in
\cite{Casalbuoni:2002pa,Casalbuoni:2002my} which is the same used
for evaluating the parameters of the NG bosons in the CFL phase.
It is enough to calculate  the self-energy of the phonons (or the
NG bosons) through one-loop diagrams due to fermion pairs. Again
the couplings of the phonons to the fermions are obtained noticing
that the gap acts as a Majorana mass for the quasi-particles.
Therefore the couplings originate from the substitutions
(\ref{subst1}) and (\ref{phonon_cube}). In this way one finds the
following results: for the single plane-wave \be v_\perp^2=\frac 1
2\left(1-\left(\frac{\delta\mu}{|\vec q\,|}\right)^2\right),~~~
v_\parallel^2=\left(\frac{\delta\mu}{|\vec q\,|}\right)^2\ee and
for the cube \be a=\frac 1 {12},~~~b=0,~~~c=\frac
1{12}\left(3\left(\frac{\delta\mu}{|\vec
q\,|}\right)^2-1\right).\ee

\section{Astrophysical implications \label{XI}}

\begin{center}
\begin{figure}[htb]
\epsfxsize=8truecm \centerline{\epsffile{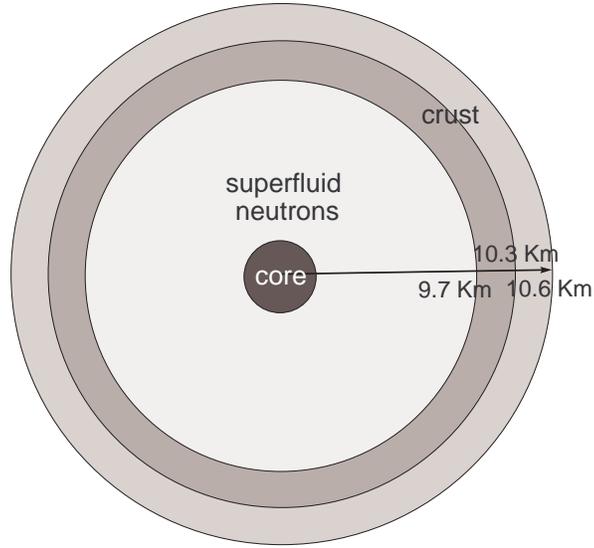}} \noindent
\caption{\it Schematic view of a neutron star as computed by an
equation of state with three nucleon interaction
\cite{Shapiro}.}\label{star2}
\end{figure}
\end{center}

Should we look for a laboratory to test color superconductivity,
we would face the problem that in the high energy physics
programmes aiming at new states of matter, such as the Quark Gluon
Plasma, the region of the $T-\mu$ plane under investigation is
that of low density and high temperature. On the contrary we need
physical situations characterized by low temperature and high
densities. These conditions are supposed to occur in the  inner
core of neutron stars, under the hypothesis that, at the center of
these compact stars, nuclear matter has become so dense as to
allow the transition to quark matter. A schematic view of a
neutron star is in Fig. \ref{star2}.
In the subsequent sections we shall give a pedagogical
introduction to the physics of compact stars and we shall review
some possible astrophysical implications of the color
superconductivity. \subsection{A brief introduction to compact
stars} To begin with, let us show that for a fermion gas high
chemical potential means  high density. To simplify the argument
we assume that the fermions are massless and not interacting, so
that $\mu=\epsilon_F=p_F$. From the grand potential in Eq.
(\ref{5.164}) \be\Omega=gV\int\frac{d^3\b
p}{(2\pi)^3}(\epsilon_{\b p}-\mu)\theta(\mu-\epsilon_{\b p}),\ee
one finds ($g$ is the number of species of fermions times 2 for
the spin)\be \Omega=-\frac{gV}{24\pi^2}\mu^4.\ee It follows that
\be\rho=-\frac{\de\Omega}{\de \mu}=\frac{gV}{6\pi^2}\mu^3,\ee
which means that the chemical potential increases as $\rho^{1/3}$:
This is the reason why we should search color superconductivity in
media with very high baryonic density.

In the general case, the equation of state is determined, for
instance, by evaluating the pressure as a function of the other
thermodynamical quantities. This can be done starting from the
grand potential. For the simple case of free massless fermions we
obtain \be P=-\Omega=\frac{gV}{24\pi^2}\mu^4,\ee and therefore the
equation of state has the form \be P=K\rho^{\frac 4 3}.\ee Clearly
this result also holds for massive fermions in the Ultra
Relativistic (UR) case where the mass is negligible. In the Non
Relativistic (NR) case one has \be P=\frac 4{15}\frac
{gV}{\pi^2}\frac{m^{3/2}}{\sqrt{2}}\mu^{5/2},~~~~\rho=\frac
2{3}\frac {g}{\pi^2}\frac{m^{3/2}}{\sqrt{2}}\mu^{3/2}\ee from
which\be P=K\rho^{\frac 5 3},\ee. More generally the equation of
state can be approximated by the expression \be P=K\rho^\gamma,
\ee and the two cases discussed above are characterized as
follows:
 \bea
&&\rho\ll 10^6 g/cm^3~ {\rm \{electrons\}}\cr
  NR\left( \gamma=\frac 5 3\right)&&\cr && \rho\ll 10^{15}
g/cm^3~{\rm \{neutrons\}} \cr&&\cr&&\cr&& \rho\gg 10^6 g/cm^3~{\rm
\{electrons\}}\cr UR\left(\gamma=\frac 4 3\right)&&\cr&& \rho\gg
10^{15} g/cm^3~{\rm \{neutrons\}}\ .\eea Let us note explicitly
that, at $T=0$, $P\neq 0$. This is a quantum-mechanical effect due
to the Pauli principle and the Fermi Dirac statistics (for
comparison, for a classical Maxwell Boltzmann gas $P\to 0$ when
$T\to 0$). In absence of other sources of outward pressure it is
the  pressure of the degenerate fermion gas  that balances the
gravity and avoids the stellar collapse.

One can see that the densities that can be reached in the compact
stars are very different depending on the nature of the fermions.
The two cases correspond to two classes of compact stars, the
white dwarfs and the neutron stars. White dwarfs (w.d.s) are stars
that have exhausted nuclear fuel;
 well known examples are Sirius B,
or 40 Eri B. In the Hertzsprung-Russel diagram w.d.s fill in a
narrow corner below the main sequence. In a w.d. stellar
equilibrium is reached through a compensation between  the inward
pressure generated by gravity and the outward  pressure of
degenerate electrons. Typical values of the central density, mass
and radius for a w.d. are $\rho= 10^6 g/cm^3$, $M\sim
 M_\odot$,  $R\sim 5,000 km$. Notice that the nuclear saturation
 density, defined as the density of a nucleon of radius 1.2 $fm$. is
 about $1.5\times 10^{14}~g/cm^3$.

Suppose now that in the star higher values of $\rho$ are reached.
  If $\rho$ increases, inverse beta decay becomes important: \be
e^-p\to n\nu\ .\ee This process fixes the chemical composition at
equilibrium \be \mu_e+\mu_p=\mu_n\ . \ee In the cases of
ultrarelativistic particles we get \be
\rho_e^{1/3}+\rho_p^{1/3}=\rho_n^{1/3}.\ee On the other hand one
has to enforce  neutrality:\be \rho_e=\rho_p\ ,\ee implying \be
\frac{\rho_p}{\rho_n}=\frac 1 8\, .\ee This number should be seen
as a benchmark value, as it is derived under simplifying
hypotheses, most notably the absence of interactions and the
neglect of masses. In any event it suggests that, for higher
densities, the star tends to have a relatively larger fraction of
neutrons and therefore it is named a {\it neutron star.} It must
be stressed that one of the relevant facts about neutron stars is
that the general relativity effects cannot be ignored and the
relevant equilibrium equations to be used are the
Oppenheimer-Volkov equations of hydrostatic equilibrium.

The following simple argument, due to Landau (1932) can be used to
evaluate the relevant parameters of white dwarfs and neutron stars
(see the textbook \cite{Shapiro}; more recent reviews of compact
stars are in \cite{Tsuruta:1998ab,Page:1998ab,Heiselberg:2000dn}).
Let us consider $N$ fermions in a sphere of radius $R$ at $T=0$;
the number of fermion per volume unit scales as $\dd
n\sim{N}/{R^3}$; the volume per fermion is therefore $\sim  1/{n}$
and the uncertainty on the position is of the order of $n^{-1/3}$;
the Fermi momentum is of the order of the uncertainty on the
fermion momentum and therefore \be p_F\sim n^{1/3}\hbar~,\ee a
result we obtained already under more stringent hypotheses
(Fermi-Dirac distribution) and derived again here using only the
uncertainty relations. The Fermi energy of the baryons is
therefore
 \be \epsilon_F\sim
\frac{\hbar cN^{1/3}}{R}\ ,\ee if $N$ is the total number of
baryons. Note that this applies both to neutron stars and to
electron stars, because also in stars where the pressure mainly
come from electrons there will be a considerable amount of protons
and neutrons and the largest part of the energy comes from the
baryons, not from the electrons.
 On the other hand the gravitational energy per baryon is
\be E_G\sim -\frac{GNm_B^2}{R}\ ,\ee and the total
 energy  can be estimated as\be
E=E_G+E_F\sim \frac{\hbar cN^{1/3}}{R}-\frac{GNm_B^2}{R}\ . \ee
Now equilibrium can exist only if $E\ge 0$. As a matter of fact if
$E<0$ ($N$ large) then $\lim_{R\to 0} E=-\infty$, which means that
the energy is unbounded from below and the system is unstable.
Therefore, $E\ge 0$ gives the maximum number of baryons as
follows: \be N\le N_{max}=\left(\frac{\hbar
c}{Gm_B^2}\right)^{3/2}\sim 2\times 10^{57}\ .\ee As a consequence
the maximum mass is \be M_{max}=N_{max}m_B=1.5 M_\odot\
.\label{chandra}\ee This mass can be estimated better and its
better determination ($\sim 1.4 M_\odot$) is known as the
Chandrasekhar limit; for our purposes the estimate (\ref{chandra})
is however sufficient. Notice that the Chandrasekhar limit is
similar for compact stars where the degeneracy pressure is mainly
supplied by electrons and also where it is supplied by baryons.

 One can  estimate the radius of a star whose mass is given by (\ref{chandra}).
 One has
\be \epsilon_F \sim\frac{\hbar c}{R}\,N_{max}^{1/3}\,\sim\,
\frac{\hbar c}{R}\left(\frac{\hbar c}{Gm_B^2}\right)^{1/2}\ee and,
therefore,
%\beas&& =5\times 10^8 \,cm\ \{m=m_e \}\cr
% R\sim\frac{\hbar
%}{mc}\left(\frac{\hbar c}{Gm_B^2}\right)^{1/2}\Big\{&&\cr
%&&=
%3\times \,10^5 \,cm\{m=m_n\}\eeas
% ......
\be R\sim\frac{\hbar }{mc}\left(\frac{\hbar c}{Gm_B^2}
\right)^{1/2}\,=\, \Big\{^{\dd{\ 5\times 10^8 \,cm\ \{m=m_e
\}}}_{\dd{\ 3\times \,10^5 \,cm\ \{m=m_n \}}.} \ee
 If a neutron star accretes its mass beyond the  Chandrasekhar
limit nothing can prevent the collapse and it becomes a black
hole\footnote{The exact determination of the mass limit depends on
the model for nuclear forces; for example in \cite{Cameron:1974ab}
the neutron star mass limit is increased beyond $1.4 M_\odot$.}.
In  Table \ref{neutronstars} we summarize our discussion; notice
that we report for the various stars also the value of the
parameter $GM/Rc^2$ i.e. the ratio of the Schwarzschild radius to
the star's radius. Its smallness
 measures the validity of the approximation of neglecting the
 general relativity effects; one can see that for the sun and the white
 dwarfs the newtonian treatment of
 gravity represents a fairly good approximation.
\begin{center}
\begin{table}[htb]
\begin{tabular}{|c|c|c|c|c|}
  \hline &&&& \\ $$ & $M$ & $R$ &$\dd\rho\left(\frac{g}{cm^3}\right)$ &
  $\dd \frac{GM}{Rc^2}$  \\ &&&&\\
  \hline &&&&\\
  Sun & $M_\odot$ & $\dd R_\odot$ & 1 & $10^{-6}$ \\ &&&&\\
  White Dwarf & $\leq M_\odot$ & $\dd 10^{-2}R_\odot$ & $\leq 10^7$&  $10^{-4}$
  \\ &&&&\\
  Neutron Star & $1-3 M_\odot$ & $\dd 10^{-5}R_\odot$ & $\leq 10^{15}$&
 $10^{-1}$  \\ &&&&\\
  Black Hole & arbitrary & $\dd\frac{2GM}{c^2}$ & $\sim \frac{M}{R^3}$&  $1$ \\ &&&&\\ \hline
\end{tabular}
\caption{\it Parameters of different stellar
objects.\label{neutronstars}}\end{table}\end{center}
 Neutron stars are the most likely candidate  for the
theoretical description of pulsars. Pulsars are rapidly rotating
stellar objects, discovered in 1967 by  Hewish and collaborators
and identified as rotating neutron stars by  \cite{gold}; so far
about 1200 pulsars have been identified.

Pulsars are characterized by the presence of strong magnetic
fields with the magnetic and rotational axis misaligned; therefore
they continuously emit electromagnetic energy (in the form of
radio waves) and constitute indeed a very efficient mean to
convert rotational energy into electromagnetic radiation. The
rotational energy loss is due to dipole radiation and is therefore
given by \be \frac{dE}{dt}= I\omega\frac{d\omega}{dt}=
-\frac{B^2R^6\omega^4\sin^2\theta}{6c^3}\ .\ee Typical values in
this formula are, for the moment of inertia $I\sim R^5\rho\sim
10^{45}$g/cm$^3$, magnetic fields $B\sim 10^{12}$ G, periods
$T=2\pi/\omega$ in the range $1.5$ msec-8.5 sec; these periods
increase slowly\footnote{Rotational period and its derivative can
be used to estimate the pulsar's age by the approximate formula
${T}/2({dT}/{dt})$, see e.g. \cite{Lorimer:1999dh}.}, with
derivatives ${dT}/{dt}\sim 10^{-12}$ - $10^{-21}$, and never
decrease except for occasional jumps (called {\it glitches}).

Glitches were first observed in the Crab and Vela pulsars in 1969;
the variations in the rotational frequency are of the order
$10^{-8}-10^{-6}$.

 This last feature is the most significant
phenomenon pointing to neutron stars as a model of pulsars in
comparison  to other form of hadronic matter, such as strange
quarks. It will be discussed in more detail in Section
\ref{astro}, where we will examine the possible role played by the
crystalline superconducting phase. In the subsequent three
paragraphs we will instead deal with other possible astrophysical
implications of color superconductivity.

\subsection{Supernovae neutrinos and cooling of neutron stars}
Neutrino diffusion is the single most important mechanism in the
cooling of young neutron stars, i.e. with an age $<10^{5}$ years;
it affects both the early stage and the late time evolution of
these compact stars. To begin with let us consider  the early
evolution of a Type II Supernova.

Type II supernovae are supposed to be born by collapse of massive
($M \sim 8 - 20  M_\odot$) stars\footnote{The other supernovae,
i.e. type I supernovae, result from the complete explosion of a
star with $4M_\odot\leq M \leq 8 M_\odot$ with no remnants.}.
These massive stars have unstable iron cores\footnote{Fusion
processes favor the formation of iron, as the binding energy per
nucleon in nuclei has a maximum for $A\sim 60$.} with masses
 of the order
 of the Chandrasekhar mass. The explosion producing the supernova
originates within the core, while the external mantle of the red
giant star produces remnants that can be analyzed by different
means, optical, radio and X rays. These studies agree with the
hypothesis of a core explosion. The emitted energy ($\sim 10^{51}
erg$) is much less than the total gravitational energy of the
star, which confirms that the remnants are produced by the outer
envelope of the massive star; the bulk of the gravitational
energy, of the order of $10^{53}$ erg, becomes internal energy of
the proto neutron star (PNS). The suggestion that neutron stars
may be formed in supernovae explosions was advanced in 1934 by
\cite{baade} and it has been subsequently confirmed by the
observation of the Crab pulsar in the remnant of the Crab
supernova observed in China in 1054 A.D.

We do not proceed in this description as it is beyond the scope of
this review and we concentrate our attention on the cooling of the
PNS\footnote{See
\cite{Colgate:1966ab,Nadyozhin:1973,Burrows:1986,Keil:1995,Pons:1999}
for further discussions.}, which is mostly realized through
neutrino diffusion. By this mechanism one passes from the initial
temperature $T\sim 20-30$ MeV to the cooler temperatures of the
neutron star at subsequent stages. This phase of fast cooling
lasts 10-20 secs and the neutrinos emitted during it have mean
energy $\sim 20$ MeV. These properties, that can be predicted
theoretically, are also confirmed by data from SN 1987A.

The role of quark color superconductivity at this stage of the
evolution of the neutron stars has been discussed in
\cite{Carter:2000xf}. In this paper the neutrino mean free path is
computed in a color super-conducting medium made up by quarks in
two flavor (2SC model). The results obtained indicate that the
cooling process by neutrino emission slows down when the quark
matter undergoes the phase transition to the superconducting phase
at the critical temperature $T_c$, but then accelerates when $T$
decreases below $T_c$. There should be therefore changes in the
neutrino emission by the PNS and they might be observed in some
future supernova event; this would produce an interesting test for
the existence of a  color superconducting phase in compact stars.

 Let us now
consider the subsequent evolution of the neutron star, which also
depends on neutrino diffusion. The simplest processes of neutrino
production are the so called direct Urca processes \bea
&&f_1+\ell\to f_2+\nu_\ell\ ,\cr &&\cr &&f_2+\ell\to f_1+\ell\to
f_2+\bar\nu_\ell\ ;\eea by these reactions, in absence of quark
superconductivity, the interior temperature $T$ of the star drops
below $10^9$ K ($\sim 100$ KeV) in a few minutes and in $10^2$
years to temperatures $\sim 10^7$ K. Generally speaking the effect
of the formation of gaps is to slow down the cooling, as it
reduces
 both the emissivity and the specific heat. However not only quarks, if present in the neutron star, but also
other fermions, such as neutrons, protons or hyperons have gaps,
 as the formation of fermion pairs is unavoidable if there is an
attractive interaction in any channel. Therefore, besides quark
color superconductivity, one has also the phenomenon of baryon
superconductivity and neutron superfluidity, which is the form
assumed by this phenomenon for neutral particles. The analysis is
therefore rather complicated; the thermal evolution of a late time
neutron star has been discussed in
\cite{Page:2000wt,Blaschke:1999qx}, but no clear signature for the
presence of color superconductivity seems to emerge from the
theoretical simulations and, therefore, one may tentatively
conclude that the late time evolution of the neutron stars does
not offer a good laboratory to test the existence of color
superconductivity in compact stars.

\subsection{$R$-mode instabilities in neutron stars and strange
stars\label{rmode}} Rotating relativistic stars are in general
unstable against the rotational mode ($r$-mode instability)
\cite{Anderson:1998ab}. The instability is due to  the emission of
angular momentum by gravitational waves from the mode. Unless it
is damped by viscosity effects, this instability would spin down
the star in relatively short times. More recently it has been
realized \cite{Bildsten:1999zn,Andersson:2000pt} that in neutron
stars there is an important viscous interaction damping the
$r$-mode i.e. that between the external metallic crust and the
neutron superfluid. The consequence of this damping is that
$r$-modes are significant only for young neutron stars, with
periods $T<2$ msec. For larger rotating periods the damping of the
$r-$mode implies that the stars slow down only due to magnetic
dipole braking.

All this discussion is relevant for the nature of pulsars: Are
they neutron stars or strange stars?

The existence of strange stars, i.e. compact stars made of quarks
$u,\,d,\,s$ in equal ratios would be a consequence of the
existence of stable strangelets. This hypothetical form of nuclear
matter, made of a large number of $u,\,d,\,s$ quarks has been
suggested by  \cite{Bodmer:1971} and  \cite{Witten:1984} as
energetically favored in comparison to other hadronic phases when
a large baryonic number is involved. The reason is that in this
way the fermions, being of different flavors, could circumvent the
Pauli principle and have a lower energy, in spite of the larger
strange quark mass. If strangelets do exist, basically all the
pulsars should be strange stars because the annihilation of
strange stars, for example from a binary system, would fill the
space around with strangelets that, in turn, would convert
ordinary  nuclear stars into strange stars.

An argument in favor of the identification of pulsars with strange
stars is the scarcity of pulsars with very high frequency ($T<2.5$
msec).

This seems to indicate that indeed the $r-$mode instability is
effective in slowing down the compact stars and
 favors strange stars, where, differently from neutron stars,
 the crust can be absent.
Even in the presence of the external crust, that in a quark star
can be formed by the gas after the supernova explosion or
subsequent accretion, the dampening of the $r-mode$ is less
efficient. As a matter of fact, since electrons are only slightly
bounded, in comparison with quarks that are confined, they tend to
form an atmosphere having a thickness of a few hundred Fermi; this
atmosphere produces a separation  between the nuclear crust and
the inner quark matter and therefore the viscosity is much
smaller.

In quark matter with color superconductivity the presence of gaps
$\Delta\gg T$  exponentially reduces the bulk and shear viscosity,
which renders the $r-$mode unstable. According to
\cite{Madsen:1999ci,Madsen:2001zh} this would rule out compact
stars entirely made of quarks in the CFL case (the 2SC model would
be marginally compatible, as there are ungapped quarks in this
case). For example for $\Delta> 1$ MeV any star having $T<10$ msec
would be unstable, which would contradict the observed existence
of pulsars with time period less than 10 msec.

However this conclusion does not rule out the possibility of
neutron stars with a quark core in the color superconducting
state, because as we have stressed already, for them the dampening
of the $r-$mode instability would be provided by the viscous
interaction between the nuclear crust and the neutron superfluid.
\subsection{Miscellaneous results} Color superconductivity is a
Fermi surface phenomenon and as such, it does not affect
significantly the equation of state of the compact star. Effects
of this phase could be seen in other astrophysical contexts, such
as those considered in the two previous paragraphs or in relation
to the pulsar glitches, which will be examined in the next
paragraph. A few other investigations have been performed in the
quest of possible astrophysical signatures of color
superconductivity; for instance in \cite{Ouyed:2001fv} it has been
suggested that the existence of a 2SC phase might be partly
responsible of the gamma ray bursts, due to the presence in the
two-flavor superconducting phase of a light glueball that can
decay into two photons. Another interesting possibility is related
to the stability of strangelets, because, as observed in
\cite{Madsen:2001fu}, CFL strangelets, i.e. lumps of strange quark
matter in the CFL phase may be significantly more stable than
strangelets without color superconductivity.

Finally we wish to mention the observation of \cite{Alford:1999pb}
concerning the evolution of the magnetic field in the interior of
neutron stars. Inside an ordinary neutron star, neutron pairs are
responsible for superfluidity, while proton pairs produce BCS
superconductivity. In this condition magnetic fields experience
the ordinary Meissner effect and are either expelled or restricted
to flux tubes where there is no pairing. In the CFL (and also 2SC)
case, as we know from  \ref{VC}, a particular $U(1)$ group
generated by \be\tilde Q=
 {\bf 1}\otimes Q-{Q}\otimes{\bf 1}\label{u1}\ee
 remains unbroken and plays the role of
electromagnetism. Instead of being totally dragged out or confined
in flux tubes, the magnetic field will partly experience Meissner
effect (the component $\tilde A_\mu$), while the remaining part
will remain free in the star. During the slowing down this
component of the magnetic field should not decay because, even
though the color superconductor is not a BCS conductor for the
group generated by (\ref{u1}), it may be a good conductor due to
the presence of the electrons in the compact star. Therefore it
has been suggested \cite{Alford:1999pb} that a quark matter core
inside a neutron star may serve as an "anchor" for the magnetic
field.

\subsection{Glitches in neutron stars\label{astro}}

\begin{center}
\begin{figure}[htb]
\epsfxsize=9truecm \centerline{\epsffile{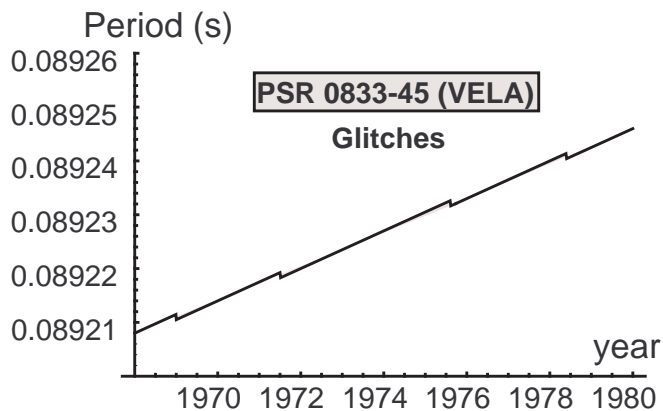}} \noindent
\caption{\it The variation of the period of the pulsar PSR 0833-45
(VELA) with the typical structure of glitches
shown.}\label{glitches}
\end{figure}
\end{center}

Glitches, that is very sudden variations in the period, are a
typical phenomenon of the pulsars, in the sense that probably all
the pulsars have glitches (for a recent review see
\cite{Link:2000mu}), see Fig. \ref{glitches}. Several models have
been proposed to explain the glitches. Their most popular
explanation is based on the idea that these sudden jumps of the
rotational frequency are due to the the angular momentum stored in
the superfluid neutrons in the inner crust (see Fig. \ref{star2}),
more precisely in vortices pinned to nuclei. When the star slows
down, the superfluid neutrons do not participate in the movement,
until the state becomes unstable and there is a release of angular
momentum to the crust, which is seen as a jump in the rotational
frequency.

The presence of glitches is one of the main reasons for the
identification of pulsars with neutron stars; as a matter of fact
neutron stars are supposed to have a dense metallic crust,
differently from quark stars where the crust is absent or, if
present, is much less dense ($\approx 10^{11}$ g cm$^{-3}$).

\begin{center}
\begin{figure}[htb]
\epsfxsize=6truecm \centerline{\epsffile{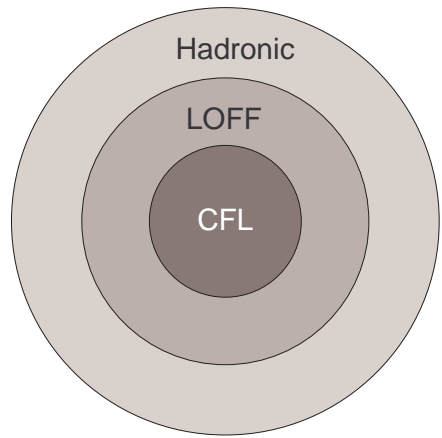}} \noindent
\caption{\it A model of a neutron star with an inner shell in the
LOFF phase.}\label{LOFFstar}
\end{figure}
\end{center}

Since the density in the inner of a star is a function of the
radius, it results that one has a sort of laboratory to study the
phase diagram of QCD at zero temperature, at least in the
corresponding range of densities. A possibility is that  a CFL
state occurs as a core of the star, then a shell in the LOFF state
and eventually the exterior part made up of neutrons (see Fig.
\ref{LOFFstar}). Since in the CFL state the baryonic number is
broken there is superfluidity. Therefore the same mechanism
explained above might work with vortices in the CFL state pinned
to the LOFF crystal.

This would avoid the objection raised in \cite{Friedman:1991qz}
that excludes the existence of strange stars. This objection is
based on the following observation: strange stars cannot have a
metallic crust, and in that case they can hardly develop vortices,
so no glitches would arise. However, if the strange matter exists,
strange stars should be rather common, as we discussed in Section
\ref{rmode}, in contrast with the widespread appearance of
glitches in pulsars. Therefore, if the color crystalline structure
is able to produce glitches, the argument in favor of the
existence of strange stars would be reinforced.

\begin{center}
\begin{figure}[htb]
\epsfxsize=9truecm \centerline{\epsffile{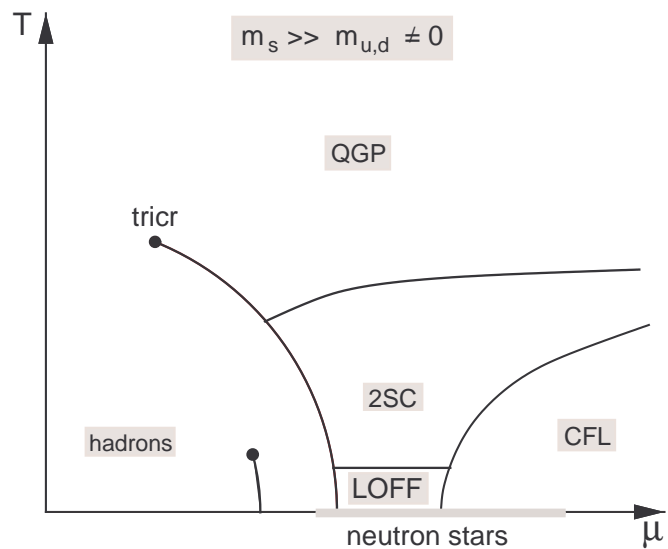}} \noindent
\caption{\it A possible QCD phase diagram showing the end point of
the first order transition phase, labelled as "tricr".  The hadron
phase with the transition to the nuclear phase, the Quark-gluon
plasma, the CFL, the 2SC and the LOFF phase are also
shown.}\label{phase}
\end{figure}
\end{center}

 Considering
the very narrow range of values for $\delta\mu$ in order to be in
the LOFF phase one can ask if the previous possibility has some
chance to be realized \cite{Casalbuoni:2003ab}. Notice that using
the typical LOFF value $\delta\mu\approx 0.75\Delta_{BCS}$ one
would need values of $\delta\mu$ around $15\div 70~MeV$. Let us
consider  a very crude model of three free quarks with
$M_u=M_d=0$, $M_s\not =0$. Assuming at the core of the star a
density around $10^{15}$ g/cm$^3$, that is from 5 to 6 times the
saturation nuclear density $0.15\times 10^{15}$ g/cm$^3$, one
finds for $M_s$ ranging between 200 and 300 $MeV$ corresponding
values of $\delta\mu$ between 25 and 50 $MeV$. We see that these
values are just in the right range for being within the LOFF
window. Therefore a possible phase diagram for QCD could be of the
form illustrated in Fig. \ref{phase}. We see that in this case a
coexistence of the CFL, LOFF and neutron matter would be possible
inside the neutron star.

%\input{Conclusions}
%\section*{Acknowledgments}
\begin{acknowledgments}
I would like to thank the CERN Theory Division, the DESY Theory
Group and the Department of Physics of the University of
Barcellona for the kind hospitality offered during the preparation
and the completion of these lecture notes. I would like to thank
also all my collaborators to the work I have done in this subject.
In particular I would like to thank G. Nardulli, R. Gatto, M.
Mannarelli and M. Ruggieri.
\end{acknowledgments}
\newpage

%%%%%%%%%%%%%%%%%%%%%%%%%%%%%%%%%%%%%%%%%%
\appendix
\section{The gap equation in the functional formalism from
HDET}\label{Appendix1}

We will now derive the gap equation by using the functional
formalism within the general formalism of Section \ref{HDET}. The
method used is a trivial generalization of the one used in Section
\ref{IIIC}. We start again from Eq. \ref{5.54} \be{\cal
L}_I=-\frac G 4\epsilon_{ab}\epsilon_{\dot c\dot
d}V_{ABCD}\psi_a^A\psi_b^B\psi_{\dot c}^{C\dagger}\psi_{\dot
d}^{D\dagger}\ee For simplicity we suppress  the notation for
fixed velocity. We will insert inside the functional integral the
following identity

\be {\rm const}=\int{\cal
D}(\Delta_{AB},\Delta^*_{AB})e^{\dd{-\frac i G\int d^4
x}F[\Delta_{AB},\Delta^*_{AB},\psi^A,\psi^{A\dagger}]}\ee where
\be F=\left[\Delta_{AB}-\frac G 2V_{GHAB}(\psi^{GT}
C\psi^H)\right]W_{ABCD}\left[\Delta^*_{CD}+\frac G
2V_{CDEF}(\psi^{E\dagger} C\psi^{F*})\right]\ee The quantity
$W_{ABCD}$ is defined in such a way that \be
W_{ABCD}V_{CDEF}=\delta_{AE}\delta_{BF},~~~V_{ABCD}W_{CDEF}=\delta_{AE}\delta_{DF}\ee
Then we get \bea F&=&\frac G 4
V_{ABCD}\psi^{CT}C\psi^B\psi^{C\dagger}C\psi^{D*}+\frac 1
2\Delta_{CD}^*\psi^{CT}C\psi^D-\frac 1
2\Delta_{AB}\psi^{A\dagger}C\psi^{B*}\nn\\&-&\frac 1 g\Delta_{AB}
W_{ABCD}\Delta_{CD}^*\eea Normalizing at the free case ($G=0$) we
get \bea \frac Z{Z_0}&=&\frac 1 {Z_0}\int{\cal
D}(\psi,\psi^\dagger){\cal
D}(\Delta,\Delta^*)e^{\dd{iS_0[\psi,\psi^\dagger]}}\nn\\ &\times&
e^{\dd{+i\int
d^4x\left[-\frac{\Delta_{AB}W_{ABCD}\Delta_{CD}^*}G-\frac 1
2\Delta_{AB}(\psi^{A\dagger} C\psi^{B*})+\frac 1
2\Delta^*_{AB}(\psi^{AT}C\psi^B)\right]}}\eea Going to the
velocity formalism and introducing the Nambu-Gor'kov field we can
write the fermionic part in the exponent inside the functional
integral as \be\int d^4x\int\frac{d\b
v}{4\pi}\chi^{A\dagger}S^{-1}_{AB}\chi^B\ee with \be
S^{-1}_{AB}=\left(\matrix{ V\cdot
\ell\,\delta_{AB}&-\Delta_{AB}\cr -\Delta_{AB}^* & \tilde V\cdot
\ell\, \delta_{AB}}\right)\ee Using again the replica trick as in
Section \ref{IIIC} we perform the functional integral over the
Fermi fields obtaining \be\frac Z{Z_0}=\frac 1{Z_0}\left[{\rm
det}S^{-1}\right]e^{\dd{-\frac i G\int d^4x
\Delta_{AB}W_{ABCD}\Delta_{CD}^*}}\equiv e^{\dd{iS_{\rm eff}}}\ee
with \be S_{\rm eff}=-\frac i 2{\rm Tr}[\log S_0S^{-1}]-\frac i
G\int d^4x \Delta_{AB}W_{ABCD}\Delta_{CD}^*\ee Differentiating
with respect to $\Delta_{AB}$ we get immediately the gap equation
(\ref{5.73}).

\section{Some useful integrals\label{appendix2}}
We list a few 2-D integrals which have been used in the text
\cite{Nardulli:2002ma}. Let us define\be I_n=\int
\frac{d^{N}\ell}{(V\cdot\ell\,\tilde V\cdot\ell -
\Delta^2+i\epsilon)^{n+1}}= \frac{i\,(-i)^{n+1}\pi^{\frac N
2}}{n!} \frac{\Gamma(n+1-\frac N 2)}{\Delta^{2n+2-N}}\ ; \ee
therefore, for $N=2-\epsilon$ and denoting by $\gamma$ the
Euler-Mascheroni constant, we get\bea
I_0&=&-\frac{2i\pi}{\epsilon}+i\pi\ln\pi\Delta^2+i\pi\gamma \ ,\cr
I_1&=&+\frac{i\pi}{\Delta^2}\ ,\ I_2=-\frac{i\pi}{2\Delta^4}\ ,\
I_3=+\frac{i\pi}{3\Delta^6}\ .\eea Moreover  defining \be
I_{n,\,m}=\int \frac{d^{2}\ell}{ (V\cdot\ell\,\tilde V\cdot\ell
-\Delta^2+i\epsilon)^n (V\cdot\ell\,\tilde V\cdot\ell
-\Delta^{\prime 2}+i\epsilon)^m}\ , \ee we find \bea I_{1,\,1}&=&
\frac{i\pi}{\Delta^2-\Delta^{\prime\, 2}}
\ln\frac{\Delta^2}{\Delta^{\prime\, 2}} \ ,\cr I_{2,\,1}&=&i\pi
\left[\frac{1}{\Delta^2(\Delta^2-\Delta^{\prime\, 2})}-
\frac{1}{(\Delta^2-\Delta^{\prime\, 2})^2}
\ln\frac{\Delta^2}{\Delta^{\prime\, 2}} \right]\ ,\cr
I_{3,\,1}&=&\frac{i\pi}2\left[
\frac{-1}{\Delta^4(\Delta^2-\Delta^{\prime\, 2})}-
\frac{2}{\Delta^2(\Delta^2-\Delta^{\prime\, 2})^2}
+\frac{2}{(\Delta^2-\Delta^{\prime\, 2})^3}\ln
\frac{\Delta^2}{\Delta^{\prime\, 2}} \right]\ ,\cr
I_{2,\,2}&=&i\pi\left[
\left(\frac{1}{\Delta^2}+\frac{1}{\Delta^{\prime\,2}}\right)
\frac{1}{(\Delta^2-\Delta^{\prime\, 2})^2}
+\frac{2}{(\Delta^2-\Delta^{\prime\, 2})^3}\ln
\frac{\Delta^2}{\Delta^{\prime\, 2}} \right]. \eea
 Of some interest is the following infrared
 divergent integral: \be \tilde I_1=\int \frac{(V\cdot\ell)^2\
d^{2}\ell}{(V\cdot\ell\,\tilde V\cdot\ell )^{2}}\ . \ee We can
regularize  the divergence  by going to finite temperature  and
then taking the limit $T\to 0$ ($\ell_0=i\omega_n,\ \omega_n=\pi
T(2n+1)$; $T\to 0,\,\mu\to \infty$): \be \tilde I_1=2\pi
Ti\int_{-\mu}^{+\mu}dx
\sum_{-\infty}^{+\infty}\frac{(i\omega_n-x)^2}{(x^2+\omega_n^2)^2}
=2\pi Ti\left(-\frac{1}T\tanh\frac{\mu}{2T}
\right)\rightarrow-2\pi i \ . \ee Other divergent integrals, such
as \be \tilde I=\int \frac{d^{2}\ell}{(V\cdot\ell\,\tilde
V\cdot\ell )(V\cdot\ell\,\tilde V\cdot\ell -\Delta^2)} \ee are
treated in a similar way.

 Finally, useful
angular integrations are: \bea \int\frac{d\b v}{4\pi}
v^jv^k&=&\frac{\delta^{jk}}{3}\ ,\\ \int\frac{d\b
v}{4\pi}v^iv^jv^kv^\ell&=& \frac 1{15}(\delta^{ij}\delta^{k\ell}+
\delta^{ik}\delta^{j\ell}+\delta^{i\ell}\delta^{jk})\ . \eea
\newpage

%%%%%%%%%%%%%%%%%%%%%%%%%%%%%%%%%%%%%%%%%%
\bibliographystyle{apsrmp}
\bibliography{barcellona}

\begin{thebibliography}{177}
\expandafter\ifx\csname natexlab\endcsname\relax\def\natexlab#1{#1}\fi
\expandafter\ifx\csname bibnamefont\endcsname\relax
  \def\bibnamefont#1{#1}\fi
\expandafter\ifx\csname bibfnamefont\endcsname\relax
  \def\bibfnamefont#1{#1}\fi
\expandafter\ifx\csname citenamefont\endcsname\relax
  \def\citenamefont#1{#1}\fi
\expandafter\ifx\csname url\endcsname\relax
  \def\url#1{\texttt{#1}}\fi
\expandafter\ifx\csname urlprefix\endcsname\relax\def\urlprefix{URL }\fi
\providecommand{\bibinfo}[2]{#2}
\providecommand{\eprint}[2][]{\url{#2}}

\bibitem[{\citenamefont{Abbott and Fahri}(1981)}]{Abbott:1980ab}
\bibinfo{author}{\bibnamefont{Abbott}, \bibfnamefont{L.~F.}}, and
  \bibinfo{author}{\bibfnamefont{E.}~\bibnamefont{Fahri}},
  \bibinfo{year}{1981}, \bibinfo{journal}{CERN-TH3015} .

\bibitem[{\citenamefont{Abrikosov}(1957)}]{abrikosov:1957jk}
\bibinfo{author}{\bibnamefont{Abrikosov}, \bibfnamefont{A.~A.}},
  \bibinfo{year}{1957}, \bibinfo{journal}{Zh. Exsp. teor. Fiz.}
  \textbf{\bibinfo{volume}{32}}, \bibinfo{pages}{1442}.

\bibitem[{\citenamefont{Abrikosov} \emph{et~al.}(1963)\citenamefont{Abrikosov,
  Gor'kov, and Dzyaloshinski}}]{abrikosov}
\bibinfo{author}{\bibnamefont{Abrikosov}, \bibfnamefont{A.~A.}},
  \bibinfo{author}{\bibfnamefont{L.~P.} \bibnamefont{Gor'kov}}, and
  \bibinfo{author}{\bibfnamefont{I.~E.} \bibnamefont{Dzyaloshinski}},
  \bibinfo{year}{1963}, \emph{\bibinfo{title}{Methods of Quantum Field Theory
  in Statistical Physics}} (\bibinfo{publisher}{Dover, New York}).

\bibitem[{\citenamefont{Alford}(2001)}]{Alford:2001dt}
\bibinfo{author}{\bibnamefont{Alford}, \bibfnamefont{M.~G.}},
  \bibinfo{year}{2001}, \bibinfo{journal}{Ann. Rev. Nucl. Part. Sci.}
  \textbf{\bibinfo{volume}{51}}, \bibinfo{pages}{131}.

\bibitem[{\citenamefont{Alford} \emph{et~al.}(2000)\citenamefont{Alford,
  Berges, and Rajagopal}}]{Alford:1999pb}
\bibinfo{author}{\bibnamefont{Alford}, \bibfnamefont{M.~G.}},
  \bibinfo{author}{\bibfnamefont{J.}~\bibnamefont{Berges}}, and
  \bibinfo{author}{\bibfnamefont{K.}~\bibnamefont{Rajagopal}},
  \bibinfo{year}{2000}, \bibinfo{journal}{Nucl. Phys.}
  \textbf{\bibinfo{volume}{B571}}, \bibinfo{pages}{269}.

\bibitem[{\citenamefont{Alford} \emph{et~al.}(2003)\citenamefont{Alford,
  Bowers, Cheyne, and Cowan}}]{Alford:2002rz}
\bibinfo{author}{\bibnamefont{Alford}, \bibfnamefont{M.~G.}},
  \bibinfo{author}{\bibfnamefont{J.~A.} \bibnamefont{Bowers}},
  \bibinfo{author}{\bibfnamefont{J.~M.} \bibnamefont{Cheyne}}, and
  \bibinfo{author}{\bibfnamefont{G.~A.} \bibnamefont{Cowan}},
  \bibinfo{year}{2003}, \bibinfo{journal}{Phys. Rev.}
  \textbf{\bibinfo{volume}{D67}}, \bibinfo{pages}{054018}.

\bibitem[{\citenamefont{Alford} \emph{et~al.}(2001)\citenamefont{Alford,
  Bowers, and Rajagopal}}]{Alford:2000ze}
\bibinfo{author}{\bibnamefont{Alford}, \bibfnamefont{M.~G.}},
  \bibinfo{author}{\bibfnamefont{J.~A.} \bibnamefont{Bowers}}, and
  \bibinfo{author}{\bibfnamefont{K.}~\bibnamefont{Rajagopal}},
  \bibinfo{year}{2001}, \bibinfo{journal}{Phys. Rev.}
  \textbf{\bibinfo{volume}{D63}}, \bibinfo{pages}{074016}.

\bibitem[{\citenamefont{Alford} \emph{et~al.}(1998)\citenamefont{Alford,
  Rajagopal, and Wilczek}}]{Alford:1998zt}
\bibinfo{author}{\bibnamefont{Alford}, \bibfnamefont{M.~G.}},
  \bibinfo{author}{\bibfnamefont{K.}~\bibnamefont{Rajagopal}}, and
  \bibinfo{author}{\bibfnamefont{F.}~\bibnamefont{Wilczek}},
  \bibinfo{year}{1998}, \bibinfo{journal}{Phys. Lett.}
  \textbf{\bibinfo{volume}{B422}}, \bibinfo{pages}{247}.

\bibitem[{\citenamefont{Alford} \emph{et~al.}(1999)\citenamefont{Alford,
  Rajagopal, and Wilczek}}]{Alford:1998mk}
\bibinfo{author}{\bibnamefont{Alford}, \bibfnamefont{M.~G.}},
  \bibinfo{author}{\bibfnamefont{K.}~\bibnamefont{Rajagopal}}, and
  \bibinfo{author}{\bibfnamefont{F.}~\bibnamefont{Wilczek}},
  \bibinfo{year}{1999}, \bibinfo{journal}{Nucl. Phys.}
  \textbf{\bibinfo{volume}{B537}}, \bibinfo{pages}{443}.

\bibitem[{\citenamefont{Andersson}(1998)}]{Anderson:1998ab}
\bibinfo{author}{\bibnamefont{Andersson}, \bibfnamefont{N.}},
  \bibinfo{year}{1998}, \bibinfo{journal}{Astrophys. J.}
  \textbf{\bibinfo{volume}{502}}, \bibinfo{pages}{708}.

\bibitem[{\citenamefont{Andersson} \emph{et~al.}(2000)\citenamefont{Andersson,
  Jones, Kokkotas, and Stergioulas}}]{Andersson:2000pt}
\bibinfo{author}{\bibnamefont{Andersson}, \bibfnamefont{N.}},
  \bibinfo{author}{\bibfnamefont{D.~I.} \bibnamefont{Jones}},
  \bibinfo{author}{\bibfnamefont{K.~D.} \bibnamefont{Kokkotas}}, and
  \bibinfo{author}{\bibfnamefont{N.}~\bibnamefont{Stergioulas}},
  \bibinfo{year}{2000}, \bibinfo{journal}{Astrophys. J.}
  \textbf{\bibinfo{volume}{534}}, \bibinfo{pages}{L75}.

\bibitem[{\citenamefont{Baade and Zwicky}(1934)}]{baade}
\bibinfo{author}{\bibnamefont{Baade}, \bibfnamefont{W.}}, and
  \bibinfo{author}{\bibfnamefont{F.}~\bibnamefont{Zwicky}},
  \bibinfo{year}{1934}, \bibinfo{journal}{Proc. Nat. Acad. Sci.}
  \textbf{\bibinfo{volume}{20}}.

\bibitem[{\citenamefont{Bailin and Love}(1984)}]{Bailin:1984bm}
\bibinfo{author}{\bibnamefont{Bailin}, \bibfnamefont{D.}}, and
  \bibinfo{author}{\bibfnamefont{A.}~\bibnamefont{Love}}, \bibinfo{year}{1984},
  \bibinfo{journal}{Phys. Rept.} \textbf{\bibinfo{volume}{107}},
  \bibinfo{pages}{325}.

\bibitem[{\citenamefont{Banks and Rabinovici}(1979)}]{Banks:1979ab}
\bibinfo{author}{\bibnamefont{Banks}, \bibfnamefont{T.}}, and
  \bibinfo{author}{\bibfnamefont{E.}~\bibnamefont{Rabinovici}},
  \bibinfo{year}{1979}, \bibinfo{journal}{Nucl. Phys.}
  \textbf{\bibinfo{volume}{B160}}, \bibinfo{pages}{349}.

\bibitem[{\citenamefont{Bardeen}(1956)}]{Bardeen:1956ab}
\bibinfo{author}{\bibnamefont{Bardeen}, \bibfnamefont{J.}},
  \bibinfo{year}{1956}, \emph{\bibinfo{title}{{\rm in} Handbuch der Physik. vol
  XV}} (\bibinfo{publisher}{Springer Verlag, Berlin}).

\bibitem[{\citenamefont{Bardeen} \emph{et~al.}(1957)\citenamefont{Bardeen,
  Cooper, and Schrieffer}}]{Bardeen:1957kj}
\bibinfo{author}{\bibnamefont{Bardeen}, \bibfnamefont{J.}},
  \bibinfo{author}{\bibfnamefont{L.~N.} \bibnamefont{Cooper}}, and
  \bibinfo{author}{\bibfnamefont{J.~R.} \bibnamefont{Schrieffer}},
  \bibinfo{year}{1957}, \bibinfo{journal}{Phys. Rev.}
  \textbf{\bibinfo{volume}{106}}, \bibinfo{pages}{162}.

\bibitem[{\citenamefont{Barrois}(1977)}]{Barrois:1977xd}
\bibinfo{author}{\bibnamefont{Barrois}, \bibfnamefont{B.~C.}},
  \bibinfo{year}{1977}, \bibinfo{journal}{Nucl. Phys.}
  \textbf{\bibinfo{volume}{B129}}, \bibinfo{pages}{390}.

\bibitem[{\citenamefont{Beane} \emph{et~al.}(2000)\citenamefont{Beane, Bedaque,
  and Savage}}]{Beane:2000ms}
\bibinfo{author}{\bibnamefont{Beane}, \bibfnamefont{S.~R.}},
  \bibinfo{author}{\bibfnamefont{P.~F.} \bibnamefont{Bedaque}}, and
  \bibinfo{author}{\bibfnamefont{M.~J.} \bibnamefont{Savage}},
  \bibinfo{year}{2000}, \bibinfo{journal}{Phys. Lett.}
  \textbf{\bibinfo{volume}{B483}}, \bibinfo{pages}{131}.

\bibitem[{\citenamefont{Bedaque and Schafer}(2002)}]{Bedaque:2001je}
\bibinfo{author}{\bibnamefont{Bedaque}, \bibfnamefont{P.~F.}}, and
  \bibinfo{author}{\bibfnamefont{T.}~\bibnamefont{Schafer}},
  \bibinfo{year}{2002}, \bibinfo{journal}{Nucl. Phys.}
  \textbf{\bibinfo{volume}{A697}}, \bibinfo{pages}{802}.

\bibitem[{\citenamefont{Benfatto and Gallavotti}(1990)}]{benfatto:1990ab}
\bibinfo{author}{\bibnamefont{Benfatto}, \bibfnamefont{G.}}, and
  \bibinfo{author}{\bibfnamefont{G.}~\bibnamefont{Gallavotti}},
  \bibinfo{year}{1990}, \bibinfo{journal}{Phys. Rev.}
  \textbf{\bibinfo{volume}{B42}}, \bibinfo{pages}{9967}.

\bibitem[{\citenamefont{Bildsten and Ushomirsky}(1999)}]{Bildsten:1999zn}
\bibinfo{author}{\bibnamefont{Bildsten}, \bibfnamefont{L.}}, and
  \bibinfo{author}{\bibfnamefont{G.}~\bibnamefont{Ushomirsky}},
  \bibinfo{year}{1999}, \eprint{astro-ph/9911155}.

\bibitem[{\citenamefont{Blaschke} \emph{et~al.}(2000)\citenamefont{Blaschke,
  Klahn, and Voskresensky}}]{Blaschke:1999qx}
\bibinfo{author}{\bibnamefont{Blaschke}, \bibfnamefont{D.}},
  \bibinfo{author}{\bibfnamefont{T.}~\bibnamefont{Klahn}}, and
  \bibinfo{author}{\bibfnamefont{D.~N.} \bibnamefont{Voskresensky}},
  \bibinfo{year}{2000}, \bibinfo{journal}{Astrophys. J.}
  \textbf{\bibinfo{volume}{533}}, \bibinfo{pages}{406}.

\bibitem[{\citenamefont{Bodmer}(1971)}]{Bodmer:1971}
\bibinfo{author}{\bibnamefont{Bodmer}, \bibfnamefont{A.~R.}},
  \bibinfo{year}{1971}, \bibinfo{journal}{Phys. Rev.}
  \textbf{\bibinfo{volume}{D4}}, \bibinfo{pages}{1601}.

\bibitem[{\citenamefont{Bowers}(2003)}]{Bowers:2003ye}
\bibinfo{author}{\bibnamefont{Bowers}, \bibfnamefont{J.~A.}},
  \bibinfo{year}{2003}, \eprint{hep-ph/0305301}.

\bibitem[{\citenamefont{Bowers and Rajagopal}(2002)}]{Bowers:2002xr}
\bibinfo{author}{\bibnamefont{Bowers}, \bibfnamefont{J.~A.}}, and
  \bibinfo{author}{\bibfnamefont{K.}~\bibnamefont{Rajagopal}},
  \bibinfo{year}{2002}, \bibinfo{journal}{Phys. Rev.}
  \textbf{\bibinfo{volume}{D66}}, \bibinfo{pages}{065002}.

\bibitem[{\citenamefont{Brown}
  \emph{et~al.}(2000{\natexlab{a}})\citenamefont{Brown, Liu, and
  Ren}}]{Brown:2000eh}
\bibinfo{author}{\bibnamefont{Brown}, \bibfnamefont{W.~E.}},
  \bibinfo{author}{\bibfnamefont{J.~T.} \bibnamefont{Liu}}, and
  \bibinfo{author}{\bibfnamefont{H.-c.} \bibnamefont{Ren}},
  \bibinfo{year}{2000}{\natexlab{a}}, \bibinfo{journal}{Phys. Rev.}
  \textbf{\bibinfo{volume}{D62}}, \bibinfo{pages}{054013}.

\bibitem[{\citenamefont{Brown}
  \emph{et~al.}(2000{\natexlab{b}})\citenamefont{Brown, Liu, and
  Ren}}]{Brown:1999aq}
\bibinfo{author}{\bibnamefont{Brown}, \bibfnamefont{W.~E.}},
  \bibinfo{author}{\bibfnamefont{J.~T.} \bibnamefont{Liu}}, and
  \bibinfo{author}{\bibfnamefont{H.-c.} \bibnamefont{Ren}},
  \bibinfo{year}{2000}{\natexlab{b}}, \bibinfo{journal}{Phys. Rev.}
  \textbf{\bibinfo{volume}{D61}}, \bibinfo{pages}{114012}.

\bibitem[{\citenamefont{Brown}
  \emph{et~al.}(2000{\natexlab{c}})\citenamefont{Brown, Liu, and
  Ren}}]{Brown:1999yd}
\bibinfo{author}{\bibnamefont{Brown}, \bibfnamefont{W.~E.}},
  \bibinfo{author}{\bibfnamefont{J.~T.} \bibnamefont{Liu}}, and
  \bibinfo{author}{\bibfnamefont{H.-c.} \bibnamefont{Ren}},
  \bibinfo{year}{2000}{\natexlab{c}}, \bibinfo{journal}{Phys. Rev.}
  \textbf{\bibinfo{volume}{D62}}, \bibinfo{pages}{054016}.

\bibitem[{\citenamefont{Buckley and Zhitnitsky}(2002)}]{Buckley:2002ur}
\bibinfo{author}{\bibnamefont{Buckley}, \bibfnamefont{K.~B.~W.}}, and
  \bibinfo{author}{\bibfnamefont{A.~R.} \bibnamefont{Zhitnitsky}},
  \bibinfo{year}{2002}, \bibinfo{journal}{JHEP} \textbf{\bibinfo{volume}{08}},
  \bibinfo{pages}{013}.

\bibitem[{\citenamefont{Burrows and Lattimer}(1986)}]{Burrows:1986}
\bibinfo{author}{\bibnamefont{Burrows}, \bibfnamefont{A.}}, and
  \bibinfo{author}{\bibfnamefont{J.~M.} \bibnamefont{Lattimer}},
  \bibinfo{year}{1986}, \bibinfo{journal}{Astrophys. J.}
  \textbf{\bibinfo{volume}{307}}.

\bibitem[{\citenamefont{Buzdin and Kachkachi}(1997)}]{buzdin:1997ab}
\bibinfo{author}{\bibnamefont{Buzdin}, \bibfnamefont{A.~I.}}, and
  \bibinfo{author}{\bibnamefont{Kachkachi}}, \bibinfo{year}{1997},
  \bibinfo{journal}{Phys. Lett.} \textbf{\bibinfo{volume}{A225}},
  \bibinfo{pages}{341}.

\bibitem[{\citenamefont{Cameron and Canuto}(1974)}]{Cameron:1974ab}
\bibinfo{author}{\bibnamefont{Cameron}, \bibfnamefont{A.~G.~W.}}, and
  \bibinfo{author}{\bibfnamefont{V.}~\bibnamefont{Canuto}},
  \bibinfo{year}{1974}, \emph{\bibinfo{title}{{\rm in} Proceedings 16th Solvay
  Conference on Astrophysics and Gravitation}}
  (\bibinfo{publisher}{Universit\'e Libre de Bruxelles}).

\bibitem[{\citenamefont{Carter and Reddy}(2000)}]{Carter:2000xf}
\bibinfo{author}{\bibnamefont{Carter}, \bibfnamefont{G.~W.}}, and
  \bibinfo{author}{\bibfnamefont{S.}~\bibnamefont{Reddy}},
  \bibinfo{year}{2000}, \bibinfo{journal}{Phys. Rev.}
  \textbf{\bibinfo{volume}{D62}}, \bibinfo{pages}{103002}.

\bibitem[{\citenamefont{Casalbuoni}
  \emph{et~al.}(2002{\natexlab{a}})\citenamefont{Casalbuoni, De~Fazio, Gatto,
  Nardulli, and Ruggieri}}]{Casalbuoni:zx}
\bibinfo{author}{\bibnamefont{Casalbuoni}, \bibfnamefont{R.}},
  \bibinfo{author}{\bibfnamefont{F.}~\bibnamefont{De~Fazio}},
  \bibinfo{author}{\bibfnamefont{R.}~\bibnamefont{Gatto}},
  \bibinfo{author}{\bibfnamefont{G.}~\bibnamefont{Nardulli}}, and
  \bibinfo{author}{\bibfnamefont{M.}~\bibnamefont{Ruggieri}},
  \bibinfo{year}{2002}{\natexlab{a}}, \bibinfo{journal}{Phys. Lett.}
  \textbf{\bibinfo{volume}{B547}}, \bibinfo{pages}{229}.

\bibitem[{\citenamefont{Casalbuoni}
  \emph{et~al.}(2000)\citenamefont{Casalbuoni, Duan, and
  Sannino}}]{Casalbuoni:2000cn}
\bibinfo{author}{\bibnamefont{Casalbuoni}, \bibfnamefont{R.}},
  \bibinfo{author}{\bibfnamefont{Z.-y.} \bibnamefont{Duan}}, and
  \bibinfo{author}{\bibfnamefont{F.}~\bibnamefont{Sannino}},
  \bibinfo{year}{2000}, \bibinfo{journal}{Phys. Rev.}
  \textbf{\bibinfo{volume}{D62}}, \bibinfo{pages}{094004}.

\bibitem[{\citenamefont{Casalbuoni}
  \emph{et~al.}(2001{\natexlab{a}})\citenamefont{Casalbuoni, Duan, and
  Sannino}}]{Casalbuoni:2000jn}
\bibinfo{author}{\bibnamefont{Casalbuoni}, \bibfnamefont{R.}},
  \bibinfo{author}{\bibfnamefont{Z.-y.} \bibnamefont{Duan}}, and
  \bibinfo{author}{\bibfnamefont{F.}~\bibnamefont{Sannino}},
  \bibinfo{year}{2001}{\natexlab{a}}, \bibinfo{journal}{Phys. Rev.}
  \textbf{\bibinfo{volume}{D63}}, \bibinfo{pages}{114026}.

\bibitem[{\citenamefont{Casalbuoni}
  \emph{et~al.}(2002{\natexlab{b}})\citenamefont{Casalbuoni, Fabiano, Gatto,
  Mannarelli, and Nardulli}}]{Casalbuoni:2002my}
\bibinfo{author}{\bibnamefont{Casalbuoni}, \bibfnamefont{R.}},
  \bibinfo{author}{\bibfnamefont{E.}~\bibnamefont{Fabiano}},
  \bibinfo{author}{\bibfnamefont{R.}~\bibnamefont{Gatto}},
  \bibinfo{author}{\bibfnamefont{M.}~\bibnamefont{Mannarelli}}, and
  \bibinfo{author}{\bibfnamefont{G.}~\bibnamefont{Nardulli}},
  \bibinfo{year}{2002}{\natexlab{b}}, \bibinfo{journal}{Phys. Rev.}
  \textbf{\bibinfo{volume}{D66}}, \bibinfo{pages}{094006}.

\bibitem[{\citenamefont{Casalbuoni and Gatto}(1981)}]{Casalbuoni:1981nd}
\bibinfo{author}{\bibnamefont{Casalbuoni}, \bibfnamefont{R.}}, and
  \bibinfo{author}{\bibfnamefont{R.}~\bibnamefont{Gatto}},
  \bibinfo{year}{1981}, \bibinfo{journal}{Phys. Lett.}
  \textbf{\bibinfo{volume}{B103}}, \bibinfo{pages}{113}.

\bibitem[{\citenamefont{Casalbuoni and Gatto}(1999)}]{Casalbuoni:1999wu}
\bibinfo{author}{\bibnamefont{Casalbuoni}, \bibfnamefont{R.}}, and
  \bibinfo{author}{\bibfnamefont{R.}~\bibnamefont{Gatto}},
  \bibinfo{year}{1999}, \bibinfo{journal}{Phys. Lett.}
  \textbf{\bibinfo{volume}{B464}}, \bibinfo{pages}{111}.

\bibitem[{\citenamefont{Casalbuoni}
  \emph{et~al.}(2001{\natexlab{b}})\citenamefont{Casalbuoni, Gatto, Mannarelli,
  and Nardulli}}]{Casalbuoni:2001gt}
\bibinfo{author}{\bibnamefont{Casalbuoni}, \bibfnamefont{R.}},
  \bibinfo{author}{\bibfnamefont{R.}~\bibnamefont{Gatto}},
  \bibinfo{author}{\bibfnamefont{M.}~\bibnamefont{Mannarelli}}, and
  \bibinfo{author}{\bibfnamefont{G.}~\bibnamefont{Nardulli}},
  \bibinfo{year}{2001}{\natexlab{b}}, \bibinfo{journal}{Phys. Lett.}
  \textbf{\bibinfo{volume}{B511}}, \bibinfo{pages}{218}.

\bibitem[{\citenamefont{Casalbuoni}
  \emph{et~al.}(2002{\natexlab{c}})\citenamefont{Casalbuoni, Gatto, Mannarelli,
  and Nardulli}}]{Casalbuoni:2002pa}
\bibinfo{author}{\bibnamefont{Casalbuoni}, \bibfnamefont{R.}},
  \bibinfo{author}{\bibfnamefont{R.}~\bibnamefont{Gatto}},
  \bibinfo{author}{\bibfnamefont{M.}~\bibnamefont{Mannarelli}}, and
  \bibinfo{author}{\bibfnamefont{G.}~\bibnamefont{Nardulli}},
  \bibinfo{year}{2002}{\natexlab{c}}, \bibinfo{journal}{Phys. Rev.}
  \textbf{\bibinfo{volume}{D66}}, \bibinfo{pages}{014006}.

\bibitem[{\citenamefont{Casalbuoni}
  \emph{et~al.}(2002{\natexlab{d}})\citenamefont{Casalbuoni, Gatto, Mannarelli,
  and Nardulli}}]{Casalbuoni:2001ha}
\bibinfo{author}{\bibnamefont{Casalbuoni}, \bibfnamefont{R.}},
  \bibinfo{author}{\bibfnamefont{R.}~\bibnamefont{Gatto}},
  \bibinfo{author}{\bibfnamefont{M.}~\bibnamefont{Mannarelli}}, and
  \bibinfo{author}{\bibfnamefont{G.}~\bibnamefont{Nardulli}},
  \bibinfo{year}{2002}{\natexlab{d}}, \bibinfo{journal}{Phys. Lett.}
  \textbf{\bibinfo{volume}{B524}}, \bibinfo{pages}{144}.

\bibitem[{\citenamefont{Casalbuoni}
  \emph{et~al.}(2001{\natexlab{c}})\citenamefont{Casalbuoni, Gatto, and
  Nardulli}}]{Casalbuoni:2000na}
\bibinfo{author}{\bibnamefont{Casalbuoni}, \bibfnamefont{R.}},
  \bibinfo{author}{\bibfnamefont{R.}~\bibnamefont{Gatto}}, and
  \bibinfo{author}{\bibfnamefont{G.}~\bibnamefont{Nardulli}},
  \bibinfo{year}{2001}{\natexlab{c}}, \bibinfo{journal}{Phys. Lett.}
  \textbf{\bibinfo{volume}{B498}}, \bibinfo{pages}{179}.

\bibitem[{\citenamefont{Casalbuoni}
  \emph{et~al.}(2001{\natexlab{d}})\citenamefont{Casalbuoni, Gatto, and
  Nardulli}}]{Casalbuoni:2000nb}
\bibinfo{author}{\bibnamefont{Casalbuoni}, \bibfnamefont{R.}},
  \bibinfo{author}{\bibfnamefont{R.}~\bibnamefont{Gatto}}, and
  \bibinfo{author}{\bibfnamefont{G.}~\bibnamefont{Nardulli}},
  \bibinfo{year}{2001}{\natexlab{d}}, \bibinfo{journal}{Phys. Lett. (B498
  Erratum)} \textbf{\bibinfo{volume}{B517}}, \bibinfo{pages}{483}.

\bibitem[{\citenamefont{Casalbuoni}
  \emph{et~al.}(2002{\natexlab{e}})\citenamefont{Casalbuoni, Gatto, and
  Nardulli}}]{Casalbuoni:2002hr}
\bibinfo{author}{\bibnamefont{Casalbuoni}, \bibfnamefont{R.}},
  \bibinfo{author}{\bibfnamefont{R.}~\bibnamefont{Gatto}}, and
  \bibinfo{author}{\bibfnamefont{G.}~\bibnamefont{Nardulli}},
  \bibinfo{year}{2002}{\natexlab{e}}, \bibinfo{journal}{Phys. Lett.}
  \textbf{\bibinfo{volume}{B543}}, \bibinfo{pages}{139}.

\bibitem[{\citenamefont{Casalbuoni}
  \emph{et~al.}(2003)\citenamefont{Casalbuoni, Gatto, Nardulli, and
  Ruggieri}}]{Casalbuoni:2003cs}
\bibinfo{author}{\bibnamefont{Casalbuoni}, \bibfnamefont{R.}},
  \bibinfo{author}{\bibfnamefont{R.}~\bibnamefont{Gatto}},
  \bibinfo{author}{\bibfnamefont{G.}~\bibnamefont{Nardulli}}, and
  \bibinfo{author}{\bibfnamefont{M.}~\bibnamefont{Ruggieri}},
  \bibinfo{year}{2003}, \eprint{hep-ph/0302077}.

\bibitem[{\citenamefont{Casalbuoni and Nardulli}(2003)}]{Casalbuoni:2003ab}
\bibinfo{author}{\bibnamefont{Casalbuoni}, \bibfnamefont{R.}}, and
  \bibinfo{author}{\bibfnamefont{G.}~\bibnamefont{Nardulli}},
  \bibinfo{year}{2003}, \eprint{hep-ph/0305069}.

\bibitem[{\citenamefont{Chandrasekhar}(1962)}]{chandrasekhar}
\bibinfo{author}{\bibnamefont{Chandrasekhar}, \bibfnamefont{B.~S.}},
  \bibinfo{year}{1962}, \bibinfo{journal}{App. Phys. Lett.}
  \textbf{\bibinfo{volume}{1}}, \bibinfo{pages}{7}.

\bibitem[{\citenamefont{Cini and Touschek}(1958)}]{Cini:1958ab}
\bibinfo{author}{\bibnamefont{Cini}, \bibfnamefont{M.}}, and
  \bibinfo{author}{\bibfnamefont{B.}~\bibnamefont{Touschek}},
  \bibinfo{year}{1958}, \bibinfo{journal}{Nuovo Cimento}
  \textbf{\bibinfo{volume}{7}}, \bibinfo{pages}{422}.

\bibitem[{\citenamefont{Clogston}(1962)}]{clogston}
\bibinfo{author}{\bibnamefont{Clogston}, \bibfnamefont{A.~M.}},
  \bibinfo{year}{1962}, \bibinfo{journal}{Phys. Rev. Lett.}
  \textbf{\bibinfo{volume}{9}}, \bibinfo{pages}{266}.

\bibitem[{\citenamefont{Colgate and White}(1966)}]{Colgate:1966ab}
\bibinfo{author}{\bibnamefont{Colgate}, \bibfnamefont{S.~A.}}, and
  \bibinfo{author}{\bibfnamefont{R.~H.} \bibnamefont{White}},
  \bibinfo{year}{1966}, \bibinfo{journal}{Astrophys. J.}
  \textbf{\bibinfo{volume}{143}}.

\bibitem[{\citenamefont{Collins and Perry}(1975)}]{collins:1975ab}
\bibinfo{author}{\bibnamefont{Collins}, \bibfnamefont{J.~C.}}, and
  \bibinfo{author}{\bibfnamefont{M.~J.} \bibnamefont{Perry}},
  \bibinfo{year}{1975}, \bibinfo{journal}{Phys. Rev. Lett.}
  \textbf{\bibinfo{volume}{34}}, \bibinfo{pages}{1353}.

\bibitem[{\citenamefont{Combescot and Mora}(2002)}]{combescot:2002ab}
\bibinfo{author}{\bibnamefont{Combescot}, \bibfnamefont{R.}}, and
  \bibinfo{author}{\bibfnamefont{C.}~\bibnamefont{Mora}}, \bibinfo{year}{2002},
  \bibinfo{journal}{Eur. P. J.} \textbf{\bibinfo{volume}{B28}},
  \bibinfo{pages}{397}.

\bibitem[{\citenamefont{Cooper}(1956)}]{cooper:1956fz}
\bibinfo{author}{\bibnamefont{Cooper}, \bibfnamefont{L.~N.}},
  \bibinfo{year}{1956}, \bibinfo{journal}{Phys. Rev.}
  \textbf{\bibinfo{volume}{104}}, \bibinfo{pages}{1189}.

\bibitem[{\citenamefont{Corak} \emph{et~al.}(1954)\citenamefont{Corak, Goodman,
  Satterthwaite, and Wexler}}]{Corak:1954ab}
\bibinfo{author}{\bibnamefont{Corak}, \bibfnamefont{W.~S.}},
  \bibinfo{author}{\bibfnamefont{B.~B.} \bibnamefont{Goodman}},
  \bibinfo{author}{\bibfnamefont{C.~B.} \bibnamefont{Satterthwaite}}, and
  \bibinfo{author}{\bibfnamefont{A.}~\bibnamefont{Wexler}},
  \bibinfo{year}{1954}, \bibinfo{journal}{Phys. Rev.}
  \textbf{\bibinfo{volume}{96}}, \bibinfo{pages}{1442}.

\bibitem[{\citenamefont{Corak} \emph{et~al.}(1956)\citenamefont{Corak, Goodman,
  Satterthwaite, and Wexler}}]{Corak:1956ab}
\bibinfo{author}{\bibnamefont{Corak}, \bibfnamefont{W.~S.}},
  \bibinfo{author}{\bibfnamefont{B.~B.} \bibnamefont{Goodman}},
  \bibinfo{author}{\bibfnamefont{C.~B.} \bibnamefont{Satterthwaite}}, and
  \bibinfo{author}{\bibfnamefont{A.}~\bibnamefont{Wexler}},
  \bibinfo{year}{1956}, \bibinfo{journal}{Phys. Rev.}
  \textbf{\bibinfo{volume}{102}}, \bibinfo{pages}{656}.

\bibitem[{\citenamefont{Daunt and Mendelssohn}(1946)}]{Daunt:1946ab}
\bibinfo{author}{\bibnamefont{Daunt}, \bibfnamefont{J.~G.}}, and
  \bibinfo{author}{\bibfnamefont{K.}~\bibnamefont{Mendelssohn}},
  \bibinfo{year}{1946}, \bibinfo{journal}{Proc. Roy. Soc.}
  \textbf{\bibinfo{volume}{A185}}, \bibinfo{pages}{225}.

\bibitem[{\citenamefont{Dimopoulos}
  \emph{et~al.}(1980{\natexlab{a}})\citenamefont{Dimopoulos, Raby, and
  Susskind}}]{Dimopoulos:1980ab}
\bibinfo{author}{\bibnamefont{Dimopoulos}, \bibfnamefont{S.}},
  \bibinfo{author}{\bibfnamefont{S.}~\bibnamefont{Raby}}, and
  \bibinfo{author}{\bibfnamefont{L.}~\bibnamefont{Susskind}},
  \bibinfo{year}{1980}{\natexlab{a}}, \bibinfo{journal}{Nucl. Phys.}
  \textbf{\bibinfo{volume}{B169}}, \bibinfo{pages}{373}.

\bibitem[{\citenamefont{Dimopoulos}
  \emph{et~al.}(1980{\natexlab{b}})\citenamefont{Dimopoulos, Raby, and
  Susskind}}]{Dimopoulos:1980cd}
\bibinfo{author}{\bibnamefont{Dimopoulos}, \bibfnamefont{S.}},
  \bibinfo{author}{\bibfnamefont{S.}~\bibnamefont{Raby}}, and
  \bibinfo{author}{\bibfnamefont{L.}~\bibnamefont{Susskind}},
  \bibinfo{year}{1980}{\natexlab{b}}, \bibinfo{journal}{Nucl. Phys.}
  \textbf{\bibinfo{volume}{B173}}, \bibinfo{pages}{208}.

\bibitem[{\citenamefont{Eguchi}(1976)}]{Eguchi:1976iz}
\bibinfo{author}{\bibnamefont{Eguchi}, \bibfnamefont{T.}},
  \bibinfo{year}{1976}, \bibinfo{journal}{Phys. Rev.}
  \textbf{\bibinfo{volume}{D14}}, \bibinfo{pages}{2755}.

\bibitem[{\citenamefont{Eliashberg}(1960)}]{Eliashberg:1960ab}
\bibinfo{author}{\bibnamefont{Eliashberg}, \bibfnamefont{G.~M.}},
  \bibinfo{year}{1960}, \bibinfo{journal}{Sov. Phys. JETP}
  \textbf{\bibinfo{volume}{11}}, \bibinfo{pages}{696}.

\bibitem[{\citenamefont{Evans} \emph{et~al.}(2000)\citenamefont{Evans,
  Hormuzdiar, Hsu, and Schwetz}}]{Evans:1999at}
\bibinfo{author}{\bibnamefont{Evans}, \bibfnamefont{N.}},
  \bibinfo{author}{\bibfnamefont{J.}~\bibnamefont{Hormuzdiar}},
  \bibinfo{author}{\bibfnamefont{S.~D.~H.} \bibnamefont{Hsu}}, and
  \bibinfo{author}{\bibfnamefont{M.}~\bibnamefont{Schwetz}},
  \bibinfo{year}{2000}, \bibinfo{journal}{Nucl. Phys.}
  \textbf{\bibinfo{volume}{B581}}, \bibinfo{pages}{391}.

\bibitem[{\citenamefont{Evans}
  \emph{et~al.}(1999{\natexlab{a}})\citenamefont{Evans, Hsu, and
  Schwetz}}]{Evans:1998ek}
\bibinfo{author}{\bibnamefont{Evans}, \bibfnamefont{N.}},
  \bibinfo{author}{\bibfnamefont{S.~D.~H.} \bibnamefont{Hsu}}, and
  \bibinfo{author}{\bibfnamefont{M.}~\bibnamefont{Schwetz}},
  \bibinfo{year}{1999}{\natexlab{a}}, \bibinfo{journal}{Nucl. Phys.}
  \textbf{\bibinfo{volume}{B551}}, \bibinfo{pages}{275}.

\bibitem[{\citenamefont{Evans}
  \emph{et~al.}(1999{\natexlab{b}})\citenamefont{Evans, Hsu, and
  Schwetz}}]{Evans:1998nf}
\bibinfo{author}{\bibnamefont{Evans}, \bibfnamefont{N.}},
  \bibinfo{author}{\bibfnamefont{S.~D.~H.} \bibnamefont{Hsu}}, and
  \bibinfo{author}{\bibfnamefont{M.}~\bibnamefont{Schwetz}},
  \bibinfo{year}{1999}{\natexlab{b}}, \bibinfo{journal}{Phys. Lett.}
  \textbf{\bibinfo{volume}{B449}}, \bibinfo{pages}{281}.

\bibitem[{\citenamefont{Fradkin and Shenker}(1979)}]{Fradkin:1979ab}
\bibinfo{author}{\bibnamefont{Fradkin}, \bibfnamefont{E.}}, and
  \bibinfo{author}{\bibfnamefont{S.~H.} \bibnamefont{Shenker}},
  \bibinfo{year}{1979}, \bibinfo{journal}{Phys. Rev.}
  \textbf{\bibinfo{volume}{D19}}, \bibinfo{pages}{3682}.

\bibitem[{\citenamefont{Frautschi}(1978)}]{Frautschi:1978rz}
\bibinfo{author}{\bibnamefont{Frautschi}, \bibfnamefont{S.~C.}},
  \bibinfo{year}{1978}, \bibinfo{note}{presented at Workshop on Hadronic Matter
  at Extreme Energy Density, Erice, Italy, Oct 13-21, 1978}.

\bibitem[{\citenamefont{Friedman and Caldwell}(1991)}]{Friedman:1991qz}
\bibinfo{author}{\bibnamefont{Friedman}, \bibfnamefont{J.~L.}}, and
  \bibinfo{author}{\bibfnamefont{R.~R.} \bibnamefont{Caldwell}},
  \bibinfo{year}{1991}, \bibinfo{journal}{Phys. Lett.}
  \textbf{\bibinfo{volume}{B264}}, \bibinfo{pages}{143}.

\bibitem[{\citenamefont{Frolich}(1950)}]{Frolich:1950ab}
\bibinfo{author}{\bibnamefont{Frolich}, \bibfnamefont{H.}},
  \bibinfo{year}{1950}, \bibinfo{journal}{Phys. Rev.}
  \textbf{\bibinfo{volume}{79}}, \bibinfo{pages}{845}.

\bibitem[{\citenamefont{Frolich}(1952)}]{Frolich:1952ab}
\bibinfo{author}{\bibnamefont{Frolich}, \bibfnamefont{H.}},
  \bibinfo{year}{1952}, \bibinfo{journal}{Proc. Roy. Soc.}
  \textbf{\bibinfo{volume}{A215}}, \bibinfo{pages}{291}.

\bibitem[{\citenamefont{Fulde and Ferrell}(1964)}]{FF}
\bibinfo{author}{\bibnamefont{Fulde}, \bibfnamefont{P.}}, and
  \bibinfo{author}{\bibfnamefont{R.~A.} \bibnamefont{Ferrell}},
  \bibinfo{year}{1964}, \bibinfo{journal}{Phys. Rev.}
  \textbf{\bibinfo{volume}{135}}, \bibinfo{pages}{A550}.

\bibitem[{\citenamefont{de~Gennes}(1989)}]{deGennes}
\bibinfo{author}{\bibnamefont{de~Gennes}, \bibfnamefont{P.~G.}},
  \bibinfo{year}{1989}, \emph{\bibinfo{title}{Superconductivity in Metals and
  Alloys}} (\bibinfo{publisher}{Addison Wesley, Reading, MA}).

\bibitem[{\citenamefont{Ginzburg and Landau}(1950)}]{ginzburg:1950xz}
\bibinfo{author}{\bibnamefont{Ginzburg}, \bibfnamefont{L.}}, and
  \bibinfo{author}{\bibfnamefont{L.~D.} \bibnamefont{Landau}},
  \bibinfo{year}{1950}, \bibinfo{journal}{Zh. Exsp. teor. Fiz.}
  \textbf{\bibinfo{volume}{20}}, \bibinfo{pages}{1064}.

\bibitem[{\citenamefont{Ginzburg}(1953)}]{Ginzburg:1953ab}
\bibinfo{author}{\bibnamefont{Ginzburg}, \bibfnamefont{V.~L.}},
  \bibinfo{year}{1953}, \bibinfo{journal}{Fortschr. Phys.}
  \textbf{\bibinfo{volume}{1}}, \bibinfo{pages}{101}.

\bibitem[{\citenamefont{Ginzburg and Andryushin}(1994)}]{Ginzburg}
\bibinfo{author}{\bibnamefont{Ginzburg}, \bibfnamefont{V.~L.}}, and
  \bibinfo{author}{\bibfnamefont{E.~A.} \bibnamefont{Andryushin}},
  \bibinfo{year}{1994}, \emph{\bibinfo{title}{Superconductivity}}
  (\bibinfo{publisher}{World Scientific}).

\bibitem[{\citenamefont{Glover and Tinkham}(1956)}]{Glover:1956ab}
\bibinfo{author}{\bibnamefont{Glover}, \bibfnamefont{R.~E.}}, and
  \bibinfo{author}{\bibfnamefont{M.}~\bibnamefont{Tinkham}},
  \bibinfo{year}{1956}, \bibinfo{journal}{Phys. Rev.}
  \textbf{\bibinfo{volume}{104}}, \bibinfo{pages}{844}.

\bibitem[{\citenamefont{Gold}(1969)}]{gold}
\bibinfo{author}{\bibnamefont{Gold}, \bibfnamefont{T.}}, \bibinfo{year}{1969},
  \bibinfo{journal}{Nature} \textbf{\bibinfo{volume}{221}},
  \bibinfo{pages}{25}.

\bibitem[{\citenamefont{Gor'kov}(1959)}]{gorkov:1959hy}
\bibinfo{author}{\bibnamefont{Gor'kov}, \bibfnamefont{L.~P.}},
  \bibinfo{year}{1959}, \bibinfo{journal}{Zh. Exsp. teor. Fiz.}
  \textbf{\bibinfo{volume}{36}}, \bibinfo{pages}{1918}.

\bibitem[{\citenamefont{Gorter and
  Casimir}(1934{\natexlab{a}})}]{Gorter:1934ab}
\bibinfo{author}{\bibnamefont{Gorter}, \bibfnamefont{C.~J.}}, and
  \bibinfo{author}{\bibfnamefont{H.~G.~B.} \bibnamefont{Casimir}},
  \bibinfo{year}{1934}{\natexlab{a}}, \bibinfo{journal}{Phys. Z.}
  \textbf{\bibinfo{volume}{35}}, \bibinfo{pages}{963}.

\bibitem[{\citenamefont{Gorter and
  Casimir}(1934{\natexlab{b}})}]{Gorter:1934lm}
\bibinfo{author}{\bibnamefont{Gorter}, \bibfnamefont{C.~J.}}, and
  \bibinfo{author}{\bibfnamefont{H.~G.~B.} \bibnamefont{Casimir}},
  \bibinfo{year}{1934}{\natexlab{b}}, \bibinfo{journal}{Z. Tech. Phys.}
  \textbf{\bibinfo{volume}{15}}, \bibinfo{pages}{539}.

\bibitem[{\citenamefont{Gross and Wilczek}(1973)}]{gross:1973ab}
\bibinfo{author}{\bibnamefont{Gross}, \bibfnamefont{D.}}, and
  \bibinfo{author}{\bibfnamefont{F.}~\bibnamefont{Wilczek}},
  \bibinfo{year}{1973}, \bibinfo{journal}{Phys. Rev. Lett.}
  \textbf{\bibinfo{volume}{30}}, \bibinfo{pages}{1343}.

\bibitem[{\citenamefont{Gusynin and Shovkovy}(2002)}]{Gusynin:2001tt}
\bibinfo{author}{\bibnamefont{Gusynin}, \bibfnamefont{V.~P.}}, and
  \bibinfo{author}{\bibfnamefont{I.~A.} \bibnamefont{Shovkovy}},
  \bibinfo{year}{2002}, \bibinfo{journal}{Nucl. Phys.}
  \textbf{\bibinfo{volume}{A700}}, \bibinfo{pages}{577}.

\bibitem[{\citenamefont{Heiselberg and
  Pandharipande}(2000)}]{Heiselberg:2000dn}
\bibinfo{author}{\bibnamefont{Heiselberg}, \bibfnamefont{H.}}, and
  \bibinfo{author}{\bibfnamefont{V.}~\bibnamefont{Pandharipande}},
  \bibinfo{year}{2000}, \bibinfo{journal}{Ann. Rev. Nucl. Part. Sci.}
  \textbf{\bibinfo{volume}{50}}, \bibinfo{pages}{481}.

\bibitem[{\citenamefont{Hong}(2000{\natexlab{a}})}]{Hong:1999ru}
\bibinfo{author}{\bibnamefont{Hong}, \bibfnamefont{D.~K.}},
  \bibinfo{year}{2000}{\natexlab{a}}, \bibinfo{journal}{Nucl. Phys.}
  \textbf{\bibinfo{volume}{B582}}, \bibinfo{pages}{451}.

\bibitem[{\citenamefont{Hong}(2000{\natexlab{b}})}]{Hong:1998tn}
\bibinfo{author}{\bibnamefont{Hong}, \bibfnamefont{D.~K.}},
  \bibinfo{year}{2000}{\natexlab{b}}, \bibinfo{journal}{Phys. Lett.}
  \textbf{\bibinfo{volume}{B473}}, \bibinfo{pages}{118}.

\bibitem[{\citenamefont{Hong}(2001)}]{Hong:2000ck}
\bibinfo{author}{\bibnamefont{Hong}, \bibfnamefont{D.~K.}},
  \bibinfo{year}{2001}, \bibinfo{journal}{Acta Phys. Polon.}
  \textbf{\bibinfo{volume}{B32}}, \bibinfo{pages}{1253}.

\bibitem[{\citenamefont{Hong}
  \emph{et~al.}(2000{\natexlab{a}})\citenamefont{Hong, Lee, and
  Min}}]{Hong:1999ei}
\bibinfo{author}{\bibnamefont{Hong}, \bibfnamefont{D.~K.}},
  \bibinfo{author}{\bibfnamefont{T.}~\bibnamefont{Lee}}, and
  \bibinfo{author}{\bibfnamefont{D.-P.} \bibnamefont{Min}},
  \bibinfo{year}{2000}{\natexlab{a}}, \bibinfo{journal}{Phys. Lett.}
  \textbf{\bibinfo{volume}{B477}}, \bibinfo{pages}{137}.

\bibitem[{\citenamefont{Hong}
  \emph{et~al.}(2000{\natexlab{b}})\citenamefont{Hong, Miransky, Shovkovy, and
  Wijewardhana}}]{Hong:1999fh}
\bibinfo{author}{\bibnamefont{Hong}, \bibfnamefont{D.~K.}},
  \bibinfo{author}{\bibfnamefont{V.~A.} \bibnamefont{Miransky}},
  \bibinfo{author}{\bibfnamefont{I.~A.} \bibnamefont{Shovkovy}}, and
  \bibinfo{author}{\bibfnamefont{L.~C.~R.} \bibnamefont{Wijewardhana}},
  \bibinfo{year}{2000}{\natexlab{b}}, \bibinfo{journal}{Phys. Rev.}
  \textbf{\bibinfo{volume}{D61}}, \bibinfo{pages}{056001}.

\bibitem[{\citenamefont{Hong} \emph{et~al.}(1999)\citenamefont{Hong, Rho, and
  Zahed}}]{Hong:1999tz}
\bibinfo{author}{\bibnamefont{Hong}, \bibfnamefont{D.~K.}},
  \bibinfo{author}{\bibfnamefont{M.}~\bibnamefont{Rho}}, and
  \bibinfo{author}{\bibfnamefont{I.}~\bibnamefont{Zahed}},
  \bibinfo{year}{1999}, \bibinfo{journal}{Phys. Lett.}
  \textbf{\bibinfo{volume}{B468}}, \bibinfo{pages}{261}.

\bibitem[{\citenamefont{'t~Hooft}(1976)}]{'tHooft:1976up}
\bibinfo{author}{\bibnamefont{'t~Hooft}, \bibfnamefont{G.}},
  \bibinfo{year}{1976}, \bibinfo{journal}{Phys. Rev. Lett.}
  \textbf{\bibinfo{volume}{37}}, \bibinfo{pages}{8}.

\bibitem[{\citenamefont{'t~Hooft}(1980)}]{Hooft:1979ab}
\bibinfo{author}{\bibnamefont{'t~Hooft}, \bibfnamefont{G.}},
  \bibinfo{year}{1980}, \emph{\bibinfo{title}{{\rm in} Recent Developments in
  Gauge Theories}} (\bibinfo{publisher}{Plenum Press}).

\bibitem[{\citenamefont{Hsu}(2000)}]{Hsu:2000sy}
\bibinfo{author}{\bibnamefont{Hsu}, \bibfnamefont{S.~D.~H.}},
  \bibinfo{year}{2000}, \eprint{hep-ph/0003140}.

\bibitem[{\citenamefont{Hsu} \emph{et~al.}(2001)\citenamefont{Hsu, Sannino, and
  Schwetz}}]{Hsu:2000by}
\bibinfo{author}{\bibnamefont{Hsu}, \bibfnamefont{S.~D.~H.}},
  \bibinfo{author}{\bibfnamefont{F.}~\bibnamefont{Sannino}}, and
  \bibinfo{author}{\bibfnamefont{M.}~\bibnamefont{Schwetz}},
  \bibinfo{year}{2001}, \bibinfo{journal}{Mod. Phys. Lett.}
  \textbf{\bibinfo{volume}{A16}}, \bibinfo{pages}{1871}.

\bibitem[{\citenamefont{Imsshennyk and Nadyozhin}(1973)}]{Nadyozhin:1973}
\bibinfo{author}{\bibnamefont{Imsshennyk}, \bibfnamefont{V.~S.}}, and
  \bibinfo{author}{\bibfnamefont{D.~K.} \bibnamefont{Nadyozhin}},
  \bibinfo{year}{1973}, \bibinfo{journal}{Sov. Phys. JETP}
  \textbf{\bibinfo{volume}{36}}.

\bibitem[{\citenamefont{Jackson and Sannino}(2003)}]{Jackson:2003dk}
\bibinfo{author}{\bibnamefont{Jackson}, \bibfnamefont{A.~D.}}, and
  \bibinfo{author}{\bibfnamefont{F.}~\bibnamefont{Sannino}},
  \bibinfo{year}{2003}, \eprint{hep-ph/0308182}.

\bibitem[{\citenamefont{Jaffe}(1977)}]{Jaffe:1977ab}
\bibinfo{author}{\bibnamefont{Jaffe}, \bibfnamefont{R.~L.}},
  \bibinfo{year}{1977}, \bibinfo{journal}{Phys. Rev. Lett.}
  \textbf{\bibinfo{volume}{38}}, \bibinfo{pages}{195, 617(E)}.

\bibitem[{\citenamefont{Jaikumar} \emph{et~al.}(2002)\citenamefont{Jaikumar,
  Prakash, and Schafer}}]{Jaikumar:2002vg}
\bibinfo{author}{\bibnamefont{Jaikumar}, \bibfnamefont{P.}},
  \bibinfo{author}{\bibfnamefont{M.}~\bibnamefont{Prakash}}, and
  \bibinfo{author}{\bibfnamefont{T.}~\bibnamefont{Schafer}},
  \bibinfo{year}{2002}, \bibinfo{journal}{Phys. Rev.}
  \textbf{\bibinfo{volume}{D66}}, \bibinfo{pages}{063003}.

\bibitem[{\citenamefont{Josephson}(1962)}]{Josephson:1962ab}
\bibinfo{author}{\bibnamefont{Josephson}, \bibfnamefont{D.}},
  \bibinfo{year}{1962}, \bibinfo{journal}{Phys. Lett.}
  \textbf{\bibinfo{volume}{1}}, \bibinfo{pages}{251}.

\bibitem[{\citenamefont{Josephson}(1965)}]{Josephson:1965ab}
\bibinfo{author}{\bibnamefont{Josephson}, \bibfnamefont{D.}},
  \bibinfo{year}{1965}, \bibinfo{journal}{Adv. Phys.}
  \textbf{\bibinfo{volume}{14}}, \bibinfo{pages}{419}.

\bibitem[{\citenamefont{Kamerlingh~Onnes}(1911)}]{Kamerlingh:1911ab}
\bibinfo{author}{\bibnamefont{Kamerlingh~Onnes}, \bibfnamefont{H.}},
  \bibinfo{year}{1911}, \bibinfo{journal}{Leiden Comm.}
  \textbf{\bibinfo{volume}{120b, 122b, 124c}}.

\bibitem[{\citenamefont{Kaplan and Reddy}(2002{\natexlab{a}})}]{Kaplan:2001qk}
\bibinfo{author}{\bibnamefont{Kaplan}, \bibfnamefont{D.~B.}}, and
  \bibinfo{author}{\bibfnamefont{S.}~\bibnamefont{Reddy}},
  \bibinfo{year}{2002}{\natexlab{a}}, \bibinfo{journal}{Phys. Rev.}
  \textbf{\bibinfo{volume}{D65}}, \bibinfo{pages}{054042}.

\bibitem[{\citenamefont{Kaplan and Reddy}(2002{\natexlab{b}})}]{Kaplan:2001hh}
\bibinfo{author}{\bibnamefont{Kaplan}, \bibfnamefont{D.~B.}}, and
  \bibinfo{author}{\bibfnamefont{S.}~\bibnamefont{Reddy}},
  \bibinfo{year}{2002}{\natexlab{b}}, \bibinfo{journal}{Phys. Rev. Lett.}
  \textbf{\bibinfo{volume}{88}}, \bibinfo{pages}{132302}.

\bibitem[{\citenamefont{Keil and Janka}(1995)}]{Keil:1995}
\bibinfo{author}{\bibnamefont{Keil}, \bibfnamefont{W.}}, and
  \bibinfo{author}{\bibfnamefont{H.~T.} \bibnamefont{Janka}},
  \bibinfo{year}{1995}, \bibinfo{journal}{Astron. Astrophys.}
  \textbf{\bibinfo{volume}{296}}.

\bibitem[{\citenamefont{Kundu and Rajagopal}(2002)}]{Kundu:2001tt}
\bibinfo{author}{\bibnamefont{Kundu}, \bibfnamefont{J.}}, and
  \bibinfo{author}{\bibfnamefont{K.}~\bibnamefont{Rajagopal}},
  \bibinfo{year}{2002}, \bibinfo{journal}{Phys. Rev.}
  \textbf{\bibinfo{volume}{D65}}, \bibinfo{pages}{094022}.

\bibitem[{\citenamefont{Landau and Lifshitz}(1996)}]{landau1}
\bibinfo{author}{\bibnamefont{Landau}, \bibfnamefont{L.}}, and
  \bibinfo{author}{\bibfnamefont{E.~M.} \bibnamefont{Lifshitz}},
  \bibinfo{year}{1996}, \emph{\bibinfo{title}{Statistical Physics, Part I}}
  (\bibinfo{publisher}{Butterworth, Heinemann}).

\bibitem[{\citenamefont{Landau} \emph{et~al.}(1980)\citenamefont{Landau,
  Lifshitz, and Pitaevskii}}]{Landau2}
\bibinfo{author}{\bibnamefont{Landau}, \bibfnamefont{L.}},
  \bibinfo{author}{\bibfnamefont{E.~M.} \bibnamefont{Lifshitz}}, and
  \bibinfo{author}{\bibfnamefont{L.~P.} \bibnamefont{Pitaevskii}},
  \bibinfo{year}{1980}, \emph{\bibinfo{title}{Statistical Physics, Part II}}
  (\bibinfo{publisher}{Oxford: Pergamon}).

\bibitem[{\citenamefont{Larkin and Ovchinnikov}(1964)}]{LO}
\bibinfo{author}{\bibnamefont{Larkin}, \bibfnamefont{A.~J.}}, and
  \bibinfo{author}{\bibfnamefont{Y.~N.} \bibnamefont{Ovchinnikov}},
  \bibinfo{year}{1964}, \bibinfo{journal}{Zh. Exsp. teor. Fiz.}
  \textbf{\bibinfo{volume}{47}}, \bibinfo{pages}{1136}.

\bibitem[{\citenamefont{Le~Bellac}(1996)}]{LeBellac:1996ab}
\bibinfo{author}{\bibnamefont{Le~Bellac}, \bibfnamefont{M.}},
  \bibinfo{year}{1996}, \emph{\bibinfo{title}{Thermal Field Theory}}
  (\bibinfo{publisher}{Cambridge University Press, Cambridge, England}).

\bibitem[{\citenamefont{Leibovich} \emph{et~al.}(2001)\citenamefont{Leibovich,
  Rajagopal, and Shuster}}]{Leibovich:2001xr}
\bibinfo{author}{\bibnamefont{Leibovich}, \bibfnamefont{A.~K.}},
  \bibinfo{author}{\bibfnamefont{K.}~\bibnamefont{Rajagopal}}, and
  \bibinfo{author}{\bibfnamefont{E.}~\bibnamefont{Shuster}},
  \bibinfo{year}{2001}, \bibinfo{journal}{Phys. Rev.}
  \textbf{\bibinfo{volume}{D64}}, \bibinfo{pages}{094005}.

\bibitem[{\citenamefont{Link} \emph{et~al.}(2000)\citenamefont{Link, Epstein,
  and Lattimer}}]{Link:2000mu}
\bibinfo{author}{\bibnamefont{Link}, \bibfnamefont{B.}},
  \bibinfo{author}{\bibfnamefont{R.~I.} \bibnamefont{Epstein}}, and
  \bibinfo{author}{\bibfnamefont{J.~M.} \bibnamefont{Lattimer}},
  \bibinfo{year}{2000}, \eprint{astro-ph/0001245}.

\bibitem[{\citenamefont{London and London}(1935)}]{london:1935cd}
\bibinfo{author}{\bibnamefont{London}, \bibfnamefont{F.}}, and
  \bibinfo{author}{\bibfnamefont{H.}~\bibnamefont{London}},
  \bibinfo{year}{1935}, \bibinfo{journal}{Proc. Roy. Soc.}
  \textbf{\bibinfo{volume}{A149}}, \bibinfo{pages}{71}.

\bibitem[{\citenamefont{Lorimer}(1999)}]{Lorimer:1999dh}
\bibinfo{author}{\bibnamefont{Lorimer}, \bibfnamefont{D.~R.}},
  \bibinfo{year}{1999}, \eprint{astro-ph/9911519}.

\bibitem[{\citenamefont{Madsen}(2000)}]{Madsen:1999ci}
\bibinfo{author}{\bibnamefont{Madsen}, \bibfnamefont{J.}},
  \bibinfo{year}{2000}, \bibinfo{journal}{Phys. Rev. Lett.}
  \textbf{\bibinfo{volume}{85}}, \bibinfo{pages}{10}.

\bibitem[{\citenamefont{Madsen}(2001)}]{Madsen:2001fu}
\bibinfo{author}{\bibnamefont{Madsen}, \bibfnamefont{J.}},
  \bibinfo{year}{2001}, \bibinfo{journal}{Phys. Rev. Lett.}
  \textbf{\bibinfo{volume}{87}}, \bibinfo{pages}{172003}.

\bibitem[{\citenamefont{Madsen}(2002)}]{Madsen:2001zh}
\bibinfo{author}{\bibnamefont{Madsen}, \bibfnamefont{J.}},
  \bibinfo{year}{2002}, \bibinfo{journal}{eConf}
  \textbf{\bibinfo{volume}{C010815}}, \bibinfo{pages}{155}.

\bibitem[{\citenamefont{Manuel and Tytgat}(2000)}]{Manuel:2000wm}
\bibinfo{author}{\bibnamefont{Manuel}, \bibfnamefont{C.}}, and
  \bibinfo{author}{\bibfnamefont{M.~H.~G.} \bibnamefont{Tytgat}},
  \bibinfo{year}{2000}, \bibinfo{journal}{Phys. Lett.}
  \textbf{\bibinfo{volume}{B479}}, \bibinfo{pages}{190}.

\bibitem[{\citenamefont{Manuel and Tytgat}(2001)}]{Manuel:2000xt}
\bibinfo{author}{\bibnamefont{Manuel}, \bibfnamefont{C.}}, and
  \bibinfo{author}{\bibfnamefont{M.~H.~G.} \bibnamefont{Tytgat}},
  \bibinfo{year}{2001}, \bibinfo{journal}{Phys. Lett.}
  \textbf{\bibinfo{volume}{B501}}, \bibinfo{pages}{200}.

\bibitem[{\citenamefont{Maxwell}(1950)}]{Maxwell:1950}
\bibinfo{author}{\bibnamefont{Maxwell}, \bibfnamefont{E.}},
  \bibinfo{year}{1950}, \bibinfo{journal}{Phys. Rev.}
  \textbf{\bibinfo{volume}{78}}, \bibinfo{pages}{477}.

\bibitem[{\citenamefont{Meissner and Ochsenfeld}(1933)}]{meissner:1933ab}
\bibinfo{author}{\bibnamefont{Meissner}, \bibfnamefont{W.}}, and
  \bibinfo{author}{\bibfnamefont{R.}~\bibnamefont{Ochsenfeld}},
  \bibinfo{year}{1933}, \bibinfo{journal}{Naturwissenschaften}
  \textbf{\bibinfo{volume}{21}}, \bibinfo{pages}{787}.

\bibitem[{\citenamefont{Miransky and Shovkovy}(2002)}]{Miransky:2001tw}
\bibinfo{author}{\bibnamefont{Miransky}, \bibfnamefont{V.~A.}}, and
  \bibinfo{author}{\bibfnamefont{I.~A.} \bibnamefont{Shovkovy}},
  \bibinfo{year}{2002}, \bibinfo{journal}{Phys. Rev. Lett.}
  \textbf{\bibinfo{volume}{88}}, \bibinfo{pages}{111601}.

\bibitem[{\citenamefont{Nambu}(1960)}]{Nambu:1960cs}
\bibinfo{author}{\bibnamefont{Nambu}, \bibfnamefont{Y.}}, \bibinfo{year}{1960},
  \bibinfo{journal}{Phys. Rev.} \textbf{\bibinfo{volume}{117}},
  \bibinfo{pages}{648}.

\bibitem[{\citenamefont{Nambu and
  Jona-Lasinio}(1961{\natexlab{a}})}]{Nambu:1961tp}
\bibinfo{author}{\bibnamefont{Nambu}, \bibfnamefont{Y.}}, and
  \bibinfo{author}{\bibfnamefont{G.}~\bibnamefont{Jona-Lasinio}},
  \bibinfo{year}{1961}{\natexlab{a}}, \bibinfo{journal}{Phys. Rev.}
  \textbf{\bibinfo{volume}{122}}, \bibinfo{pages}{345}.

\bibitem[{\citenamefont{Nambu and
  Jona-Lasinio}(1961{\natexlab{b}})}]{Nambu:1961fr}
\bibinfo{author}{\bibnamefont{Nambu}, \bibfnamefont{Y.}}, and
  \bibinfo{author}{\bibfnamefont{G.}~\bibnamefont{Jona-Lasinio}},
  \bibinfo{year}{1961}{\natexlab{b}}, \bibinfo{journal}{Phys. Rev.}
  \textbf{\bibinfo{volume}{124}}, \bibinfo{pages}{246}.

\bibitem[{\citenamefont{Nardulli}(2002)}]{Nardulli:2002ma}
\bibinfo{author}{\bibnamefont{Nardulli}, \bibfnamefont{G.}},
  \bibinfo{year}{2002}, \bibinfo{journal}{Riv. Nuovo Cim.}
  \textbf{\bibinfo{volume}{25N3}}, \bibinfo{pages}{1}.

\bibitem[{\citenamefont{Nielsen and Chada}(1976)}]{Chada:1976cn}
\bibinfo{author}{\bibnamefont{Nielsen}, \bibfnamefont{H.~B.}}, and
  \bibinfo{author}{\bibfnamefont{S.}~\bibnamefont{Chada}},
  \bibinfo{year}{1976}, \bibinfo{journal}{Nucl. Phys.}
  \textbf{\bibinfo{volume}{B105}}, \bibinfo{pages}{445}.

\bibitem[{\citenamefont{Ouyed and Sannino}(2001)}]{Ouyed:2001fv}
\bibinfo{author}{\bibnamefont{Ouyed}, \bibfnamefont{R.}}, and
  \bibinfo{author}{\bibfnamefont{F.}~\bibnamefont{Sannino}},
  \bibinfo{year}{2001}, \bibinfo{journal}{Phys. Lett.}
  \textbf{\bibinfo{volume}{B511}}, \bibinfo{pages}{66}.

\bibitem[{\citenamefont{Page}(1998)}]{Page:1998ab}
\bibinfo{author}{\bibnamefont{Page}, \bibfnamefont{D.}}, \bibinfo{year}{1998},
  \emph{\bibinfo{title}{{\rm in} The Many Faces of Neutron Stars}}
  (\bibinfo{publisher}{Kluwer Academic Publishers}).

\bibitem[{\citenamefont{Page} \emph{et~al.}(2000)\citenamefont{Page, Prakash,
  Lattimer, and Steiner}}]{Page:2000wt}
\bibinfo{author}{\bibnamefont{Page}, \bibfnamefont{D.}},
  \bibinfo{author}{\bibfnamefont{M.}~\bibnamefont{Prakash}},
  \bibinfo{author}{\bibfnamefont{J.~M.} \bibnamefont{Lattimer}}, and
  \bibinfo{author}{\bibfnamefont{A.}~\bibnamefont{Steiner}},
  \bibinfo{year}{2000}, \bibinfo{journal}{Phys. Rev. Lett.}
  \textbf{\bibinfo{volume}{85}}, \bibinfo{pages}{2048}.

\bibitem[{\citenamefont{Pines}(1958)}]{Pines:1958ab}
\bibinfo{author}{\bibnamefont{Pines}, \bibfnamefont{D.}}, \bibinfo{year}{1958},
  \bibinfo{journal}{Phys. Rev.} \textbf{\bibinfo{volume}{109}},
  \bibinfo{pages}{280}.

\bibitem[{\citenamefont{Pippard}(1953)}]{Pippard:1953ab}
\bibinfo{author}{\bibnamefont{Pippard}, \bibfnamefont{A.~B.}},
  \bibinfo{year}{1953}, \bibinfo{journal}{Proc. Roy. Soc.}
  \textbf{\bibinfo{volume}{A216}}, \bibinfo{pages}{547}.

\bibitem[{\citenamefont{Pisarski and Rischke}(1999)}]{Pisarski:1999av}
\bibinfo{author}{\bibnamefont{Pisarski}, \bibfnamefont{R.~D.}}, and
  \bibinfo{author}{\bibfnamefont{D.~H.} \bibnamefont{Rischke}},
  \bibinfo{year}{1999}, \bibinfo{journal}{Phys. Rev.}
  \textbf{\bibinfo{volume}{D60}}, \bibinfo{pages}{094013}.

\bibitem[{\citenamefont{Pisarski and
  Rischke}(2000{\natexlab{a}})}]{Pisarski:1999tv}
\bibinfo{author}{\bibnamefont{Pisarski}, \bibfnamefont{R.~D.}}, and
  \bibinfo{author}{\bibfnamefont{D.~H.} \bibnamefont{Rischke}},
  \bibinfo{year}{2000}{\natexlab{a}}, \bibinfo{journal}{Phys. Rev.}
  \textbf{\bibinfo{volume}{D61}}, \bibinfo{pages}{074017}.

\bibitem[{\citenamefont{Pisarski and
  Rischke}(2000{\natexlab{b}})}]{Pisarski:1999bf}
\bibinfo{author}{\bibnamefont{Pisarski}, \bibfnamefont{R.~D.}}, and
  \bibinfo{author}{\bibfnamefont{D.~H.} \bibnamefont{Rischke}},
  \bibinfo{year}{2000}{\natexlab{b}}, \bibinfo{journal}{Phys. Rev.}
  \textbf{\bibinfo{volume}{D61}}, \bibinfo{pages}{051501}.

\bibitem[{\citenamefont{Polchinski}(1993)}]{Polchinski:1992ed}
\bibinfo{author}{\bibnamefont{Polchinski}, \bibfnamefont{J.}},
  \bibinfo{year}{1993}, \emph{\bibinfo{title}{{\rm in} Recent directions in
  particle theory: from superstrings and black holes to the standard model
  (TASI - 92)}} (\bibinfo{publisher}{World Scientific}),
  \eprint{hep-th/9210046}.

\bibitem[{\citenamefont{Politzer}(1973)}]{politzer:1973ab}
\bibinfo{author}{\bibnamefont{Politzer}, \bibfnamefont{H.~D.}},
  \bibinfo{year}{1973}, \bibinfo{journal}{Phys. Rev. Lett.}
  \textbf{\bibinfo{volume}{30}}, \bibinfo{pages}{1346}.

\bibitem[{\citenamefont{Pons and et~al}(1999)}]{Pons:1999}
\bibinfo{author}{\bibnamefont{Pons}, \bibfnamefont{J.}}, and
  \bibinfo{author}{\bibnamefont{et~al}}, \bibinfo{year}{1999},
  \bibinfo{journal}{Astrophys. J.} \textbf{\bibinfo{volume}{513}}.

\bibitem[{\citenamefont{Rajagopal and Shuster}(2000)}]{Rajagopal:2000rs}
\bibinfo{author}{\bibnamefont{Rajagopal}, \bibfnamefont{K.}}, and
  \bibinfo{author}{\bibfnamefont{E.}~\bibnamefont{Shuster}},
  \bibinfo{year}{2000}, \bibinfo{journal}{Phys. Rev.}
  \textbf{\bibinfo{volume}{D62}}, \bibinfo{pages}{085007}.

\bibitem[{\citenamefont{Rajagopal and Wilczek}(2001)}]{Rajagopal:2000wf}
\bibinfo{author}{\bibnamefont{Rajagopal}, \bibfnamefont{K.}}, and
  \bibinfo{author}{\bibfnamefont{F.}~\bibnamefont{Wilczek}},
  \bibinfo{year}{2001}, \emph{\bibinfo{title}{{\rm in} At the frontier of
  particle physics, vol. 3}} (\bibinfo{publisher}{World Scientific}),
  \eprint{hep-ph/0011333}.

\bibitem[{\citenamefont{Rapp} \emph{et~al.}(1998)\citenamefont{Rapp, Schafer,
  Shuryak, and Velkovsky}}]{Rapp:1998zu}
\bibinfo{author}{\bibnamefont{Rapp}, \bibfnamefont{R.}},
  \bibinfo{author}{\bibfnamefont{T.}~\bibnamefont{Schafer}},
  \bibinfo{author}{\bibfnamefont{E.~V.} \bibnamefont{Shuryak}}, and
  \bibinfo{author}{\bibfnamefont{M.}~\bibnamefont{Velkovsky}},
  \bibinfo{year}{1998}, \bibinfo{journal}{Phys. Rev. Lett.}
  \textbf{\bibinfo{volume}{81}}, \bibinfo{pages}{53}.

\bibitem[{\citenamefont{Rapp} \emph{et~al.}(2000)\citenamefont{Rapp, Schafer,
  Shuryak, and Velkovsky}}]{Rapp:1999qa}
\bibinfo{author}{\bibnamefont{Rapp}, \bibfnamefont{R.}},
  \bibinfo{author}{\bibfnamefont{T.}~\bibnamefont{Schafer}},
  \bibinfo{author}{\bibfnamefont{E.~V.} \bibnamefont{Shuryak}}, and
  \bibinfo{author}{\bibfnamefont{M.}~\bibnamefont{Velkovsky}},
  \bibinfo{year}{2000}, \bibinfo{journal}{Annals Phys.}
  \textbf{\bibinfo{volume}{280}}, \bibinfo{pages}{35}.

\bibitem[{\citenamefont{Reddy} \emph{et~al.}(2003)\citenamefont{Reddy,
  Sadzikowski, and Tachibana}}]{Reddy:2002xc}
\bibinfo{author}{\bibnamefont{Reddy}, \bibfnamefont{S.}},
  \bibinfo{author}{\bibfnamefont{M.}~\bibnamefont{Sadzikowski}}, and
  \bibinfo{author}{\bibfnamefont{M.}~\bibnamefont{Tachibana}},
  \bibinfo{year}{2003}, \bibinfo{journal}{Nucl. Phys.}
  \textbf{\bibinfo{volume}{A714}}, \bibinfo{pages}{337}.

\bibitem[{\citenamefont{Reynolds} \emph{et~al.}(1950)\citenamefont{Reynolds,
  Serin, Wright, and Nesbitt}}]{Reynolds:1950}
\bibinfo{author}{\bibnamefont{Reynolds}, \bibfnamefont{C.~A.}},
  \bibinfo{author}{\bibfnamefont{B.}~\bibnamefont{Serin}},
  \bibinfo{author}{\bibfnamefont{W.~H.} \bibnamefont{Wright}}, and
  \bibinfo{author}{\bibfnamefont{L.~B.} \bibnamefont{Nesbitt}},
  \bibinfo{year}{1950}, \bibinfo{journal}{Phys. Rev.}
  \textbf{\bibinfo{volume}{78}}, \bibinfo{pages}{487}.

\bibitem[{\citenamefont{Rho}
  \emph{et~al.}(2000{\natexlab{a}})\citenamefont{Rho, Shuryak, Wirzba, and
  Zahed}}]{Rho:2000ww}
\bibinfo{author}{\bibnamefont{Rho}, \bibfnamefont{M.}},
  \bibinfo{author}{\bibfnamefont{E.~V.} \bibnamefont{Shuryak}},
  \bibinfo{author}{\bibfnamefont{A.}~\bibnamefont{Wirzba}}, and
  \bibinfo{author}{\bibfnamefont{I.}~\bibnamefont{Zahed}},
  \bibinfo{year}{2000}{\natexlab{a}}, \bibinfo{journal}{Nucl. Phys.}
  \textbf{\bibinfo{volume}{A676}}, \bibinfo{pages}{273}.

\bibitem[{\citenamefont{Rho}
  \emph{et~al.}(2000{\natexlab{b}})\citenamefont{Rho, Wirzba, and
  Zahed}}]{Rho:1999xf}
\bibinfo{author}{\bibnamefont{Rho}, \bibfnamefont{M.}},
  \bibinfo{author}{\bibfnamefont{A.}~\bibnamefont{Wirzba}}, and
  \bibinfo{author}{\bibfnamefont{I.}~\bibnamefont{Zahed}},
  \bibinfo{year}{2000}{\natexlab{b}}, \bibinfo{journal}{Phys. Lett.}
  \textbf{\bibinfo{volume}{B473}}, \bibinfo{pages}{126}.

\bibitem[{\citenamefont{Rischke}(2000)}]{Rischke:2000qz}
\bibinfo{author}{\bibnamefont{Rischke}, \bibfnamefont{D.~H.}},
  \bibinfo{year}{2000}, \bibinfo{journal}{Phys. Rev.}
  \textbf{\bibinfo{volume}{D62}}, \bibinfo{pages}{034007}.

\bibitem[{\citenamefont{Rischke} \emph{et~al.}(2001)\citenamefont{Rischke, Son,
  and Stephanov}}]{Rischke:2000cn}
\bibinfo{author}{\bibnamefont{Rischke}, \bibfnamefont{D.~H.}},
  \bibinfo{author}{\bibfnamefont{D.~T.} \bibnamefont{Son}}, and
  \bibinfo{author}{\bibfnamefont{M.~A.} \bibnamefont{Stephanov}},
  \bibinfo{year}{2001}, \bibinfo{journal}{Phys. Rev. Lett.}
  \textbf{\bibinfo{volume}{87}}, \bibinfo{pages}{062001}.

\bibitem[{\citenamefont{Sakita}(1985)}]{Sakita}
\bibinfo{author}{\bibnamefont{Sakita}, \bibfnamefont{B.}},
  \bibinfo{year}{1985}, \emph{\bibinfo{title}{Quantum theory of many-variable
  systems and fields}} (\bibinfo{publisher}{World Scientific}).

\bibitem[{\citenamefont{Sannino}(2000)}]{Sannino:2000kg}
\bibinfo{author}{\bibnamefont{Sannino}, \bibfnamefont{F.}},
  \bibinfo{year}{2000}, \bibinfo{journal}{Phys. Lett.}
  \textbf{\bibinfo{volume}{B480}}, \bibinfo{pages}{280}.

\bibitem[{\citenamefont{Sarma}(1963)}]{sarma:1963ab}
\bibinfo{author}{\bibnamefont{Sarma}, \bibfnamefont{G.}}, \bibinfo{year}{1963},
  \bibinfo{journal}{J. Phys. Chem. Solids} \textbf{\bibinfo{volume}{24}},
  \bibinfo{pages}{1029}.

\bibitem[{\citenamefont{Schafer}(2000{\natexlab{a}})}]{Schafer:2000ew}
\bibinfo{author}{\bibnamefont{Schafer}, \bibfnamefont{T.}},
  \bibinfo{year}{2000}{\natexlab{a}}, \bibinfo{journal}{Phys. Rev. Lett.}
  \textbf{\bibinfo{volume}{85}}, \bibinfo{pages}{5531}.

\bibitem[{\citenamefont{Schafer}(2000{\natexlab{b}})}]{Schafer:1999fe}
\bibinfo{author}{\bibnamefont{Schafer}, \bibfnamefont{T.}},
  \bibinfo{year}{2000}{\natexlab{b}}, \bibinfo{journal}{Nucl. Phys.}
  \textbf{\bibinfo{volume}{B575}}, \bibinfo{pages}{269}.

\bibitem[{\citenamefont{Schafer}(2000{\natexlab{c}})}]{Schafer:2000tw}
\bibinfo{author}{\bibnamefont{Schafer}, \bibfnamefont{T.}},
  \bibinfo{year}{2000}{\natexlab{c}}, \bibinfo{journal}{Phys. Rev.}
  \textbf{\bibinfo{volume}{D62}}, \bibinfo{pages}{094007}.

\bibitem[{\citenamefont{Schafer}(2002)}]{Schafer:2001za}
\bibinfo{author}{\bibnamefont{Schafer}, \bibfnamefont{T.}},
  \bibinfo{year}{2002}, \bibinfo{journal}{Phys. Rev.}
  \textbf{\bibinfo{volume}{D65}}, \bibinfo{pages}{074006}.

\bibitem[{\citenamefont{Schafer}(2003)}]{Schafer:2003ab}
\bibinfo{author}{\bibnamefont{Schafer}, \bibfnamefont{T.}},
  \bibinfo{year}{2003}, \eprint{hep-ph/0304281}.

\bibitem[{\citenamefont{Schafer and Shuryak}(1998)}]{Schafer:1998wv}
\bibinfo{author}{\bibnamefont{Schafer}, \bibfnamefont{T.}}, and
  \bibinfo{author}{\bibfnamefont{E.~V.} \bibnamefont{Shuryak}},
  \bibinfo{year}{1998}, \bibinfo{journal}{Rev. Mod. Phys.}
  \textbf{\bibinfo{volume}{70}}, \bibinfo{pages}{323}.

\bibitem[{\citenamefont{Schafer} \emph{et~al.}(2001)\citenamefont{Schafer, Son,
  Stephanov, Toublan, and Verbaarschot}}]{Schafer:2001bq}
\bibinfo{author}{\bibnamefont{Schafer}, \bibfnamefont{T.}},
  \bibinfo{author}{\bibfnamefont{D.~T.} \bibnamefont{Son}},
  \bibinfo{author}{\bibfnamefont{M.~A.} \bibnamefont{Stephanov}},
  \bibinfo{author}{\bibfnamefont{D.}~\bibnamefont{Toublan}}, and
  \bibinfo{author}{\bibfnamefont{J.~J.~M.} \bibnamefont{Verbaarschot}},
  \bibinfo{year}{2001}, \bibinfo{journal}{Phys. Lett.}
  \textbf{\bibinfo{volume}{B522}}, \bibinfo{pages}{67}.

\bibitem[{\citenamefont{Schafer and
  Wilczek}(1999{\natexlab{a}})}]{Schafer:1998ef}
\bibinfo{author}{\bibnamefont{Schafer}, \bibfnamefont{T.}}, and
  \bibinfo{author}{\bibfnamefont{F.}~\bibnamefont{Wilczek}},
  \bibinfo{year}{1999}{\natexlab{a}}, \bibinfo{journal}{Phys. Rev. Lett.}
  \textbf{\bibinfo{volume}{82}}, \bibinfo{pages}{3956}.

\bibitem[{\citenamefont{Schafer and
  Wilczek}(1999{\natexlab{b}})}]{Schafer:1998na}
\bibinfo{author}{\bibnamefont{Schafer}, \bibfnamefont{T.}}, and
  \bibinfo{author}{\bibfnamefont{F.}~\bibnamefont{Wilczek}},
  \bibinfo{year}{1999}{\natexlab{b}}, \bibinfo{journal}{Phys. Lett.}
  \textbf{\bibinfo{volume}{B450}}, \bibinfo{pages}{325}.

\bibitem[{\citenamefont{Schafer and
  Wilczek}(1999{\natexlab{c}})}]{Schafer:1999pb}
\bibinfo{author}{\bibnamefont{Schafer}, \bibfnamefont{T.}}, and
  \bibinfo{author}{\bibfnamefont{F.}~\bibnamefont{Wilczek}},
  \bibinfo{year}{1999}{\natexlab{c}}, \bibinfo{journal}{Phys. Rev.}
  \textbf{\bibinfo{volume}{D60}}, \bibinfo{pages}{074014}.

\bibitem[{\citenamefont{Schafer and
  Wilczek}(1999{\natexlab{d}})}]{Schafer:1999jg}
\bibinfo{author}{\bibnamefont{Schafer}, \bibfnamefont{T.}}, and
  \bibinfo{author}{\bibfnamefont{F.}~\bibnamefont{Wilczek}},
  \bibinfo{year}{1999}{\natexlab{d}}, \bibinfo{journal}{Phys. Rev.}
  \textbf{\bibinfo{volume}{D60}}, \bibinfo{pages}{114033}.

\bibitem[{\citenamefont{Schrieffer}(1964)}]{Schrieffer}
\bibinfo{author}{\bibnamefont{Schrieffer}, \bibfnamefont{J.~R.}},
  \bibinfo{year}{1964}, \emph{\bibinfo{title}{Theory of Superconductivity}}
  (\bibinfo{publisher}{The Benjamin/Cummings Publishing Company, Inc.}).

\bibitem[{\citenamefont{Shankar}(1994)}]{Shankar:1994pf}
\bibinfo{author}{\bibnamefont{Shankar}, \bibfnamefont{R.}},
  \bibinfo{year}{1994}, \bibinfo{journal}{Rev. Mod. Phys.}
  \textbf{\bibinfo{volume}{66}}, \bibinfo{pages}{129}.

\bibitem[{\citenamefont{Shapiro and Teukolski}(1983)}]{Shapiro}
\bibinfo{author}{\bibnamefont{Shapiro}, \bibfnamefont{S.~L.}}, and
  \bibinfo{author}{\bibfnamefont{S.~A.} \bibnamefont{Teukolski}},
  \bibinfo{year}{1983}, \emph{\bibinfo{title}{Black holes, White Dwarfs and
  Neutron Stars: The Physics of Compact Objects}}
  (\bibinfo{publisher}{Wiley-Interscience}).

\bibitem[{\citenamefont{Shifman} \emph{et~al.}(1980)\citenamefont{Shifman,
  Vainshtein, and Zakharov}}]{Shifman:1980uw}
\bibinfo{author}{\bibnamefont{Shifman}, \bibfnamefont{M.~A.}},
  \bibinfo{author}{\bibfnamefont{A.~I.} \bibnamefont{Vainshtein}}, and
  \bibinfo{author}{\bibfnamefont{V.~I.} \bibnamefont{Zakharov}},
  \bibinfo{year}{1980}, \bibinfo{journal}{Nucl. Phys.}
  \textbf{\bibinfo{volume}{B163}}, \bibinfo{pages}{46}.

\bibitem[{\citenamefont{Shimahara}(1998)}]{Shimahara:1998ab}
\bibinfo{author}{\bibnamefont{Shimahara}, \bibfnamefont{H.}},
  \bibinfo{year}{1998}, \bibinfo{journal}{J. Phys. Soc. Japan}
  \textbf{\bibinfo{volume}{67}}, \bibinfo{pages}{736}.

\bibitem[{\citenamefont{Shovkovy and Ellis}(2002)}]{Shovkovy:2002kv}
\bibinfo{author}{\bibnamefont{Shovkovy}, \bibfnamefont{I.~A.}}, and
  \bibinfo{author}{\bibfnamefont{P.~J.} \bibnamefont{Ellis}},
  \bibinfo{year}{2002}, \bibinfo{journal}{Phys. Rev.}
  \textbf{\bibinfo{volume}{C66}}, \bibinfo{pages}{015802}.

\bibitem[{\citenamefont{Shovkovy and Wijewardhana}(1999)}]{Shovkovy:1999mr}
\bibinfo{author}{\bibnamefont{Shovkovy}, \bibfnamefont{I.~A.}}, and
  \bibinfo{author}{\bibfnamefont{L.~C.~R.} \bibnamefont{Wijewardhana}},
  \bibinfo{year}{1999}, \bibinfo{journal}{Phys. Lett.}
  \textbf{\bibinfo{volume}{B470}}, \bibinfo{pages}{189}.

\bibitem[{\citenamefont{Son}(1999)}]{Son:1998uk}
\bibinfo{author}{\bibnamefont{Son}, \bibfnamefont{D.~T.}},
  \bibinfo{year}{1999}, \bibinfo{journal}{Phys. Rev.}
  \textbf{\bibinfo{volume}{D59}}, \bibinfo{pages}{094019}.

\bibitem[{\citenamefont{Son and Stephanov}(2000{\natexlab{a}})}]{Son:1999cm}
\bibinfo{author}{\bibnamefont{Son}, \bibfnamefont{D.~T.}}, and
  \bibinfo{author}{\bibfnamefont{M.~A.} \bibnamefont{Stephanov}},
  \bibinfo{year}{2000}{\natexlab{a}}, \bibinfo{journal}{Phys. Rev.}
  \textbf{\bibinfo{volume}{D61}}, \bibinfo{pages}{074012}.

\bibitem[{\citenamefont{Son and Stephanov}(2000{\natexlab{b}})}]{Son:2000tu}
\bibinfo{author}{\bibnamefont{Son}, \bibfnamefont{D.~T.}}, and
  \bibinfo{author}{\bibfnamefont{M.~A.} \bibnamefont{Stephanov}},
  \bibinfo{year}{2000}{\natexlab{b}}, \bibinfo{journal}{Phys. Rev. (D61
  Erratum)} \textbf{\bibinfo{volume}{D62}}, \bibinfo{pages}{059902}.

\bibitem[{\citenamefont{Son} \emph{et~al.}(2001)\citenamefont{Son, Stephanov,
  and Zhitnitsky}}]{Son:2000fh}
\bibinfo{author}{\bibnamefont{Son}, \bibfnamefont{D.~T.}},
  \bibinfo{author}{\bibfnamefont{M.~A.} \bibnamefont{Stephanov}}, and
  \bibinfo{author}{\bibfnamefont{A.~R.} \bibnamefont{Zhitnitsky}},
  \bibinfo{year}{2001}, \bibinfo{journal}{Phys. Rev. Lett.}
  \textbf{\bibinfo{volume}{86}}, \bibinfo{pages}{3955}.

\bibitem[{\citenamefont{Tinkham}(1995)}]{Tinkham}
\bibinfo{author}{\bibnamefont{Tinkham}, \bibfnamefont{M.}},
  \bibinfo{year}{1995}, \emph{\bibinfo{title}{Introduction to
  Superconductivity}} (\bibinfo{publisher}{McGraw-Hill}).

\bibitem[{\citenamefont{Tsuruta}(1998)}]{Tsuruta:1998ab}
\bibinfo{author}{\bibnamefont{Tsuruta}, \bibfnamefont{S.}},
  \bibinfo{year}{1998}, \bibinfo{journal}{Phys. Rep.}
  \textbf{\bibinfo{volume}{292}}, \bibinfo{pages}{1}.

\bibitem[{\citenamefont{Wang and Rischke}(2002)}]{Wang:2001aq}
\bibinfo{author}{\bibnamefont{Wang}, \bibfnamefont{Q.}}, and
  \bibinfo{author}{\bibfnamefont{D.~H.} \bibnamefont{Rischke}},
  \bibinfo{year}{2002}, \bibinfo{journal}{Phys. Rev.}
  \textbf{\bibinfo{volume}{D65}}, \bibinfo{pages}{054005}.

\bibitem[{\citenamefont{Weinberg}(1996)}]{WeinbergII}
\bibinfo{author}{\bibnamefont{Weinberg}, \bibfnamefont{S.}},
  \bibinfo{year}{1996}, \emph{\bibinfo{title}{The Quantum theory of Fields,
  Vol. II}} (\bibinfo{publisher}{Cambridge University Press}).

\bibitem[{\citenamefont{Witten}(1984)}]{Witten:1984}
\bibinfo{author}{\bibnamefont{Witten}, \bibfnamefont{E.}},
  \bibinfo{year}{1984}, \bibinfo{journal}{Phys. Rev.}
  \textbf{\bibinfo{volume}{D30}}, \bibinfo{pages}{272}.

\bibitem[{\citenamefont{Zarembo}(2000)}]{Zarembo:2000pj}
\bibinfo{author}{\bibnamefont{Zarembo}, \bibfnamefont{K.}},
  \bibinfo{year}{2000}, \bibinfo{journal}{Phys. Rev.}
  \textbf{\bibinfo{volume}{D62}}, \bibinfo{pages}{054003}.

\bibitem[{\citenamefont{Ziman}(1964)}]{Ziman}
\bibinfo{author}{\bibnamefont{Ziman}, \bibfnamefont{J.~M.}},
  \bibinfo{year}{1964}, \emph{\bibinfo{title}{Principles of the Theory of
  Solids}} (\bibinfo{publisher}{Cambridge University Press, New York}).

\end{thebibliography}
\end{document}